\documentclass[twoside, a4paper, 10pt]{book}

\usepackage{amsmath}
\usepackage{amssymb}
\usepackage{amsbsy}
\usepackage{mathrsfs}
\usepackage{graphicx}
\usepackage{fancyhdr}
\usepackage{array}
\usepackage{algpseudocode}
\usepackage{algorithm}

\RequirePackage[plainpages=true]{hyperref}
\usepackage{color}

\newcommand{\de}{\partial}
\newcommand{\eq}{\Leftrightarrow}
\DeclareMathOperator*{\extremum}{extremum} 

\newcommand{\beq}{\begin{equation}}
\newcommand{\eeq}{\end{equation}}
\newcommand{\bea}{\begin{eqnarray}}
\newcommand{\eea}{\end{eqnarray}}

\newcommand{\ksat}[1][$k$]{{#1}-\textsc{sat}}
\newcommand{\kxorsat}[1][$k$]{{#1}-\textsc{xorsat}}
\newcommand{\sat}{\textsc{sat}}
\newcommand{\unsat}{\textsc{unsat}}
\newcommand{\true}{\textsc{true}}
\newcommand{\false}{\textsc{false}}
\newcommand{\walksat}{Walk\textsc{sat}}
\newcommand{\uecsp}[1][$d,k$]{(#1)-\textnormal{UE-CSP}}
\newcommand{\wprop}{\textsc{Warning Propagation}}
\newcommand{\typ}{\textsc{typical}}
\newcommand{\notyp}{\textsc{not typical}}
\newcommand{\hyp}[1]{Hypothesis~{#1}}

\newcommand{\journal}[4]{\textit{#1} {\bf #2} #3 (#4)}
\newcommand{\jsp}{J. Stat. Phys.}
\newcommand{\nat}{Nature}
\newcommand{\sci}{Science}
\newcommand{\prl}{Phys. Rev. Lett.}
\newcommand{\pr}{Phys. Rev.}
\newcommand{\jphys}{J. Phys.}

\begin{document}


\begin{titlepage}

 \begin{center}
     \vspace{2em}
     \includegraphics[height=.15\textheight]{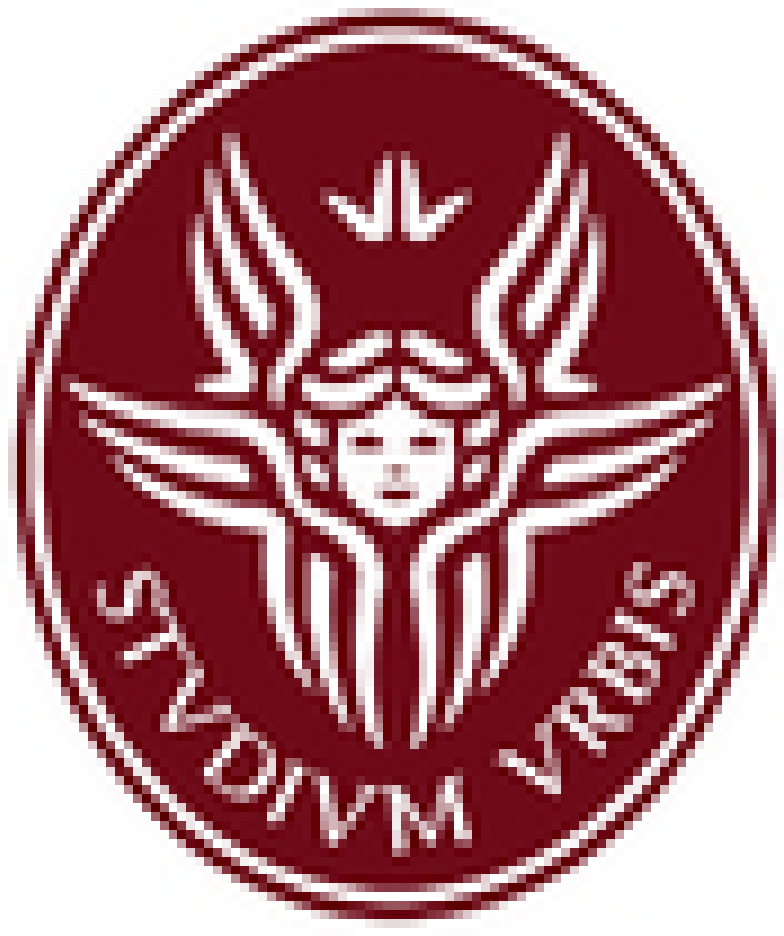}\\
     \vspace{1em}
     {\textsc{Sapienza Universit\`{a} di Roma}}\\
     {\textsc{Dottorato di Ricerca in Fisica}}\\
     {\textsc{Scuola di dottorato ``Vito Volterra''}}\\
     \vspace{8em}
     {\huge \textbf{Theoretical analysis of optimization problems}}\\
     \vspace{8em}
     {\textsc{Thesis submitted to obtain the degree of}}\\
     {\textsc{\textit{"Dottore di Ricerca" - Doctor Philosophi\ae}}}\\
     {\textsc{PhD in Physics - XX cycle - October 2007}}\\
     \vspace{3em}
     {\textsc{by}}\\
     \vspace{1em}
     {\Large \textbf{Fabrizio Altarelli}}\\
     \vspace{10em}
 \end{center}

  \begin{center}
    \begin{tabular}{c c c c c c c c c c}
      \textbf{Program Coordinator} & & & & ~~~~~~~~~~~~~~~~~~~~~~~~ & & & & & \textbf{Thesis Advisors} \\[0.2cm]
      \large{Prof. Enzo Marinari} & & & & ~~~~~~~~~~~~~~~~~~~~~~~~ & & & & & \large{Prof. Giorgio Parisi}\\[0.2cm]
      & & & & ~~~~~~~~~~~~~~~~~~~~~~~~ & & & & & \large{Dr. Nicolas Sourlas} \\[0.2cm]
      & & & & ~~~~~~~~~~~~~~~~~~~~~~~~ & & & & & \large{Dr. R\'emi Monasson} \\
    \end{tabular}
  \end{center}

\end{titlepage}

\pagestyle{empty}
\newpage
\mbox{}
\newpage
\mbox{}
\vskip 5cm
\hspace{10cm}
\emph{Alla dottoressa Federici}

\pagenumbering{roman}
\setcounter{page}{1}
\tableofcontents

\pagestyle{fancy}
\newpage
\chapter*{Introduction}
\addcontentsline{toc}{chapter}{Introduction}

The mean field theory of disordered systems is a well established topic in statistical mechanics, developed in the past thirty years with remarkable success (see \cite{Mezard87} for the classical reference). Originally, interest in this field was motivated by the experimental discovery of \emph{spin glasses}, metallic compounds formed by diluting a ferromagnetic metal in a diamagnetic host, and which exhibit peculiar magnetic properties: on one hand, the low dilution ensures that the locations of the ferromagnetic atoms (and therefore their interactions) are random, so that their structure presents no order; on the other hand, evidence is found at low temperature for a transition to a phase in which the local magnetization is frozen (in a direction variable from point to point) and which therefore displays some properties  characteristic of the presence of order. The phenomenology of spin glasses is indeed very rich, and the development of theoretical models able to explain and reproduce it in full has been a major challenge (and achievement) of statistical mechanics in the past thirty years, requiring the introduction of innovative concepts and techniques.

During the same time span, a large number of interesting results have been obtained in the understanding of \emph{combinatorial optimization problems}, and in the development of \emph{computational complexity theory} (see \cite{Papadimitriou98} for an introduction). Combinatorial optimization problems (that is to say, problems in which the \emph{optimal} configuration in a large and discrete set of candidates has to be found) are of the greatest interest for practical applications, and turn out to be general enough to deserve considerable attention from the theoretical point of view as well. Moreover, they are the cornerstone of complexity theory, the purpose of which is to characterize the intrinsic ``hardness'' of solving problems, and also the efficiency of algorithms used to solve them.

Very early, it was recognized that these two fields, apparently far from each other, and studied by different communities of researchers, actually have very much in common. It was soon realized that random distributions of some well known and very important (both theoretically and in view of applications) combinatorial optimization problems were formally equivalent to diluted spin glass models, and could be treated with such powerful tools as the replica method and (somewhat later) the cavity approach. This has led, in the past decade, to a very fecund transfer of problems and ideas across the two fields, leading to significant advances in our understanding of both.

A first area of interest is the characterization of the different phases of models that are relevant from both the optimization and statistical mechanics points of view. These models consist of a collection of $N$ Ising spins that interact with $k$-body couplings ($k=2,3,\dots$) with random strenghts (the exact form of the interactions defines each model). The number of interactions to which individual spins participate, also called \emph{connectivity}, plays the role of the control parameter, analogous to the pressure in thermodynamic systems. As the connectivity is varied, the free energy ``landscape'' undergoes a series of dramatic structural changes, that correspond to the onset different ``macroscopic'' properties of the system, such as the presence of an exponential (in $N$) number of local minima in the landscape, or the value of the energy density of the global ground state being of order 1 rather than order $N$.

Another area of interest is the analysis of the algorithms that can be used in order to solve the optimization problem represented by each model, or in equivalent but more physical terms, to find its ground state configurations. There is a huge variety in these algorithms: some of them define a \emph{dynamical} process which modifies the configuration, in a manner similar to the well known Metropolis algorithm; some others perform a \emph{sequential assignment} of the values of spins trying to minimize the number of positive contributions to the hamiltonian; others still do not act on the spins themselves, but rather on some \emph{effective variables}, such as the magnetic field conjugated to each spin. Accordingly, a full ``taxonomy'' of algorithms can be constructed, and the average behaviour of whole classes of algorithms, with similar structure but different attributes, can be characterized, allowing both to identify those algorithms that are of interest from the point of view of actual applications, and also to reach a better understanding in the properties of the models themselves.

Even though the list of the topics that have been studied in this field, and which are of interest for current research, includes many more, I shall limit the discussion to the previous ones: in this thesis, I have worked on problems that stem from these two lines of research. In the first Part, I shall therefore introduce the models I have studied and give an overview of the most relevant known results. Chapter 1 is an introduction to the physics of disordered systems: the main concepts of the statistical mechanics of spin glasses are introduced, with a discussion of the main phenomenological features that characterize them, and of the replica method, which allows to study them analytically. In Chapter 2, I shall introduce combinatorial optimization problems, and some important results from the theory of computational complexity that are relevant to my work; in particular, the two boolean satisfiability problems that I have been interested in, called \ksat\ and \kxorsat, are defined, and their properties are discussed. Finally, in Chapter 3 I shall review the results obtained by applying the methods and concepts developed for spin glasses to these two problems, and what their interpretation as spin glasses can teach us about the physics of these and similar systems.

In the second Part of the thesis, I shall present some of the original results that I have obtained in collaboration with R\'emi Monasson, Giorgio Parisi and Francesco Zamponi.

The first problem we have studied is motivated by a well known (but not as well understood) empirical observation: a large variety of systems present a phase in which the ground states form \emph{clusters} and the spins are \emph{frozen}; in this phase, no \emph{local} search algorithm is capable to find the ground states in an \emph{efficient} manner. In this context (and very losely speaking), clusters are sets of configurations which all have the ground state energy and which are connected, while different clusters are well separated (two configurations are considered \emph{adjacent} if they differ by a number of spins which is of order 1 in $N$, \emph{connected} if one can be reached from the other with a series of adjacent steps, and \emph{separated} if this is not possible); a frozen spin is a spin that takes the same value in all the configurations of a cluster; a local search algorithm is an algorithm that only uses information about the values of a number of variables which is of order 1 in $N$; and efficient means that the time (or number of elementary computations) required to find a ground state configuration with this algorithm grows faster than any polynomial in $N$.

The simplest model presenting such a ``clustered-frozen'' phase is \kxorsat, which I mentioned before and I shall discuss in Chapter 2; on the other hand, one of the most studied (and useful in practical applications) local algorithms is DPLL, which works by assigning variables in sequence according to some simple rule called heuristics, and which I shall also discuss in Chapter 2. In order to gain a better understanding of the failure of local search algorithms in the clustered-frozen phase, we have studied a farly general class of DPLL heuristics for \kxorsat, obtaining some results that I shall present in Chapter 4. Most notably, we have obtained the first proof (to the best of my knowledge) that any heuristic in this class fails to find a ground state in polynomial time in $N$ with probability 1 as $N$ goes to infinity. Moreover, we have obtained an argument that supports the claim that in the large $k$ limit, one of the heuristics belonging to the class we have studied (and which was previously introduced and called GUC) is capable of finding ground states efficiently with probability 1 up to the onset of the clustered-frozen phase, while all the other heuristics previously studied were known to fail well before this phase transition.

The second problem we have considered concerns the most studied and celebrated combinatorial optimization problem: \ksat. There are many reasons motivating the interest for \ksat, notably that it is the first problem for which NP-completeness (which is the key concept in computational complexity theory) was proven, and that it is so general that a huge number of other problems (many of which are relevant in view of applications) can be expressed as particular instances of \ksat. As a result of these extended studies, a very rich phase structure has emerged, with a multitude of transitions determined by temperature and connectivity. The aim of our work was to study the phase which is obtained at zero temperature when the connectivity goes to infinity. Apart from the intrinsic interest of studying one of the phases of the system, this problem is very interesting due to some recent results in computational complexity theory that establish a link between the \emph{average} case complexity of \ksat\ at large connectivity and the \emph{worst} case complexity of several other problems. No relation between these two measures of complexity was previously known, and the complexity class of the problems considered depends on the properties of \ksat\ at large connectivity.

The main result we have obtained is that this phase of the system is characterized by the presence of a single cluster of ground states in which the fraction of spins that are not frozen goes exponentially to 0 as the connectivity is increased, and that the field conjugated to frozen spins is of the same order of the connectivity. I shall present these results in Chapter 5, together with a discussion of their interest and consequences for computational complexity theory.

Moreover, during the past year I have engaged in the study of yet another algorithm for boolean satisfiability problems, going under the name of \walksat. This work, which consists in a numerical characterization of the average behavior of the algorithm, and in elucidating the properties of \ksat\ that this behavior imply, is still in progress, and will constitute the object of a future publication.

\newpage
\mbox{}

\newpage
\pagenumbering{arabic}
\setcounter{page}{1}

\begin{part}{Statistical mechanics of optimization problems}
\chapter{Statistical mechanics of disordered systems}
\label{Chap_statmech}

\section{Statistical mechanics and phase transitions}

In this section I shall introduce some notation and briefly review some fundamental concepts of statistical mechanics, illustrating them with the example of the Ising ferromagnet.

\subsection{The Gibbs distribution}

A general system studied in statistical mechanics will have a large number $N$ of degrees of freedom $\{x_i \in \mathbb X |\ i=1,\dots,N\}$. A configuration $\mathscr C \in \mathbb X^N$ of the system is determined by specifying the value taken by each $x_i$. The hamiltonian of the system will be an extensive function of the configuration, $H(\mathscr C)$. 

The statistical properties of the system are determined by the probability distribution of the configurations. If a system can exchange energy with its surrounding at temperature\footnote{I shall always use ``natural'' units, in which the Boltzmann constant is equal to 1.} $T \equiv 1/\beta$, this probability is given by the Gibbs distribution:
\begin{equation}
 \mathbb P[\mathscr C] = \frac 1 {Z(\beta)} e^{-\beta H(\mathscr C)}
\end{equation}
where the partition function $Z(\beta)$ is a normalization. In fact, it is much more than a normalization, since all the equilibrium properties of the system can be computed from it. For example, the average moments of the energy are given by its derivatives:
\begin{eqnarray}
 E(\beta) \equiv \mathbb E[H(\mathscr C)] &=& - \frac \de {\de \beta} \log Z(\beta), \\
\mathbb E[H(\mathscr C)^2] - \mathbb E[H(\mathscr C)]^2 &=& \frac {\de^2} {\de \beta^2} \log Z(\beta), \ \dots
\end{eqnarray}

The entropy and the free energy can be introduced in two equivalent ways. The ``microcanonical'' entropy is the logarithm of the number of configurations with energy $E$:
\begin{equation}
 S_\mathrm m(E) = \log \left| \{\mathscr C \in \mathbb X^N : H(\mathscr C) = E \} \right|.
\end{equation}
We can expect it to be an extensive quantity and define the entropy density $s_\mathrm m(e) = S_\mathrm m(Ne)/N$. Since the Gibbs measure depends on the configurations only through the energy, we can greatly simplify the description of the system by considering the probability to find it in \emph{any} configuration of energy~$E$:
\begin{equation}
 \mathbb P[E] = \sum_{\{\mathscr C \in \mathbb X^N | H(\mathscr C) = E\}} \frac 1 {Z(\beta)} e^{-\beta H(\mathscr C)} = \frac 1 {Z(\beta)} e^{-\beta E + S_\mathrm m(E)}
\equiv \frac 1 {Z(\beta)} e^{-\beta F_\mathrm m(E)}
\label{prob_E}
\end{equation}
where we have introduced the free energy $F_\mathrm m(E) \equiv N f_\mathrm m(E/N) \equiv E - S_\mathrm m(E)/\beta$.

On the other hand, the ``canonical'' entropy is defined in terms of the Gibbs distribution as	
\begin{equation}
 S_\mathrm c(\beta) \equiv - \mathbb E [\log \mathbb P[\mathscr C] ] = -\sum_{\mathscr C} \mathbb P[\mathscr C] \log \mathbb P[\mathscr C] \,,
\end{equation}
(notice that $\mathbb P[\mathscr C]$ depends on $\beta$) while the free energy is defined
\begin{equation}
 F_\mathrm c(\beta) \equiv - \frac 1 \beta \log Z(\beta) \,.
\label{free_can}
\end{equation}
Notice that these definitions imply that
\begin{equation}
 S_\mathrm c(\beta) 
= -\mathbb E \left[ \log \frac {e^{-\beta H(\mathscr C)}} {Z(\beta)} \right]
= \beta E(\beta) - \beta F_\mathrm c(\beta)
\end{equation}
which formally corresponds to the similar microcanonical relation.

The relationship between the microcanonical and canonical approaches becomes evident in the thermodynamic limit $N \rightarrow \infty$. In this limit, we can compute the canonical free energy with the Laplace method:
\begin{eqnarray}
 f_\mathrm c(\beta) 
	&=& - \lim_{N \to \infty} \frac 1 N \frac 1 \beta \log Z(\beta) \\
        &=& - \lim_{N \to \infty} \frac 1 N \frac 1 \beta \log \int de \ e^{-N \beta f_\mathrm m(e)} \\
        &=& - \lim_{N \to \infty} \frac 1 N \frac 1 \beta \log e^{-N \beta f_\mathrm m(\bar e)} \\
        &=& f_\mathrm m(\bar e)		\label{equiv_can_mic}
\end{eqnarray}
where $\bar e$ is the value that maximizes the exponent, i.e. $f_\mathrm m'(\bar e) = 0
\eq s_\mathrm m'(\bar e) = \beta$. But in the thermodynamic limit the energy is concentrated, so that
\begin{equation}
 e(\beta) = \lim_{N \to \infty} \int de \ e \ \frac {e^{-N \beta e + N s_\mathrm c(e)}} {Z(\beta)}
= \bar e \ e^{-\beta \bar e + s_\mathrm m(\bar e) - \beta f_\mathrm m(\bar e)}
= \bar e
\end{equation}
from (\ref{free_can}) and (\ref{equiv_can_mic}). Therefore $f_\mathrm c(\beta) = f_\mathrm m(e(\beta))$ and $s_\mathrm c(\beta) = s_\mathrm m(e(\beta))$.

The physical interpretation of the free energy becomes clear by observing that (\ref{prob_E}) can be rewritten as
\begin{equation}
 \mathbb P[e] = \frac 1 {Z(\beta)} e^{-N \beta f_\mathrm m(e)} 
= e^{-N \beta [f_\mathrm m(e) - f_\mathrm m(e(\beta))]} \,,
\end{equation}
i.e. the probability that $e$ takes a value which is different from the expected value is exponentially small in $N$ and the corresponding large deviations function is the free energy itself.

Also notice that if the energy of a configuration $\mathscr C$ only depends on some extensive observable $O$, i.e. $H(\mathscr C) = \mathscr E(O(\mathscr C))$ where $\mathscr E$ is some function, then the expected value and the distribution of the large deviations of $O$ can be expressed in a similar way in terms of the free energy, by writing it as a function of $o \equiv O/N$.

\subsection{Phase transitions and ergodicity breaking}

Let us now discuss a specific example: the infinite range Ising ferromagnet. The degrees of freedom are $N$ Ising spins $\sigma_i \in \{-1,1\}$. We consider that each spin interacts with all the others and with a homogeneous external field $h^\mathrm{ext}$:
\begin{equation}
 H(\mathscr C) = -\sum_{i<j}^{1,N} J_N \sigma_i \sigma_j -\sum_i h^\mathrm{ext} \sigma_i \,.
\label{h_ising}
\end{equation}
In order for the energy to be extensive, $J_N$ must scale with the number of spins as $J/N$, and we set the energy units so that the factor $J$ is 1.

It is easy to solve this model with the trick discussed in the last paragraph of the previous section: the energy (\ref{h_ising}) depends on the configuration only through the total magnetization $M(\mathscr C) = \sum_i \sigma_i$, which is an extensive quantity. In terms of densities
\begin{equation}
 e(m) = - \frac 1 2 m^2 - h^\mathrm{ext} m.
\end{equation}
The number of configurations with magnetization $M$ is just $\binom{N}{N_+}$ where $N_+ = (N+M)/2$ is the number of up spins, so that the (microcanonical) entropy is obtained by Stirling's approximation:
\begin{equation}
 s(m) = - \frac {1+m} 2 \log \frac {1+m} 2 - \frac {1-m} 2 \log \frac {1-m} 2 \,.
\end{equation}
The equilibrium magnetization $\bar m$ is obtained introducing $f(m) = e(m) - s(m)/\beta$ from the condition
\begin{equation}
 f'(\bar m) = 0 \eq - \bar m - h^\mathrm{ext} - \frac 1 {2\beta} [ \log(1+\bar m) - \log(1- \bar m) ] = 0 \,,
\end{equation}
from which the self-consistent equation
\begin{equation}
 \bar m = \tanh [\beta(\bar m + h^\mathrm{ext})]
\end{equation}
 is found. 

We see that for $\beta > 1$ this equation admits a solution with $\bar m \neq 0$ even if $h^\mathrm{ext}=0$, i.e. there is a spontaneous magnetization, while for $\beta < 1$ this is not the case. This is one of the simplest examples of phase transition, in which the magnetization has the role of the \emph{order parameter} characterizing the phases. Notice that the existence of a spontaneous magnetization is a very striking phenomenon: in the absence of an external field, the energy is an even function of the magnetization, and the Gibbs weight of the configurations with magnetization $m$ is the same as that corresponding to magnetization $-m$, so that the expected value of the magnetization is 0 \emph{at all temperatures}.

The solution of this apparent contradiction can be understood by a more careful consideration the free energy of the problem. In the absence of field, $f(m)$ is an even function of $m$. It can be easily seen that the sign of $f''(0)$ is the same as that of $1-\beta$: at high temperature $m=0$ is the absolute minimum of $f$, while at low temperature $f$ has two equal minima $f(m_+) = f(m_-)$. In this line of reasoning, we are implicitly assuming that the external field is \emph{exactly} 0 when we take the thermodynamic limit. However, this is not a satisfactory assumption: the magnetic field is a physical parameter, while the thermodynamic limit is an idealization, so that the description of the physical ferromagnet should be obtained by considering a finite size system in the presence of a (possibly small) magnetic field, and computing the thermodynamic limit of the system in the presence of the field, which can then be taken to 0. The expected magnetization in the absence of field is then
\begin{equation}
 m_0 = \lim_{h^\mathrm{ext} \to 0} \lim_{N \to \infty} \frac 1 N \mathbb E[M|\beta,h^\mathrm{ext}] \,.
\end{equation}
As a consequence, the degeneracy between the two minima of $f$ present when $\beta > 1$ is removed before we take the limit of zero field, and only one 
of the two minima will contribute to the Gibbs measure. Loosely speaking, in the presence of spontaneous magnetization $m_0 > 0$, in order to reach a configuration of magnetization $m < 0$ the system must cross a free energy barrier of order $O(N)$, which cannot occur in the thermodynamic limit: the configuration space then breaks in two distinct regions, one containing all the configurations with positive magnetization and the other those with a negative one, and the two regions are dynamically disconnected. This is an example of \emph{ergodicity breaking} (for a clarifying discussion of ergodicity breaking in magnetic systems, see Chapter 2 of \cite{Parisi88}).

A final remark concerning the nature of the phase transition. We can compute the magnetization as a function of the external field by looking at the positions of the minima of $f$. In the absence of field, when $\beta = 1 + \epsilon$ the two minima are separated by a distance of order $o(1)$ (as $\epsilon \to 0$), and the value of the spontaneous magnetization grows continuously from 0 to a finite value with $\beta-1$. However, a different situation can occur, in which at the critical temperature the free energy has two \emph{well separated} minima, such that one is favored for $\beta = \beta_c + \epsilon$ and the other for $\beta = \beta_c - \epsilon$. In this case, when the temperature crosses the critical value, the order parameter undergoes a discontinuous change. This kind of discontinuous phase transitions is called of \emph{first order}, while continuous ones are called of \emph{second order}.

\section{Disordered systems and spin glasses}

Disorder is ubiquitous in nature: amorphous materials are infinitely more common than crystals; biological systems sometimes manifest order in the form of regular behavior, but rarely of structure; the distribution of matter in the universe is irregular at any scale... Countless more examples show that, in fact, disorder is the rule of nature, and order is the exception.

However, the apparent lack of order and structure is not a sufficient criterion to consider a system as properly \emph{disordered}. After all, a snapshot of the positions of molecules in a gas shows no sign of order, and yet gasses have a perfectly regular behavior under most conditions. On the other hand, a system as simple as a double pendulum can have an incredibly complicated dynamical evolution, with no signs of regularity at all, but would hardly be considered disordered.

In this section I shall try to give some examples of systems in which disorder plays a crucial role in determining their behavior, and which can be understood in terms of some very general concepts, in order to obtain a better characterization of what ``proper'' disordered systems are. I shall also introduce a formalism that has proven extremely powerful to describe them in a quantitative way.

\subsection{Origins of disorder}

In general, a disordered system can be characterized as having two distinct sets of parameters. The first one corresponds to the degrees of freedom of the system that have a dynamical evolution during the observation of the system. The second set corresponds to some parameters that influence the dynamics of the degrees of freedom, but that do not change during the observation, and which have ``random'' or irregular values. 

In some cases the distinction between the two sets of variables will be purely dynamical. Glasses are a prototypical example of this kind of systems. They lack any long-range order, but locally the positions of atoms are very constrained. As a result, the motion of an atom typically requires the rearrangement of a number of neighbors that varies widely, and some degrees of freedom are effectively ``frozen'' over the experimental time scales, while others undergo a fast dynamical evolution. Another example of this class of system is provided by kinetically constrained models, which are a simplification and generalization of glasses. These models generally study particles on lattices that undergo some simple dynamics, e.g. each site can be either empty or occupied by one particle, and particles can hop from one site to the next under some conditions that are specific to the model and which typically include that the site be empty. Depending on the boundary conditions and on the specific dynamical rules a rich phenomenology can be produced.

In other cases the distinction between dynamical variables and ``frozen'' parameters is explicit: some parameters (e.g. the interaction strength between pairs of particles) take constant random values, extracted from some known distribution. This kind of disorder is said to be \emph{quenched}\footnote{Notice, however, that there is no fundamental difference between quenched and dynamically induced disorder: in both cases, a large number of parameters is effectively frozen in random values. The difference is mainly related to the description, rather than the physics of the system.}. The most celebrated example is that of magnetic impurities diluted in noble metal alloys, in which the positions of the impurities, and therefore the strengths of their magnetic interactions, are in fact random, giving rise to a very peculiar phenomenology. The theoretical models  introduced to study these materials and to reproduce their behavior go under the name of spin glasses. The rest of this section will be devoted to introduce the most widely studied models of spin glasses, while their phenomenology and the analytical techniques used to solve them will be discussed in the latter sections of this Chapter.

\subsection{Spin glass models}

The simplest models for spin glasses has the following hamiltonian (for the classical introduction to the field, see \cite{Mezard87}):
\begin{equation}
 H_J = \sum_{i,j} J_{ij} \sigma_i \sigma_k
\label{spin_glass_H}
\end{equation}
where the $J \equiv \{J_{ij}\}$ are random couplings and $\sigma \equiv \{\sigma_i\}$ are Ising spins. Depending on the geometry of the interaction, several models can be obtained:
\begin{description}
 \item {\bf Edwards-Anderson (EA) |} The interactions involve only nearest neighbors on a lattice of dimension $D$, and their strengths are random variables extracted from a Gaussian distribution with zero average and finite variance. This was the first model introduced to describe magnetic alloys \cite{Edwards75}.
 \item {\bf Sherrington-Kirkpatrick (SK) |} Each $J_{ij}$ (for each distinct couple of indices) is extracted from a Gaussian distribution. In order for the energy to be extensive, the standard deviation of the distribution must be of order $O(N^{-1/2})$ \cite{Sherrington75}.
 \item {\bf Bethe lattice |} The interactions between spins are described by a Bethe lattice (i.e. a random graph with a finite connectivity $k$ and with no loops), and their strength has a standard deviation proportional to $k^{-1/2}$.
\end{description}

A simple generalization is obtained by allowing the interaction to involve a number of spins $p>2$:
\begin{equation}
 H_J = \sum_{i_1, i_2, \dots, i_p} J_{i_1 i_2 \dots i_p} \sigma_{i_1} \sigma_{i_2} \dots \sigma_{i_p} \,.
\end{equation}
In such $\boldsymbol p${\bf-spin} models the spins can be either Ising or real ($\sigma_i \in \mathbb R$). In the latter case a spherical constraint $\sum_i \sigma_i^2 = 1$ is imposed. Many more models have been proposed and studied, which I shall not describe.

\subsection{Mean field theory and diluted models}

Even though the Edwards-Anderson model was the first spin glass model to be proposed, in 1975, it still waits for a general solution. In fact, most of the progress made in spin glasses has been obtained on the basis of \emph{mean field theory}. Mean field theory can be defined as in the case of the Ising ferromagnet by writing the hamiltonian (\ref{spin_glass_H}) in terms of local fields, 
\begin{equation}
 H_J(\sigma) = \sum_i h_i(\sigma) \sigma_i \,, \ \ \ h_i(\sigma) = \sum_j J_{ij} \sigma_j
\end{equation}
and replacing the configuration-dependent value of $h_i$ with its thermal average, which depends on magnetizations rather than spin values. This approach can be generalized (and made much more powerful) by writing directly an expression for the free energy which depends on the local magnetizations $\{m_i\}$ and looking for the values of $\{m_i\}$ that satisfy the set of equations $\de f / \de m_i = 0$, an approach that goes under the names of Thouless, Anderson and Palmer (TAP) \cite{Thouless77}. However much care should be exercised in deriving the expression for the free energy, and it should be kept in mind that since this doesn't (usually) come from a variational principle, there is no requirement for the solutions to the TAP equations to be minima of the free energy. As we shall see, the mean field results can be derived in a more transparent, but more complicated, analytical way. 

A very important point to stress is that mean field results are in general exact for infinite range models, such as SK (and this has been recently rigorously proved), but are only approximations for large (but finite) range models, which become poor approximations if the range of interaction is short. This is due to the fact that in long range models, local fluctuations of thermodynamic quantities have no global effects, while in short range models they become crucial. However, finite range models have proven themselves very elusive so far. This raises the question of how to include local fluctuation effects in more tractable models.

A step towards this direction is provided by \emph{diluted} models, of which the Bethe lattice model introduced in the previous subsection is an example. A more general case is obtained when the geometry of the model is an Erd\H{o}s-R\'enyi random graph, in which each pair of spins has the same probability of being connected, and the average connectivity is finite. In these models, the corrections to mean field theory arise from loops, which are typically of length $O(\log N)$, and their magnitude is small and can be dealt with (as we shall see when I will introduce the cavity method). On the other hand, local fluctuations are present in diluted models, and they can be studied in this context.

\subsection{Frustration, local degeneracies, complexity}
\label{Par Frustration}

A very general and important feature of the spin glass hamiltonian (\ref{spin_glass_H}) is that its global minima, which govern the low temperature behavior of the system, cannot be found by \emph{local} optimization. This fact has two causes, and very deep implications. 

The first cause is \emph{frustration}, which can be most simply illustrated by an example: if $J_{12}, J_{13} > 0$ while $J_{23} < 0$ there is no possible assignment of $\sigma_1, \sigma_2, \sigma_3$ that will make all three terms in $J_{12} \sigma_1 \sigma_2 + J_{13} \sigma_1 \sigma_3 + J_{23} \sigma_2\sigma_3$ negative. Some of the addends in the hamiltonian will have to be positive, and the minimization of the hamiltonian requires a global approach. 

Also, once it is clear that some interactions will have to give positive contributions, it is also clear that a large number of choices are possible for \emph{which} terms to make positive: in general a large number of configurations will have the ground state energy density. But this local degeneracy, which is the second obstacle to local optimization, can occur independently of frustration. If we consider (only for the sake of this argument) an Ising $p$-spin model with large $p$ and all the $J$'s positive, we see that the number of assignments that minimize each term in the hamiltonian (separately) is $2^{p-1}$. Each many-spin interaction term poses a very weak constraint on the individual spins.

The consequence of frustration and local degeneracy is that in general the ground state of a spin glass will be highly degenerate. Not only the number of minimal energy configurations will be exponential in the size $N$ of the system, but often, due to disorder, the Gibbs measure will decompose in a large number 
$\mathcal N$ of pure states. In some cases this number will be exponential: $\mathcal N \sim e^{\Sigma N}$ where $\Sigma > 0$ is called complexity; in other cases $\mathcal N$ will be sub-exponential in $N$, but still large.

\subsection{The order parameter of disordered systems}

The most striking feature of spin glasses is that there is order hidden in their disorder. If one looks at a ``typical'' configuration of a spin glass, it will look the same at any temperature: each spin points in an apparently random direction. However, as the temperature is lowered, each spin becomes more and more ``frozen'' in a particular direction, which will depend on the site and which will ``look'' as disordered as the typical high temperature configuration. At sufficiently low temperatures, even though the site-averaged magnetization is zero, the local average magnetization is not. A convenient measure of this hidden order was introduced by Edwards and Anderson \cite{Edwards75}, and goes under their names:
\begin{equation}
 q_{\mathrm{EA}} = \frac 1 N \sum_i m_i^2
\end{equation}
where $m_i$ is the thermal average of $\sigma_i$. In the following I shall denote thermal averages with angled brackets, e.g. $m_i = \left< \sigma_i \right>$.

Of course, since the hamiltonian is dependent on the specific values of the random couplings, the value of $m_i$ will also depend on them. However, for many physical observables the average over sites is equal to the average over disorder:
\begin{equation}
 \lim_{N \to \infty} \frac 1 N \sum_i O_J(i) = \overline{O_J(\cdot)} \equiv \int d \mu(J) \ O_J(\cdot)
\end{equation}
where $\mu(J)$ is the distribution of disorder. Such observables are said to be \emph{self-averaging}, and the Edwards-Anderson order parameter $q_\mathrm{EA}$ is one of them. On the other hand, if physically relevant observables were to be dependent on the realization of disorder, i.e. on the specific sample, there would be very little to say about them, and very little interest in their study.

The Edwards-Anderson order parameter is very closely related to a more general quantity, the overlap, which can be defined on two different contexts. The overlap between microscopic configurations $\sigma$ and $\tau$ can be defined as 
\begin{equation}
 q_{\sigma \tau} = \frac 1 N \sum_i \sigma_i \tau_i
\end{equation}
which will be in the interval $[-1,1]$. The value 1 will correspond to perfectly correlated configurations, -1 to perfectly anti-correlated ones, and 0 to uncorrelated $\sigma$ and $\tau$. The concept of overlap can be extended to thermodynamic states, and is particularly interesting in the presence of ergodicity breaking. If we consider two different thermodynamic states $\alpha$ and $\beta$, we can compute 
\begin{equation}
 q_{\alpha \beta} = \frac 1 N \sum_i \left< \sigma_i \right>_\alpha \left< \sigma_i \right>_\beta
\end{equation}
which will measure how different the two states are. 

When a single state is present, the Edwards-Anderson order parameter is just $q_\mathrm{EA} = q_{\alpha \alpha}$, the self-overlap of the state with itself. However, in presence of ergodicity breaking, the Gibbs measure decomposes in a sum over pure states,
\begin{equation}
 \left< O \right> = \sum_\sigma \frac 1 Z \ O(\sigma) \ e^{-\beta H(\sigma)} = \sum_\alpha \frac {Z_\alpha} Z \sum_{\sigma \in \alpha} \frac 1 {Z_\alpha} \ O(\sigma) \ e^{-\beta H(\sigma)} = \sum_\alpha w_\alpha \left< O \right>_\alpha
\end{equation}
where $Z_\alpha \equiv \sum_{\sigma \in \alpha} \exp(-\beta H(\sigma))$ and $w_\alpha \equiv Z_\alpha / Z$ is the relative weight of the state $\alpha$ in the decomposition. In this case, the Edwards-Anderson parameter is given by
\begin{equation}
 q_\mathrm{EA} = \overline{ \frac 1 N \sum_i \left< \sigma_i \right >^2 } 
= \overline{ \frac 1 N \sum_i \left( \sum_{\alpha} w_\alpha \left< \sigma_i \right>_\alpha \right)^2} 
= \overline{ \sum_{\alpha, \beta} w_\alpha w_\beta \ q_{\alpha \beta} }
\end{equation}
in which not just the self-overlaps of the states are considered, but also the overlaps among different states.

A very powerful characterization of the structure of the thermodynamic states is provided by the distribution of overlaps between states,
\begin{equation}
 \mathcal P(q) = \overline {\sum_{\alpha, \beta} w_\alpha w_\beta \ \delta(q - q_{\alpha \beta}) }
\end{equation}
which gives the probability that two configurations picked at random from the Gibbs distribution have overlap $q$. In terms of $\mathcal P(q)$ we will have
\begin{equation}
 q_\mathrm{EA} = \int dq \ \mathcal P(q) \ q \,.
\end{equation}

\section{Phenomenology of disordered systems}

As I have tried to explain in the previous section, disordered systems share three characteristic features: first, the presence of quenched disorder; second, the effects of frustration and local degeneracy, which lead to the existence of many thermodynamic states at low temperature; third, the ``freezing'' of the dynamical degrees of freedom in a disordered configuration at low temperature. From the phenomenological point of view, the two latter characteristics are the most relevant ones.

In this section I shall briefly review the phenomenology of disordered systems that support this picture, and which is common to a very wide class of systems, regardless of the specificities of different models.

\subsection{Spin glass susceptibilities}

The first clear observation of a ``hidden'' order in disordered systems came from measures of the low-field AC magnetic susceptibility in diluted solutions of iron in gold. The magnetic susceptibility $\chi$  is directly related to the Edwards-Anderson order parameter $q_\mathrm{EA}$. It is defined locally as $\chi_{ii} = \de m_i / \de h_i^\mathrm{ext}$, where $h_i^\mathrm{ext}$ is the applied external field. Since the contribution of the external field to the hamiltonian is always a linear term $-\sum_i h_i^\mathrm{ext} \sigma_i$, it is easy to see that the following fluctuation-response relation must hold:
\begin{equation}
 \chi_{ii} = \frac {\de m_i} {\de h_i^\mathrm{ext}} 
= \frac {\de^2} {\de (h_i^\mathrm{ext})^2 } \frac 1 \beta \log Z(\beta, \{h_i^\mathrm{ext}\})
= \beta \left< \left( \sigma_i - \left< \sigma_i \right> \right) ^ 2 \right> 
= \beta (1 - m_i^2) \,.
\end{equation}
The measured local susceptibility is the average of $\chi_{ii}$ over the sites:
\begin{equation}
 \chi_\mathrm{loc} = \frac 1 N \sum_i \chi_{ii} = \beta ( 1 - q_\mathrm{EA} ) \,.
\end{equation}
In the absence of magnetic ordering at low temperatures, $\chi_\mathrm{loc}$ should diverge as $1/T$. The measured susceptibility shows a sharp cusp instead of a divergence, which indicates that below a certain temperature $q_\mathrm{EA} > 0$ (Fig.~\ref{Fig_magnetic}).

\begin{figure}
 \includegraphics[height=10.5cm]{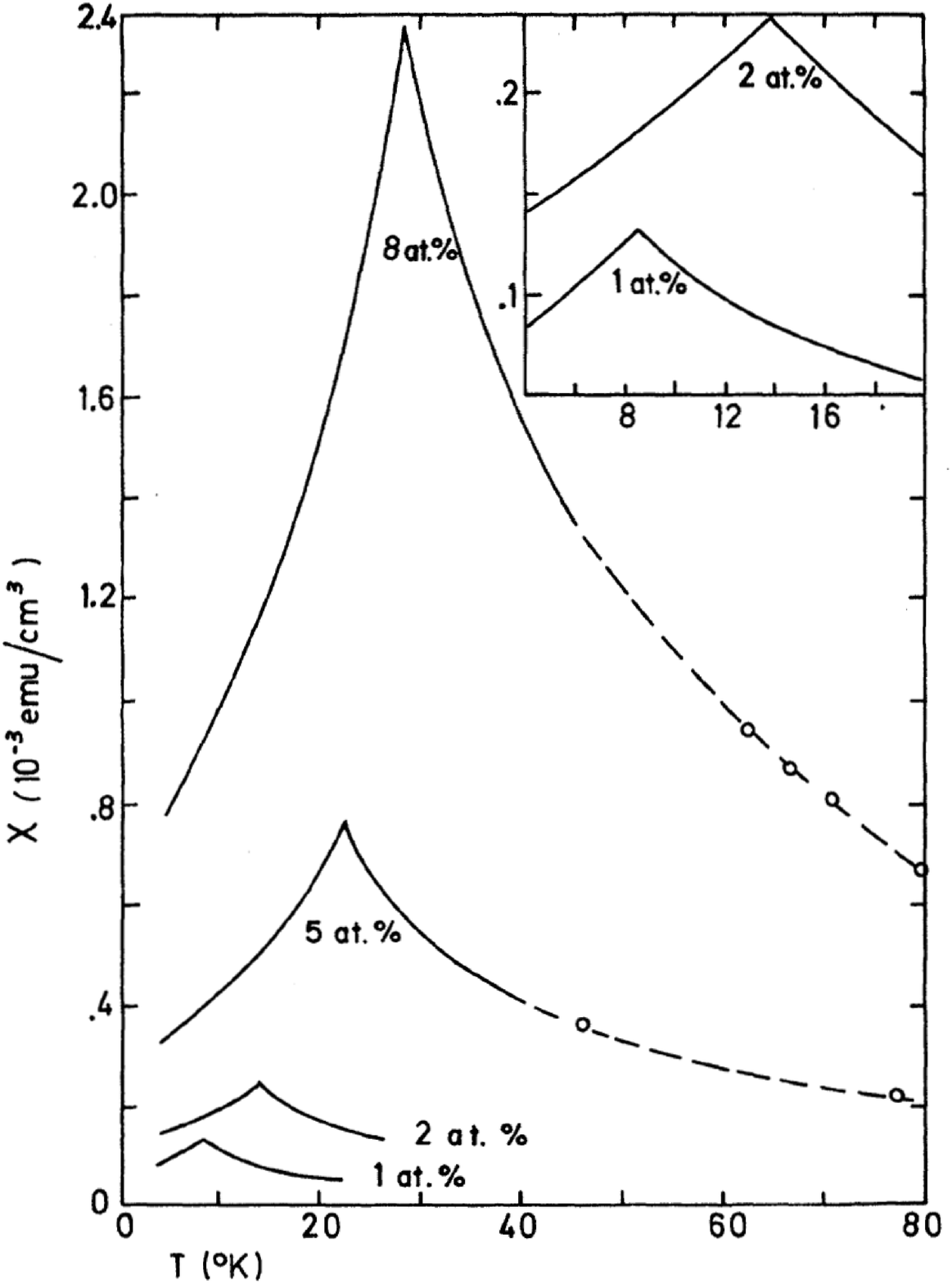}
 \includegraphics[height=10.75cm]{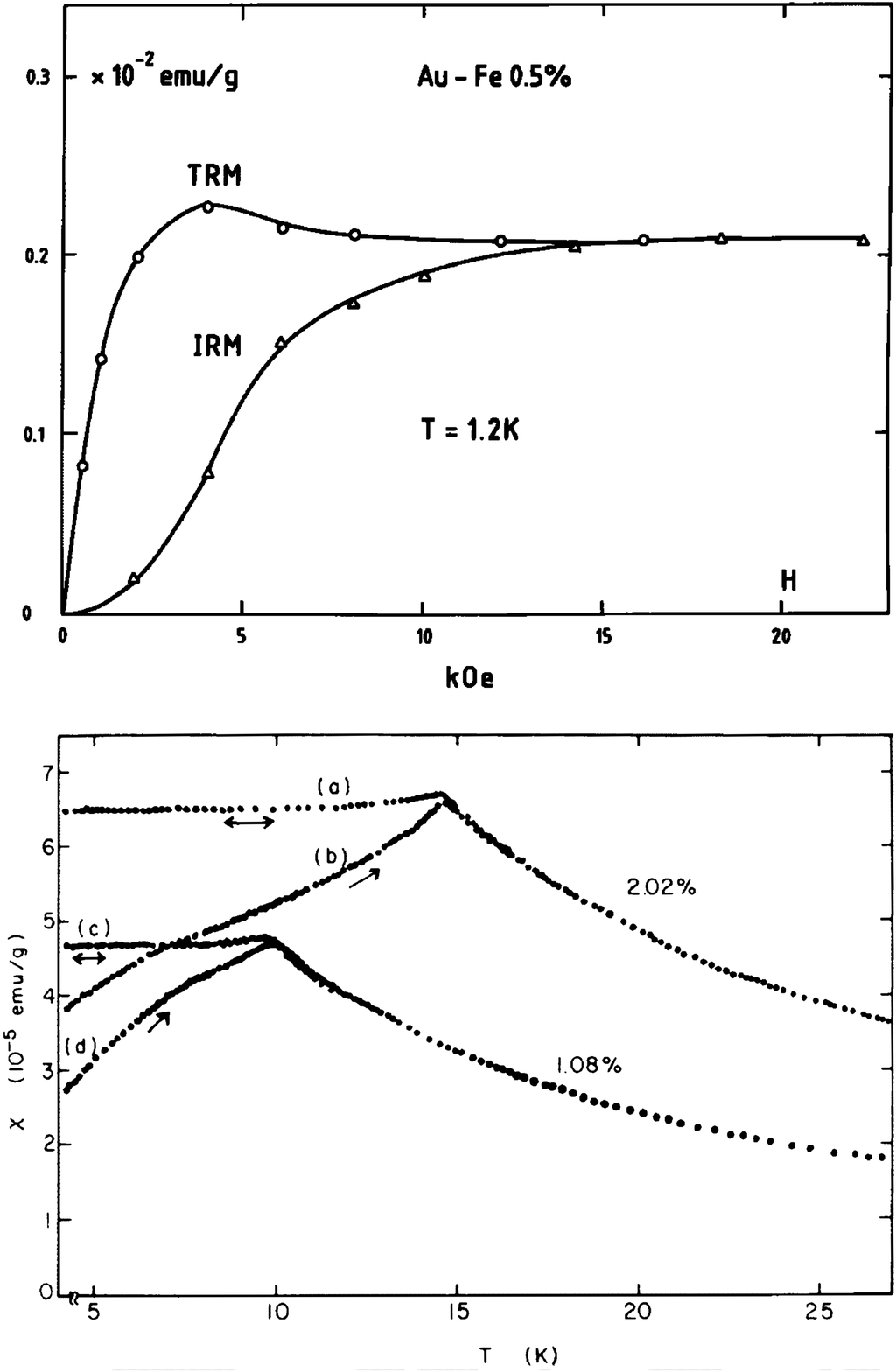}
 \caption{
 Magnetic properties of spin glasses.
 \emph{Left}~The AC susceptibility of \underline{Au}Fe alloys at different Fe concentrations for low field  ($\simeq 5$ G) and $\nu = 155 \ \mathrm{Hz}$ (from \cite{Cannella72}).
\emph{Right bottom}~The DC susceptibility of \underline{Cu}Mn for two Mn concentrations. Curves (a) and (c) were obtained by cooling in the measurement field (FC),(b) and (d) are the results of zero-field-cooled (ZFC) experiments (from \cite{Tholence74}).
\emph{Right top}~Remanent magnetization in \underline{Au}Fe (from \cite{Nagato79}).
 }
 \label{Fig_magnetic}
\end{figure}

A more detailed analysis of the frequency dependence of the measured AC susceptibility suggests the existence of a \emph{glassy} magnetic phase, i.e. a phase characterized by the existence of many metastable states. This is clearly confirmed by measures of DC magnetic susceptibility and of remanent magnetization, which both display a very strong dependence of the response on the details of the preparation of the sample. In DC susceptibility measures it can be seen that below a critical temperature, which coincides with the extrapolation to zero frequency of the position of the cusps in AC measurements, two different values of susceptibility can be measured: if the sample is cooled in the absence of field one obtains $\chi_\mathrm{zfc}$, which is lower than $\chi_\mathrm{fc}$, the value which is obtained when the sample is cooled in the presence of field. Moreover, if the external field is strong, a ``remanent'' magnetization is observed after it is switched off. The value of the remanent magnetization again depends on whether the field was applied during the cooling of the sample or only later. In the first case, the so called Thermo-Remanent Magnetization (TRM) is larger than the Isothermal Remanent Magnetization (IRM) (Fig.~\ref{Fig_magnetic}). This dependence on preparation of the sample properties clearly demonstrate that many different low temperature thermodynamic states are accessible to the system, and that they are well separated from each other, in the sense that the free energy barriers between states are extensive. 


\subsection{Divergence of relaxation times}

The main characteristic of glassy behavior is the divergence of the relaxation time at finite temperature. For structural glasses, the relaxation time $t_\alpha$ is defined as the decay time of density fluctuations, and it is accessible experimentally both directly and through the Maxwell relation
\begin{equation}
 \eta = G_\infty t_\alpha
\end{equation}
where $\eta$ is the viscosity and $G_\infty$ is the infinite-frequency shear modulus of the liquid. Experiments show that super-cooled liquids have a viscosity which can vary by as much as 15 orders of magnitude when the temperature varies by a factor of two above the glass forming temperature (Fig.~\ref{Fig_relaxation}). Similar results are obtained from direct measurements.

\begin{figure}
 \includegraphics[width=9cm]{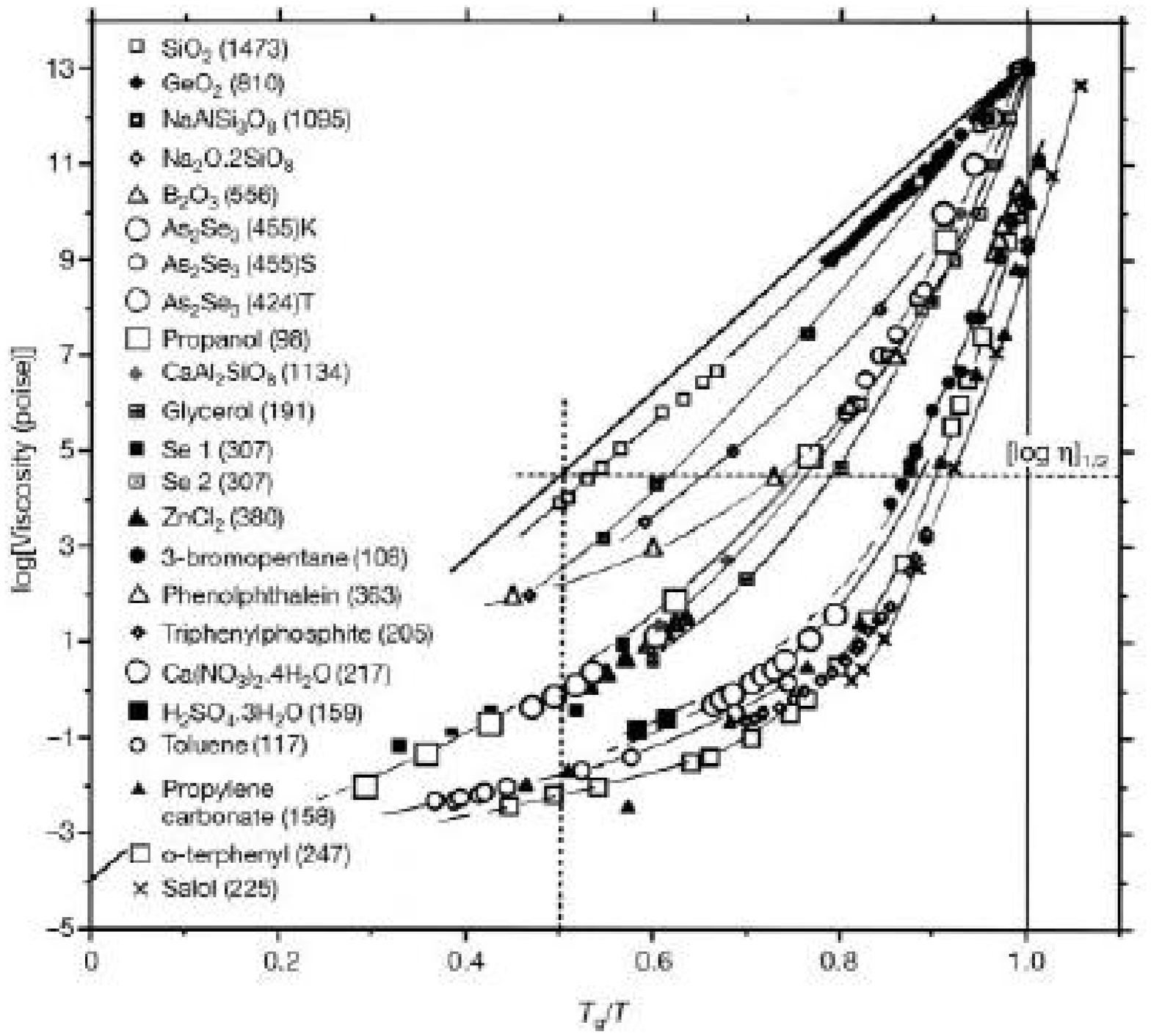}
 \includegraphics[width=6.4cm]{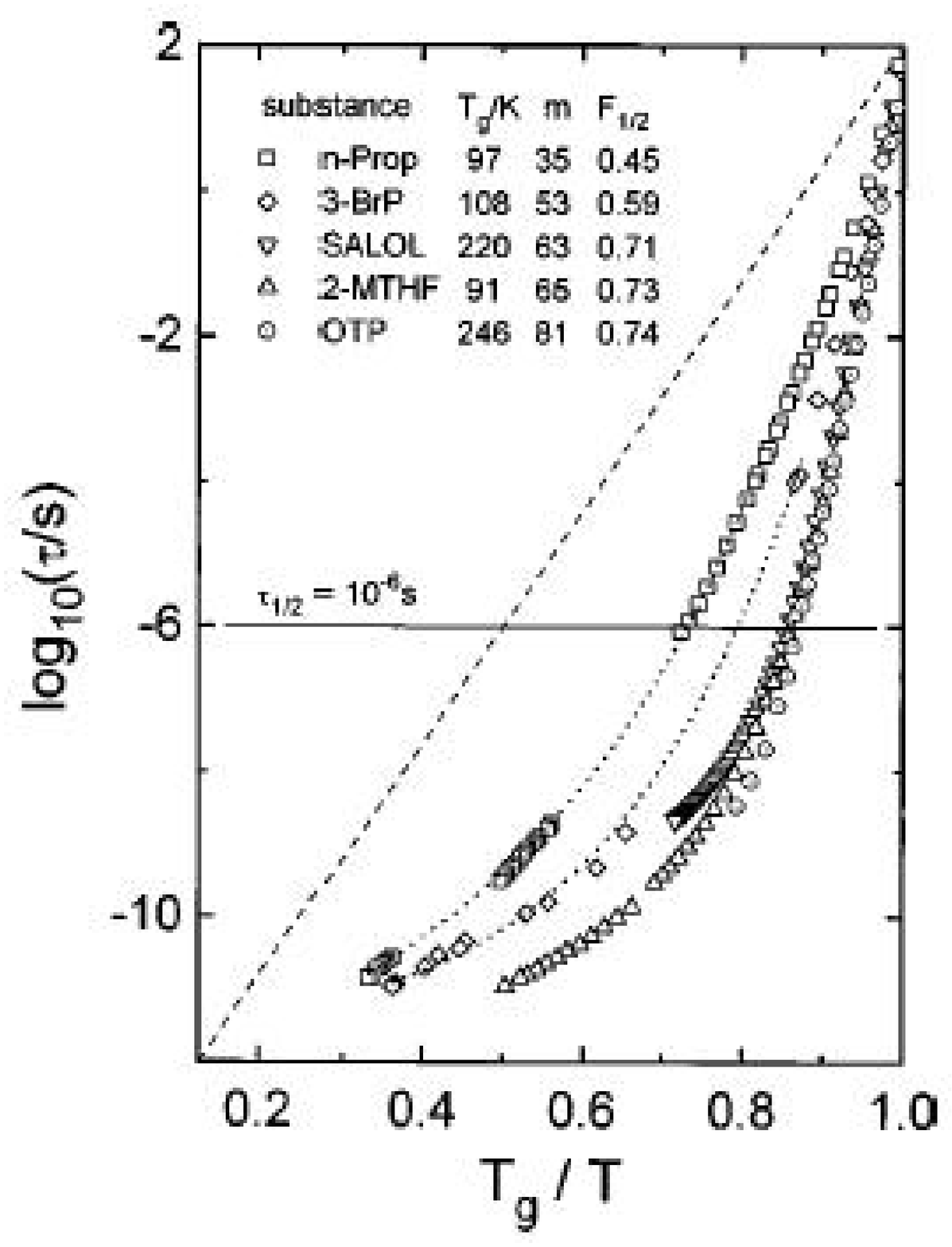}
 \caption{
 \emph{Left}~Viscosity measures for many glass forming liquids (from \cite{Martinez01}). The glass forming temperature $T_g$ is reported in parenthesis in the legend for each liquid.
 \emph{Right}~Structural relaxation times from dielectric relaxation measurements (from \cite{Richert98})
 }
 \label{Fig_relaxation}
\end{figure}

Spin glass models also show a divergence in relaxation times. A good example is provided by the $p$-spin spherical model (for $p \geq 3$). At high temperatures, the Fluctuation-Dissipation Theorem (FDT) holds, and the correlation $C(t,t')$ is related to the response $F(t,t')$ by the relation
\begin{equation}
 \frac \de {\de t} C(t,t') = -T \ F(t,t') \,.
\end{equation}
If the system equilibrates, the correlation function becomes invariant under time translations, $C(t, t+\tau) = C_\mathrm{eq}(\tau)$ and it is possible to derive a differential equation for $C_\mathrm{eq}(\tau)$, whose numerical solution for $p=3$ is shown in figure \ref{Fig_correlation}.

\begin{figure}
 \includegraphics[angle=-90,width=7.7cm]{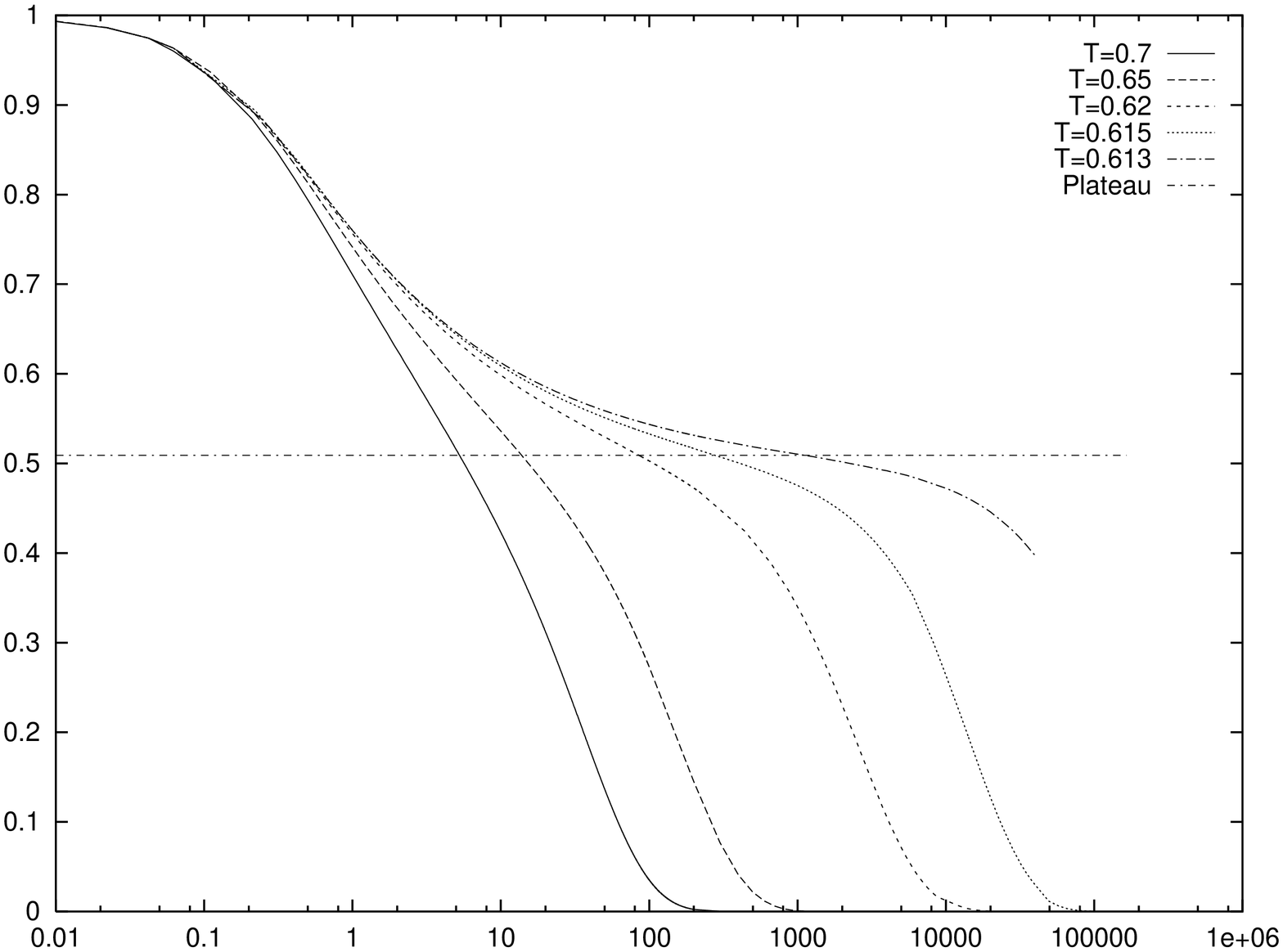}
 \includegraphics[angle=-90,width=7.7cm]{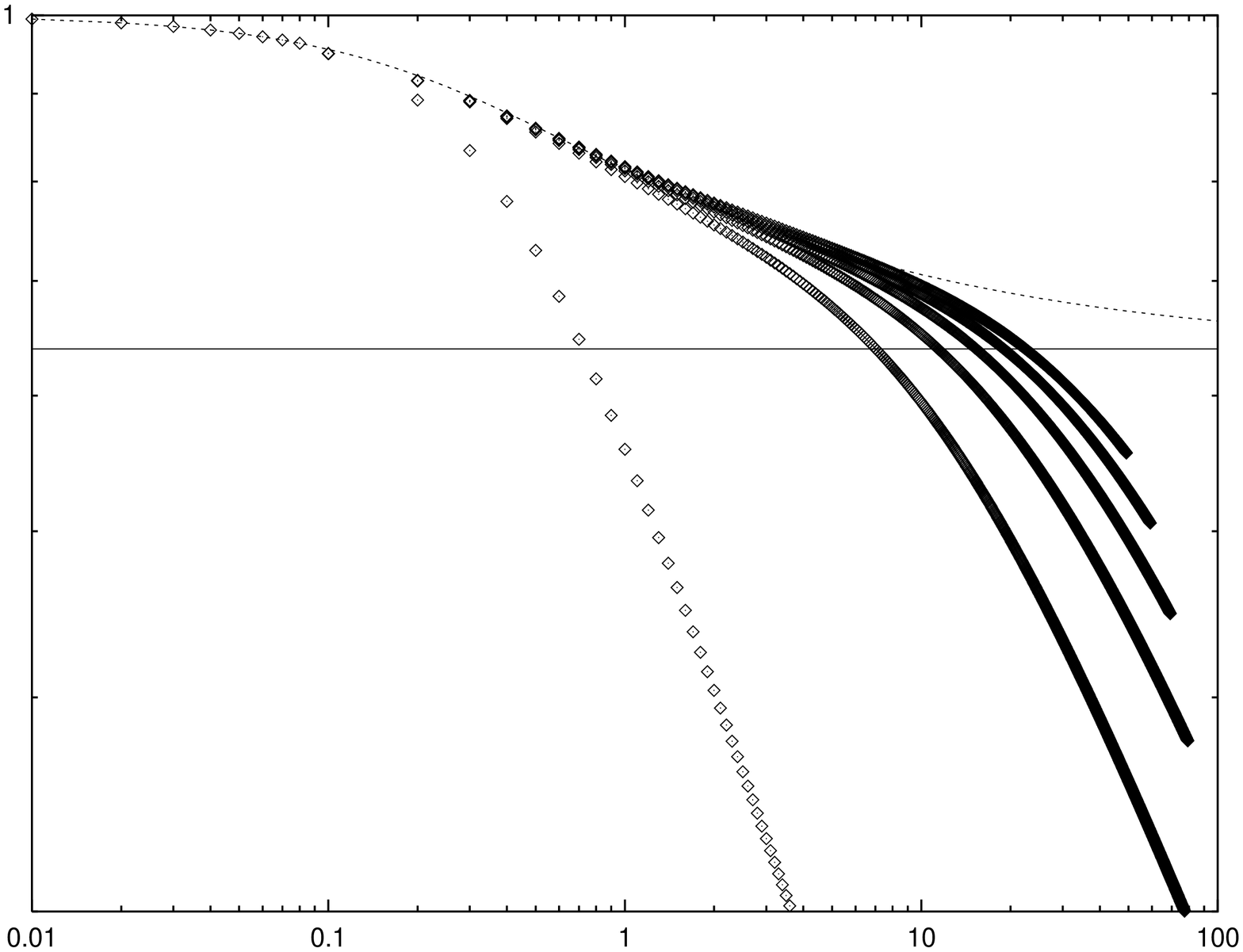}
 \caption{
 \emph{Left}~The translationally invariant correlation function $C_\mathrm{eq}(\tau)$ as a function of $\tau$, for different temperatures $T$. The horizontal line is the value of $q_\mathrm{EA}$.
 \emph{Right}~The out of equilibrium correlation function $C(t_\mathrm{w}, t+t_\mathrm{w})$ as a function of $t$ for different waiting times $t_\mathrm{w}$ at temperature $T=0.5$. The dotted line is computed in the limit $t_\mathrm{w} \to \infty$ and the horizontal line is its limiting value for $t \to \infty$. Both figures are from \cite{Barrat97}.
 }
 \label{Fig_correlation}
\end{figure}

What one sees is that as the temperature is decreased, a \emph{plateau} forms. The length of the plateau diverges as $T \to T_d$. The analysis of the model shows that $T_d$ is the temperature at which the free energy becomes dominated by an exponential number of metastable states with energy higher than the ground state. The value of the plateau coincides with $q_\mathrm{EA}$.

\subsection{Ageing}

If the temperature is lowered below $T_d$, a striking break-down of the translational invariance of the correlation function occurs, signalling that the system becomes unable to equilibrate.
In this regime, the correlation function $C(t_\mathrm{w},t_\mathrm{w}+t)$ depends separately on the waiting time $t_\mathrm{w}$ and on the duration of the observation $t$. Only in the limit $t_\mathrm{w} \to \infty$ the validity of the FDT is recovered and the system finally equilibrates.

This is an example of a very general phenomenon, observed in structural glasses as well as in spin glasses, which goes under the name of \emph{ageing}. Many observables for disordered systems maintain a time dependence for very long times under stable external conditions, indicating that they cannot equilibrate. This again confirms the existence of many metastable states which ``trap'' the dynamics of the system.

\section{The replica method}
\label{Sec Replica method}

In this section I shall briefly review one of the two equivalent analytical methods that can be used to investigate the equilibrium properties disordered systems: the replica method \cite{Mezard87}.

\subsection{The replica trick}

As I mentioned in the second section of this chapter, many physically relevant quantities are self-averaging, which is to say that their thermodynamic average is independent on the specific sample. A most notable example of a self-averaging quantity is the free energy density,
\begin{equation}
 f_J(\beta) = - \lim_{N \to \infty} \frac 1 {\beta N} \log Z_J(\beta)
\end{equation}
where the subscript $J$ denotes the dependence on the disorder. Because of the self-averageness of $f$, the free energy of any sample will be the same, and will be equal to the average over the distribution of $J$ of $f_J$:
\begin{equation}
 f(\beta) = \overline{ f_J(\beta) } 
= - \lim_{N \to \infty} \frac 1 {\beta N} \overline{ \log Z_J(\beta) } 
\equiv - \lim_{N \to \infty} \frac 1 {\beta N} \int d\mu(J) \ \log \sum_\sigma e^{-\beta H_J(\sigma)} \,.
\label{quenched_f}
\end{equation}
Unfortunately, the presence of the logarithm in the integral over the disorder makes it impossible to calculate it directly. However, one can use the following identity
\begin{equation}
 \log x = \lim_{n \to 0} \frac {x^n-1} n
\end{equation}
and write
\begin{equation}
 \overline{ \log Z_J(\beta) } 
= \overline {\lim_{n \to 0} \frac { Z_J(\beta)^n -1 } n }
= \lim_{n \to 0} \frac { \overline{ Z_J(\beta)^n } -1} n
= \lim_{n \to 0} \log \overline{ Z_J(\beta)^n } \,.
\end{equation}
By doing this, instead of $\overline{ \log Z_J(\beta) }$ one has to compute $\overline{Z_J(\beta)^n}$, which turns out to be much simpler. Notice that $Z_J(\beta)^n$ is the partition function of a system in which the dynamical degrees of freedom are replicated $n$ times and the quenched parameters are the same in each replica (hence the name, \emph{replica trick}).

\subsection{Solution of the $p$-spin spherical model}

As an example of the replica method, I am going to sketch its application to the $p$-spin spherical model. The hamiltonian is given by
\begin{equation}
 H_J(\sigma) = \sum_{i_1, \dots, i_p} J_{i_1 \cdots i_p} \ \sigma_{i_1} \cdots \sigma_{i_p} \,.
\end{equation}
The disorder $J$ has a gaussian distribution with average 0, and in order for the hamiltonian to be extensive its variance must scale as $N^{p-1}$:
\begin{equation}
 \mathbb P[J_{i_1 \cdots i_p} = J] = \mu(J) 
= \sqrt{ \frac {2 N^{p-1}} {2 \pi p!} } \ \exp \left\{ - \frac 1 2 J^2 \frac {2 N^{p-1} } {p!} \right\}\,.
\end{equation}

The starting point is to compute the gaussian integral over the gaussian distribution of disorder:
\begin{eqnarray}
 \overline {Z_J(\beta)^n} 
&=& \prod_{i_1 \cdots i_p} \int d\mu(J_{i_1 \cdots i_p}) \ \int d\sigma^1 \cdots d\sigma^n \ \exp \left\{ -\beta J_{i_1 \cdots i_p} \sum_{a=1}^n \sigma_{i_1}^a \cdots \sigma_{i_p}^a \right\} \\
&=& \prod_{i_1 \cdots i_p} \int d\sigma^1 \cdots d\sigma^n \ \exp \left\{ \frac {\beta^2 p!} {4 N^{p-1}} \sum_{a,b}^{1,n} \sigma_{i_1}^a \sigma_{i_1}^b \cdots \sigma_{i_p}^a \sigma_{i_p}^b \right\}
\end{eqnarray}
where I have dropped an overall normalization constant which doesn't give an extensive contribution. Here and in the following, I shall always denote by $i,j,k,\dots$ site indices running from 1 to $N$ and with $a,b,c,\dots$ replica indices running from 1 to $n$. Notice that after the integral we are left with a system in which \emph{sites are independent} and \emph{replicas are coupled}, which we can rewrite:
\begin{eqnarray}
  \overline {Z_J(\beta)^n} 
&=& \int d\sigma^1 \cdots d\sigma^n \ \exp \left\{ \frac {\beta^2} {4 N^{p-1}} \sum_{a,b}^{1,n} \left( \sum_i \sigma_i^a \sigma_i^b \right)^p \right\}
\end{eqnarray}

We can now introduce the overlaps between replicas,
\begin{equation}
 Q_{ab} = \frac 1 N \sum_i \sigma_i^a \sigma_i^b
\end{equation}
and multiply by
\begin{equation}
 1 = \int dQ_{ab} \int d\lambda_{ab} \exp \left\{i \lambda_{ab} \left[ NQ_{ab} - \sum_i \sigma_i^a \sigma_i^b \right] \right\}
\end{equation}
to obtain:
\begin{eqnarray}
 \overline{ Z_J(\beta)^n } 
&=&  \int d\sigma^a \cdots d\sigma^n \ \int dQ \ \int d\lambda \ \exp \left\{ \frac {\beta^2 N} 4 \sum_{ab} Q_{ab}^p + N \sum_{ab} \lambda_{ab} Q_{ab} - \sum_i \sum_{ab} \sigma_i^a \lambda_{ab} \sigma_i^b \right\} \nonumber \\ \ 
\end{eqnarray}
(where $Q \equiv \{Q_{ab}\}$ and $\lambda \equiv \{ \lambda_{ab} \}$). This integral is now gaussian in $\sigma$, and can be performed to obtain:
\begin{equation}
 \overline{ Z_J(\beta)^n } = \int dQ \ d\lambda e^{-N S(Q,\lambda)}
\end{equation}
where the action is
\begin{equation}
 S(Q,\lambda) = - \frac {\beta^2} 4 \sum_{ab} Q_{ab}^p - \sum_{ab} \lambda_{ab} Q_{ab} + \frac 1 2 \log \det(2 \lambda) \,.
 \label{action}
\end{equation}

This integral can be done using the Laplace method, in order to obtain
\begin{equation}
 f = - \lim_{N \to \infty} \frac 1 {\beta N} \lim_{n \to 0} \frac 1 n \log \int dQ \ d\lambda e^{-N S(Q,\lambda)}
= - \lim_{n \to 0} \frac 1 n \lim_{N \to \infty} \frac 1 {\beta N} \log e^{-N S(\bar Q, \bar \lambda)}
= \lim_{n \to 0} \frac 1 {\beta n} S(\bar Q, \bar \lambda)
\end{equation}
where $\bar Q$ and $\bar \lambda$ extremize the action. Notice however that we had to invert the order in which the limits over $N$ and $n$ are taken, which is not \emph{a priori} a legitimate manipulation.
Assuming it to be correct, the saddle point equations one obtains are the following:
\begin{eqnarray}
 \lambda_{ab} &=& \frac 1 2 (Q^{-1})_{ab} \,, \\
 \frac {\de f} {\de Q_{ab}} &=& 0 = \frac {\beta^2 p} 2 Q_{ab}^{p-1} + (Q^{-1})_{ab}
 \label{saddlepoint}
\end{eqnarray}

As we see, the parameter space over which one has to minimize $f$ is the space of symmetric matrices $Q$. The dimension of these matrices is $n$, which is assumed to go to 0: the only way to obtain a meaningful result is to write an expression for $f$ which is valid for \emph{any} finite $n$ and then do an analytic continuation of this expression for $n \to 0$. However, this requires that the matrix $Q_{ab}$ be parameterized in such a way that the matrix elements will depend on $n$ and on a fixed number $r$ of parameters $\{p_1, p_2, \dots, p_r\}$, which will be set to the values that satisfy the saddle point equations and which will be functions of $n$.

This rather intricate procedure raises three issues. The first is related to the fact that the whole procedure is far from rigorous from the mathematical point of view. Second, the parameterization of $Q$ in the particular form I've described limits the scope for the extremalization of $f$: it is not at all clear \emph{a priori} that the absolute extremum of $f$ corresponds to a matrix of the ``right'' form, and we may end up with an extremum that is not the ``true'' one. Finally, the stability of the free energy which is obtained in the end should be carefully checked \emph{a posteriori}. I shall return on these issues later.

A ``naive'' hypothesis would be to assume that since the replicas are just a formal expedient to compute $f$, the physical quantities should be independent of the replica index, and the overlap matrix $Q$ should be invariant under permutations of the replica indices. This would lead to the very simple parameterization $Q_{ab} = q_0 + (1-q_0) \delta_{ab}$ (the diagonal elements of $Q$ are determined by the spherical constraint to be 1). However, as already noted, the replicas \emph{do} have a physical interpretation: the replicated partition function, which is the proper self-averaging quantity to compute, corresponds to a composite system consisting of $n$ replicas of the original one. There is no reason why, in the presence of many states, different replicas should find themselves in the same state. Quite on the contrary, one should expect the breaking of the replica symmetry to be the signature of the presence of many states. It turns out that this intuition is correct.

The solution of the $p$-spin model \cite{Crisanti92} can be obtained by applying the replica-symmetry breaking (RSB) scheme introduced by Parisi to solve the SK model \cite{Parisi79, Parisi80, Parisi83}. The following parameterization is assumed for $Q$:
\begin{equation}
 Q_{ab} = \delta_{ab} + q_1 (1-\delta_{ab}) \mathbb I \left( a \div m = b \div m \right) + q_0 \mathbb I \left( a \div m \neq b \div m \right)
\end{equation}
where the free parameters are $\left\{m, q_0, q_1\right\}$, $a \div m$ represents the integer division of $a$ by $m$, and $\mathbb I(\mathsf{event})$ is the indicator function of $\mathsf{event}$ (i.e. it is 1 if $\mathsf{event}$ is true and 0 otherwise). The parameters are subject to the conditions $0 < m < n$, with $m$ such that $n$ is a multiple of $m$, and $0 < q_0 < q_1 < 1$. This parameterization corresponds to a matrix $Q$ which is made of $n/m$ identical blocks of size $m$ covering the main diagonal, with 1 on the main diagonal and $q_1$ outside of it in each block, and $q_0$ outside the blocks (notice that the case $m=n$ and $q_0 = q_1$ would correspond to the replica symmetric solution). This parameterization is known as \emph{one-step replica-symmetry breaking}, or 1RSB for short.

If this parameterization is substituted in the expression of the action $S(Q,\lambda)$ (\ref{action}), and the limit $n \to 0$ is computed, the following expression for the free energy is obtained:
\begin{eqnarray}
 f_\mathrm{1RSB} &=& - \frac \beta 4 \left[ 1 + (m-1)q_1^p - m q_0^p \right]
- \frac 1 {2 \beta} \left\{ \frac{m-1}{m} \log(1-q_1) + \right. \nonumber \\
&& \left. + \frac 1 m \log \left[ m(q_1 - q_0) + (1 - q_1) \right] + \frac{q_0}{m(q_1-q_0)-(1-q_1)} \right\} \,.
\end{eqnarray}
This expression can then be minimized to obtain the values of $m$, $q_1$ and $q_0$. What one sees is that for high temperature, a solution with $m = 1$ exists and is stable. However (for $p\geq3$) as the temperature is lowered to $T_s$ the solution with $m=1$ becomes \emph{unstable} and a new solution with $m < 1$ appears, which is stable and has a lower free energy than the solution with $m=1$. The value of $m$ undergoes a discontinuity as $T$ crosses $T_s$, jumping from 1 to a value which is at a finite distance from 1. 
As I have already mentioned, the existence of a replica-symmetry breaking solution is the signature of a glassy phase in which many different thermodynamic states coexist. The $p$-spin model undergoes a phase transition at $T_s$ from a paramagnetic to a glassy phase.

I would like to conclude this section with three remarks. The first concerns the issues I mentioned regarding the validity of the replica method. As I wrote, in general the procedure is not mathematically rigorous. However, one should note that in the case of the SK model the Parisi solution has been recently proved to be exact. Moreover the method has been applied to a large number of fairly different models, and in each case the results obtained are sensible: it appears safe to conjecture its validity, with the proviso that the stability of the solution it gives should be checked \emph{a posteriori} and that one cannot rule out the existence of other solutions, possibly with lower free energy.

Second, the example of the $p$-spin is particularly simple. In other models, including SK, one needs to consider a more complicated parameterization of the overlap matrix, which consists in applying the procedure I described recursively: one starts with a ``block'' of size $m_i$ which has 1 on the main diagonal and $q_i$ outside of it, and introduces blocks of size $m_{i+1} = m_i \div p$ (for some integer $p$) on the diagonal, with the same structure as the starting block, but a new value $q_{i+1}$ for the off-diagonal elements. This procedure can be repeated for any number of \emph{steps}. The solution of the $p$-spin is \emph{one-step replica-symmetry breaking}, denoted 1RSB. In the case of the SK model one needs an infinite number of steps, and the solution is said to be \emph{full replica-symmetry breaking} (FRSB).

Finally, the parameters over which one needs to extremize the free energy are the matrix elements of $Q$ (through the Parisi parameterization), which are \emph{scalar} quantities. This is a general feature of fully-connected models. However, as we shall see in the following section, the parameters to be minimized become much more complicated in the case of diluted models.

\subsection{Replica formalism for diluted models}
\label{Par Replica method for diluted}

In order to apply the replica method to diluted systems, one needs to generalize the approach that I have outlined for the case of the $p$-spin \cite{Mezard85, Orland85, DeDominicis87, Monasson99}. The starting point is the same: the average over disorder of the $n$-replicated partition function. For a system of Ising spins $\sigma_i \in \{-1,1\}$, with $\sigma \equiv \{\sigma_1,\dots,\sigma_N\}$ and with hamiltonian $H_J(\sigma)$, we have:
\begin{equation}
 \overline{ Z_J(\beta)^n } 
= \int d\mu(J) \sum_{\sigma^1} \cdots \sum_{\sigma^n} \exp \left\{- \beta \sum_{a=1}^n H_J(\sigma^a) \right\}
= \sum_{\sigma^1} \cdots \sum_{\sigma^n} \overline{ \exp \left\{- \beta \sum_{a=1}^n H_J(\sigma^a) \right\} }
\label{replicated_Z}
\end{equation}
where $\sigma^a$ is the $N$-spin configuration of the $a^\mathrm{th}$ replica. In fact, the $n$-replicated spin configuration is a matrix $\boldsymbol \sigma$ with $N$ rows corresponding to the sites and $n$ columns corresponding to the replicas. The $i^\mathrm{th}$ row is the $n$-component vector $\vec \sigma_i$ in which the component $\sigma_i^a$ is the value of the spin on the site $i$ for replica $a$, and the $a^\mathrm{th}$ column is the $N$-component configuration of replica $a$.

As an example of hamiltonian, we can consider the diluted version of the Ising $p$-spin model, which we shall discuss more in detail in the following:
\begin{equation}
 H_J(\sigma) = \sum_{m=1}^M \frac 1 2 \left( 1 - J_m \, \sigma_{i^m_1} \cdots \sigma_{i^m_p} \right)
\end{equation}
where the sum is over $M$ terms, each consisting of the product of $p$ spin, with indices $i^m_j$ with $j=1,\dots,p$ selected uniformly at random between 1 and $N$, and where the couplings $J_m$ are $\pm 1$ uniformly at random. The additive constant present in each term of the sum is such that the energy is positive or null. The factor $1/2$ is such that the value of the energy is equal to the number of terms in the sum which have a $J_m$ with a different sign relative to the product of the spins. On a random configuration, half the terms will be equal to 1 and the other half to 0, so that the energy will be extensive if $M = O(N)$.

We can interpret the right hand side of (\ref{replicated_Z}) as the partition function of an effective hamiltonian $\mathscr H$ depending on the full replicated configuration $\boldsymbol \sigma$:
\begin{equation}
 \overline{ Z_J(\beta)^n }
= \sum_{\boldsymbol \sigma} \exp \left\{ -\beta \mathscr H (\boldsymbol \sigma) \right\}
\label{effective_H}
\end{equation}
Since the distribution of disorder is independent on the site, the averaged quantity in the right hand side of (\ref{replicated_Z}) must be invariant under permutations of site indices. This implies that the effective hamiltonian (\ref{effective_H}) can depend on $\boldsymbol \sigma$ only through 
\begin{equation}
 c(\vec \tau) \equiv \frac 1 N \sum_i \mathbb I (\vec \sigma_i = \vec \tau) 
\end{equation}
which is the fraction of sites that have replicated configuration $\vec \tau$.
Even though $c(\vec \tau)$ actually depends on the replicated configuration $\boldsymbol \sigma$, we are going to assume it to be fixed and avoid its appearance in the notation. Also, notice that $\sum_{\vec \tau} c(\vec \tau) = 1$.

The overlap between replica configurations $Q_{ab}$ can also be expressed in terms of $c(\sigma)$:
\begin{equation}
 Q_{ab} = \frac 1 N \sum_i \sigma_i^a \sigma_i^b
= \sum_{\vec \tau} c(\vec \tau) \ \tau^a \tau^b \,.
\label{overlap_ct}
\end{equation}
This was to be expected: in the calculation for the $p$-spin, the free energy we obtained depended only on $Q$, and (\ref{overlap_ct}) implies that what we obtained was actually dependent on $c(\vec \tau)$ only. This is a general feature of fully connected models: their free energies (or rather, the actions whose extrema are equal to the free energy) depend only on the overlaps between replicas. However, for diluted models one needs to generalize (\ref{overlap_ct}) to include higher moments:
\begin{equation}
 Q_{a_1\cdots a_k} \equiv \sum_{\vec \tau} c(\vec \tau) \ \tau^{a_1} \cdots \tau^{a_k} \,.
\end{equation}
The crucial point is that even though these quantities are more complicated than the overlaps, they are still conceptually equivalent to $c(\vec \tau)$, which provides the full description of the structure of the states of the system, be it fully connected or diluted.


To see more in details how it is possible to write the free energy in terms of $c(\vec\tau)$, we can go back to (\ref{effective_H}) where we recall that $\mathscr H(\boldsymbol \sigma) = \mathscr H[c(\vec \tau)]$:
\begin{equation}
 \overline{ Z_J(\beta)^n } = \sum_{\boldsymbol \sigma} e^{-\beta \mathscr H(\boldsymbol \sigma)}
= \sum_{\left\{c(\vec\tau)\right\}}^{0,N} \frac{N!}{\prod_{\vec\tau}[Nc(\vec \tau)]!} \ e^{-\beta \mathscr H[c(\vec\tau)]} \ \mathbb I \left[ \sum_{\vec\tau} c(\vec\tau) = 1 \right]
\end{equation}
where the sum is over $2^n$ variables, each variable being the value of $c$ for one of the possible $2^n$ $n$-component spin configurations, that take values between 0 and $N$, where the multinomial factor is just the number of replicated configurations $\boldsymbol \sigma$ that give rise to the same distribution $c(\vec\tau)$, and where the last indicator function ensures the normalization of $c(\vec\tau)$.

In the limit $N \to \infty$ the sum becomes an integral and the multinomial coefficient can be approximated with Stirling's formula to obtain
\begin{eqnarray}
f(\beta) &=& -\lim_{N\to\infty} \frac 1 {\beta N} \lim_{n\to0} \frac 1 n \overline{ Z_J(\beta)^n } \\
&=& -\lim_{n\to0} \frac 1 n \lim_{N\to\infty} \frac 1 {\beta N} \int_0^1 \left( \prod_{\vec\tau} d c(\vec\tau) \right) \ \exp \left\{ N \left[ - \left( \sum_{\vec\tau} c(\vec\tau) \log c(\vec\tau) \right) - \beta \mathscr H[c(\vec\tau)] \right] \right\} 
\times \nonumber \\
&&\times \mathbb I \left[ \sum_{\vec\tau} c(\vec\tau) = 1 \right]  \\
&=& \lim_{n\to0} \frac 1 n \frac 1 \beta \extremum_{\{c(\vec\tau) : \sum_{\vec\tau} c(\vec\tau) = 1\}} \left\{\sum_{\vec\tau} c(\vec\tau) \log c(\vec\tau) + \beta \mathscr H[c(\vec\tau)] \right\}
\label{f_functional}
\end{eqnarray}
where (as before) we have exchanged the order of the limits $N\to\infty$ and $n\to0$.

With this formalism, the problem of computing the free energy of a (possibly diluted) disordered Ising model is decomposed into three tasks:
\begin{enumerate}
 \item Find the effective hamiltonian $\mathscr H[c(\vec\tau)]$
 \item Compute, for each value of $n$, the extremum of the free energy functional in $c(\vec\tau)$ appearing on the right hand side of (\ref{f_functional})
 \item Perform the analytic continuation of the result to $n=0$
\end{enumerate}

In Chapter \ref{Chap_ksat} I shall use this formalism to derive some properties of the solutions of an optimization problem which is formally equivalent to a diluted Ising spin glass.


\chapter{Optimization problems and algorithms}
\label{Chap_optimization}

In the previous Chapter, I have given a very brief overview of the physics of disordered systems. In this Chapter, I shall introduce a different kind of disordered systems, which arise from the study of combinatorial optimization problems, and I shall discuss some aspects specific to them, and what they have in common with the disordered systems studied in physics.

In the first Section, I shall give some examples of combinatorial optimization problems; in Section \ref{Sec Satisfiability} I shall introduce the two specific problems that have been the subject of my research, \ksat\ and \kxorsat; then I shall introduce some notions from complexity theory, in Section \ref{Sec Complexity}; finally, in \ref{Sec Algorithms} I shall present some families of algorithm that are useful for finding solutions to optimization problems, and whose properties also shed some light on the underlying structure of the problems themselves.

Most of the material discussed in this Chapter can be found in \cite{Papadimitriou98}.

\section{Some examples of combinatorial optimization problems}
\label{Sec Examples of comb opt prob}

Optimization problems are concerned with finding the ``best'' (or \emph{optimal}) allocation of finite resources to achieve some purpose. It is clearly a very general and important class of problems. An early example of optimization problem is narrated in Virgil's Aeneid: Dido, a Phenician princess, is obliged to flee Tyre, her hometown, after her husband is murdered by her brother, a cruel tyrant. She embarks with a small group of refugees, and lands in Lybia, where she asks the king Iarbas to purchase some land to found a new city, Carthage. Iarbas, in love with Dido but rejected by her, has no intention to allow the settlement, and offers only as much land as can be enclosed in a bull's hide. He is, however, outwitted by Dido, who cuts the hide in thin stripes, which she joins to form a long string. With that, she encloses an area shaped as a semi-circle, delimited by the sea, and sufficient to build Carthage.
In this legendary tale, Dido not only had the brilliant idea of cutting the hide, but also solved a non-trivial optimization problem: what is the curve of given perimeter that encloses the largest area?

Combinatorial optimization problems are, in a way, simpler: the set of possible solutions is \emph{discrete}. This restriction might appear severe in view of practical applications, but in fact it is not: many resources, such as industrial machines, skilled workers or computer chips are indeed indivisible.
Let us begin with an example, which I shall use to illustrate a general, formal definition, and after which I shall give some more examples of different families of combinatorial optimization problems.

Consider the following
\begin{description}
 \item[Knapsack Problem (KP)] Given a set $\mathcal S$ of items $i=1,\dots,N$, each having a value $v_i \in \mathbb R^+$ and a weight $w_i \in \mathbb R^+$, what is the subset $\mathcal S' \subset \mathcal S$ with the largest total value $V = \sum_{i \in \mathcal S'} v_i$ and such that the total weight $W = \sum_{i \in \mathcal S'} w_i$ is $W \leq W^*$ for some given $W^*$?
\end{description}
The possible solutions (or configurations) are all the subsets that can be formed with elements from $\mathcal S$, which are a discrete set of cardinality $2^N$ (corresponding to the two choices ``present'' or ``not-present'' for each item in $\mathcal S$). A specific instance of the general problem is defined by the pairs $\{(v_i,w_i), \ i=1,\dots,N\}$, and by the maximum allowed weight $W^*$.

In general, an instance of the problems I shall consider will be defined by specifying the following three characteristics:
\begin{enumerate}
 \item A set $\mathcal C$ of possible \emph{configurations} $\mathscr C$;
 \item A \emph{cost function} $F: \mathcal C \rightarrow \mathbb R$ that associates a cost $F(\mathscr C)$ to every configuration $\mathscr C \in \mathcal C$, and which can be computed in polynomial time;
 \item An \emph{objective}, that is to say a condition on $F(\mathscr C)$ which must be satisfied.
\end{enumerate}
In the knapsack example, $\mathcal C$ is the set of all subsets of $\mathcal S$, the cost function $F$ is 
\begin{equation}
 F(\mathscr C) = \mathbb I \left[ \sum_{i \in \mathscr C} w_i \leq W^* \right] \times \sum_{i \in \mathscr C} v_i
\end{equation}
and the objective is of the form $F(\mathscr C) > F^*$.

In general, for a given instance, one can ask the following questions:
\begin{description}
 \item[Decision] Does a configuration that realizes the objective exist?
 \item[Optimization] What is the ``tightest'' objective which can be realized? For example, the largest value of $F^*$.
 \item[Search] Which configuration realizes the objective?
 \item[Enumeration] How many configurations realize the objective?
 \item[Approximation] Which configuration realizes a weaker form of the objective, for example $F(\mathscr C) > \gamma F^*$ for some constant $\gamma < 1$?
\end{description}
The knapsack example above is a combination of an optimization problem (finding the largest possible value which can be realized) and a solution one (finding the corresponding configuration). Of course, one could ask many more questions. These are just the ones I shall be interested in in the following.

Let me cite a few more examples of problems:
\begin{description}
 \item[Number Partitioning] Given a set of $N$ positive integers $S = \{n_i \in \mathbb N, \ i=1,\dots,N\}$, find a subset $S' \subset S$ such that $\sum_{i \in S'} n_i = \sum_{i \in S \backslash S'} n_i$.
 \item[Subset Sum] Given a positive integer $K$ and a set of $N$ positive integers $S = \{n_i \in \mathbb N, \ i=1,\dots,N\}$, find a subset $S' \subset S$ such that $\sum_{i \in S'} n_i = K$. 
 \item[Integer Linear Programming (ILP)] Given a $n$-component real vector $\mathbf c$, a $n \times m$ real matrix $\mathbf A$, and a $m$-component real vector $b$, find a $n$-component vector $x$ with non-negative integer components and which maximizes $\mathbf c \cdot \mathbf x$ subject to the constraints $\mathbf A \mathbf x \leq \mathbf b$.
 \item[Is Prime] Given a positive integer $N$, determine if $N$ is prime.
\end{description}

Many combinatorial optimization problems are defined on \emph{graphs}. A graph $\mathcal G$  is a double set of points, called \emph{vertices}, $v \in \mathcal V$, and of distinct segments connecting pairs of points in $\mathcal V$, called \emph{edges}, $e \in \mathcal E$: $\mathcal G = (\mathcal V, \mathcal E)$. Three special kinds of graphs are \emph{cycles}, i.e. loops; \emph{trees}, which are connected graphs that contain no cycles; and \emph{bipartite} graphs, in which the set of vertices is divided in two, $\mathcal V = \mathcal V_1 \cup \mathcal V_2$, and all edges have an endpoint in $\mathcal V_1$ and the other in $\mathcal V_2$. Let me just mention a few important problems defined on graphs:
\begin{description}
 \item[Hamiltonian Cycle (HC)] Given a graph $\mathcal G = (\mathcal V, \mathcal E)$, find a cycle $\mathcal G' \subset \mathcal G$ containing all the vertices of $\mathcal G$.
 \item[Traveling Salesman Problem (TSP)] Given a graph $\mathcal G = (\mathcal V, \mathcal E)$ and a weight $w(e) \in \mathbb R^+$ associated to each edge, find a HC with minimum total weight.
 \item[Minimum Spanning Tree (MST)] Given a graph $\mathcal G = (\mathcal V, \mathcal E)$ and a weight $w(e) \in \mathbb R^+$ associated to each edge, find a tree $\mathcal G' \subset \mathcal G$ containing all the vertices of $\mathcal G$ with minimum total weight.
 \item[Vertex covering (VC)] Given a graph $\mathcal G = (\mathcal V, \mathcal E)$, find a subset $\mathcal V' \subset V$ of the vertices of $\mathcal G$ such that each edge $e \in \mathcal E$ has at least one of its endpoints in $\mathcal V'$, and minimizing $|\mathcal V'|$.
 \item[$\boldsymbol q$-Coloring ($\boldsymbol q$-COL)] Given a graph $\mathcal G = (\mathcal V, \mathcal E)$, assign to each vertex a color $c \in \{1,2,3,\dots,q\}$ such that no edge in $\mathcal V$ has two endpoints of the same color.
 \item[Matching] Given a graph $\mathcal G = (\mathcal V, \mathcal E)$ and a weight $w(e) \in \mathbb R^+$ associated to each edge, find a subgraph $\mathcal G' \subset \mathcal G$ such that each vertex in $\mathcal V'$ has one and only one edge in $\mathcal E'$, and which maximizes the total weight. Often $\mathcal G$ is bipartite, in which case the problem is called \emph{bipartite matching}.
 \item[Max Clique] Given a graph $\mathcal G = (\mathcal V, \mathcal E)$ , find its largest \emph{clique}, i.e. fully connected subgraph.
 \item[Min (or Max) Cut] Given a graph $\mathcal G = (\mathcal V, \mathcal E)$ and a weight $w(e) \in \mathbb R^+$ associated to each edge, find a partition $(\mathcal V_1, \mathcal V_2)$ of $\mathcal V$ such that the total weight of the edges that have an edge in $\mathcal V_1$ and the other in $\mathcal V_2$ is minimized (or maximized).
\end{description}

All these problems are interesting from the theoretical point of view, and relevant for their practical applications. A further family of problems concerns \emph{boolean satisfiability}, which I shall introduce in the next Section. The importance of boolean satisfiability problems and their connection to the other problems will be discussed in Section \ref{Sec Complexity}.

\section{Boolean satisfiability: \ksat\ and \kxorsat}
\label{Sec Satisfiability}

Boolean satisfiability problems are concerned with the following general question: given a boolean function $\mathcal F(x)$ over $N$ boolean variables $x \equiv (x_1,\dots,x_N) \in \{\true,\false\}^N$, is there an assignment of the variables which makes the function evaluate to $\true$?
The different problems of the family correspond to specific choices of the form of the function $\mathcal F$.

\subsection{Introduction to \ksat}

The prototype of satisfiability problems is the following. Given a $N$-tuple of boolean variables $x = (x_1,\dots,x_N)$, a \emph{literal} is defined as a variable or its negation, e.g. $x_3$ and $\bar x_7$; a \emph{$k$-clause} (or simply clause, of length $k$) is defined as the disjunction of $k$ literals, e.g. for $k=3$: $x_2 \vee \bar x_4 \vee x_7$; finally, a \emph{formula} is defined as the conjunction of $M$ clauses. For example, for $N=7, M=3$:
\begin{equation}
 \mathcal F(x) = \left( \bar x_3 \vee x_5 \vee \bar x_6 \right) \wedge \left( x_2 \vee x_3 \right) \wedge (x_1 \vee x_3 \vee \bar x_5 \vee x_7) \,.
\end{equation}
Such a formula is said to be in \emph{conjunctive normal form} (CNF), which is defined as
\begin{equation}
 \mathcal F(x) = \bigwedge_{m=0}^M \left[ \left( \bigvee_{j \in \mathcal I_m} x_j \right) \vee \left( \bigvee_{j' \in \mathcal I'_m} \bar x_{j'} \right) \right]
\end{equation}
where $\mathcal I_m$ and $\mathcal I'_m$ are subsets of $\{1,\dots,N\}$ such that $I_m \cap I'_m = \oslash$ for each $m=1,\dots,M$.

The \emph{satisfiability} problem (\sat) is the problem of determining if a given CNF formula admits at least one satisfactory assignment (also called a \emph{solution}) or not.
An interesting special case is that in which all the clauses have the same 
length $k$, in which case the problem is known as \ksat. If the answer is ``yes'', the formula is said to be \emph{satisfiable}, which I shall denote by \sat\footnote{The use of \sat\ to designate both the general satisfiability problem and the satisfiable property of a formula should not lead to confusion, since in the future I shall be concerned exclusively with \ksat.}, otherwise it is \emph{unsatisfiable} which I shall denote \unsat. 

The same questions apply to \ksat\ as to any other combinatorial optimization problem, namely the decision, optimization, solution, enumeration, and approximation problems, where the quantity to be minimized is the number of violated clauses.

A lot of attention has been devoted to \ksat, principally for three reasons: first, for its theoretical relevance; many problems, from theorem proving procedures in propositional logic (the original motivation for \ksat), to learning models in artificial intelligence, to inference and data analysis, can all be expressed as CNF formul\ae. Second, because it is directly involved in a large number of practical problems, from VLSI circuits design to cryptography, from scheduling to communication protocols, all of which actually require solving or optimizing real instances of \ksat\ formul\ae. Third, and probably most notably, because of its central role in complexity theory, which I shall discuss in the next Section.

The questions of interest in the study of \ksat\ can  be divided in two broad families: on one hand those regarding the general properties of CNF formul\ae\ and of their solutions (when they exist); on the other hand, those concerning the algorithms capable of answering the different questions one may ask (decision, optimization, \dots); and of course, the intersection of the two (for example, proving that a certain algorithm succeeds in finding a solution under some assumptions also proves that a formula verifying those same assumptions must be \sat).

Also the answers that one can seek can be divided in two (or rather, their qualitative types): on one hand the results that are true in general and for any instance of \ksat\ (under certain conditions), and on the other hand results that are true in a \emph{probabilistic} way. Let me clarify this last case with an example. Suppose one considers the ensemble of all possible \ksat\ formul\ae\ with given $N$ and $M$, with uniform weight. The total number $\mathcal N_\mathrm C$ of $k$-clauses that one can form with $N$ variables is given by the number of choices of $k$ among $N$ indices times the number of choices for the $k$ negations, i.e.
\begin{equation}
 \mathcal N_\mathrm C = \binom N k 2^k \,.
\end{equation}
The number of formul\ae\ $\mathcal N_\mathrm F$ that can be made with $k$ independently chosen clauses is then
\begin{equation}
 \mathcal N_\mathrm F = (\mathcal N_\mathrm C)^M \,.
\end{equation}
Consider now a clause $\mathcal C$ in the formula, for simplicity $\mathcal C = x_1 \vee \cdots \vee x_k$. This clause will be satisfied by any of the $2^k$ possible values of $(x_1,\dots,x_k)$ \emph{except} the one corresponding to $x_i = \false$ for $i=1,\dots,k$: out of all the possible assignments, only a fraction $1-1/2^k$ will satisfy any given clause. Since the formula contains $M \equiv \alpha N$ clauses (where $\alpha$ is defined as the ratio $M/N$), the \emph{average} number of satisfying assignments will be
\begin{equation}
 \mathcal N_\mathrm S = 2^N \times \left(1 - \frac 1 {2^k} \right)^M =
 \left[ 2 \left( 1 - \frac 1 {2^k} \right) ^ \alpha \right]^N \,.
\end{equation}
If we consider \emph{large} formul\ae, i.e. the limit $N \to \infty$, we see that the average number of solutions tends to 0 if
\begin{equation}
 \alpha > - \frac {\log 2} {\log \left( 1 - 2^{-k} \right)} \,.
\end{equation}
Notice that the average number of solutions is larger than or equal to the probability that a formula is \sat, since
\begin{equation}
 \mathcal N_\mathrm S = \sum_{n=0}^{2^N} n \times \mathbb P[\text{The number of solutions is }n] \geq \sum_{n=1}^{2^N} \mathbb P[\text{The number of solutions is }n]
\end{equation}
and the sum on the right hand side is the probability that a formula is \sat.
Therefore, we see that \emph{in the limit $N \to \infty$ a random \ksat\ formula chosen with uniform weight among all those with $M = \alpha N$ clauses is \unsat\ with probability $1$ if $\alpha > - \log 2 / \log (1-2^{-k})$.}

This kind of statement is very useful to characterize the \emph{typical} properties of \ksat\ formul\ae\ under some given conditions. In many cases, the typical behavior is the interesting one, as it dominates the observable phenomena. The problem of studying \ksat\ formul\ae\ extracted from some distribution is often called Random-\ksat. If the distribution is not specified, the uniform one is assumed.

Many interesting properties are easily proved for Random-\ksat. For example, for $\alpha \to 0$ the probability $P_\mathrm{Sat}(\alpha)$ that a random formula is \sat\ tends to 1. And it must be a decreasing function of $\alpha$, since the property of being \sat\ is \emph{monotone}: in order for a formula to be \sat, any sub-formula (made with a subset of its clauses) has to be satisfiable as well. In other words, adding clauses to a formula can only decrease its chances of being \sat, and adding random clauses to a random formula can only decrease its probability of being \sat.

From the physicist's point of view, probabilistic results are most interesting, because a random distribution of formul\ae\ can be treated as a disordered system with some distribution of disorder. Indeed, one can represent Random-\ksat\ as a spin glass. Each variable $x_i$ will correspond to an Ising spin $\sigma_i$, which will be 1 if $x_i = \true$ and $-1$ otherwise. For a given configuration, the number of violated clauses will play the role of the energy:
\begin{equation}
 E(\sigma) = \sum_{m=1}^M \prod_{j=1}^k \frac{1 + J^m_j \sigma_{i^m_j}} 2
 \label{ksat hamiltonian}
\end{equation}
where $i^m_j$ is the index of the $j^\text{th}$ variable appearing in the $m^\text{th}$ clause, and $J^m_j$ is 1 if the variable appears negated and $-1$ otherwise. The set of $\{J^m_j\}$ and $\{i^m_j\}$ defines some random couplings which involve terms with $1,2,\dots,k$ spins, have unit strength, and are attractive or repulsive with equal probability. As usual with statistical mechanics systems, we shall be interested in the thermodynamic limit $N \to \infty$. Since a random configuration violates a random clause with probability $2^{-k}$, the energy is extensive (i.e. proportional to $N$) if $\alpha$ is of order $O(1)$ as $N \to \infty$. This is a perfectly legitimate \emph{diluted }spin glass model. In fact, in Chapters \ref{Chap_phase_diagrams} and \ref{Chap_ksat} I shall present some results on Random-\ksat\ obtained applying the replica method of Paragraph \ref{Par Replica method for diluted} to \ref{ksat hamiltonian}.

\subsection{Introduction to \kxorsat}

Another interesting boolean satisfiability problem goes under the name of \kxorsat, and is obtained when the boolean function $\mathcal F(x)$ is the boolean equivalent of a linear system of equations:
\begin{equation}
 \mathcal F(x) = \bigwedge_{m=1}^M \left[ \left( \bigoplus_{j=1}^k x_{i^m_j} \right) \oplus y_m \right]
\end{equation}
where the symbol $\oplus$ denotes the logical operation XOR, and where $i^m_j \in \{1,\dots,N\}$ for $m = 1,\dots,M$ and $j = 1,\dots,k$ are some variable indices, and where $y \equiv (y_1,\dots,y_M)$ is some constant boolean vector.
If we make the correspondence $\true = 1$ and $\false = 0$, this formula is equivalent to the linear system
\begin{equation}
 \left\{
  \begin{aligned}
   x_{i^1_1} \oplus \cdots \oplus x_{i^1_k} &= y_1 \,, \\
   x_{i^2_1} \oplus \cdots \oplus x_{i^2_k} &= y_2 \,, \\
   \cdots \\
   x_{i^M_1} \oplus \cdots \oplus x_{i^M_k} &= y_M \,. \\
  \end{aligned}
 \right.
\end{equation}

An immediate consequence of this remark is that a very efficient algorithm is available to find if a given \kxorsat\ formula is \sat, which assignments are solutions, and what is their number: the Gauss elimination procedure. One may even wonder why such a problem is interesting at all, given that it is equivalent to linear boolean algebra. The reasons are threefold: first, \kxorsat\ is less easy that it seems. For example, if one determines with the Gauss elimination procedure that a \kxorsat\ instance is not satisfiable, he could be interested in finding an approximate optimal configuration, i.e. an assignments which is guaranteed to satisfy a fraction $1-\epsilon$ of the maximum possible number of clauses, for some given $\epsilon > 0$. Such an approximation algorithm, however, is not known (or rather, no such algorithm is known to work \emph{efficiently}, the meaning of which will become clear in the next Section). Second, many questions regarding the \emph{dynamics} of algorithms that can be applied to both \ksat\ and \kxorsat\ are interesting, difficult to answer for \ksat, more manageable for \kxorsat, and \emph{a priori} should have at least qualitatively similar answers for the two problems. In these cases, \kxorsat\ constitutes an excellent starting point to understand what happens in \ksat. Finally, and foremost from the point of view of physicists, because \kxorsat\ is a legitimate, and very interesting, spin glass model in its own. In fact, the diluted Ising $p$-spin model with couplings $\pm 1$ \emph{is} \kxorsat: defining the energy as the number of violated clauses (as for \ksat) and using the correspondence between boolean variables and Ising spins, we have
\begin{equation}
 E(\sigma) = \sum_{m=1}^M \frac {1 - y_m \, \sigma_{i^m_1} \cdots \sigma_{i^m_k} } 2 \,.
\end{equation}

As in the case of \ksat, the spin glass model is defined for some distribution of disorder, corresponding to an ensemble of possible \kxorsat\ formul\ae\ with a given measure, and we shall consider the thermodynamic limit $N \to \infty$ with some finite $\alpha = M / N$.

\section{Computational complexity}
\label{Sec Complexity}

Introducing \kxorsat, I made the following implicit statement: that since an efficient algorithm for solving it was known, it could possibly be regarded as a less interesting problem than \ksat. Is such a statement reasonable? Not really: whether a problem is ``harder'' than another or not should be an intrinsic property of the problem, if it is meaningful at all, and should not be related to our knowledge (or lack thereof) of algorithms.

The question of what makes a problem intrinsically ``hard'', and how to compare the ``hardness'' of different problems without introducing contingent dependencies (on the techniques and tools actually available to solve them) is the subject of computational complexity theory. It is a branch or rigorous mathematics, and it involves highly abstract (and quite complicated) models of computation. With no pretense in this direction, I shall only aim at giving the ``flavor'' of the most relevant concepts and results. An excellent (rigorous) introduction to the field is provided by the already cited reference \cite{Papadimitriou98}.

\subsection{Algorithms and computational resources}

The first issue to be addressed is how to measure computational complexity. Let us consider that we have some decision problem, and an algorithm which can solve any instance of the problem. In order to compute the solution to the problem, the algorithm will use some \emph{computational resources}. The most important of them is the \emph{time} it will take to complete the computation. Other examples are the memory required to store the intermediate steps of the computation (usually referred to as \emph{space}); some algorithms are probabilistic (we shall discuss them later), and require a supply of \emph{random numbers}; in order to save space, some intermediate results may have to be erased, which has an \emph{energy} cost (the loss of information corresponds to a decrease in entropy). There are several other relevant resources that one can consider. However, I shall consider only time.

In order to eliminate the dependency of the running time on such practical aspects as the hardware used to perform the computation or the actual code used to implement the algorithm, time will be defined as the number of elementary operations (such as arithmetic operations on single digit numbers, or comparisons between bits, \emph{et c\ae tera}) needed to complete the calculation. This will depend on the particular instance of the problem considered, and general results are obtained considering the \emph{worst} possible instance for any given size $n$ of the problem, and then taking the asymptotic behavior for large $n$. For example, if two different algorithms are available to solve the same problem, with times that scale as $t_1 \sim O(n^2)$ and $t_2 \sim O(n^3 \log n)$ respectively, then for large enough $n$ it is sure that algorithm 1 will perform better than algorithm 2, regardless of the details of the dependency of $t$ on $n$, and therefore of the specificities of the implementation.
	
Clearly, the main \emph{theoretical} distinction will be between algorithm that have running times that increase as polynomials of the input size, and algorithms for which $t$ increases as an exponential of the input size. This is easily seen by considering what happens to the ``accessible'' size of the input if the speed at which elementary operations are performed is increased by some constant factor, for different scaling behaviors of $t$ versus $n$. This is done in Table \ref{Tab Complexity}. Notice, however, that \emph{in practice} an algorithm running in time scaling as $10^3 n^3$ will take much longer than one scaling as $2^{10^{-3} n}$ for $n$ up to $\simeq 10^4$. The point is that in the analysis of known algorithms, such ``extreme'' coefficients never occur.

\begin{table}
 \begin{center}
  \begin{tabular}{|c|c|c|c|}
   \hline
   $t$ & $n_\mathrm a(1)$ & $n_\mathrm a(100)$ & $n_\mathrm a(10000)$ \\
   \hline
   $O(n)$      & $n_1$ & $100 \times n_1$   & $10000 \times n_1$  \\ 
   $O(n^2)$    & $n_2$ & $10 \times n_2$    & $100 \times n_2$    \\
   $O(n^3)$    & $n_3$ & $4.6 \times n_3$   & $21.5 \times n_3$   \\
   $O(2^n)$    & $n_4$ & $n_4 + 6.6$        & $n_4 + 13.3$        \\
   $O(2^{2n})$ & $n_5$ & $n_5 + 3.3$        & $n_5 + 6.6$         \\
   \hline
  \end{tabular}
 \end{center}
 \caption{
  Increase of the ``accessible'' problem sizes for different scalings of running time, and for different increases in the computer speed. The first column reports the scaling of $t$ as a function of $n$ for different algorithms; the second column is the size of problems that can be computed in some given maximum time, which is denoted by $n_i$; the third column reports the value of $n_i$ obtained if the computer speed is increased by a factor 100; the last column corresponds to a factor of 10000. Notice that while polynomial algorithms have accessible sizes that increase by a constant \emph{factor}, for exponential algorithms the increase is an \emph{additive} constant. 
 }
 \label{Tab Complexity}
\end{table}

\subsection{Computation models and complexity classes}

The analysis of algorithms provides (constructive) upper bounds on the computational resources required by the algorithm to solve some problem. A more interesting (and challenging) question would be to find some \emph{lower} bound on the resources needed to perform some computation, independently on the algorithm used, which would then be a property of the problem itself. The theory of computational complexity tries to answer this question.

In order to do that, \emph{computation models} are introduced, which define what can (and cannot) be done in a computation. The most celebrated example of computation model is the \emph{Turing machine} \cite{Turing36}, which consists of the following: a \emph{tape}, made of an unlimited number \emph{squares}, each of which can contain a symbol $s$ from some finite alphabet $\Sigma$; a \emph{head} which reads the tape and can perform some action $a$ on it, such as ``write $s$ in this empty square'', ``move right one square'', ``erase this square'', ``halt'' \emph{et caetera}; an internal \emph{state} of the head, which is an element $q_i$ of a finite set $\{q_1,\dots,q_r\}$; finally, a \emph{computation rule}, which associates to any pair $(s,q_i)$ a pair $(a,q_{i'})$, where $s$ is the symbol on the square currently under the head and $q_i$ its internal state, depending on which, $a$ is an action performed by the head and $q_{i'}$ is the new internal state of the head.

The computation begins with some input written on the tape, and proceeds according to the computation rule, until the computation ends (i.e. the head halts). The result of the computation is what is written on the tape at the end. Different computation rules will compute different quantities, i.e. solve different problems.

Notice that \emph{any} decision problem can be expressed in such a way that the instance is a string written in the alphabet $\Sigma$ and the output is \textsc{yes} or \textsc{no}, and therefore can be addressed by a suitable Turing machine. For example, a graph can be represented by a string over the alphabet $\{0,1,2,3,4,5,6,7,8,9,(,-,)\}$ by specifying the number of vertices and then for each edge, the pair of vertices it connects, for example: $5(1-3)(1-5)(2-4)(2-5)(3-5)$. The decision problem is then equivalent to identifying which strings correspond to instances for which the answer is \textsc{yes}, that is to say whether the input string is or not an element of the subset of possible strings for which the answer is \textsc{yes}. Since subsets of possible strings are often called \emph{languages}, decision problems are also referred to as languages, or as set recognition problems.

There are many variants of the Turing machine, such as \emph{binary} machines, working on the alphabet {0,1}; or \emph{multi-tape} machines (which have a finite number of tapes and heads, and for which the computation rule specifies the joint action of all of them); or \emph{universal} machines, for which the computation rule is provided as an input on the tape (which can always be done, since the rule can be represented as a string). For most of them, it can be proved that they are equivalent to a simple Turing machine, with an overhead on running time which is at most polynomial in the input size. Moreover, many other computation models, sometimes drastically different from the Turing machine, have been proved to be equivalent to it. It is a well established belief (but far from provable), going under the name of \emph{Church-Turning thesis}, that any computation which can physically be performed can be represented by a Turing machine.

Another very important variant is the \emph{non-deterministic} Turing machine, which is a Turing machine with a computation rule which is not single valued: the machine is able to ``split'' (creating an identical copy of itself) and perform different actions on different tapes. One can either interpret this as a Turing machine with an infinite number of heads and tapes and which can transfer an infinite amount of information from one tape to another, or as ``the most lucky'' Turing machine, which at each split only executes one of the possible actions prescribed by the computation rule, and such that it leads to the ``best'' answer for the problem. Such a computation model is not feasible in practice, but we shall see that is very important from the theoretical point of view. In the following, by \emph{polynomial time} I shall always mean on a deterministic Turing machine, unless differently specified.

Since the Turing machine is such a general paradigm for computations, it can be used to define \emph{complexity classes}, i.e. classes of problems that have similar complexity. There are many different complexity classes that are relevant, but we shall focus on two of them:
\begin{description}
 \item[Deterministic Polynomial Time (P)] The class P is defined as the class of all decision problems that can be solved in polynomial time by a deterministic (i.e. ``normal'') Turing machine.
 \item[Non-deterministic Polynomial Time (NP)] The class NP is defined as the class of all decision problems that can be solved in polynomial time by a non-deterministic Turing machine.
\end{description}

Some comments are in order. First, notice that these class definitions do not refer to any specific algorithm: it is the fact that it is \emph{possible} to solve them under certain conditions which matters, not that we are able to do it. Notably, no polynomial time algorithm is known for \emph{any} NP problem, so the possibility to solve them in polynomial time on non-deterministic Turing machines is a mere definition. 

However, and this is the second point, it is a very \emph{meaningful} definition: for most problems, it is clear whether a problem is in P, in NP, or in none of the two. For example, for \ksat\ an obvious algorithm is polynomial on a non-deterministic Turing machine: proceed in steps, and assign a variable at each step, splitting between the assignments \true\ and \false, then simplify the formula, and verify that there are no contradictions (i.e. clauses which cannot be satisfied); if this happens, halt the corresponding head; if some head achieves to assign all the variables, then it has find a satisfying assignment and the answer is \sat; on the contrary, if all the heads halt before they have assigned all the variables, there is no satisfying assignment and the answer is \unsat. This procedure is obviously polynomial, so \ksat\ is in NP. On the other hand, we have seen that the Gauss elimination procedure is polynomial (on a normal computer, and therefore on a Turing machine as well), and so \kxorsat\ is in P.

Third, notice that any problem which is in P is also, \emph{a fortiori}, in NP. In fact, the question of whether P and NP are \emph{equal} (i.e. if there exist polynomial time algorithms to solve any NP problem) is one of the central open problems in complexity theory. It is strongly believed that the answer is no, but no proof (or disproof) of this is known.

Fourth, an equivalent, and more ``practical'' definition of NP is the following:
NP is the class of all problems for which it is possible to issue a \emph{certificate} in (deterministic) polynomial time. A certificate is the the answer \textsc{yes} or \textsc{no} for a specific configuration, provided as input together with the instance of the problem. In other words, NP problems are such that a candidate solution can be verified in polynomial time.
Again, it is obvious that \ksat\ is in NP, and that any problem in P is also in NP. The equivalence of the two definitions is easy to verify: if a certificate to a problem can be issued in polynomial time, a non-deterministic Turing machine can test in parallel all the possible configurations and find if some of them has answer \textsc{yes}. On the other hand, if a non-deterministic Turing machine can solve in polynomial time a problem, it can also check if any of the configurations for which the answer is \textsc{yes} coincides with the configuration submitted for the certificate.

Finally, notice that these definitions, given for decision problems, actually extend to search and optimization problems, so that if a decision problem belongs to NP (or P), then all of them are in the same class. For example, the optimization problem of \ksat\ consists in finding the smallest value of $E$ such that the decision problem ``An assignment which satisfies $M-E$ clauses exists'' gives answer \textsc{yes}. One can solve in (non-deterministic) polynomial time for $E=0$, then for $E=1$ and so on, and find in (non-deterministic) polynomial time the smallest $E$. However, the complexity classes of \emph{enumeration} problems are often different.

\subsection{Reductions, hardness and completeness}

A \emph{reduction} is a polynomial time algorithm which maps an instance of some decision problem into an instance of some other decision problem, such that the two instances always have the same answer. More formally, let us consider two decision problems $A$ and $B$. Recall that $A$ (and also $B$) can be viewed as the subset of the strings over the alphabet $\{0,1\}$ which describe the instances of the problem that give answer \textsc{yes}. Then, we can write $x \in A$ to mean that the string $x$ represents an instance of problem $A$ for which the answer is \textsc{yes}, denote by $|x|$ the length of the string $x$, and define functions that associate a string to another string, i.e. $f: \, \{0,1\}^* \rightarrow \{0,1\}^*$ (the superscript $*$ denotes the set of all the possible strings in the alphabet). A formal definition of reduction is the following:
\begin{description}
 \item[Reduction] A decision problem $A$ \emph{reduces} to the decision problem $B$, denoted by $A \leq_p B$, if there exists a function $f: \, \{0,1\}^* \rightarrow \{0,1\}^*$, computable in polynomial time $p(|x|)$, such that $x \in A \Leftrightarrow f(x) \in B$.
\end{description}
Notice that since the function is computable in time bounded by $p(|x|)$, we must have
\begin{equation}
 |f(x)| \leq p(|x|) \,.
\end{equation}

The concept of reduction is very powerful, since it permits to relate the complexity of different problems. In particular, one can define problems that are ``at least as difficult'' as \emph{any} problem in some class:
\begin{description}
 \item[Hardness] A decision problem $A$ is C-\emph{hard} for some computational complexity class C if for any problem $B \in \text C$, $B \leq A$.
 \item[Completeness] A decision problem $A$ is C-\emph{complete} for some computational complexity class C if $A \in \text C$ and for any problem $B \in \text C$, $B \leq A$.
\end{description}
Loosely speaking, C-complete problems are the most difficult problems to solve in class C, and if an efficient (i.e. polynomial) algorithm is found for a C-complete algorithm, it can solve efficiently any problem in C (for which a reduction is known).

The importance of \ksat\ in complexity theory is due to the following
\begin{description}
 \item[Cook-Levin Theorem] \ksat[3] is NP-complete \cite{Cook71, Levin73}.
\end{description}
This was the first result on NP-completeness, introduced the concept, and proved that several other problems, to which \sat\ can be reduced, where also NP-complete.

The proof of the Cook-Levin theorem is surprisingly simple, and emphasizes the (conceptual) importance of the non-deterministic Turing machine: it is simply a mapping of the time evolution of the Turing machine into a \sat\ formula, in which the interpretation of boolean variables is ``The cell $i$ contains the symbol $j$ at time $k$ in the computation'', or ``The head is over cell $i$ at step $k$ in the computation'', and ``The head is in state $q_i$ at the step $k$ of the computation'' (where $i,j,k$ act as variable indices). The proof shows how to form a legitimate \sat\ formula for any given non-deterministic Turing machine, and then that any \sat\ formula can be reduced to \ksat[3].

\ksat\ proves a very powerful tool for reductions, because of its generality and simple structure. The following problems are easily proven NP-complete, by reducing \ksat\ to them: Integer Linear Programming, Hamiltonian Cycle, Traveling Salesman, Max Clique, Max Cut, Vertex Covering, 3-Coloring,~\dots\ . The list is very, very long.

The fact that so many important problems are in NP, and that no efficient algorithms are known (and probably exist) to solve them, seems very discouraging in view of applications. However, this need not be the case, as I shall point out in the following Paragraph.

\subsection{Other measures of complexity}

The complexity classes P and NP are defined in terms of the asymptotic behavior of the running time for the \emph{worst} possible instance of any size $n$. In many practical problems, one can be satisfied if some much less stringent requirements are met:
\begin{itemize}
 \item If the \emph{typical} running time over some distribution of instances is polynomial.
 \item If an approximate optimal solution can be found in polynomial time for any approximation factor $\epsilon$.
\end{itemize}
Average-case complexity theory studies the first question; the theory of complexity of approximation studies the second.

Many average-case complexity results analyze the average time that some given algorithm takes to solve an instance of a problem, for a given distribution of instances. It is often the case that a NP problem is solved in polynomial time \emph{on average} over some ``natural'' distribution of instances. For example, for problems defined on graphs, one can form the uniform distribution over all graphs containing $n$ vertices and with some average connectivity. Then, one can prove that 3-\textsc{col} can be solved in \emph{linear} time on average. Often, however, all the algorithms known for some NP problem take exponential time on average. Alternatively, one can study the probability with which an algorithm finds an answer in polynomial time.

A crucial point in average-case complexity theory is the choice of the distribution. For example, the best known algorithm for Subset Sum take exponential time if the $n$ numbers in the set are taken uniformly in the range $[1,2^n]$. However, if this range is extended to $[1,2^{n \log^2 n}]$,  the average time for the best algorithm becomes polynomial. Even in cases when the dependency on the distribution is less dramatic, it remains a crucial point. For example, the reductions that map many NP problems on \ksat\ introduce a very peculiar structure in the \ksat\ formul\ae\ they generate, so that even though the distribution of the instances of the original problem is a natural one, the distribution of \ksat\ formul\ae\ that are obtained is almost \emph{never} a natural one. Thus, even though \ksat\ can be solved efficiently on average in many cases under natural distributions, these results do \emph{not} extend to the problems that can be reduced to \ksat. On the other hand, even when it is possible to characterize the distribution of \ksat\ formul\ae\ generated by some reduction, it is usually either impossible to find an algorithm that is efficient on average on them, or even to perform the analysis of the average case. This poses a severe limitation to the applicability of average-case complexity results.

On the other hand, complexity of approximation results are very interesting in view of applications. They are, however, usually more technical than the results I have discussed, and beyond the level of this introduction. I shall only cite the Probabilistically Checkable Proof (PCP) theorem and its consequences on the approximability of \textsc{max}-\ksat[3] \cite{Arora92}, which is the optimization problem of 3-CNF formul\ae.

One of the two equivalent definitions of the class NP requires that NP problems can be certified in polynomial time. The following definition extends the same concept:
\begin{description}
 \item[Probabilistically Checkable Proof (PCP)] Given two functions $r,q: \mathbb N \rightarrow \mathbb N$, a problem $L$ belongs to the class $\text{PCP}(r,q)$ if there is a polynomial time probabilistic function (called \emph{verifier}) $V: \, \{0,1\}^* \rightarrow \{0,1\}$ which, given as an input: a string $x$; a string $\pi$ (called \emph{proof}); a sequence of $r(|x|)$ random bits; and which uses a substring of $\pi$, of size $q(|x|)$ and chosen at random, to compute $V^\pi(x)$, and is such that
 \begin{equation}
  \left\{
   \begin{aligned}
    x \in L &\Rightarrow \exists \pi : \ \mathbb P[V^\pi(x) = 1] = 1 \,; \\
  x \notin L &\Rightarrow \forall \pi : \ \mathbb P[V^\pi(x) = 1] \leq \frac 1 2 \,.
   \end{aligned}
  \right.
 \label{PCP}
 \end{equation}
\end{description}
In this definition, the proof $\pi$ is the analogous of the candidate configuration in NP: it is some string which is provided as an input, and which, if well chosen, can prove that $x \in L$ (i.e. that the answer to the instance represented by $x$ of the decision problem $L$ is \textsc{yes}). The verifier $V(x)$ is the analogous of the algorithm which issues the certificate, i.e. it gives, in polynomial time, an answer which is \textsc{yes} or \textsc{no} and which is related to the answer to the instance represented by $x$. However, $V(x)$ is probabilistic, that is to say, it is a random variable. The source of the randomness is provided by the $r(|x|)$ random bits used to compute $V(x)$. For the problems in PCP$(r,q)$, the distribution of the values of $V(x)$ verifies the condition (\ref{PCP}). Finally, notice that only a number $q(|x|)$ of symbols in $\pi$ is actually used in the computation of $V^\pi(x)$, and these symbols are chosen at random.

At first sight, the class PCP seems very unnatural, and of little interest. The following theorem proves this impression very much wrong:
\begin{description}
 \item[PCP Theorem] NP = PCP$\bigl(O(\log n),O(1)\bigr)$ \,.
\end{description}

Again, several remarks. First, notice that any mathematical statement can represented by a string, and that any mathematical proof can be represented by another string. Mathematical statements can be divided in two: right ones (i.e. theorems), and wrong ones. One can consider the following decision problem, called \textsc{theorem}: given a mathematical statement, is it a theorem? It is clear enough that it is possible to verify if a proof provided to support a statement is correct or not in a time which is polynomial in the length of the proof. Therefore, \textsc{theorem} is in NP.

What this theorem states is that any theorem represented by a string $x$ can be recognized by looking at a \emph{finite} number of randomly chosen bits of some suitable proof, represented by a string $\pi$, and evaluating some polynomial time function $V$. Then, if $V^\pi(x) = 0$ the statement is not a theorem with probability 1, while if $V^\pi(x) = 1$ it may or may not be. Conversely, if the statement $x$ \emph{is} a theorem, then there must be some proof $\pi$ such that $V^\pi(x) = 1$ with probability 1, and if the statement is \emph{not} a theorem, the probability that $V^\pi(x) = 1$ is less than or equal to 1/2 for \emph{any} proof string $\pi$.
One can therefore check if the proof of any theorem (of any length) is correct just by looking at a \emph{finite} number of bits in the proof, provided it is put in a suitable form, and obtain a probabilistic result which is correct with probability 1 if the answer is \textsc{no}, and correct with probability $p$ if the answer is \textsc{yes}, for any $p<1$.

Second, the same reasoning applies to \emph{any} NP decision problem, not just \textsc{theorem}. For example, if, instead of providing a candidate solution to check if an instance of \ksat\ is satisfiable, one provided a PCP proof $\pi$, then it would be possible to check it in \emph{constant} time, rather than polynomial, obtaining a probabilistic result.

Third, even though the PCP theorem is very surprising in itself, the following corollary is also remarkable:
\begin{description}
 \item[Hardness of approximation of MAX-3-SAT] The PCP Theorem implies that there exists $\epsilon > 0$ such that $(1-\epsilon)$-approximation of \textsc{max}-\ksat[3] is NP-hard.
\end{description}
In other words, it is at least as difficult to find an approximation to the optimal assignment as it is to find the optimal assignment itself (if the approximation has to be good enough).

The theory of complexity of approximation is very rich and well established. However, I shall not discuss it any further.

\subsection{Connections to the work presented in Part II}

In Chapter \ref{Chap_xorsat}, I shall present some results about what a certain class of algorithms can and cannot do on average for \kxorsat, and also for an extension of \kxorsat\ which is NP-complete.

The motivation for the work in presented in Chapter \ref{Chap_ksat} is a recent result which establishes a relation between the \emph{average-case} complexity for \ksat[3] on the uniform distribution, and  the \emph{worst-case} complexity of approximation for several problems. The results I shall present provide an indication that some hypothesis, on which the previous relation is based, might be wrong.

\section{Search algorithms}
\label{Sec Algorithms}

In the previous Section, I have introduced the concept of computational complexity, which measures how difficult it is to solve a problem. In this Section, I shall introduce several algorithms that attempt to do it in practice for the search problems associated to \ksat\ and \kxorsat, that is to say algorithms which try to find satisfying assignments for a given formula.

There is a huge variety of approaches and ``strategies'' to solve combinatorial optimization problems, and notably \ksat. It is important to notice that, due to their formal similarity, the vast majority of the algorithms that can solve \ksat\ can also solve \kxorsat\ and \emph{vice versa}, although with different performances (sometimes dramatically). I shall therefore discuss the two problems jointly, specifying the cases in which there are notable differences.

This introduction will be far from exhaustive: I shall focus on those algorithms of interest in view of the discussion of Part II. They can be divided in broadly
two \emph{families}:
\begin{description}
 \item[Random-walks] In random walks, all the variables are assigned at the first step of execution, typically at random, or following some more refined rule. In the following steps, single variables or groups of variables are selected and ``flipped'' (i.e. their value is changed), according to some stochastic rule which depends on the configuration. The algorithm stops when a solution is found, or when an upper bound to the number of steps has been reached. An algorithm in this family is specified by the rule according to which variables are flipped.
 \item[DPLL Procedure] In the DPLL procedure, variables are assigned \emph{sequentially}: at each step, a variable is selected according to some \emph{heuristic} rule, and its value set according to some \emph{strategy}. Once a variable is assigned, the formula is simplified by replacing it with its value. Under this process, the formula therefore evolves into a shorter and mixed one (i.e. including clauses of different lengths). Two events are especially important in the DPLL procedure: the generation of \emph{unit clauses} and of \emph{contradictions}. An algorithm in the DPLL family is specified by these four characteristics: the heuristic, the strategy, the action taken in the presence of unit clauses, and that in presence of contradictions.
\end{description}

The rest of this Section is organized in
two Paragraphs, one for each family of algorithms. In each case, I shall consider the average case performance over the uniform distribution of instances, for either \ksat\ or \kxorsat.

\subsection{Random-walk algorithms}

The most familiar random-walk algorithm for physicists is the Metropolis Monte-Carlo procedure, which is capable of sampling configuration with probability equal to their Gibbs weight. In particular, the zero temperature version of the Metropolis algorithm consists in picking at each step a variable at random and flipping it if this decreases the number of violated clauses, and is a very simple example of ``greedy'' algorithm, i.e. an algorithm which tries to perform a local optimization of the configuration at every move.

Based on the qualitative arguments about frustration presented in Paragraph \ref{Par Frustration}, such a \emph{local optimization} procedure is bound to fail in disordered systems. The following arguments shows that this is the case with probability 1 for uniformly drawn random instances of \kxorsat[3]. Consider the subformula represented in Figure \ref{Fig Blocked island}, which I shall call a ``blocked island''. It is clear that if such a subformula is present in the formula, and if it is found in a configuration such as one of those depicted in the figure, a greedy algorithm will not be able to reach a satisfying assignment. In \cite{Cocco04} it is shown that in the limit $N \to \infty$ this situation occurs with finite probability
\begin{equation}
 p = \frac{729}{1024} \alpha ^ 7 e^{-45 \alpha}
\end{equation}
where $\alpha$ is the clause to variable ratio, $\alpha \equiv M/N$. The average number of blocked islands in a random \kxorsat[3] formula is $p N = O(N)$, and it is a lower bound to the minimum number of violated clauses of configurations that greedy algorithms are able to find.

\begin{figure}
 \includegraphics[width=\textwidth]{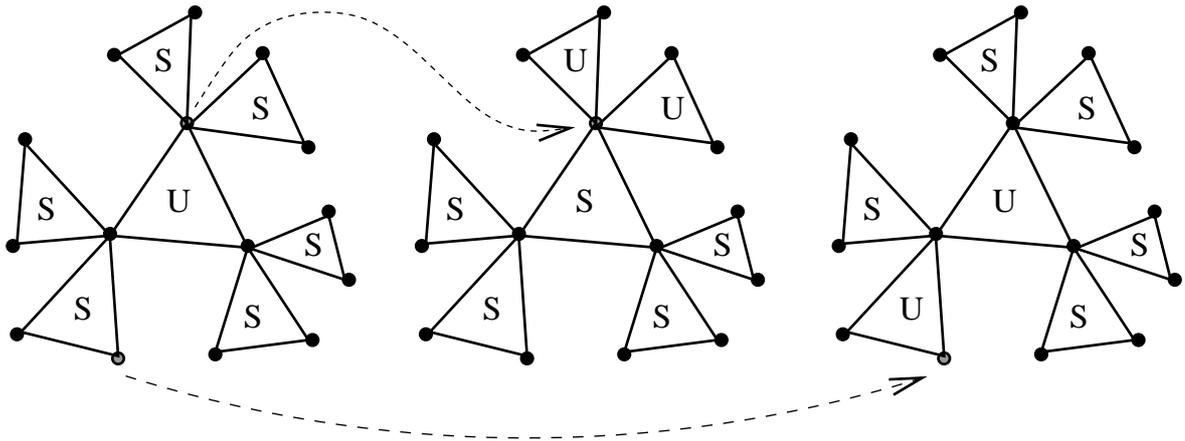}
 \caption{Representation of a ``blocked island''. Each dot in the diagram corresponds to a variable, and triangles represent 3-clauses containing the variables at the vertices. The left-most diagram shows an isolated subformula; the variables in the subformula are all assigned, in such a way that the clauses marked with the letter S are satisfied, those with the letter U are unsatisfied. If one of the variables appearing in the central clause is flipped, the second configuration is obtained; if one of the variables which do not appear in the central clause is flipped, the third configuration is obtained. In both cases, the number of unsatisfied clauses increases by 1 (From \cite{Cocco04}).}
 \label{Fig Blocked island}
\end{figure}

More interesting are ``less greedy'' algorithms. A simple example is provided by Pure Random Walk Sat (PR\walksat), which was introduced in \cite{Papadimitriou91}, and is defined as follows: initially, assign all the variables uniformly at random; then, at each step pick uniformly at random a clause among those that are violated, and a variable among those appearing in it, and flip it; repeat, until a satisfying assignment is found, or a number $T_\mathrm{max}$ of steps has been performed. 
Notice that by flipping a variable which appears in a violated clause, that clause becomes satisfied; however, if that variable also appear in other clauses that were satisfied before the flip, they might become unsatisfied after. This is why this algorithm is ``less'' greedy.
The possible outcomes of the algorithm are two: either a satisfying assignment is produced, or the output is \textsc{undetermined}. 

In \cite{Papadimitriou91}, it was shown that PR\walksat\ finds a solution with probability 1 for \emph{any} satisfiable instance of \ksat[2] in a number of steps (i.e. time) of order $O(N^2)$. An interesting  extension of this result to \ksat[3] was obtained in \cite{Schoening02}, where it is shown that if $T_\mathrm{max} = 3N$ and the procedure is repeated for a number $R$ of times without obtaining a satisfying assignment, then the probability that the instance is \sat\ is upper-bounded by $\exp[-R \, (3/4)^N]$. By taking $R$ sufficiently larger than $(4/3)^N$, the probability that an instance for which no satisfying assignment has been found is nonetheless satisfiable can be made arbitrarily small. Also, notice that, even though the running time of such a procedure (for any fixed probability bound) is exponential, it is still exponentially smaller than $2^N$, which would be required by exhaustive search.

The previous results hold for \emph{any} instance, and the probabilities mentioned are over the choices of the algorithm. Another interesting question is to analyze the average-case behavior over the uniform distribution of \ksat\ instances. This was done in \cite{Alekhnovich03, Semerjian03, Barthel03}. In the first of these papers, a rigorous bound is found for the values of the clause-to-variable ratio $\alpha \equiv M/N$ for which PR\walksat\ finds a solution in polynomial time with probability 1: $\alpha < \alpha_\mathrm{PRWalkSAT} \simeq 1.63$ (for $k=3$). This is the first example I mention of an algorithmic bound on $\alpha$, i.e. a threshold value separating two different behaviors of the same algorithm. Many more will follow. Also, notice that since with probability 1 PR\walksat\ finds a solution for random \ksat[3] formul\ae\ with $\alpha < \alpha_\mathrm{PRWalkSAT}$, this implies that these formul\ae\ are satisfiable with probability 1. 

In \cite{Semerjian03, Barthel03} the same problem was studied with ``physical'' methods. In particular, a numerical study indicates that random instances are solvable with probability 1 in polynomial time if $\alpha \lesssim 2.7$, while for larger values the time becomes exponential. The analysis of the master equation performed in \cite{Semerjian03} shows that the average fraction of unsatisfied clauses, $\varphi(t)$, after $tN$ steps of the algorithm, is a deterministic function which depends on $\alpha$ and goes to 0 in finite $t$ if $\alpha \lesssim 2.7$, while for larger $\alpha$ it tends asymptotically to a finite value, which is 0 for $\alpha \simeq 2.7$ and then increases. In this second regime, it can happen that solutions are found because of \emph{fluctuations}, but the time which this requires is exponential in $N$.

A somewhat more complicated variant of this algorithm goes under the name of \walksat, and is defined as follows:

\algblockx[MIf]{mif}{mend}[1]{\textbf{if} #1 \textbf{then}}{\textbf{end if}}
\algcblockx[MElse]{MIf}{melse}{mend}[1]{\textbf{else} #1\textbf{:}}{\textbf{end if}}
\algcblockx[MElse]{MElse}{melse}{mend}[1]{\textbf{else} #1\textbf{:}}{\textbf{end if}}

\begin{algorithmic}
 \Procedure{\textnormal{Walk}sat}{$p,T_\mathrm{max}$}
 \State{Assign uniformly at random each variable}
 \Repeat
  \State{Select uniformly at random a clause $\mathcal C$ which is \unsat}
  \State{For each variable $x_i$ in $\mathcal C$, compute the \emph{break-count} $b(x_i)$, defined as the number of clauses currently satisfied that will be violated if $x_i$ is flipped}
  \mif{A variable $x_j$ in $\mathcal C$ has break-count $b(x_j)=0$}
   \State{Flip $x_j$}
  \melse{With probability $p$ }
    \State{Select the variable in $\mathcal C$ with the lowest break-count (or select uniformly at random one of the variables with the lowest break-count, if there are more than one), and flip it}
  \melse{With probability $1-p$ }
    \State{Select uniformly at random a variable in $\mathcal C$ and flip it}
  \mend
 \Until{There are no \unsat\ clauses, or the number of steps exceeds $T_\mathrm{max}$}
 \If{A solution $X$ has been found}
   \Return{$X$}
 \Else{ }
   \Return{\textsc{undecided}}
 \EndIf
\EndProcedure
\end{algorithmic}

As in the case of PR\walksat, variables to be flipped are selected only in clauses that are currently \unsat. However, instead of picking a variable at random, \walksat\ looks for a variable which can be flipped \emph{without making any clause \unsat\ which is currently \sat}. Notice that in doing this the total number of \unsat\ clauses must decrease of at least 1 (i.e. the selected clause becoming \sat). On the other hand, if some clauses currently \sat\ have to become \unsat, the variable which minimizes their number is selected, with probability $p$, or otherwise any variable in the clause uniformly at random. In both of these cases, the total number of \unsat\ clauses can increase.

The average case performance of \walksat\ is astonishingly good. Numerical experiments suggest that its typical running time (e.g. the median over a series of runs) remains linear for $\alpha$ up to $4.15$ (for $k=3$) \cite{Aurell05}. Interestingly, this value coincides with the threshold for the stability of the 1RSB solution \cite{Montanari04}.

For larger values of $\alpha$, the behavior of \walksat\ becomes more complicated. The average running time becomes exponential, with a peculiar structure in the average fraction of unsatisfied clauses as a function of the number of steps (divided by $N$). A detailed analysis of this behavior is the object of current work in collaboration with Giorgio Parisi.

\subsection{DPLL algorithms}

The DPLL procedure is a firmly established complete algorithm for \ksat\ and similar constraint satisfaction problems. For concreteness, and for future reference in Chapter \ref{Chap_xorsat}, I shall consider the case of \kxorsat. DPLL was introduced by Davis and Putnamm in 1960 \cite{Davis60} and developed by Davis, Logemann and Loveland in 1962 \cite{Davis62}, and has many variants. 

The basic principle is to assign the variables in sequential order, and simplify the formula after each assignment. This generates a sub-formula in which clauses that are satisfied are eliminated, and clauses in which the assigned variable appears decrease in length of one unit. If a unit clause is generated (i.e. a clause of length 1), this clause determines the value of the variable appearing in it, and it is assigned accordingly. This event is called \emph{Unit Propagation} (UP). The rule according to which the variable to be assigned is selected is called \emph{heuristic}. Most often, the value assigned is selected uniformly at random, but sometimes a rule, called \emph{strategy}, determines it. The simplest example of heuristic consists in selecting the variable uniformly at random among those not yet assigned, as well as the value, but giving priority to UP; it is called Unit Clause (UC).

A crucial distinction between DPLL variants is the action taken if a \emph{contradiction} arises, i.e. in the case of \kxorsat, a pair of unit clauses for the same variable with conflicting assignments. If this occurs, no value of the variable in question will satisfy the subformula, and therefore the original one. This event signals that some of the previous assignments were wrong. Two possible actions can then be taken: either modify some of the previous assignments, or output \textsc{undetermined} and possibly restart the procedure. In the first case, the algorithm \emph{backtracks} to the last variable which was set by a ``free'' step (as opposed to a UP or a backtrack), and inverts it. In the second case, the algorithm is no longer complete, but we shall see that it can still be interesting in the average case.

Formally, we can describe the DPLL procedure with and without backtracking with the two following procedures, in which $\mathcal F$ is the formula and $H$ is the heuristic, i.e. a function which associates an index of a variable not yet assigned to a subformula. With no backtracking, 

\algloopdefx[SimpleIf]{SimpleIf}[1]{\textbf{if }#1\textbf{ then }}
\algloopdefx[SimpleFor]{SimpleFor}[1]{\textbf{for }#1\textbf{ do }}

\begin{algorithmic}
 \Procedure{DPLL without backtracking}{$\mathcal F, H$}
 \Repeat
   \SimpleFor{every unit clause $\mathcal U$ in $\mathcal F$}
     \State{$\mathcal F \leftarrow \mathrm{Simplify}(\mathcal F,\mathcal U)$}
   \State{$i \leftarrow H(\mathcal F)$}
   \State{$\mathcal F \leftarrow \mathrm{Simplify}[\mathcal F, x_i = S(\mathcal F)]$}
   \SimpleIf{a contradiction is present}
     \State{\Return{\textsc{undetermined}}}
 \Until{all the variables are assigned}
 \State{\Return{\true}}
 \EndProcedure
\end{algorithmic}
where $S(\mathcal F)$ is the strategy according to which values for assignments are decided.
With backtracking the procedure is somewhat more complicated, and it is more conveniently expressed in a \emph{recursive} form:
\begin{algorithmic}
 \Procedure{DPLL with backtracking}{$\mathcal F, H$}
 \SimpleIf{all the the clauses are satisfied}
   \State{\Return{\true}}
 \SimpleIf{a contradiction is present}
   \State{\Return{\false}}
 \SimpleFor{every unit clause $\mathcal U$ in $\mathcal F$}
   \State{$\mathcal F \leftarrow \mathrm{Simplify}(\mathcal F,\mathcal U)$}
 \State{$i \leftarrow H(\mathcal F)$}
 \State{\Return{$\mathrm{DPLL}[\mathrm{Simplify}(\mathcal F, x_i=\true), H] \ \bigvee \ \mathrm{DPLL}[\mathrm{Simplify}(\mathcal F, x_i=\false), H]$} }
  \EndProcedure
\end{algorithmic}
The complete variant of DPLL (i.e. the one with backtracking) has been extensively studied (see for example \cite{Cocco01, Cocco04a} and references therein). In the following, I shall concentrate on DPLL without backtracking.

Many different heuristics for DPLL have been studied, in view of both theoretical studies and applications. In the following, an important role will be played by the Generalized Unit Clause (GUC), introduced and studied in \cite{Chao86, Chao90, Frieze96}, which is defined as follows: at each step, select uniformly at random a clause among those of \emph{shortest length}, and then uniformly at random a variable in it. This generalizes the UP rule to clauses of length larger than unit, hence the name. 

The analysis of the average case behavior of DPLL heuristics can be simplified considerably using the following approach, introduced in \cite{Achlioptas01}. Consider the state of the formula after $T$ variables have been set. It will contain a number $C_j(T)$ of clauses of length $j=1,2,\dots,k$ (for some values of $T$, some unit clauses will not have been removed yet, hence the term $j=1$). The formula can be described as a table in which each row represents a clause, and each ``slot'' in it represents a variable. Initially, there are $M$ rows, each of length $k$, which then become shorter as the algorithm proceeds. If the heuristics we consider consist in the selection of either a variable uniformly at random, or of a slot in the table according to some rule which does not depend on the content of the slots, then the subformul\ae\ that are generated are uniformly random conditioned on their lengths. This is the case of both UC (which always selects the variable uniformly at random) and of GUC (which selects, uniformly at random, first a row in the table among those of shortest length, and then a slot in the row).

In the case of UC, at each step a variable is selected uniformly at random. Because of the statistical independence of the subformul\ae, each slot has a probability $1/(N-T)$ of containing the selected variable, and a clause of length $j$ has probability $j/(N-T)$ of containing it. Since the clauses of length $j$ that contain the selected variable become of length $j-1$, the \emph{average} variation in the number of clauses is
\begin{equation}
 \mathbb E[C_j(T+1) - C_j(T) | \{C_j(T)\}] = \frac {(j+1) C_{j+1}(T) - j C_j(T)} {N-T} 
 \label{Delta C_j UC}
\end{equation}
where, for notational simplicity, we set $C_{k+1}(T) \equiv 0$.
Notice that this is the same equation one obtains for steps in which UP is applied, when instead of selecting the variable uniformly at random it is selected among those appearing in unit clauses.

A theorem by Wormald \cite{Wormald95}, the statement of which is rather technical and I shall omit, ensures that (under some very general assumptions which are satisfied by all the heuristics we shall consider) the clause \emph{densities} are concentrated in the thermodynamic limit,
\begin{equation}
 \mathbb E[C_j(T)] = N c_j(T/N)
\end{equation}
where $c_j(t)$ is a function determined by the differential equation obtained dividing (\ref{Delta C_j UC}) by $\Delta T = 1$.
\begin{equation}
 \frac {d c_j(t)} {dt} = \lim_{N \to \infty} \frac {\mathbb 	E[\Delta C_j(T) | \{C_j(T)\}]} {\Delta T} = \frac {(j+1) c_{j+1}(t) - j c_j(t)}{1-t} \hspace{.5cm} (j=2,\dots,k)\,.
 \label{Differential UC}
\end{equation}

Since the initial formula contains $M = \alpha N$ clauses of length $k$, the initial condition for this system of equations is $c_j(0) = \delta_{j,k} \alpha$. Notice that, if at any time, $c_1(t) > 0$, i.e. the formula contains an extensive number of unit clauses, each of them has a probability of order $1/N$ of containing any given variable, so that there is a finite probability that two unit clauses will contain the same variable. If this happens, a contradiction is generated with finite probability at each step of the algorithm, so that over a finite interval of time $\Delta t$ this will happen with probability 1. Therefore, if at any time during the evolution of the formula $c_1(t)$ becomes positive, the algorithm will generate a contradiction and will stop. This is the reason why the range of values of $j$ starts with 2.
Since the rate at which unit clauses are generated is
\begin{equation}
 \frac {2 c_2(t)}{1-t}
\end{equation}
and the rate at which they are removed is at most 1 (because one variable is set at each time step, and therefore at most one unit clause is removed), the condition for the onset of contradictions is
\begin{equation}
 \frac {2 c_2(t)}{1-t} = 1 \,.
 \label{Condition for contra}
\end{equation}

The system of equations (\ref{Differential UC}) is easily solved:
\begin{equation}
 c_j(t) = \alpha \binom k j (1-t)^j t^{k-j} \hspace{.5cm} (j=2,\dots,k) \,.
\label{Sol UC}
\end{equation}
The algorithm will provide a solution with probability 1 if all the variables are set without generating contradictions, i.e. if $2 c_2(t) / (1-t) < 1$ for all $t \in [0,1]$. For $c_2(t)$ given by (\ref{Sol UC}), this function reaches a maximum for $t = t^* \equiv (k-2)/(k-1)$, in which its value is
\begin{equation}
 \max_{t \in [0,1]} \frac {2 c_2(t)}{1-t} = \alpha k \left( \frac{k-2}{k-1} \right)^{k-2}
\end{equation}
which is equal to 1 if
\begin{equation}
 \alpha = \alpha^\mathrm{UC}_\mathrm h \equiv \frac 1 k \left( \frac{k-1}{k-2} \right)^{k-2} \,.
\end{equation}
Notice that this implies that for $\alpha \leq \alpha^\mathrm{UC}_\mathrm h$, random \kxorsat\ formul\ae\ from the uniform distribution are satisfiable with probability 1, and UC is capable in finding a satisfactory assignment in linear time with probability 1.

A similar analysis can be performed for GUC. Initially, the formula contains $M$ clauses of length $k$. As variables are set, some clauses become shorter: let us suppose that after $T$ steps the number of clauses of length $j$ is $C_j(T)$ for $j=j^*,\dots,k$ with $j^* > 1$, and 0 for $j < j^*$, and let us consider what happens starting from there. When a variable is set by GUC, it is selected among the shortest clauses, i.e. those of length $j^*$. A clause of length $j^*-1$ is generated, and the other numbers of clauses vary only if the same variable appears in other equations. That is to say, after the first variable has been set the average variations in $C_j(T)$ are:
\begin{eqnarray}
 \left< \Delta^{(1)} C_j(T) \right> &\equiv& \mathbb E[C_j(T+1) - C_j(T) | \{C_j(T)\}] \nonumber \\
 &=& \frac {(j+1) C_{j+1}(T) - j C_j(T)} {N-T} \hspace{.5cm} (j = j^*+1, \dots, k) \,, \\
 \left< \Delta^{(1)} C_{j^*}(T) \right> &\equiv& \mathbb E[C_{j^*}(T+1) - C_{j^*}(T) | \{C_j(T)\}]
 \nonumber \\
 &=& -1 + \frac {(j^*+1) C_{j^*+1}(T) - j^* C_{j^*}(T)} {N-T} \,, \\
 \left< \Delta^{(1)} C_{j^*-1}(T) \right> &\equiv& \mathbb E[C_{j^*-1}(T+1)| \{C_j(T)\}]
 \nonumber \\
 &=& 1 + \frac {j^* C_{j^*}(T)} {N-T} \,.
\end{eqnarray}
where the superscript $(n)$ indicates that $n$ variables have been set (here, $n=1$).

Notice that the average number of clauses of length $j^*-1$ is now of order $O(1)$, and not smaller than 1. GUC will then select a clause from one of the clauses of length $j^*-1$, giving:
\begin{eqnarray}
 \left< \Delta^{(2)} C_j(T) \right> 
 &=& 2 \frac {(j+1) C_{j+1}(T) - j C_j(T)} {N-T} + O(N^{-1}) \hspace{.5cm} (j = j^*+1, \dots, k)\,, \\
 \left< \Delta^{(2)} C_{j^*}(T) \right> 
 &=& -1 + 2 \frac {(j^*+1) C_{j^*+1}(T) - j^* C_{j^*}(T)} {N-T}  + O(N^{-1})\,, \\
 \left< \Delta^{(2)} C_{j^*-1}(T) \right> 
 &=& 2 \frac {j^* C_{j^*}(T)} {N-T} + O(N^{-1})\,, \\
 \left< \Delta^{(2)} C_{j^*-2}(T) \right> 
 &=& 1  + O(N^{-1}) \,.
\end{eqnarray}
In this equations, the terms $O(N^{-1})$ come from the fact that we are considering the initial $T$ for evaluating the functions, which results in a variation of $O(1)$ in the values of the $C_j$. Notice that UP do not contribute to values of $j$ that are smaller than $j^*-1$, because the number of clauses of such lengths are not extensive.

It will then take (on average) $j^*-1$ steps (after the first one) to ``empty'' one of the clauses of length $j^*-1$ that have been generated: 
\begin{eqnarray}
 \left< \Delta^{(j^*)} C_j(T) \right> 
 &=& j^* \frac {(j+1) C_{j+1}(T) - j C_j(T)} {N-T} + O(N^{-1}) \hspace{.5cm} (j = j^*+1, \dots, k)\,, \\
 \left< \Delta^{(j^*)} C_{j^*}(T) \right> 
 &=& -1 + j^* \frac {(j^*+1) C_{j^*+1}(T) - j^* C_{j^*}(T)} {N-T}  + O(N^{-1})\,, \\
 \left< \Delta^{(j^*)} C_{j^*-1}(T) \right> 
 &=& j^* \frac {j^* C_{j^*}(T)} {N-T} + O(N^{-1})\,, \\
 \left< \Delta^{(j^*)} C_{j^*-2}(T) \right> 
 &=& O(N^{-1}) \,.
\end{eqnarray}

Let us call a \emph{round} the sequence of steps starting with the assignment of a variable in a clause of length $j^*-1$ and ending when there are no more clauses shorter that $j^*-1$, such as the steps from 2 to $j^*$ in the previous argument. Each round has the same duration: $j^*-1$ steps. During such a round, the variation of the average number of clauses of length $j^*-1$ is
\begin{equation}
  \left< \Delta^{(\mathrm{round})} C_{j^*-1}(T) \right> 
 = -1 + (j^*-1) \frac {j^* C_{j^*}(T)} {N-T} + O(N^{-1})\,,
\end{equation}
so that after $r \geq 1$ rounds the average variations will be
\begin{eqnarray}
 \left< \Delta^{(1 + r(j^*-1))} C_j(T) \right> 
 &=& [1 + r(j^*-1)] \frac {(j+1) C_{j+1}(T) - j C_j(T)} {N-T} + O(N^{-1}) \hspace{.5cm} (j = j^*+1, \dots, k)\,, \nonumber \\
 \\
 \left< \Delta^{(1 + r(j^*-1))} C_{j^*}(T) \right> 
 &=& - 1 + [1 + r(j^*-1)] \frac {(j^*+1) C_{j^*+1}(T) - j^* C_{j^*}(T)} {N-T} + O(N^{-1})\,, \\
 \left< \Delta^{(1 + r(j^*-1))} C_{j^*-1}(T) \right> 
 &=& 1 + \frac {j^* C_{j^*}(T)} {N-T} + r\left[ -1 + (j^*-1) \frac {j^* C_{j^*}(T)} {N-T} \right]  + O(N^{-1})\,.
\end{eqnarray}

There are two possible cases: either after a finite average number $R$ of rounds the average number of clauses of length $j^*-1$ returns to 0, or not. In the first case, $R$ is obtained from the condition:
\begin{eqnarray}
 && \left< \Delta^{(1 + R(j^*-1))} C_{j^*-1}(T) \right> = 0 \\
 &\Leftrightarrow& R = \frac { 1 + \frac {j^* C_{j^*}(T)}{N-T} } {1 - (j^*-1) \frac {j^* C_{j^*}(T)}{N-T} } + O(N^{-1}) \,.
\end{eqnarray}
Notice that, since $R$ is an \emph{average} number, it needs not be integer, and also that the condition for $R$ to be finite is
\begin{equation}
 \frac {j^* C_{j^*}(T)}{N-T} < \frac 1 {j^*-1} \,.
\end{equation}
After $R$ rounds, the number of steps that have been taken is
\begin{equation}
 \Delta T = 1 + R \times (j^*-1) = \frac {j^*} { 1 - (j^*-1) \frac {j^* C_{j^*}(T)}{N-T} } + O(N^{-1})
\label{Delta T for GUC}
\end{equation}
and the total average variations will be:
\begin{eqnarray}
 \left< \Delta^{(\Delta T)} C_j(T) \right> 
 &=& \Delta T \frac {(j+1) C_{j+1}(T) - j C_j(T)} {N-T} + O(N^{-1}) \hspace{.5cm} (j = j^*+1, \dots, k)\,, \nonumber \\
 \\
 \left< \Delta^{(\Delta T)} C_{j^*}(T) \right> 
 &=& - 1 + \Delta T \frac {(j^*+1) C_{j^*+1}(T) - j^* C_{j^*}(T)} {N-T} + O(N^{-1})\,, \\
 \left< \Delta^{(1 + r(j^*-1))} C_{j^*-1}(T) \right> 
 &=& O(N^{-1})\,. 
\label{Finite GUC}
\end{eqnarray}

Wormald's theorem can be applied, ensuring that in the thermodynamic limit the contributions of order $O(N^{-1})$ are ininfluential, and that the average densities are concentrated around the functions $c_j(t)$ that are solutions of the differential equations obtained by dividing (\ref{Finite GUC}) by $\Delta T$, given by (\ref{Delta T for GUC}). The equations we obtain are the following:
\begin{eqnarray}
 \frac {d c_j}{dt} &=& \frac {(j+1) c_{j+1} - j c_j}{1-t} \hspace{.5cm} (j = j^*+1,\dots,k) \,, \\
 \frac {d c_{j^*}}{dt} &=& \frac {(j^*+1) c_{j^*+1} - j^* c_{j^*}}{1-t} - \frac 1 {j^*} \left[ 1 - (j^*-1) \frac {j^* c_{j^*}} {1-t}\right]
\end{eqnarray}
which we can rewrite as a single equation
\begin{equation}
 \frac {d c_j} {dt} =  \frac {(j+1) c_{j+1} - j c_j}{1-t} + \delta_{j,j^*} \left[ \frac 1 {j^*} - \frac {(j^*-1) c_{j^*}}{1-t} \right] \hspace{.5cm} (j=j^*,\dots,k)\,.
\end{equation}

We still have to analyze what happens when $R$ diverges. In that case, the rate at which clauses of length $j^*-1$ accumulate is larger than the rate at which they can be removed, and their number becomes extensive. This signals that the value of $j^*$ must decrease by one unit.

In Paragraph \ref{Par_UC and GUC} I shall give a detailed study of the solution to these equations for $k=3$, showing that GUC finds solutions in linear time with probability 1 for random formul\ae\ from the uniform distribution for $\alpha \leq \alpha_\mathrm h ^\mathrm{GUC}(3) \simeq 0.750874$, which is therefore a lower bound for the value up to which random formul\ae\ are \sat\ with probability 1.

%
%

\chapter{Phase transitions in random optimization problems}
\label{Chap_phase_diagrams}

In the previous Chapter I have introduced two random optimization problems, \ksat\ and \kxorsat, which are equivalent to some spin glass models. In this chapter I am going to review the rich phenomenology displayed by these models, consisting of several phase transitions regarding different order parameters.

I shall first make a very brief introduction to the discovery of sharp transitions in numerical experiments, mostly concerning \ksat, in Section \ref{Sec Numerical evidence for PT}; then, I shall give a rigorous derivation the phase diagram of \kxorsat\ in Section \ref{Sec Phase diagram of kxorsat}; 
finally, in Section \ref{Sec Phase diagram ksat} I shall sketch the main results on the phase diagram of \ksat.

\section{Evidence of phase transitions from numerical experiments}
\label{Sec Numerical evidence for PT}

Phase transitions are a common and well understood concept in statistical mechanics. In the context of random combinatorial optimization problems, it is far less obvious what this can mean. I shall therefore start with a definition and a simple example.

Let us consider a random problem defined over some distribution of instances, and a property $\mathcal P$ which might be true or false for each instance. I shall denote by $N$ the size of the problem, by $c$ some control parameter, and by $P(N,c)$ the probability over the distribution of instances that $\mathcal P$ is true. Then, a \emph{sharp transition} in $\mathcal P$ is defined by the following condition:
\begin{equation}
 \lim_{N \to \infty} P(N,c) =
  \begin{cases}
   0 & \text{if } c < c^* \\
   1 & \text{if } c > c^*
  \end{cases}
\label{Sharp transition}
\end{equation}
where $c^*$ is a constant \emph{threshold} independent on $N$. 

For example, we might consider random graphs with $N$ vertices and $M = cN$ edges, and ask what is the probability $P(N,c)$ that the largest connected component in the graph has size $\gamma N$ with $\gamma > 0$ and independent on $N$. This problem, called random graph percolation, has been studied by Erd\H{o}s and R\'enyi in \cite{Erdos59, Erdos60}. They have proved that the percolation indeed undergoes a sharp transition, with threshold value $c^* = 1/2$.

In numerical studies the definition (\ref{Sharp transition}) is of little use, as the size of samples has to be finite. Some method to extrapolate results to the $N \to \infty$ limit is needed. For large but finite $N$, $P(N,c)$ will be a smooth function of $c$ varying from 0 to 1, whose form will in general depend on $N$. The \emph{transition region}, defined as the range of values of $c$ in which $\epsilon < P(N,c) < 1-\epsilon$ for some finite $\epsilon$ independent on $N$, will have a width $\Delta(N)$ which will become smaller and smaller as $N$ grows. If $\Delta(N)$ scales as a power of $N$, $\Delta(N) \sim N^{-\nu}$ for some constant $\nu$, one can rescale
\begin{equation}
 P(N,c) = \phi_N \bigl( (c-c^*) N^{\nu} \bigr)
\label{Finite Size Scaling}
\end{equation}
and hope that the function $\phi_N(\cdot)$ becomes independent of $N$ for large (but experimentally accessible) $N$. If this is the case, the values of $\nu$ and $c^*$ can be obtained by fitting numerical data so that they ``collapse'' on $\phi(c)$. This is one of the simplest applications of a general method which goes under the name of \emph{finite size scaling} (see for example \cite{Kirkpatrick85}).
In the case of percolation on random graphs, a finite size scaling of the type of (\ref{Finite Size Scaling}) holds, with $\nu = 1/3$.

Finite size scaling was applied in \cite{Kirkpatrick94} to \ksat, providing the first numerical evidence for a sharp transition between a \sat\ phase where random formul\ae\ are satisfiable with probability 1 and a \unsat\ phase there they are not satisfiable with probability 1. The threshold value $\alpha_\mathrm s(k)$ was measured for $k=2,3,4,5,6$, together with the exponent $\nu(k)$. For example, for $k=3$ the values found were $\alpha_\mathrm{s}(3) \simeq 4.17$ and $\nu(3) \simeq 0.67$. However, due to the relatively small size of the formul\ae\ considered ($N \approx 100$), these values were later proved to be inaccurate (most notably the exponents).

Previous studies, for example \cite{Mitchell92}, had measured the probability of a random formula being satisfiable, pointing out that it was $1/2$ for $\alpha \simeq 4.25$ for $k=3$ and $N$ sufficiently large, but without discussing the $N$ dependence of the transition width. In fact, the main purpose of that study was to analyze a different phenomenon: the variation of the running times of the complete DPLL procedure on random formul\ae\ as a function of $\alpha$. What the authors had noticed, and motivated their work, was that formul\ae\ were ``hardest'' to solve in a region centered on the value of $\alpha$ corresponding to $\mathbb P[\text{Sat}|N,\alpha] = 1/2$.

This problem was analyzed again in \cite{Selman96}, in which finite size scaling techniques were applied to the median running time as a function of $N$ and $\alpha$. Even though the maximum of the running time is reached for $\alpha \simeq \alpha_\mathrm s(k)$ for large $N$, this is a very different phenomenon from the \sat/\unsat\ transition, since it is related to the dynamical properties of an algorithm (while the \sat/\unsat\ transition is a property of the ensemble of formul\ae\ themselves).

These two problems, the phase transitions of random constraint satisfaction problems, and the dependency on $\alpha$ of the performance of algorithms, as well as their connection to the properties of typical random formul\ae\ and of their solutions, will be the main topic of the rest of this Chapter, in which I shall present some well known results, and of the second Part of this thesis, presenting some original ones.

\section{Rigorous derivation of the phase diagram of \kxorsat}
\label{Sec Phase diagram of kxorsat}

In this section I shall present a some rigorous results on the phase diagram of \kxorsat. The cases $k=1$ and $k=2$ are much simpler than the general case $k \geq 3$. On the other hand, all values of $k \geq 3$ give rise to the same behavior (at least qualitatively), while the behavior for $k=1$ and 2 is different. For these reasons I shall restrict $k\geq3$ in this Chapter.

As in the case of \ksat, it is intuitive to expect that as the ratio $\alpha = M/N$ between the number of clauses $M$ and the number of variables $N$ increases, the probability that a random formula be satisfiable will decrease. And numerical experiments confirm that (as was the case for \ksat) the transition between the \sat\ and the \unsat\ phases becomes sharp as $N\to\infty$. 

However we shall see that the phase diagram of \kxorsat\ presents a richer structure than just a \sat/\unsat\ transition, and that the geometrical properties of the set of solutions in the \sat\ phase present a second phase transition, which can be related to the performance of search algorithm, as I shall discuss in Chapter \ref{Chap_xorsat}.

\subsection{Bounds from first and second moments}

In this paragraph I shall derive a rigorous bound for the threshold value $\alpha_\mathrm s(k)$ of the \sat/\unsat\ transition, first proved in \cite{Creignou99}.

The number of solutions $\mathcal N$ of \kxorsat\ formul\ae\ with fixed $M$ and $N$ can be regarded as a random variable whose distribution $\mathcal P(\mathcal N)$ will depend on the distribution of the formul\ae\ considered. Since this random variable only takes integer values, the following identity must hold:
\begin{equation}
 \left< \mathcal N \right> \equiv \sum_{\mathcal N = 0}^{2^N} \mathcal P(\mathcal N) \ \mathcal N \geq \sum_{\mathcal N = 1}^{2^N} \mathcal P (\mathcal N) = \mathbb P [\sat]
\end{equation}
which means that the probability of having at least a solution is smaller than or equal to the average number of solutions. This bound for the probability that a formula is satisfiable is called \emph{first moment inequality}.

Let us denote by $X \equiv \{x_i \ | \ i = 1, \dots , N\}$ a configuration of $N$ boolean variables. In order to compute the average number of solutions of a random formula drawn from the uniform distribution, we introduce the indicator function $\varepsilon_l(X)$ which is equal to 1 if the configuration $X$ verifies clause $l$ and 0 otherwise. Then:
\begin{equation}
 \left< \mathcal N \right> = \left< \sum_X \prod_{l=1}^M \varepsilon_l(X) \right> \,. 
\end{equation}
Since the clauses are extracted independently of one another, the average over the choices of the formula can be computed as an average over the choices of each clause appearing in it:
\begin{equation}
 \left< \mathcal N \right> = \sum_X \prod_{l=1}^M \left< \varepsilon_l(X) \right> \,. 
\end{equation}
Moreover, the probability that \emph{any} configuration $X$ satisfies a uniformly drawn random clause is $1/2$, since for any choice of the indices appearing in the clause (and therefore, for fixed $X$, for any left hand side of the clause), the two choices \true\ and \false\ for the right hand side have equal probability. We obtain the very simple result:
\begin{equation}
 \left< \mathcal N \right> = 2^N \times 2^{-M} = 2^{N(1-\alpha)} \,.
\label{tot_entropy}
\end{equation}
and therefore from the first moment inequality:
\begin{equation}
 \mathbb P[\sat] \leq \left< \mathcal N \right> = 2^{N(1-\alpha)}
\end{equation}
which goes to zero for $N \to \infty$ if $\alpha \geq 1$. 

A lower bound for $\mathbb P[\sat]$ can be obtained from the \emph{second moment inequality}, which is derived from the Cauchy-Schwarz inequality of the scalar product
\begin{equation}
 \mathbf u \cdot \mathbf v \equiv \sum_{\mathcal N} \mathcal P (\mathcal N) \, u_\mathcal N \, v_\mathcal N \,,
\end{equation}
which ensures that
\begin{equation}
 (\mathbf u \cdot \mathbf v)^2 \leq (\mathbf u \cdot \mathbf u) \times (\mathbf v \cdot \mathbf v)
\end{equation}
for any vector $\mathbf u$ and $\mathbf v$. In particular, by choosing $u_\mathcal N = \mathcal N$ for any $\mathcal N$ and $v_\mathcal N = 1$ for $\mathcal N \geq 1$ and $v_0 = 0$ one obtains:
\begin{equation}
 \left< \mathcal N \right> ^ 2 = \left( \sum_{\mathcal N \geq 1} \mathcal P (\mathcal N) \, \mathcal N \right)^2
\leq \left[ \sum_{\mathcal N \geq 0} \mathcal P (\mathcal N) \, \mathcal N^2 \right] \times 
\left[ \sum_{\mathcal N \geq 1} \mathcal P(\mathcal N) \, 1^2 \right]
= \left< \mathcal N^2 \right> \times \mathbb P[\sat] \,.
\end{equation}

The crucial point is to compute
\begin{equation}
 \left< \mathcal N^2 \right> = \left< \left( \sum_X \prod_{l=1}^M \varepsilon_l(X) \right) ^2 \right>
= \sum_{X,Y} \prod_{l=1}^M \left< \varepsilon_l(X) \varepsilon_l(Y) \right>
= \sum_{X,Y} \left< \varepsilon(X) \varepsilon(Y) \right> ^M
\label{N^2}
\end{equation}
where again we made use of the independence of clauses in the extraction of a random formula to write the result in terms of $\left< \varepsilon(X)\varepsilon(X) \right>$ which is the probability that \emph{both} $X$ and $Y$ satisfy a random clause. This quantity will obviously depend on how different $X$ and $Y$ are: if $X$ satisfies the clause, $Y$ will also satisfy it if and only if the number of variables appearing in the clause that are different in $X$ and $Y$ is even. When averaging over the choice of the clause, this will depend on the Hamming distance $d(X,Y)$ between $X$ and $Y$,
\begin{equation}
 d(X,Y) \equiv \frac 1 N \sum_i \mathbb I (x_i = y_i) \,.
\end{equation}

For example, for $k=3$ the probability that two configurations at distance $d$ satisfy a random clause is
\begin{equation}
 p_3(d) = \frac 1 2 \left[ (1-d)^3 + 3d^2(1-d) \right] + O(N^{-1})
\end{equation}
where the factor $1/2$ comes from the probability that $X$ satisfies the clause to begin with; the term $(1-d)^3$ is the probability that the 3 variables appearing in the clause take the same value in $X$ and $Y$; the term $3d^2(1-d)$ is the probability that two variables are different and one is equal (among those appearing in the clause) in $X$ and $Y$; and finally, we are neglecting a term of order $N^{-1}$ arising from the correlations in the choices of the variables appearing in a single clause (which must be different).

The general form will be
\begin{equation}
 p_k(d) = \frac 1 2 \sum_{l = 0, 2, \dots, k} \binom k l d^l(1-d)^{k-l} + O(N^{-1})
\end{equation}
in which only the \emph{even} terms in the binomial expansion are taken. Notice that (contrary to the upper bound obtained from the first moment inequality), the lower bound derived from the second moment inequality will therefore depend on $k$.

Going back to (\ref{N^2}) we obtain:
\begin{equation}
 \left< \mathcal N ^2 \right> = \sum_{X,Y} p_k(d(X,Y))^M = \sum_{d \ = \ 0,\, 1/N,\, 2/N,\, \cdots} p_k(d)^M \mathcal M(d)
\end{equation}
where $\mathcal M (d)$ is the number of pairs of configurations at distance $d$, i.e. 
\begin{equation}
 \mathcal M(d) = 2^N \binom N {Nd} \,,
\end{equation}
so that for large $N$
\begin{equation}
 \left< \mathcal N ^2 \right> 
= \sum_{d \ = \ 0,\, 1/N,\, 2/N,\, \cdots} \exp \left\{ N \log 2 \left[ 1 - (1-d) \log_2 (1-d) - d \log_2 d + \alpha \log_2 p_k(d) \right] \right\}
\label{average_N^2}
\end{equation}
which we evaluate with the Laplace method:
\begin{equation}
  \left< \mathcal N ^2 \right> 
= 2^{N \gamma_k(\alpha,\bar d)}
\end{equation}
where $\gamma_k(\alpha,d)$ is the function multiplying $N \log 2$ in (\ref{average_N^2}) and $\bar d$ is the value of $d$ that maximizes it in the interval $[0,1]$. The result of the second moment calculation is:
\begin{equation}
 \mathbb P[\sat] 
\geq \frac {\left< \mathcal N \right> ^2} {\left< \mathcal N ^2 \right>} 
= \frac {2^{2N(1-\alpha)}} {2^{N \gamma_k(\alpha, \bar d)}}
= 2^{N\left[2(1-\alpha) - \gamma_k(\alpha, \bar d) \right]} \,.
\label{Psat}
\end{equation}

For $k=3$ one obtains that if $\alpha \leq \alpha_0 (3) \simeq 0.889$ the function $\gamma_3(\alpha, d)$ has a global maximum in $\bar d = 1/2$ where $\gamma_3(\alpha,1/2) = 2 (1-\alpha) + o(1)$ (the asymptotics are for $N \to \infty$); if $\alpha > \alpha_0 (3)$ a maximum located at $\bar d < 1/2$ becomes larger than the local one at $d = 1/2$ and $\gamma_3(\alpha,\bar d) > 2 (1-\alpha) + o(1)$. Comparing with (\ref{Psat}) one sees that, in the limit $N \to \infty$, $\mathbb P[\sat] > 0$ if $\alpha \leq \alpha_0 (3)$.

The same analysis can be performed for larger values of $k$, leading to similar results. In fact, one can prove a stronger statement: not only $\mathbb P[\sat] > 0$ if $\alpha \leq \alpha_0 (k)$, but the lower bound is equal to 1 in the thermodynamic limit, so that random \kxorsat\ formul\ae\ are \sat\ with probability 1 if $\alpha \leq \alpha_0 (k)$.	

The conclusion of the first and second moment calculations is that, if there is a sharp transition between the \sat\ and the \unsat\ phases in \kxorsat, it must occur for $\alpha = \alpha_\mathrm s(k)$ such that
\begin{equation}
 \alpha_0 (k) \leq \alpha_\mathrm s(k) \leq 1 \,.
\label{moments}
\end{equation}
Since these bounds are not tight, one cannot conclude whether such a sharp transition exists on the basis of the first and second moment inequalities.

\subsection{Leaf removal procedure}
\label{Sec_pure-leaf-removal}

The leaf removal procedure allows to prove that a sharp transition between the \sat\ and \unsat\ phases indeed exists, to compute the value of $\alpha_\mathrm s(k)$ at which it occurs, and to characterize the geometry of the solutions \cite{Cocco03, Mezard03}.

The main idea behind this powerful argument is the following: if the formula contains a variable $x_1$ which has a unique occurrence, the value of $x_1$ is constrained only by the clause in which it appears. Given the values of the other variables that appear in it, one is free to set the value of $x_1$ so as to satisfy the clause. This means that a clause which contains a single-occurrence variable does not constrain the values of the other variables that appear in it. One can then set it apart, and look for a solution of the reduced formula in which neither the single-occurrence variable nor the clause it belongs to are present. Moreover, when a clause is set apart, it is possible that some variable that appears in it becomes a single-occurrence variable (relative to the rest of the formula), so that the removal of single-occurrence variables (\emph{leaves}) is an iterative procedure. In the following I shall give a quantitative description of this process.

Let us consider a random \kxorsat\ formula with $M$ clauses and $N$ variables. It is easy to show that the distribution of the number of occurrences $\ell$ of the variables in the formula will be a poissonian with parameter $\alpha k$:
\begin{equation}
 \mathbb P[\ell] = e^{-\alpha k} \frac{(\alpha k)^\ell}{\ell!} \,.
\label{poisson_t=0}
\end{equation}
A finite fraction $\alpha k e^{-\alpha k}$ of variables will therefore have a single occurrence. Let us assume that we proceed by removing them one at a time, in successive ``steps''.

I shall denote by $n_\ell(T)$ the average number (divided by $N$) of variables that have $\ell$ occurrences after $T$ steps. At that point, the total number of variables in the system is $N' = N-T$, and the total number of clauses is $M' = M-T$, since at each step one variable and one clause are removed from the system. During a step, the number of occurrences of the other variables that appear in the removed clause will also be decreased by one.
What is the probability that one of these other variables has $\ell$ occurrences? One might be tempted to say that it is just proportional to $n_\ell$, since that is the probability that a variable has $\ell$ occurrences. However this is wrong, for the following reason. We can regard the formula as a table with $M'$ rows and $k$ columns, where the ``slot'' in row $i$ and column $j$ contains the index of the $j^\mathrm{th}$ variable in the $i^\mathrm{th}$ clause. A variable which has $\ell$ occurrences in the formula will appear in $\ell$ slots of the table. So the number of slots in the table that contain variables that have $\ell$ occurrences is $\ell \times N \times n_\ell$, and the probability that a randomly chosen slot contains a variable with $\ell$ occurrences is $\ell n_\ell / \sum_{\ell'} \ell' n_{\ell'}$. Since the number of variables in the removed clause is $k$, the average number of variables that appear in it (apart from the single-occurrence variable that we have chosen to eliminate) and that have $\ell$ occurrences is therefore $(k-1)\ell n_\ell/\sum_{\ell'} \ell' n_{\ell'}$.

We can use Wormald's theorem, which I introduced in Chapter \ref{Chap_optimization}, to write a differential equation\footnote{A detailed derivation is provided in Section \ref{Sec_leaf-removal} for a more general case.} for $n_\ell(t)$, where $t \equiv T/N$, in the limit $N \to \infty$:
\begin{equation}
 \frac{dn_\ell}{dt} = (k-1) \frac{(\ell+1)n_{\ell+1}(t)-\ell n_\ell(t)}{k(\alpha-t)}
\hspace{1cm} (\ell>1)
\label{evolution_nl2}
\end{equation}
where $k(\alpha-t) = \sum_\ell \ell n_\ell(t)$ is the total number of slots divided by $N$ (remember that exactly $k$ slots are removed at each step). The first term corresponds to the variables that have $\ell+1$ occurrences before the clause is removed, which afterwards have $\ell$ occurrences, while the second term corresponds to the variables that have $\ell$ occurrences before the clause is removed and which afterwards have $\ell-1$ occurrences. It is easy to check that this equation can be extended to $\ell=0$ and $\ell=1$ as follows:
\begin{equation}
 \frac{dn_\ell}{dt} = (k-1) \frac{(\ell+1)n_{\ell+1}(t)-\ell n_\ell(t)}{k(\alpha-t)}
+ \delta_{\ell,0} - \delta_{\ell,1} \,.
\label{evolution_nl}
\end{equation}
The initial condition that must be imposed is (\ref{poisson_t=0})
\begin{equation}
 n_\ell (0) = e^{-\alpha k} \frac{(\alpha k)^\ell}{\ell!} \,.
\end{equation}

It is easy to see that, for $\ell \geq 2$, $n_\ell$ remains poissonian  even for $t>0$, with some time dependent parameter which is $\lambda(t)$. To prove it, one just needs to replace the ansatz
\begin{equation}
 n_\ell (t) = e^{-\lambda(t)} \frac{\lambda(t)^\ell}{\ell!}
\end{equation}
into (\ref{evolution_nl2}) to obtain
\begin{equation}
 \frac{dn_\ell}{dt} = - \frac{d\lambda}{dt} \left[ n_\ell(t) - n_{\ell-1}(t) \right]
= \frac{k-1}{k(\alpha-t)} \lambda(t) \left[ n_\ell(t) - n_{\ell-1}(t) \right]
\end{equation}
from which one obtains an equation for $\lambda(t)$ independent of $\ell$:
\begin{equation}
  \frac d {dt} \lambda(t) = - \frac{k-1}{k(\alpha-t)} \lambda(t) \, .
\end{equation}
Solving it with the initial condition $\lambda(0) = \alpha k$ gives
\begin{equation}
 \lambda(t) = \alpha k \left(1 - \frac t \alpha \right)^\frac{k-1}{k} \,.
\label{lambda}
\end{equation}

However, $n_1(t)$ is \emph{not} poissonian, because of the extra $\delta_{\ell,1}$ term in (\ref{evolution_nl}) compared to (\ref{evolution_nl2}), and to compute it we use the following trick:
\begin{equation}
 n_1(t) 
= \sum_{\ell=1}^\infty \ell n_\ell(t) - \sum_{\ell=2}^\infty \ell n_\ell(t) 
= k(\alpha-t) - \sum_{\ell=2}^\infty \ell e^{-\lambda(t)} \frac{\lambda(t)^\ell}{\ell!} = k(\alpha-t) - \left[ \lambda(t) - \lambda(t) e^{-\lambda(t)} \right]
\label{n1}
\end{equation}
which can be conveniently expressed in terms of the parameter $b \equiv (1-t/\alpha)^{1/k}$:
\begin{equation}
 n_1(b) = \lambda(b) \left[b + e^{-\lambda(b)} - 1 \right]
\label{b_leafremoval}
\end{equation}
with $\lambda(b) = \alpha k b^{k-1}$. The interval of variation of $t$ is $[0,\alpha]$ (since after $\alpha N = M$ steps all the clauses are eliminated from the system), and correspondingly $b$ varies between the initial value 1 and 0.

There are now two possibilities, depending on the value of $\alpha$: either $n_1(b) > 0$ for all $b \in [0,1]$, or for some value $b^* \in [0,1]$ one has $n_1(b^*) = 0$. In the first case the algorithm stops when all the clauses have been removed from the system. In the second case, one is left with an irreducible sub-formula containing $N(\alpha-t^*) = N \alpha (b^*)^k$ clauses and $N\sum_{\ell=2}^\infty n_\ell(b^*) = N - N(1-b^*)[1+\alpha k (b^*)^{k-1}]$ variables. Note that the sub-formula is still uniformly random, conditioned on the distribution $n_\ell(b^*)$.

It is easy to check that the first case occurs for $\alpha < \alpha_\mathrm c(k)$ where $\alpha_\mathrm c(k)$ is a constant, while for $\alpha \geq \alpha_\mathrm c(k)$ the value of $n_1$ vanishes for $b^*>0$, which is the largest solution of (\ref{b_leafremoval}). I shall denote $b_\mathrm c(k)$ the value of $b^*$ corresponding to $\alpha = \alpha_\mathrm c(k)$. Numerical values of these constants (and their asymptotics for $k \to \infty$) are shown in Table \ref{Tab_threshold}, in the following paragraph.

Let us now turn to the implications of these results on the original formula. If the first case occurs (i.e. if $\alpha < \alpha_\mathrm c(k)$), one can ``invert'' the procedure, and reinsert the clauses into the formula one at a time, in the reverse order with which they were removed. When the first clause is reinserted, one can chose freely the values of $k-1$ variables, and set the last variable to the value which satisfies the clause. In general, when one reinserts a clause containing $j$ ``new'' variables, the value of $j-1$ of them is set arbitrarily, and the last one is set to the value which satisfies the clause. Notice that since each removed clause contained a variable which had a single occurrence at the time when it was removed, each reinserted clause will contain at least one new variable. One can then obtain a solution to the original formula in this manner.

What is the number of solutions that one obtains? Not counting the variable which has been selected for removal, the average number of single-occurrence (i.e. ``new'') variables present in the clause removed at time $t$ is $(k-1) n_1(t) / [k(\alpha-t)]$. For each of them two values can be chosen. The number of solutions $\mathcal N$ is therefore
\begin{equation}
 \mathcal N \equiv 2^{Ns} \,, \hspace{.5cm} s = \int_0^{t^*} \frac {(k-1)n_1(t)}{k(\alpha-t)} dt + e^{-\alpha k}
 \label{entropy_inside}
\end{equation}
where the last term comes from the variables which do not appear in the system.
The integral is easily done recalling that for $\alpha < \alpha_\mathrm c(k)$ one has $t^*=\alpha$ and substituting (\ref{lambda}) and (\ref{n1}) to obtain $s=1-\alpha$, as expected from (\ref{tot_entropy}).

In the second case, for $\alpha_\mathrm c(k) < \alpha$, the leaf removal procedure ends with a sub-formula with a clause to variables ratio $\alpha'$ given by 
\begin{equation}
 \alpha' = \frac{\alpha (b^*)^k}{1-(1-b^*)\left[1 + \alpha k (b^*)^{k-1} \right]} \,,
 \label{alpha'}
\end{equation}
which is an increasing function of $\alpha$.
The original formula is \sat\ if and only if the sub-formula is also \sat, and we would like to know if this is the case, depending on the value of $\alpha'$. As we have seen in the calculation of the bound from the first and second moments (\ref{moments}), the upper bound $\alpha_\mathrm s(k) \leq 1$ is independent on the distribution of random instances, while the lower bound $\alpha_0(k) \leq \alpha_\mathrm s(k)$ depends on it. The computation of the lower bound must therefore be adapted to a distribution of instances which is uniform conditioned on the average numbers of occurrences $\{n_\ell(b^*)\}$ which is $0$ for $\ell=1$ and poissonian with parameter $\lambda(b^*)$ for $\ell \geq 2$. This is done in a detailed manner in \cite{Mezard03}. The result is remarkable: in the absence of single-occurrence variables, the average number of solutions becomes a concentrated quantity and $\left< \mathcal N^2 \right> = \left< \mathcal N \right> ^2$, so that the bounds from first and second moments inequalities become tight: $\alpha'_\mathrm s (k) = 1$. 

This proves that there is, indeed, a sharp transition between the \sat\ and the \unsat\ phases, and the transition value of $\alpha$ is obtained from the condition
\begin{equation}
 1 = \alpha' = \frac{\alpha (b^*)^k}{1-(1-b^*)[1+\alpha k (b^*)^{k-1}]}
\end{equation}
(notice that $b^*$ is itself a function of $\alpha$, determined by (\ref{b_leafremoval})).

The average number of solutions of the sub-formula will be
\begin{equation}
 \mathcal N' = 2^{N'(1-\alpha')} 
= 2^{N \left\{ b^* - \alpha (b^*)^k + \alpha k \left[(b^*)^k - (b^*)^{k-1} \right] \right\} } \,.
\label{N'}
\end{equation}

For each solution of the sub-formula, which I shall call ``seed'', the number of solutions of the original formula that can be obtained is still given by (\ref{entropy_inside}), where now $t^* = \alpha [ 1 - (b^*)^k ]$:
\begin{equation}
 \mathcal N_1 = 2^{N\left\{1-b^* + \alpha k \left[(b^*)^{k-1} - (b^*)^k\right] -\alpha \left[1 - (b^*)^k\right] \right\}}
\label{N1}
\end{equation}
where the subscript 1 is a reminder that this is for a fixed seed. Since for different seeds one necessarily obtains different solutions of the original formula, the total number of solutions is 
\begin{equation}
 \mathcal N = \mathcal N' \times \mathcal N_1 = 2^{N(1-\alpha)}
\end{equation}
as expected.

It is possible to prove the following properties (or at least, to give some non-rigorous arguments supporting them, see \cite{Mezard03, Mora06}):

\begin{enumerate}
 \item The average distance $d_0$ between different solutions corresponding to the same seed is
   \begin{equation}
     d_0 = \frac {1-b^*} 2
   \end{equation}
 \item The average distance $d_1$ between solutions corresponding to different seeds is
    \begin{equation}
     d_1 = \frac 1 2
    \end{equation}
 \item The maximum distance between solutions corresponding to a same seed is smaller than the minimum distance between solutions corresponding to different seeds
 \item For any two solutions $X$ and $X'$ corresponding to the same seed, there exists a sequence of solutions $X_1, \dots, X_P$ such that $X_1 = X$, $X_P = X'$ and the (intensive) distance between $X_i$ and $X_{i+1}$ is of order $o(1)$ as $N \to \infty$.
\end{enumerate}

\subsection{Phase diagram of \kxorsat}

Based on the previous analysis, the following phase diagram can be determined. Each statement is valid with probability 1 in the limit $N \to \infty$ for random \kxorsat\ formul\ae\  extracted from the uniform distribution and with $k \geq 3$.

The phase diagram of \kxorsat\ consists of three phases, dependent on the ratio $\alpha$ of clauses per variable, separated by sharp transitions located at $\alpha_\mathrm c(k)$ (for clustering) and $\alpha_\mathrm s(k)$ (for \sat/\unsat). Numerical values of the thresholds for finite $k$, and their asymptotics for $k \to \infty$ are shown in Table~\ref{Tab_threshold}.

For $\alpha < \alpha_\mathrm c(k)$ the formula is \sat\, and the solutions are homogeneously distributed in the space of configurations. Two random solutions are at an (intensive) distance $d = 1/2$, and they are connected by a sequence of solutions separated by a distance of order $o(1)$. The total number of solutions is given by (\ref{tot_entropy}),
\begin{equation}
 \mathcal N = 2^{N(1-\alpha)} \,.
\end{equation}
The value of the threshold $\alpha_\mathrm c (k)$ is the smallest value of $\alpha$ such that the equation
\begin{equation}
 b = 1-e^{-\alpha k b^{k-1}}
\label{b_clustering}
\end{equation}
has a solution with $b>0$. 

\begin{table}
\begin{center}
\begin{tabular}{|c|c|c|c|}
 \hline
 $k$ & $\alpha_\mathrm c(k)$ & $\alpha_\mathrm s(k)$ & $b_\mathrm c(k)$ \\
 \hline
 $3$ & $0.81846916$          & $0.91793528$          & $0.71533186$ \\
 $4$ & $0.77227984$          & $0.97677016$          & $0.85100070$ \\
 $5$ & $0.70178027$          & $0.99243839$          & $0.90335038$ \\
 $6$ & $0.63708113$          & $0.99737955$          & $0.93007969$ \\
 \hline
 $\infty$ & $ \log k / k$    & $1 - e^{-k}$          & $ 1 - 1 / k \log k$ \\
 \hline
\end{tabular}
\end{center}
\caption{Threshold values for the clustering and \sat/\unsat\ transitions and backbone size $b_\mathrm c$ (at the clustering transition) for various values of $k$ and (to the leading order) for $k \to \infty$.}
\label{Tab_threshold}
\end{table} 

For $\alpha_\mathrm c(k) < \alpha < \alpha_\mathrm s(k)$, the formula is \sat\ and the solutions are clustered. Each cluster is identified by a particular solution of the sub-formula generated by the leaf-removal algorithm, called a seed. The solutions belonging to a same cluster are connected, the average distance between two of them is $(1-b^*)/2$ and their number is given by (\ref{N1}):
\begin{equation}
 \mathcal N_1 = 2^{N\left\{1-b^* + \alpha k \left[(b^*)^{k-1} - (b^*)^k\right] -\alpha \left[1 - (b^*)^k\right] \right\}}
\end{equation}
where $b^*$ is the largest solution of (\ref{b_clustering}), which represents the fraction of variables that take the same value in each solution of a given cluster and is called \emph{back-bone} size.
The solutions belonging to different clusters are well separated, the average distance between two of them is $1/2$ and the number of clusters is given by (\ref{N'}):
\begin{equation}
 \mathcal N' = 2^{N'(1-\alpha')} 
= 2^{N \left\{ b^* - \alpha (b^*)^k + \alpha k \left[(b^*)^k - (b^*)^{k-1} \right] \right\} } \,.
\end{equation}
The threshold value $\alpha_\mathrm s(k)$ is given by the condition (\ref{alpha'}):
\begin{equation}
 \frac{\alpha (b^*)^k}{1-(1-b^*)\left[1 + \alpha k (b^*)^{k-1} \right]} = 1 \,.
\end{equation}

For $\alpha_\mathrm s(k) < \alpha$ the formula is \unsat. Note that as $\alpha \to \alpha_\mathrm s(k)$ from below, the entropy of the number of cluster goes to 0, i.e. the number of clusters becomes sub-exponential in $N$, while the number of solutions (inside each cluster) remains exponential in $N$, and discontinuously jumps to 0 as $\alpha$ crosses $\alpha_\mathrm s(k)$.

The entropies (i.e. $\log \mathcal N / N$) of the number of clusters and of the (total) number of solutions are shown in Figure \ref{Fig_entropy} for $k=3$.

\begin{figure}
 \includegraphics[width=7.7cm]{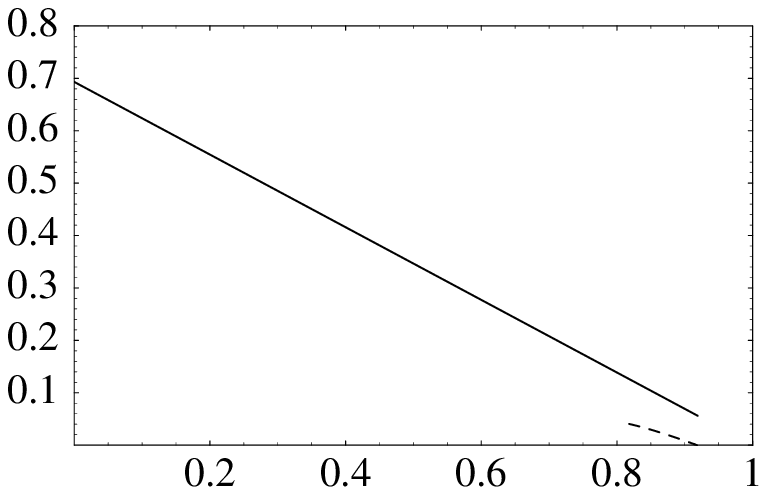}
 \includegraphics[width=7.7cm]{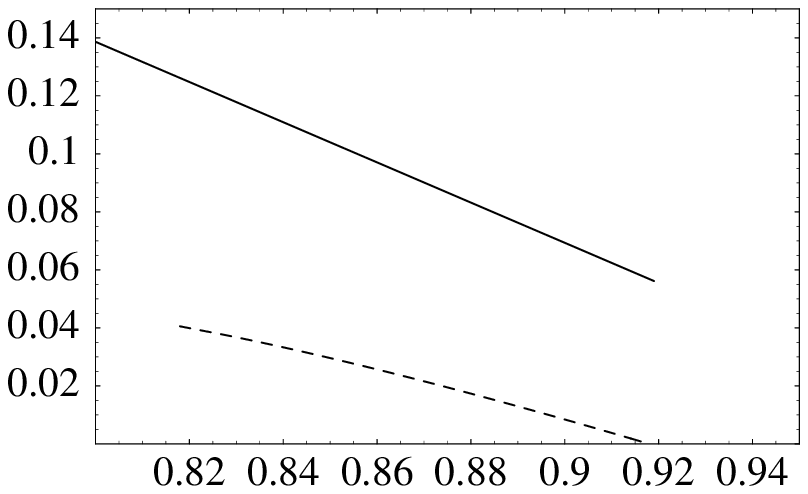}
 \caption{Total entropy $s(\alpha)$ (full line) and entropy of the number of clusters $\sigma(\alpha)$ (dashed line) as functions of $\alpha$ for \kxorsat[3]. The curve for $\sigma(\alpha)$ starts at $\alpha = \alpha_\mathrm c(3) \simeq 0.818$. The curve for $s(\alpha)$ ends at $\alpha = \alpha_\mathrm s(3) \simeq 0.918$, where $\sigma(\alpha) = 0$. The right hand panel is an inset of the full figure, on the left.}
 \label{Fig_entropy}
\end{figure}

%

\section{Heuristic results on the phase diagram of \ksat}
\label{Sec Phase diagram ksat}

The simple graph-theoretical arguments that allow the complete and rigorous  characterization of the phase diagram of \kxorsat\ do not apply in the case of \ksat. Not only the methods required to derive it are more complicated (and not rigorous), but the phase diagram itself is more complicated.

\subsection*{Sat/Unsat transition}

The existence of a \sat/\unsat\ transition in \ksat\ has been proved rigorously, but the proof of its sharpness remains an open problem. In fact, the following was proved by Friedgut \cite{Friedgut99}:
\begin{description}
 \item[Theorem] For each $k \geq 2$, there exists a sequence $\alpha_N(k)$ such that, for all $\epsilon > 0$,
 \begin{equation}
  \lim_{N \to \infty} \mathbb P[\text{Sat}|N,\alpha] =
  \begin{cases}
   1 & \text{if } \alpha = (1-\epsilon) \alpha_N(k) \,,\\
   0 & \text{if } \alpha = (1+\epsilon) \alpha_N(k)
  \end{cases}
\end{equation}
\end{description}
where $\mathbb P[\text{Sat}|N,\alpha]$ is the probability that a uniformly random \ksat\ formula with $N$ variables and $\alpha N$ clauses is satisfiable.

Note that this theorem proves a non-uniform convergence: the threshold value is a function of $N$, which does not necessarily converge to a constant.
This theorem doesn't imply that the \sat/\unsat\ transition is sharp, but it proves that it exists. The sharpness of the transitions remains a conjecture.

Rigorous upper and lower bounds have been proved for the threshold $\alpha_\mathrm s(k)$ for finite $k$ and asymptotically as $k \to \infty$ (for a review and latest results, see \cite{Achlioptas05}). Some values are listed in Table \ref{Tab Sat/unsat ksat}.

Finally, the best available estimates of the value of $\alpha_\mathrm s(k)$ have been obtained with methods derived from statistical mechanics: the analysis of a message passing procedure called Survey Propagation (SP), which is based on the cavity method \cite{Mertens06}. Some values obtained from the analysis of SP are reported in Table \ref{Tab Sat/unsat ksat}.

\begin{table}
 \begin{center}
 \begin{tabular}{|c|c|c|c|}
 \hline
 $k$ & $\alpha_\mathrm s^-(k)$ & $\alpha_\mathrm s^*(k)$ & $\alpha_\mathrm s^-(k)$ \\
 \hline
 3 & 3.52 & 4.267 & 4.51 \\
 4 & 7.91 & 9.931 & 10.23 \\
 5 & 18.79 & 21.117 & 21.33 \\
 $\infty$ & $2^k \log 2 - k$ & $ 2^k \log 2 - b_k$ & $2^k \log 2$ \\
 \hline
 \end{tabular}
 \end{center}
 \caption{Threshold values for the \sat/\unsat\ transition in \ksat. $\alpha_\mathrm s^-(k)$ is a rigorous lower bound, $\alpha_\mathrm s^*(k)$ is the prediction from the cavity method, and $\alpha_\mathrm s^+(k)$ is a rigorous upper bound. For $k \to \infty$ the rigorous bounds are exact, while in the result from the cavity computation, $b_k$ is a positive function of $k$ which converges to $(1+\log 2)/2$ as $k \to \infty$. From \cite{Achlioptas05, Mertens06}}
\label{Tab Sat/unsat ksat}
\end{table}

\subsection*{Clustering transition}

The satisfiable phase of \ksat\ has a very rich structure, presenting \emph{several} phase transitions that concern the geometry of the satisfying assignments. The first such transition is the clustering one.

The definition of the clustering phenomenon itself is much more complicated for \ksat\ than for \kxorsat. As we have seen, clustering in \kxorsat\ has a geometrical origin: the set of variables of a formula can be decomposed in two: the backbone, made of variables that are determined by solving a sub-formula of the original problem; and the leaves, that are free to take any value in any solution. This structure naturally implies the clustering of solutions, and also two properties of the clusters: first, that all clusters contain the same number of solutions; second, that the variables that are frozen inside a cluster are the same for all clusters.

In \ksat, these two properties do not hold. The fact that the variables that freeze in different clusters are not the same requires a definition of clusters independent on the backbone. This can be done by defining the clusters as a partition of the solutions such that:
\begin{enumerate}
 \item The distance between any pair of solutions belonging to different clusters is larger than the distance between any pairs of solutions belonging to the same cluster;
 \item For any pair of solutions $(X,Y)$ belonging to the same cluster, a sequence of solutions $\{X_1,\dots,X_n\}$ can be made such that $X_1 = X$, $X_n = Y$ and the distance between $X_i$ and $X_{i+1}$ is of order $O(1)$ (as $N \to \infty$).
\end{enumerate}
This approach is followed in \cite{Mezard05, Daude07}, where rigorous results are obtained for $k \geq 8$. Moreover, non-rigorous results based on the cavity method are available for any $k$ \cite{Mertens06}, and are reported in Table \ref{Tab Clustering cond ksat}.

\begin{table}
 \begin{center}
 \begin{tabular}{|c|c|c|c|}
 \hline
 $k$ & $\alpha_\mathrm c(k)$ & $\alpha_\mathrm {Cond}(k)$ & $\alpha_\mathrm s(k)$ \\
 \hline
 3 & 3.86 & 3.86 & 4.267 \\
 4 & 9.38 & 9.547 & 9.931 \\
 5 & 19.16 & 20.80 & 21.117 \\
 \hline
 \end{tabular}
 \end{center}
 \caption{
  Threshold values for the clustering $(\alpha_\mathrm c)$ and condensation $(\alpha_\mathrm{Cond})$ transitions in \ksat. The values of $\alpha_\mathrm s(k)$ from Tab \ref{Tab Sat/unsat ksat} are repeated for comparison. From \cite{Krzakala07}.
 }
\label{Tab Clustering cond ksat}
\end{table}

Notice, however, that the clustering phenomenon was first suggested for \ksat\ in \cite{Biroli00}, where a ``variational'' replica calculation was performed: based on physical intuition, a simple trial function with few free parameters was used as the functional order parameter for the free energy, as in (\ref{f_functional}), and the values of the parameters where set by finding the extremum of the corresponding entropy. With this method, an approximation to the RSB solution which describes the clustered phase was found. This led to the calculation of approximate values of the clustering threshold $\alpha_\mathrm c(k)$. In the same paper, the other difficulty arising in \ksat, i.e. the fact that different clusters have different sizes, was pointed out.

It is a very important fact, as it gives rise to two more phase transitions.

\subsection*{Condensation and freezing transitions}

Let us denote, as usual, the entropy of the number of clusters by $\Sigma$, the internal entropy of a cluster as $s_i$ and the total entropy as $s$. Each of them is defined as the logarithm of the corresponding number of objects divided by $N$. When different clusters have different sizes, a convenient way of accounting for them is to write $\Sigma$ as a function of $s_i$: $\Sigma(s_i)$ is the entropy of the number of clusters that have internal entropy $s_i$. The total entropy is then
\begin{equation}
 s = \int \left[ \Sigma(s_i) + s_i \right] \, d s_i \,.
\end{equation}
The measure of the number of solutions will be dominated by the maximum of the integrand, i.e. by the value
\begin{equation}
 s_i^* \, : \ \Sigma'(s_i^*) = -1 \,.
\end{equation}

At the clustering transition $\alpha_\mathrm c(k)$, the complexity $\Sigma(s_i^*)$ becomes positive: the space of solutions splits into an exponential number of well separated clusters, each containing an exponential number of solutions. As $\alpha$ grows, the number of solutions decreases. More specifically, it is $\Sigma(s_i^*)$ which decreases, and for $\alpha = \alpha_\mathrm{Cond}(k) < \alpha_\mathrm s(k)$, it vanishes. When this happens, both the number of solutions and the number of cluster are still exponential; however, the measure of the number of solutions is dominated by a sub-exponential number of clusters, corresponding to the largest $s_i$. As $\alpha$ increases further, the value of the \emph{maximum} of $\Sigma(s_i)$ decreases, until it vanishes at $\alpha = \alpha_\mathrm s(k)$, the \sat/\unsat\ transition. When this happens, the number of solutions vanishes abruptly, with a discontinuity in $s_i$.

Still another phase transition occurs for intermediate values of $\alpha$, corresponding to the freezing of variables within a cluster. For $\alpha_\mathrm c(k) < \alpha < \alpha_\mathrm f(k)$, there are no frozen variables (even within a cluster), while for $\alpha_\mathrm f(k) < \alpha$ frozen variables are present \cite{Semerjian07}.

\end{part}
\begin{part}{Some properties of random $k$-SAT and random $k$-XORSAT}
\chapter{Study of poissonian heuristics for DPLL in $k$-XORSAT}
\label{Chap_xorsat}

In this chapter I shall present some new results on the relationship between the clustering transition of \kxorsat\ and the performance of DPLL algorithms, obtained with R\'emi Monasson and Francesco Zamponi and published in \cite{Altarelli06}.

It is generally believed that \emph{local} algorithms cannot succeed (in finding solutions) in the clustered phase of random CSP. In this context ``local'' means that the algorithm decides assignments based on local information, such as the values of variables within a finite subset of clauses. Local algorithms therefore include, for example, search algorithms such as Metropolis or \walksat, and also the DPLL procedure. The basic argument supporting this belief is that in the clustered phase an extensive back-bone of frozen variables exists, which requires an extensive number of variables to take values that are strongly correlated. An optimization procedure which only takes into account a finite portion of the problem will not be able to find a correct assignment for the back-bone, and therefore for the problem.

An alternative argument is directly derived from spin glass theory: the free energy landscape of random CSP in the clustered phase is characterized by a large number of states, most of which have positive energy, separated by extensive barriers. In order to go from a random configuration to a ground state the system must cross these barriers, which a local optimization procedure cannot do. If this argument is plausible for search procedures, which perform a random walk in the space of configurations while trying to minimize some cost function, and which therefore can indeed remain trapped in local minima of the free energy, it is not at all clear why it should apply to the DPLL procedure. Indeed, the only evidence supporting this claim for DPLL is that no heuristic is known to succeed in the clustered phase. 

The main result that I shall present in this chapter is that no DPLL heuristic which preserves the poissonian distribution of occurrences in the sub-formul\ae\ it generates can find solutions in the clustered phase. The essence of the argument, as we shall see, is related to the geometrical properties of the graph underlying the formula (which allow the use of the leaf-removal procedure to characterize the phases), and to the very basic fact that a Unit Propagations cannot remove more than one clause for each variable that they assign.

This result is valid for random \kxorsat\ formul\ae\ extracted from the uniform distribution (with probability 1 as $N \to \infty$, as usual). It is worth noting that it can be extended to a generalization of \kxorsat\ which goes under the name of Uniquely Extensible Constraint Satisfaction Problems, or UE-CSP. In these problems, variables can take values in a set of cardinality $d$, and the form of the constraints is such that $k$ variables appear in them, and that if any $k-1$ variables appearing in a constraint are assigned, then the value of the $k^\mathrm{th}$ variable is determined. It is very interesting that $(d,k)$-UE-CSP is NP-Complete for $\{d \geq 4, k \geq 3\}$.  \kxorsat\ is a special case of $(d,k)$-UE-CSP with $d=2$. However, as far as the DPLL procedure is concerned, the class of $(d,k)$-UE-CSP is equivalent to \kxorsat\ for \emph{any} $k$ and $d$, since the only relevant feature for the sake of DPLL is that Unit Propagations be possible, and the characteristic property of UE-CSP's ensures that it is. As a consequence that there will be a sharp transition between a phase with a back-bone and a phase without it, which will occur for some $\alpha_\mathrm c(d,k)$, and that all the results that we shall derive concerning the performance of DPLL will be valid for \uecsp\ as well.

The structure of this chapter is the following: in Section \ref{Sec_leaf-removal} I shall introduce a generalization of the leaf-removal procedure to mixed formul\ae; this will allow me to introduce a \emph{potential} function that characterizes the phases of mixed formul\ae, in Section \ref{Sec_potential}; in Section \ref{Sec_trajectories} I shall characterize the trajectories that poissonian heuristics generate in the space of the density of clauses $\{c_j\}$; then I shall derive an upper bound for the values of $\alpha$ for which poissonian heuristics for DPLL can find solutions, in Section \ref{Sec_bounds}; in Section \ref{Sec_optimality GUC} I shall present an argument supporting that GUC saturates the previous bound in the limit $k \to \infty$. ; finally, in Section \ref{Sec Conclusions 4} I shall discuss the results obtained and indicate some possible directions of further investigation.

\section{Leaf-removal for mixed formul\ae}
\label{Sec_leaf-removal}

In paragraph \ref{Sec_pure-leaf-removal} I described the leaf-removal procedure applied to a \emph{pure} \kxorsat\ formula, that is to say a formula in which all the clauses involve exactly $k$ variables, as was introduced in \cite{Cocco03, Mezard03}. The leaf-removal procedure is extremely powerful, as it provides a full characterization of the phase diagram of \kxorsat. In this Section I shall generalize the analysis of the leaf-removal procedure to the case of \emph{mixed} formul\ae, containing clauses of \emph{different lengths} (where length stands for the number of variables in the clause), in order to allow the characterization of the sub-formul\ae\ generated by DPLL heuristics.

\subsection{Leaf-removal differential equations}

Let us consider a random \textsc{xorsat} formula with $N$ variables and a total of $M$ clauses of different lengths $j = 2, 3, \dots, k$. We don't consider clauses of length 1 since they are trivial, and we denote by $k$ the maximum clause length. The number of clauses of length $j$ will be denoted by $C_j(0)$, where the 0 indicates that this is the initial formula (relative to the action of the leaf-removal), and we will have $M = \sum_{j=2}^k C_j(0) = \alpha N$ for some finite $\alpha$. We shall also denote by $N_\ell(0)$ the number of variables with $\ell$ occurrences, and therefore $\sum_{\ell=0}^\infty N_\ell(0) = N$. We assume that the formula is formed by selecting uniformly at random the index of the variable appearing in each ``slot'' of each clause (with no repetitions within a clause). The distribution of the number of occurrences of the variables in the formula is then a poissonian with parameter $\lambda(0)$. Notice that the distribution of occurrences is \emph{independent} on the clause lengths (i.e. the distribution of occurrences in clauses of length $j$ is the same for all $j$).

The leaf-removal proceeds in steps. Let us denote by $T$ the number of steps that have been performed. At each step, a single-occurrence variable is selected, and the clause in which it appears is removed. What is the probability $p(j)$ that a single occurrence variable appears in a clause of length $j$? By definition, a single-occurrence variable occupies a unique slot in the formula. Since each slot can contain any variable with uniform probability, $p(j)$ will be proportional to the fraction of slots that belong to clauses of length $j$:
\begin{equation}
 p(j) = \frac{j C_j}{\sum_j j C_j} \,.
\end{equation}
If we denote by $C_j(T)$ the number of clauses of length $j$ after $T$ steps of leaf-removal, we shall have
\begin{equation}
 \mathbb E [ C_j(T+1) - C_j(T) ] = -p(j) = - \frac{j C_j(T)}{\sum_j j C_j(T)} \,.
 \label{extensive_c}
\end{equation}
Moreover, if the removed clause has length $j$, the average number of variables (excluding the one to be eliminated) with $\ell$ occurrences that appear in it will be $(j-1)\ell N_\ell(T) / \sum_{\ell'} \ell' N_{\ell'}$, and therefore
\begin{equation}
 \mathbb E [ N_\ell(T+1) - N_\ell(T) | j ] = (j-1) \frac {(\ell+1) N_{\ell+1}(T) - \ell N_\ell(T)} {\sum_\ell \ell N_\ell(T)} + \delta_{\ell,0} - \delta_{\ell,1}
\label{extensive n for fixed j}
\end{equation}
where the Kronecker deltas come from the single occurrence variable being eliminated. Multiplying by $p(j)$ and summing over $j$,
\begin{eqnarray}
 \mathbb E [ N_\ell(T+1) - N_\ell(T) ] 
&=& \sum_{j=2}^k p(j) \mathbb E [ N_\ell(T+1) - N_\ell(T) | j ] \\
&=& \sum_{j=2}^k \frac{j C_j(T)}{\sum_j j C_j(T)} (j-1) \frac {(\ell+1) N_{\ell+1}(T) - \ell N_\ell(T)} {\sum_\ell \ell N_\ell(T)} + \delta_{\ell,0} - \delta_{\ell,1} \,.
 \label{extensive_n}
\end{eqnarray}
In the limit $N \to \infty$ the variations in (\ref{extensive_c}) and (\ref{extensive_n}) are of $O(1)$, and we can apply Wormald's theorem to obtain the following differential equations for $n_\ell(t) \equiv \mathbb E [N_\ell(Nt)/N]$ and $c_j(t) \equiv \mathbb E [C_j(Nt)/N]$:
\begin{equation}
 \left\{
 \begin{aligned}
  \frac {d c_j(t)} {dt} 
	&= - \frac {j c_j(t)}{\sum_{j'=2}^k j' c_{j'}(t)} \,, \\
  \frac{d n_\ell(t)}{dt} &= \sum_{j=2}^k \frac{j(j-1)c_j(t)}{\sum_{j'=2}^k j' c_{j'}(t)} \frac {(\ell+1)n_{\ell+1}(t) - \ell n_\ell(t)} {\sum_{\ell'=0}^\infty \ell' n_{\ell'}(t)} + \delta_{\ell,0} - \delta_{\ell,1} \,.
 \end{aligned}
 \right.
 \label{gen_leaf-removal}
\end{equation}

The initial conditions for $c_j(t)$ are trivial (i.e. $c_j(0) = C_j(0) / N$), while those for $n_l$ are:
\begin{equation}
 n_\ell(0) = e^{-\lambda(0)} \frac{\lambda(0)^\ell}{\ell!} \,.
\end{equation}
Since the parameter of the poissonian coincides with its average, we shall have
\begin{equation}
 \lambda(0) = \sum_{\ell=0}^\infty \ell n_\ell(0) = \sum_{j=2}^k j c_j(0)
 \label{lambda_0}
\end{equation}
where the last equality comes from the fact that the two sums in (\ref{lambda_0}) give the number of slots in the formula, and therefore are equal. 

\subsection{Solution for $c_j(t)$}

In order to solve (\ref{gen_leaf-removal}), we observe two things: first, that the equations for $\{c_j(t)\}$ are independent on $\{n_\ell(t)\}$; second, that the equation for $c_k(t)$ implies that, as long as $c_k(t) > 0$, it is a strictly decreasing function of $t$, and therefore $c_k$ can be used as an independent variable instead of $t$. We then divide the equations for $c_j$ with $j=2,3,\dots,k-1$ by the equation for $c_k$, obtaining
\begin{equation}
 \frac{dc_j}{dc_k} = \frac j k \frac {c_j} {c_k}
\end{equation}
which admits the solution
\begin{equation}
 c_j = c_j^0 \left(\frac{c_k}{c_k^0}\right)^{j/k}
 \label{sol c_j vs c_k}
\end{equation}
where $c_j^0$ is the value of $c_j$ at $t=0$. It is convenient to introduce
\begin{equation}
 b(t) \equiv \left( \frac {c_k(t)} {c_k(0)} \right) ^ {1/k}
\end{equation}
so that (\ref{sol c_j vs c_k}) becomes
\begin{equation}
 c_j(t) = c_j(0) b(t)^j \,.
\end{equation}
Notice that $b$ is an invertible function of $c_k$, and therefore of $t$.

Let us also introduce the generating function $\gamma(b)$ of the $c_j(0)$, which will play a very important role in the following:
\begin{equation}
 \gamma(b) \equiv \sum_{j=2}^k c_j(0) b^j 
 \label{def gamma}
\end{equation}
so that
\begin{equation}
 \gamma(b(t)) = \sum_{j=2}^k c_j(0) b(t)^j = \sum_{j=2}^k c_j(t)
\end{equation}
is the total number of clauses at time $t$ and
\begin{equation}
 b(t) \gamma'(b(t)) \equiv b(t) \left. \frac{d\gamma}{db} \right|_{b=b(t)} = \sum_{j=2}^k j c_j(t) = \sum_{\ell=1}^\infty \ell n_\ell(t)
\end{equation}
is the number of slots in the formula at time $t$. Since exactly one equation is removed at each step, one must have
\begin{equation}
 \gamma(b(t)) = \alpha - t
 \label{gamma(b(t))}
\end{equation}
which implicitly defines $b(t)$ through (\ref{def gamma}):
\begin{equation}
 t(b) = \alpha - \sum_{j=2}^k c_j(0) b^j \,.
\label{t(b)}
\end{equation}

\subsection{Solution for $n_\ell(t)$}

We can now write the equations for $\{n_\ell(t) \, | \, \ell \geq 2\}$ in (\ref{gen_leaf-removal}) as
\begin{equation}
 \frac{dn_\ell}{dt} = \left. \frac {\gamma''(b)b^2}{[\gamma'(b)b]^2} \right|_{b=b(t)} \left[(\ell+1)n_{\ell+1} - \ell n_\ell\right]
= \left. \frac{\gamma''(b)}{\gamma'(b)^2} \right|_{b=b(t)} \left[(\ell+1)n_{\ell+1} - \ell n_\ell\right] \,.
\label{dnell/dt}
\end{equation}
As in the case of pure \kxorsat\ formul\ae\ the distribution of occurrences (for $\ell \geq 2$) remains poissonian at all times:
\begin{equation}
 n_\ell(t) = e^{-\lambda(t)} \frac {\lambda(t)^\ell}{\ell!} \hskip .5cm (\ell \geq 2) \,.
\end{equation}
This is easily seen by substituting this expression in (\ref{dnell/dt}), which gives an equation for $\lambda$ which is independent on $\ell$:
\begin{equation}
 \frac{d\lambda}{dt} = - \frac{\gamma''(b)}{\gamma'(b)^2} \lambda
\end{equation}
where $b=b(t)$. This is solved by noticing from (\ref{t(b)}) that 
\begin{equation}
 \frac{dt}{db} = -\gamma'(b)
\end{equation}
so that
\begin{equation}
 \frac{d\lambda}{db} = \frac{\gamma''(b)}{\gamma'(b)} \lambda
\end{equation}
with the initial condition that for $t=0$, which corresponds to $b=1$, $\lambda$ must be equal to $\lambda(0) = \sum_j j c_j(0)$. The solution is then:
\begin{equation}
 \lambda(b) = \gamma'(b)
\end{equation}
and we obtain
\begin{equation}
 n_\ell(b) = e^{-\gamma'(b)} \frac{\gamma'(b)^\ell}{\ell!} \hspace{.5cm} (\ell \geq 2)
\end{equation}
with $b=b(t)$ obtained by inverting $t = \alpha - \gamma(b)$.

For $\ell = 1$ we write
\begin{equation}
 n_1(b) 
= \sum_{\ell = 1}^\infty \ell n_\ell(b) - \sum_{\ell = 2}^\infty \ell n_\ell(b)
= b \gamma'(b) - e^{-\gamma'(b)} \gamma'(b) \left[ e^{\gamma'(b)} - 1 \right] \,.
\end{equation}
The leaf-removal will end when $n_1(b) = 0$ for some $b \in [0,1]$, which gives:
\begin{equation}
 b = 1 - e^{-\gamma'(b)} \,.
\label{b clustering}
\end{equation}
Let us denote by $b^*$ the largest solution of this equation. If $b^* = 0$ the leaf-removal removes all the clauses from the formula, which is \sat\ (with probability 1), and the solutions are unclustered. If $b^* > 0$ the leaf-removal ends with an irreducible sub-formula. The number of clauses in the sub-formula is $N \sum_{j=2}^k c_j(t^*) = N \gamma(b^*)$ and the number of variables is $N \sum_{\ell=2}^\infty n_\ell(b^*) = N e^{-\gamma'(b^*)} \left[ e^{\gamma'(b^*)} - 1 - \gamma'(b^*) \right]$. The sub-formula is \sat\ if and only if the number of variables is smaller than or equal to the number of clauses:
\begin{equation}
 \gamma(b^*) \leq b^* + (1-b^*) \log(1-b^*)
\label{sat/unsat transition}
\end{equation}
where we have used (\ref{b clustering}). The \sat/\unsat\ transition occurs when this bound is saturated.

\section{Characterization of the phases in terms of a potential}
\label{Sec_potential}

\subsection{Definition and properties of the potential $V(b)$}
\label{Par_def potential}

Let us define the following \emph{potential}, which is a function of $b$:
\begin{equation}
 V(b) = -\gamma(b) + b + (1-b) \log (1-b) \,.
\label{V(b)}
\end{equation}
The derivative of $V(b)$ is
\begin{equation}
 V'(b) = -\gamma'(b) - \log(1-b) \,,
\end{equation}
and we see that for $b=b^*$, which verifies (\ref{b clustering}) $\Leftrightarrow \gamma'(b^*) = - \log (1-b^*)$, we have
\begin{equation}
 V'(b^*) = 0 \,.
\end{equation}
The value of $b^*$ can therefore be obtained from $V(b)$, looking for the largest value in $[0,1]$ where the derivative of $V$ vanishes.

In the unclustered phase, $V(b)$ has a unique minimum at $b^* = 0$. As $\alpha$ grows, a secondary minimum develops for $b^* > 0$. The clustering transition occurs when this secondary minimum forms, and when this happens one must have $V''(b^*) = 0$.  On the other hand, from (\ref{sat/unsat transition}) and (\ref{V(b)}) one sees that the \sat/\unsat\ transition occurs when $V(b^*) = 0$. As in the pure case of paragraph \ref{Sec_pure-leaf-removal}, $b^*$ is the size of the back-bone, i.e. the fraction of variables that take the same value in each solution of a given cluster.

It is therefore possible to characterize the phase to which the formula belongs in terms of $V(b)$:
\begin{align}
 & \text{Back-bone size: } && \hspace{-3cm} b^* = \max_{b \in [0,1]} \{ b : V'(b) = 0 \} \label{cond b*} \\
 & \text{Clustering transition: } && \hspace{-3cm} V''(b^*) = 0 \label{cond clustering} \\
 & \text{\sat/\unsat\ transition: } && \hspace{-3cm} V(b^*) = 0 \label{cond sat/unsat}
\end{align}
An example of potential is provided in Figure \ref{Fig_potential}. The formul\ae\ considered for each curve are generated by the UC heuristic applied to a \kxorsat[3] formula with $\alpha = 0.8$. Each curve corresponds to a different time during the evolution under UC (more detailed explanations are given in Section \ref{Sec_trajectories}).

\begin{figure}
 \centering
 \includegraphics[width=9cm]{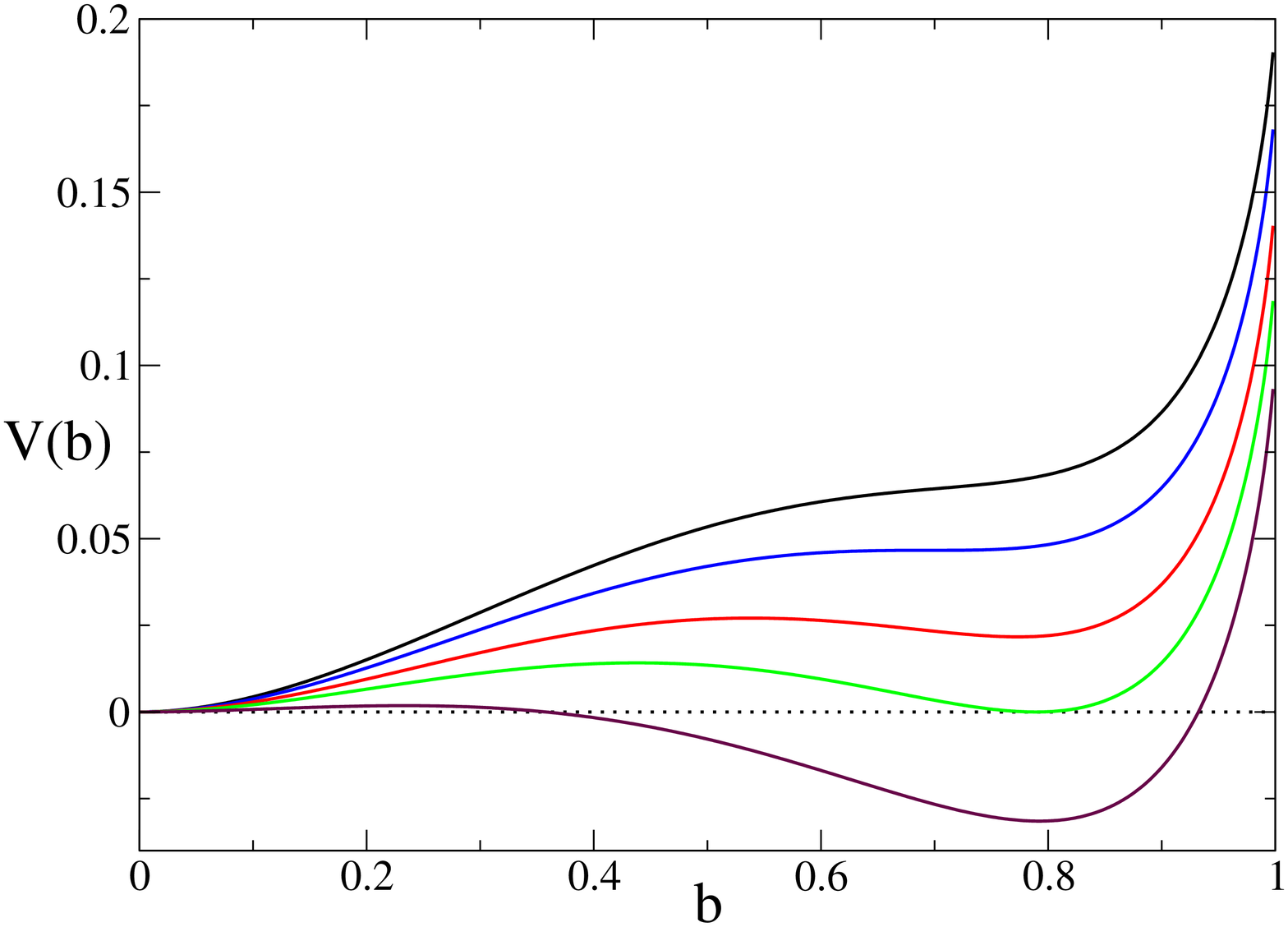}
 \caption{
  Potential $V(b)$ for different formul\ae. Each was obtained by applying the UC heuristic to a \kxorsat[3] formula with $\alpha=0.8$ for different times: from top to bottom $t = \{0, t_\mathrm c = 0.02957, 0.07327, t_\mathrm s = 0.11697, 0.20642 \}$. The first curve shows that the formula is in the unclustered phase; the second curve corresponds to the clustering transition; the third to a clustered formula; the fourth to the \sat/\unsat\ transition; finally, the formula is \unsat.
 }
 \label{Fig_potential}
\end{figure}

Notice that, given an arbitrary set of clause densities $\{c_2,\dots,c_k\}$, it is not \emph{a priori} a trivial task to determine whether random formul\ae\ conditioned by $\{c_j\}$ are \sat\ or not, and if they are \sat, whether their solutions are clustered or not. However, it suffices to compute $V(b)$ for the given set of $c_j$'s and, from its ``shape'', the answers to the previous questions become immediately clear. This is what makes the potential $V(b)$ such a powerful tool in the study of the phase transitions of \kxorsat\ (and of \uecsp).

Interestingly, a ``potential method'' was already well known in mean field theory of spin glasses. It was originally introduced by Parisi \cite{Parisi94} and developed by him and Franz \cite{Franz97, Franz97a, Franz98, Franz98a} and by Monasson \cite{Monasson95}. ``Their'' potential is derived in a completely different way: it is defined by considering two \emph{real} replicas of the system (i.e. two identical samples), with an interaction term that depends on the overlap $q$ between their configurations. The first replica is allowed to equilibrate at temperature $T$, without ``feeling'' the effect of the coupling, while the second replica equilibrates at the same temperature but is subject to the interaction. The effect of the interaction is to constrain the configurations of the second replica to those that have a fixed overlap with the equilibrium configurations of the first one. The potential $V(q)$ is then defined as the free energy of the second replica as a function of $q$, in the limit in which the interaction strength vanishes.

Even though the potential $V(q)$ is defined in a completely different way from the potential $V(b)$ defined in this Section, the two share many common features. First, both are functions of the overlap $q$, or equivalently of the fraction of frozen spins $b$; second, the properties of their minima determine the phase transitions of the system (in this regard, the definition of $V(q)$ as a free energy is much more transparent); third, the value of the potential corresponding to the secondary minimum (when it is present) is equal to the complexity. In fact, it should be possible to prove that the two potentials are actually identical by computing the full expression of the 1RSB free energy of \kxorsat\ in the case of a mixed system, and deriving the explicit expression of $V(q)$ in the most general case. The fact that the same potential can be obtained following two approaches that are so different is a very interesting fact in itself.

\subsection{Phase diagram for mixed \kxorsat\ formul\ae}
\label{Par_phase diagram}

For pure formul\ae\ the phase diagram depends on a single parameter, $\alpha$. For mixed formul\ae, the phase diagram is more complicated, as the space of parameters is $\mathcal C = \{c_2, \dots, c_k\}$ which has dimension $k-1$. Each one of the $c_j$'s varies in $[0,1]$, because if some $c_{j'} > 1$ then the sub-formula containing only the clauses of length $j'$ is \unsat (and therefore so is the complete formula). For any point $\mathbf c \in \mathcal C$ we can compute the potential $V(b)$, which depends on $\mathbf c$ through $\gamma(b) = \sum_{j=2}^k c_j b^j$, and we can define $b^*$ as the largest solution of $b = 1 - e^{-\gamma'(b)}$ in $[0,1]$.

The phase transitions are characterized by the conditions (\ref{cond clustering}) and (\ref{cond sat/unsat}). The boundary between the unclustered and the clustered phase will be the $(k-2)$-dimensional surface $\Sigma_\mathrm c$ defined by:
\begin{equation}
 \Sigma_\mathrm c = \left\{ \mathbf c \in \mathcal C : (b^* > 0) \wedge (V''(b^*) = 0) \right\} \,.
\label{Sigma_c}
\end{equation}
The boundary between the \sat\ and the \unsat\ regions in $\mathcal C$ will be the $(k-2)$-dimensional surface $\Sigma_\mathrm s$ defined by:
\begin{equation}
 \Sigma_\mathrm s = \left\{ \mathbf c \in \mathcal C : (b^* > 0) \wedge (V(b^*) = 0) \right\} \,.
\label{Sigma_s}
\end{equation}

\begin{figure}[t]
 \includegraphics[width=.55\textwidth]{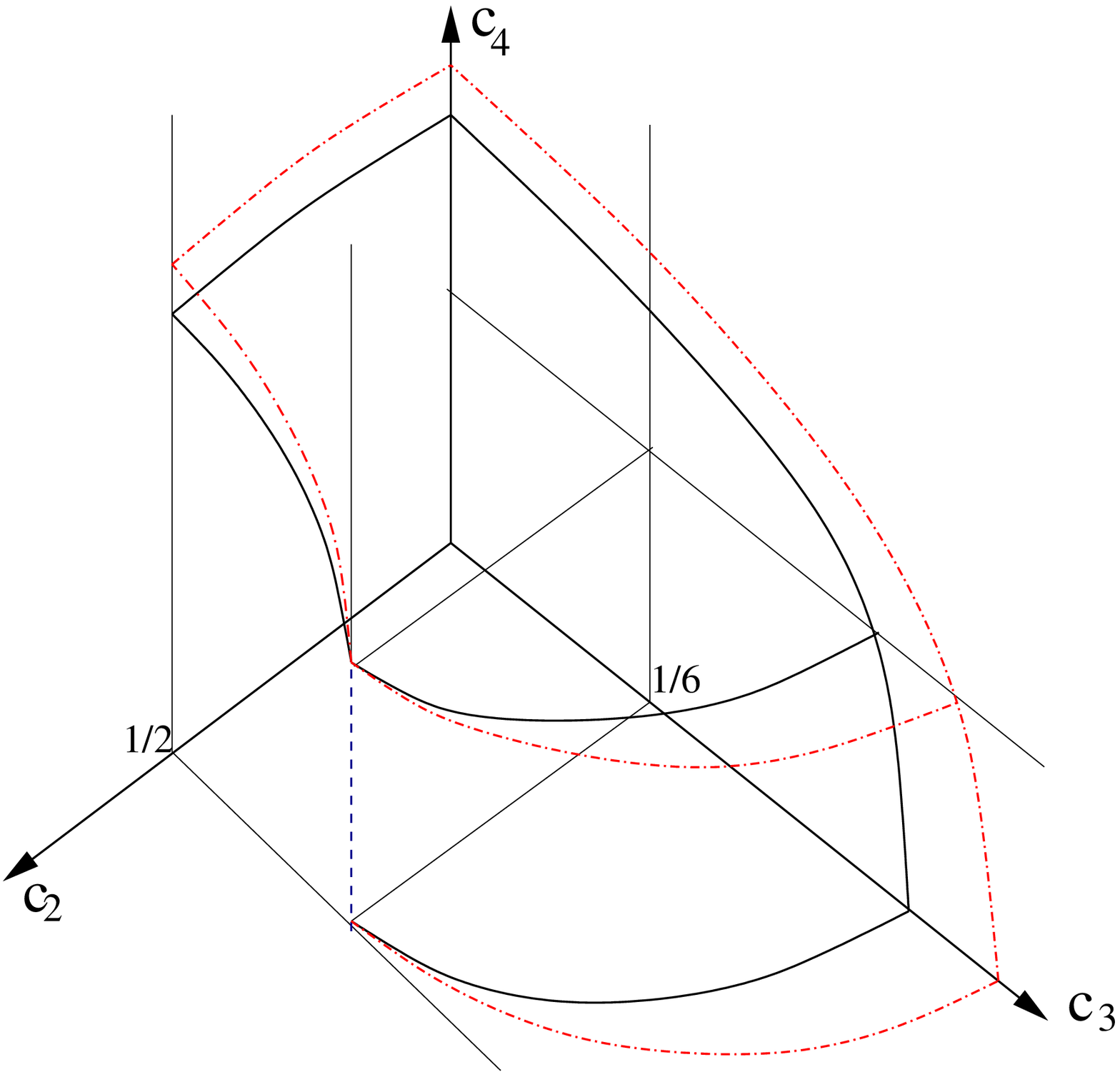}
 \includegraphics[width=.45\textwidth]{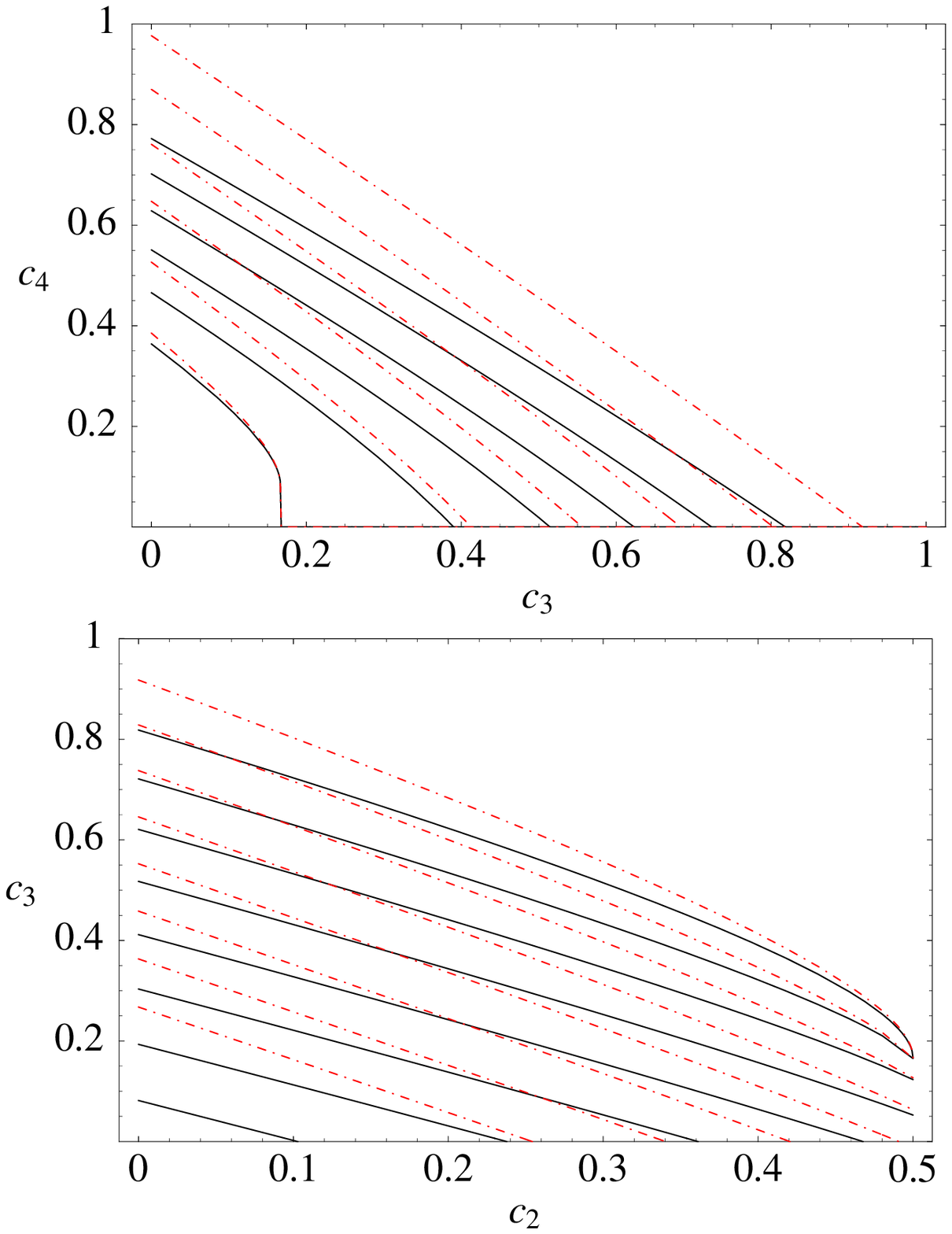}
 \caption{
  Phase diagram of mixed \kxorsat[4]. 
  \emph{Left}~A pictorial view of the surfaces $\Sigma_\mathrm c$ (full black) and $\Sigma_\mathrm s$ (dot-dashed red), intersecting on the segment $\Sigma_\mathrm k$ (dashed blue), where they are tangent to each other. Going from the origin out, the formul\ae\ are first unclustered, then clustered (after $\Sigma_\mathrm c$ is crossed) and finally \unsat\ (after $\Sigma_\mathrm s$ is crossed). 
  \emph{Right}~The sections of $\Sigma_\mathrm c$ (full black) and $\Sigma_\mathrm s$ (dot-dashed red) at constant $c_2 = \{0,0.1,0.2,0.3,0.4,0.5\}$ from top to bottom (top panel) and at constant $c_4 = \{0,0.1,0.2,0.3,0.4,0.5,0.6,0.7\}$ from top to bottom (bottom panel).
   The phase diagram for pure formul\ae\ with $k=3$ is formed by the $c_3$ axis of the bottom panel, which the two curves corresponding to $c_4=0$ intersect at $c_3 = \alpha = 0.818$ and $c_3 = \alpha = 0.918$ respectively.
 }
 \label{Fig_phases k=4}
\end{figure}

Notice that in $b=0$ one always has $V(0) = V'(0) = 0$, because the first term in $\gamma(b)$ is $c_2 b^2$ and $b + (1-b)\log(1-b) = b^2/2 + O(b^3)$ for small $b$. Also, for $c_2 = 1/2$ one has $V''(0) = 0$ (irrespectively of the values of $c_j$ for $j > 2$). Therefore, for $c_2 = 1/2$, $b=0$ is formally a solution of $V(b) = 0$ and of $V''(b) = 0$. Even though the surfaces $\Sigma_\mathrm c$ and $\Sigma_\mathrm s$ are defined with $b^* > 0$, it is possible that $b^* \to 0$ if the local minimum at $b^* > 0$ merges into the global minimum of $V$ in $b=0$. This can happen if and only if $V'''(0) = 0$ (so that $b=0$ becomes a ``flat'' saddle of $V$), which is obtained for $c_3 = 1/6$ (as seen by taking the term in $b^3$ in the above expansions). This implies that the two surfaces $\Sigma_\mathrm c$ and $\Sigma_\mathrm s$ intersect on the $(k-3)$-dimensional surface $\Sigma_\mathrm k$ of equation:
\begin{equation}
 \Sigma_\mathrm k = \left\{ \mathbf c \in \mathcal C : \left(c_2 = \frac 1 2\right) \wedge \left(c_3 = \frac 1 6\right) \right\}
\label{Sigma_k}
\end{equation}
The suffix `k' stands for \emph{critical} (the `c' being used for \emph{clustering}), because $\Sigma_\mathrm k$ is the surface where the discontinuous phase transitions (both the clustering and the \sat/\unsat\ ones) become continuous, which is traditionally called \emph{critical point} in statistical mechanics. 

The surfaces $\Sigma_\mathrm c$, $\Sigma_\mathrm s$ and $\Sigma_\mathrm k$ are tangent to each other. This can be seen by verifying that $c_2 = 1/2 - \epsilon^2$ and $c_3 = 1/6 + \epsilon$ verify (\ref{Sigma_c}), (\ref{Sigma_s}) and (\ref{Sigma_k}) with $b^* = \epsilon$ (to the leading order in $\epsilon \to 0$). The fact that $\Sigma_\mathrm c$ and $\Sigma_\mathrm s$ have an intersection where they are tangent to each other is not at all clear \emph{a priori} (as we have seen, it depends on the specific form of $V(b)$), but it will turn out to be extremely important in the following.

As an illustration, the phase diagram for $k=4$ is shown in Figure \ref{Fig_phases k=4}.

\section{Trajectories generated by poissonian heuristics}
\label{Sec_trajectories}

In Section \ref{Sec Algorithms} I introduced the DPLL procedure and discussed some properties of two specific heuristics, called respectively Unit Clause (UC) and Generalized Unit Clause (GUC), for the problem of \kxorsat[$(2+p)$]. In this section I shall extend the same kind of analysis to more general heuristics and to mixed formul\ae\ of any maximum length $k$.

I shall first define the class of heuristics considered, then derive some general properties of poissonian heuristics that will be useful in Section \ref{Sec_bounds}, and finally analyze the special cases of UC and GUC to illustrate them.

\subsection{Poissonian heuristics for DPLL}
\label{Par_def poissonian}

Let us consider a DPLL procedure without back-tracking acting on some pure \kxorsat\ formula. For the class of heuristics I want to introduce, it is convenient to modify the description of the procedure I gave in Section \ref{Sec_bounds} in such a way that unit propagations are performed by the heuristic. The modified procedure is described by the following pseudocode:
\algblockx[Switch]{switch}{swend}[1]{\textbf{switch} #1\textbf{:}}{\textbf{end switch}}
\algcblockx[Swelse]{Switch}{swelse}{swend}[1]{\textbf{otherwise} #1\textbf{:}}{\textbf{end switch}}
\begin{algorithmic}
 \Procedure{Modified DPLL}{$\{C_2(0),\dots,C_k(0)\}$}
 \Repeat
   \State{Select and assign a variable $x$ according to \textsc{Heuristic}}
   \State{Simplify the formula}
 \Until{A contradiction is generated {\bf or} All the variables are assigned}
 \EndProcedure
\end{algorithmic}
with the heuristic:
\begin{algorithmic}
 \Procedure{Poissonian Heuristic}{$\{p_j(C_1,\dots,C_k) \, | \, j = 0, \dots, k\}$}
 \switch{With probability $p_0(C_1,\dots,C_k)$}
  \State{Select uniformly at random a variable $x$}
  \State{Assign $x$ to \true\ or \false\ uniformly at random}
 \swelse{With probability $1-p_0(C_1,\dots,C_k)$}
  \State{Select at random a clause length $j \in \{1,\dots,k\}$ with probability $p_j(C_1,\dots,C_k)$}
  \State{Select uniformly at random a clause $\mathcal C$ of length $j$}
  \State{Select uniformly at random a variable $x$ appearing in $\mathcal C$}
  \State{Assign $x$ to \true\ or \false\ uniformly at random}
 \swend
 \EndProcedure
\end{algorithmic}
where $p_j(C_1,\dots,C_k)$ with $j=0,\dots,k$ are functions that characterize the heuristic.
The Unit Propagation rule then simply requires that $p_j(\{C_j\}) = \delta_{j,0}$ if $C_1 > 0$. 
Notice that $\{C_j\}$ are the extensive numbers of clauses of length $j$ in the specific formula we are considering (they are not averaged over the distribution of formul\ae). Moreover, since the alternatives corresponding to different values of $j=0,\dots,k$ are independent, it is possible to normalize the probabilities so that
\begin{equation}
 \sum_{j=0}^k p_j(C_1,\dots,C_k) = 1 \,. 
 \label{normalization p_j}
\end{equation}

It is easy to see that UC and GUC are special cases of this class of heuristics:
\begin{eqnarray}
 p_j^\mathrm{UC}(\{C_j\}) &=& 
  \begin{cases}
   \delta_{j,1} & \text{if $C_1 > 0$;} \\
   \delta_{j,0} & \text{otherwise.}
  \end{cases} 
  \label{p_j UC}\\
 p_j^\mathrm{GUC}(\{C_j\}) &=&
  \begin{cases}
   \delta_{j,1} & \text{if $C_1 > 0$;} \\
   \mathbb I [j\text{ is the lenght of the shortest clause in the formula}]
    & \text{otherwise.}
  \end{cases}
  \label{p_j GUC}
\end{eqnarray}

A very important property of this class of heuristics is that the sub-formul\ae\ that it generates are uniformly distributed, conditioned on the numbers $\{C_j\}$ of clauses of length $j$. As a consequence, the distribution of the number of occurrences of variables will remain poissonian under the action of these heuristics, even though the parameter of the poissonian may vary. This is the reason why I call this class of heuristics \emph{poissonian}.
In fact, I believe this to be the most general class of heuristics which preserve the uniform distribution of the sub-formul\ae\ it generates (even though I am unable to support this claim). 

Because of this property of the heuristics it is possible to analyze them in terms of differential equations, as we did for UC and GUC in Section \ref{Sec Algorithms}. We define the \emph{time} $t = T/N$ where $T$ is the number of variables that have been assigned, and the average clause densities $c_j(t) = \mathbb E[C_j(Nt)/N]$. The initial condition for the equations will be $c_j(0) = \alpha \delta_{j,k}$. Under the action of the heuristic, the formula will trace a trajectory in the space $\{c_j\} \subset [0,1]^{k-1}$. The dimension of the space is $k-1$ instead of $k$ because if at any time $c_1(t) > 0$ the procedure generates a contradiction with probability 1 and it fails. 

For notational convenience, I shall introduce $c_{k+1}(t) \equiv 0$ and $p_{k+1}(\{C_j\}) \equiv 0$. An analysis similar to that carried out in Section \ref{Sec Algorithms} for GUC then shows that the differential equations that determine $\{c_j\}$ are the following:
\begin{equation}
 \frac{dc_j}{dt} = \frac {(j+1)c_{j+1}(t) - j c_j(t)}{1-t} - \rho_j(t)
 \hspace{.5cm} (j=1,\dots,k)
 \label{differential equation for heuristic}
\end{equation}
where
\begin{equation}
 \rho_j(t) \equiv \lim_{\Delta T \to \infty} \frac 1 {\Delta T} \lim_{N \to \infty} \sum_{T = tN}^{tN + \Delta T - 1} \left[ p_j \bigl( \{C_{j'}(T)\} \bigr) - p_{j+1} \bigl( \{C_{j'}(T)\} \bigr) \right] \hspace{.5cm} (j=1,\dots,k)
\label{def rho_j}
\end{equation}
is (minus) the average variation of $c_j$ due to the the algorithm selecting $j+1$ or $j$ as the length for the clause from which to pick the variable to be assigned. In this equation $\Delta T$ is a number of steps of order $o(N)$, so that $c_j(t)$ can be considered constant over $\Delta T$, and which is a generalization of the ``round'' I introduced in the analysis of GUC. Notice that $\rho_j(t)$ depends on $t$ only through $\{c_j(t)\}$.

The first term in (\ref{differential equation for heuristic}) is due to the  other clauses of the formula in which the selected variable appears: on average, there will be $(j+1)c_{j+1}/(1-t)$ of them of length $j+1$ (which will become of length $j$) and $j c_j(t)/(1-t)$ of length $j$ (which will become of length $j-1$).

Since the density of unit clauses in the formula is always 0, for $j=1$ (\ref{differential equation for heuristic}) reduces to
\begin{equation}
 \frac{dc_1}{dt} = \frac{2c_2(t)}{1-t} - \rho_1(t) = 0
\end{equation}
which gives the explicit expression of $\rho_1(t)$ required to ensure Unit Propagation. The condition that signals the appearance of contradictions with probability 1 is
\begin{equation}
 \rho_1(t) = \frac{2 c_2(t)}{1-t} = 1 \,.
\end{equation}
I shall define one more $(k-2)$-dimensional surface in the phase diagram:
\begin{equation}
 \Sigma_\mathrm q = \left\{ \mathbf {\tilde c} \in [0,1]^{k-1} \, : \, \tilde c_2 = \frac 1 2 \right\}
\end{equation}
where the `q' stands for \emph{contradiction} (the `c' being very much in demand...) and where the tilde reminds us that these clause densities are normalized to the number of variables in the sub-formula, i.e. $\tilde c_j = c_j / (1-t)$.

A final remark to conclude this paragraph: since the distribution of occurrences remains poissonian at all times, the results of the previous section allow to characterize the phase to which the sub-formul\ae\ generated by the heuristics belong. The only difference is that the clause densities $c_j(t)$ are normalized to the number of variables $N$ in the initial formula, so the definition potential must be modified as follows:
\begin{eqnarray}
 V(b,t,\alpha) &\equiv& - \frac {\gamma'(b,t,\alpha)} {1-t} + b + (1-b) \log (1-b) \,, 
  \label{def V per euristica} \\
 \gamma(b,t,\alpha) &\equiv& \sum_{j=1}^k c_j(t) b^j \,.
  \label{def gamma per euristica} 
\end{eqnarray}
where the sum over $j$ can be extended to include 1 because $c_1(t) \equiv 0$. $V$ depends on $t$ and $\alpha$ through $\gamma$ and therefore through the $\{c_j(t)\}$ (which depend on $\alpha$ because of the initial condition). One should be careful not to confuse the time $t$ which appears in these equations with that introduced in the description of the leaf-removal of Section \ref{Sec_leaf-removal}: $t$ is the fraction of variables appearing in the original formula that have been assigned to obtain the sub-formula, to which the leaf-removal can then be applied.

In equation \ref{def V per euristica} the prime in $\gamma'(b,t,\alpha)$ denotes the partial derivative with respect to $b$. In the following I shall always denote derivatives with respect to $b$ with primes, and derivatives with respect to $t$ with dots (e.g. $\dot \gamma(b,t,\alpha)$). Derivatives with respect to $\alpha$ will be written explicitly.

It is convenient to supplement (\ref{def V per euristica}) and (\ref{def gamma per euristica}) with the generating function of the $\{\rho_j(t)\}$:
\begin{equation}
 \phi(b,t,\alpha) = \sum_{j=1}^k \rho_j(t) b^j 
\end{equation}
which will play an important role in the following.

\subsection{General properties of poissonian heuristics}
\label{Par_properties poissonian}

The rate at which clauses are removed from the formula is given by
\begin{equation}
 - \sum_{j=1}^k \dot c_j(t) = - \dot \gamma(1,t,\alpha) = \sum_{j=1}^k \rho_j(t)
\label{minus gamma dot}
\end{equation}
where the ``telescopic'' terms $(j+1)c_{j+1}-jc_j$ in $\dot c_j$ cancel each other. Since at each time step at most one clause is removed from the formula, one must have
\begin{equation}
 -\dot \gamma(1,t,\alpha) \leq 1 \,.
 \label{-gamma dot}
\end{equation}
This bound is saturated when $\rho_1(t) = 1$ which is the condition for the onset of contradictions.

Moreover, we can multiply (\ref{def rho_j}) by $j$ and sum over $j$ to obtain
\begin{equation}
 \sum_{j=1}^k j \rho_j(t) = \phi'(1,t,\alpha) = \lim_{\Delta T \to \infty} \frac 1 {\Delta T} \lim_{N \to \infty} \sum_{T = Nt}^{Nt+\Delta t-1} \sum_{j=1}^k p_j \left( \{C_j(T)\} \right) \leq 1
 \label{sum j rho_j}
\end{equation}
because of the normalization condition (\ref{normalization p_j}).

More generally, if we denote the average over $\Delta T$ which appears in (\ref{def rho_j}) and (\ref{sum j rho_j}) with angled brackets $\left< \cdot \right>$, we have
\begin{equation}
 \rho_j(t) = \left<p_j\right> - \left<p_{j+1}\right>
\end{equation}
where each $\left< p_j \right>$ is non-negative and they are normalized so that $\sum_{j=0}^k \left< p_j \right> = 1$ (because each term in the sum defining the average over $\Delta T$ has these properties). Then we have
\begin{equation}
 \phi(b,t,\alpha) = \sum_{j=1}^k \rho_j(t) b^j = \sum_{j=1}^k \left<p_j\right> b^j - \sum_{j=2}^k \left< p_j \right> b^{j-1} \leq b \sum_{j=1}^k \left< p_j \right> b^{j-1} \leq b
\label{bound on phi(b,t,alpha)}
\end{equation}
since $b \in [0,1]$. Moreover,
\begin{equation}
 \phi'(b,t,\alpha) = \sum_{j=1}^k j \rho_j(t) b^{j-1} = \left< p_1 \right> + \sum_{j=2}^k b^{j-2} \left[ 1 - j(1-b) \right] \left< p_j \right> \,.
\end{equation}
The coefficient in front of $\left< p_j \right>$ in the terms of the sum is maximum for $b=1$, independently of $j$, and is then equal to 1, so that
\begin{equation}
 \phi'(b,t,\alpha) \leq \left< p_1 \right> + \sum_{j=2}^k \left< p_j \right> = 1 - \left< p_0 \right> \leq 1 \,.
\label{bound on phi'(b,t,alpha)}
\end{equation}

These two bounds will be extremely useful in order to characterize the trajectories traced by poissonian heuristics. To do that, for each value of $\alpha$ in the original formula, we can define the three times $t_\mathrm c(\alpha)$, $t_\mathrm s(\alpha)$ and $t_\mathrm q(\alpha)$ at which the reduced sub-formul\ae\ cross respectively the clustering transition surface $\Sigma_\mathrm c$, the \sat/\unsat\ transition surface $\Sigma_\mathrm s$ and the contradiction surface $\Sigma_\mathrm q$ defined at the end of Section \ref{Sec_potential}. \emph{A priori} we could expect the trajectories to cross each surface more than once, and in this case we shall consider the times of first crossing. By doing this, we ensure that the three functions $t_\mathrm x(\alpha)$ (where `x' is `c', `s' or `q') are invertible, and we can define $\alpha_\mathrm x(t)$ as the value of $\alpha$ such that $t_\mathrm x = t$. On the other hand, it is possible that the trajectory never cross some (or all) of these surfaces, in which case the corresponding $t_\mathrm x(\alpha)$ will be undetermined.

Since the phase transitions are completely characterized by the potential $V$, these crossing times will be determined by the conditions (\ref{cond clustering}) and (\ref{cond sat/unsat}) on $V(b,t,\alpha)$: for given $\alpha$ and $t$ (and therefore for given $\{c_j\}$) we define $b^*$ by (\ref{cond b*}) as the largest solution of the equation $V'(b,t,\alpha) = 0$; then the clustering time $t_\mathrm c(\alpha)$ will be such that $V''(b^*,t_\mathrm c, \alpha) = 0$ and the \sat/\unsat\ time $t_\mathrm s(\alpha)$ will be such that $V(b^*, t_\mathrm s, \alpha) = 0$. As for the contradiction time $t_\mathrm q(\alpha)$, it is determined by the condition $2 c_2(t_\mathrm q) / (1-t_\mathrm q) = 1$.

Let us take the total time derivative of the condition that determines $b^*$, i.e. $V'(b^*,t,\alpha) = 0$:
\begin{equation}
 \frac d {dt} V'(b^*,t,\alpha) = V''(b^*,t,\alpha) \frac {db^*}{dt} + \frac {\de}{\de t} V'(b^*,t,\alpha) + \frac {\de}{\de \alpha} V'(b^*,t,\alpha) \frac {d \alpha}{dt} = 0 \,.
\end{equation}
The term in $db^* / dt$ is present because when $t$ changes, so do the values of $c_j(t)$ and therefore the coefficients in the power series that defines $V$, and the point where its derivative vanishes moves. In the same manner, if $b^*$ is held fixed, then as $t$ varies the only remaining parameter must vary as well, and this is $\alpha$, which gives rise to the term in $d \alpha / dt$.

At the clustering transition, $\alpha = \alpha_\mathrm c(t)$ and $b^* = b^*_\mathrm c$, the condition $V''(b^*_\mathrm d, t, \alpha_\mathrm c(t)) = 0$ is verified, so that the previous equation becomes
\begin{equation}
 \frac {d \alpha_c (t)} {dt} = - \frac {\dot V'(b^*_\mathrm c, t, \alpha_\mathrm c)} {\de_\alpha V'(b^*_\mathrm c, t, \alpha_\mathrm c)}
\label{de alpha / de t clustering}
\end{equation}
where, let me stress it again, $\alpha_\mathrm c \equiv \alpha_\mathrm c(t)$ is the value of $\alpha$ such that the trajectory crosses $\Sigma_\mathrm c$ at time $t$, and where the dot denotes a \emph{partial} time derivative.

From the definition (\ref{def V per euristica}) we have:
\begin{eqnarray}
 \dot V'(b,t,\alpha) &=& - \frac{1}{1-t} \left[ \dot \gamma'(b,t,\alpha) + \frac{\gamma'(b,t,\alpha)} {1-t} \right] \,, \\
 \de_\alpha V'(b,t,\alpha) &=& - \frac 1 {1-t} \de_\alpha \gamma'(b,t,\alpha) 
\,.
\end{eqnarray}
We can substitute these two expressions into (\ref{de alpha / de t clustering}) to obtain:
\begin{equation}
 \frac {d \alpha_c (t)} {dt} = - \left. \frac {\dot \gamma'(b,t,\alpha) + \gamma'(b,t,\alpha) / (1-t)} {\de_\alpha \gamma'(b,t,\alpha)} \right|_{b = b^*_\mathrm c(t), \alpha = \alpha_\mathrm c(t)} \,.	
 \label{d alpha / dt clustering}
\end{equation}

From the equations of motion of the heuristic (\ref{differential equation for heuristic}) we obtain the following equation for $\dot \gamma$:
\begin{eqnarray}
 \dot \gamma(b,t,\alpha) &=& \sum_{j=1}^k \frac {dc_j}{dt} b^j = \sum_{j=1}^k \left[ \frac {(j+1)c_{j+1}-j c_j}{1-t} - \rho_j \right] b^j \nonumber \\
&=& \frac{1-b}{1-t}\gamma'(b,t,\alpha) - \phi(b,t,\alpha) \,.
\label{dot gamma}
\end{eqnarray}
Differentiating it with respect to $b$ we have:
\begin{equation}
 \dot \gamma'(b,t,\alpha) = - \frac 1 {1-t} \gamma'(b,t,\alpha) + \frac{1-b}{1-t} \gamma''(b,t,\alpha) - \phi'(b,t,\alpha) \,.
\end{equation}
For $b = b^*_c(t)$ we shall have $V' = V'' = 0$, and since
\begin{equation}
 V''(b,t,\alpha) = - \frac 1 {1-t} \gamma''(b,t,\alpha) + \frac 1 {1-b}
\end{equation}
we get
\begin{equation}
 \left[ \frac{1-b}{1-t} \gamma''(b,t,\alpha) = 1 \right]_{b=b^*_\mathrm c(t), \alpha = \alpha_\mathrm c(t)}
\end{equation}
so that the numerator in (\ref{d alpha / dt clustering}) becomes $1 - \phi'(b^*_\mathrm c, t, \alpha_\mathrm c)$ and we obtain:
\begin{equation}
 \frac {d \alpha_\mathrm c (t)} {d t} = - \left. \frac{1-\phi'(b,t,\alpha)}{\de_\alpha \gamma'(b,t,\alpha)} \right|_{b = b^*_\mathrm c(t), \alpha = \alpha_\mathrm c(t)} \,.
 \label{d alpha / dt clustering finale}
\end{equation}

This is where the bounds (\ref{bound on phi(b,t,alpha)}) and (\ref{bound on phi'(b,t,alpha)}) are important (actually, it's only the second of the two which is used here): since $\phi'(b,t,\alpha) \leq 1$ for any $b,t,\alpha$, the numerator is surely positive or null. Moreover, the denominator is positive at $t=0$, when $\gamma'(b,0,\alpha) = \alpha k b^{k-1}$ independently of the heuristic. We then have to cases:
\begin{description}
 \item[Case 1]~The denominator remains positive at all times, in which case $d \alpha_\mathrm c(t) / dt$ is always negative and $\alpha_c(t)$ is a decreasing function of $t$, which implies that $t_\mathrm c(\alpha)$ is a decreasing function of $\alpha$; \\
 \item[Case 2]~If $\de_\alpha \gamma'(b,t,\alpha)$ vanishes for some value of $t$ (for a given $\alpha$), the denominator in (\ref{d alpha / dt clustering finale}) vanishes. Then $\de_\alpha t_\mathrm c(\alpha) = 0$ and $t_\mathrm c(\alpha)$ has either an extremum or an inflection point. After that, the curve will continue (with decreasing values of $\alpha$). The curve of $t_\mathrm c(\alpha)$ cannot reach the axis $\alpha = 0$ (because for $\alpha = 0$ the formula is surely unclustered, and there is no $t_\mathrm c$), and neither can it reach the $t=0$ axis (because at $t=0$ we have a pure \kxorsat\ formula, and we know that it has a unique clustering transition), so it will end at some terminal point.
\end{description}
In both cases, $t_\mathrm c(\alpha)$ is a single valued function of $\alpha$. It is the numerator of (\ref{d alpha / dt clustering finale}), not the denominator, which should change sign in order for $t_\mathrm c(\alpha)$ to take multiple values. But this cannot happen because of (\ref{bound on phi'(b,t,alpha)}). An illustration of the possible shapes of the curves for $t_\mathrm x(\alpha)$ is given in Figure~\ref{Fig possible shapes of t(alpha)}.

\begin{figure}
 \centering
 \includegraphics[width=0.6\textwidth]{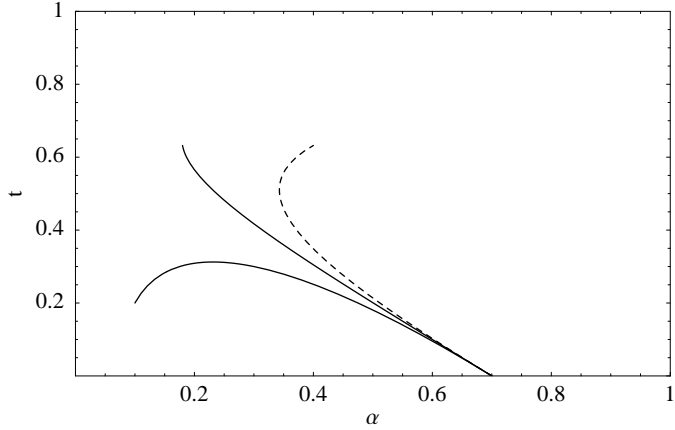}
 \caption{
  Possible shapes for the curves $t_\mathrm x(\alpha)$ (`x' being `c' or `s'). $t_\mathrm x$ is a strictly decreasing function of $\alpha$ if the denominator in (\ref{d alpha / dt clustering finale}) or (\ref{d alpha / dt sat/unsat final}) never vanishes (middle full curve). If instead it does vanish and then changes sign, $t_\mathrm x$ will develop a maximum and then continue to the left with positive derivative, but it will remain a single-valued function of $\alpha$ (bottom full curve). What cannot occur (top dashed curve) is that $t_\mathrm x'(\alpha)$ diverges and then changes sign, making $t_\mathrm x$ a multiple-valued function of $\alpha$: this would require the numerator in (\ref{d alpha / dt clustering finale}) or (\ref{d alpha / dt sat/unsat final}) to become negative, which cannot occur because of the bounds (\ref{bound on phi(b,t,alpha)}) and (\ref{bound on phi'(b,t,alpha)}). In Section~\ref{Sec_bounds} I shall prove that actually the curve representing $t_\mathrm x (\alpha)$ must end at a point where its derivative is infinite, as in the case of the middle full curve.
 }
 \label{Fig possible shapes of t(alpha)}
\end{figure}

Notice that, even though we considered initially the possibility that the trajectory cross several times $\Sigma_\mathrm c$, and defined $t_\mathrm c$ as the time of the \emph{first} crossing, the argument I just exposed shows that there can be at most one crossing. We shall see that this fact has profound implications for the performance of poissonian heuristics. Before doing that, however, let me derive an analogous argument for $t_\mathrm s(\alpha)$.

We start by taking the total time derivative of the potential,
\begin{equation}
 \frac d {dt} V(b,t,\alpha) = V'(b,t,\alpha) \frac{db}{dt} + \dot V(b,t,\alpha) + \de_\alpha V(b,t,\alpha) \frac{d\alpha}{dt} \,.
\end{equation}
At the \sat/\unsat\ transition, $b = b^*_\mathrm s(t)$, $\alpha = \alpha_\mathrm s(t)$ and $V = V' = 0$ from (\ref{cond b*}) and (\ref{cond sat/unsat}), so that we obtain:
\begin{equation}
 0 = \left[ \dot V(b,t,\alpha) + \de_\alpha V(b,t,\alpha) \frac{d \alpha}{dt} \right]_{b = b^*_\mathrm s(t), \alpha = \alpha_\mathrm s(t)}
\end{equation}
from which
\begin{equation}
 \frac{d \alpha_\mathrm s}{dt} = - \left. \frac{\dot V(b,t,\alpha)}{\de_\alpha V(b,t,\alpha)} \right|_{b = b^*_\mathrm s(t), \alpha = \alpha_\mathrm s(t)} \,.
\label{d alpha / dt sat/unsat}
\end{equation}

We can now substitute (\ref{dot gamma}) in the partial time derivative of the potential (\ref{def V per euristica}) to obtain:
\begin{eqnarray}
 \dot V(b,t,\alpha) &=& - \frac{1}{1-t} \left[\dot \gamma(b,t,\alpha) + \frac{\gamma(b,t,\alpha)}{1-t} \right] \nonumber \\
&=& - \frac 1 {1-t} \left[ \frac{1-b}{1-t}\gamma'(b,t,\alpha) - \phi(b,t,\alpha) + \frac{\gamma(b,t,\alpha)}{1-t} \right] \,.
\label{dot V sat/unsat}
\end{eqnarray}
At the \sat/\unsat\ transition we have
\begin{eqnarray}
 V(b^*_\mathrm s, t, \alpha_\mathrm s) = 0 &\Rightarrow& \frac{\gamma(b^*_\mathrm s, t, \alpha_\mathrm s)}{1-t} = b^*_\mathrm s + (1-b^*_\mathrm s)\log(1-b^*_\mathrm s) \\
 V'(b^*_\mathrm s, t, \alpha_\mathrm s) = 0 &\Rightarrow& \frac{\gamma'(b^*_\mathrm s, t, \alpha_\mathrm s)}{1-t} = - \log(1-b^*_\mathrm s)
\end{eqnarray}
so that (\ref{dot V sat/unsat}) reduces to 
\begin{equation}
 \dot V(b^*_\mathrm s, t, \alpha_\mathrm s) = - \frac 1 {1-t} \left[b^*_\mathrm s - \phi(b^*_\mathrm s, t, \alpha_\mathrm s) \right] \,.
\end{equation}

By substituting this in the numerator of (\ref{d alpha / dt sat/unsat}) we obtain:
\begin{equation}
 \frac {d \alpha_\mathrm s(t)}{dt} = -\left. \frac{b - \phi(b,t,\alpha)}{\de_\alpha \gamma(b,t,\alpha)} \right|_{b = b^*_\mathrm s(t), \alpha = \alpha_\mathrm s(t)} \,.
\label{d alpha / dt sat/unsat final}
\end{equation}
The argument now goes as for $\alpha_\mathrm c(t)$: the bound (\ref{bound on phi(b,t,alpha)}) ensures that the numerator is non-negative, and the denominator is positive at $t=0$, so that $t_\mathrm s(\alpha)$ must be single valued.

To summarize, in this paragraph I have shown that the trajectories described by poissonian heuristics can cross the clustering transition surface $\Sigma_\mathrm c$ and the \sat/\unsat\ transition surface $\Sigma_\mathrm s$ only once. Moreover, it is clear that if they reach the contradiction surface $\Sigma_\mathrm q$ the algorithm stops, and the crossing of $\Sigma_\mathrm q$ must also be unique.

\subsection{Analysis of UC and GUC}
\label{Par_UC and GUC}

In this paragraph I shall give some examples of the results of the previous paragraph based on two poissonian heuristics that are particularly simple to analyze: UC and GUC.

\subsubsection*{Analysis of UC}

The equations of motion for UC are obtained from (\ref{p_j UC}) and (\ref{differential equation for heuristic}):
\begin{equation}
 \frac{d c_j}{dt} = \frac{(j+1)c_{j+1} - jc_j}{1-t}  \hspace{0.5cm} (j \geq 2)
\end{equation}
with the initial condition $c_j(0) = \alpha \delta_{j,k}$. The solution is straightforward:
\begin{equation}
 c_j^\mathrm{UC}(t) = \alpha \binom k j (1-t)^j t^{k-j} \hspace{0.5cm} (j \geq 2) \,.
\end{equation}
As usual $\rho_1 = 2 c_2^\mathrm{UC} / (1-t)$ and for all $j > 1$ the corresponding $\rho_j = 0$. This is a direct consequence of (\ref{p_j UC}): the requirement for Unit Propagation is that the above expression of $\rho_1$ be true, and if there are no unit clauses $p_0 = 1$ and all the other $p_j$'s are 0.

We can explicitly compute $\gamma$, $V$ and $\phi$:
\begin{eqnarray}
 \gamma^\mathrm{UC}(b,t,\alpha) &=& \sum_{j=2}^k c_j^\mathrm{UC}(t) b^j = \alpha \left[t + b(1-t) \right]^k - \alpha k (1-t) t^{k-1} b - \alpha t^k
\label{gamma UC} \,, \\
 V^\mathrm{UC}(b,t,\alpha) &=& - \frac{\gamma'(b,t,\alpha)}{1-t} + b + (1-b) \log(1-b)
\nonumber \\
&=& \alpha k t^{k-1} - \alpha k \left[t + b(1-t)\right]^{k-1} + b + (1-b) \log(1-b) 
\label{V UC} \,, \\
 \phi^\mathrm{UC}(b,t,\alpha) &=& \sum_{j=1}^k \rho_j(t) b^j = \rho_1(t) b = \frac{2 c^\mathrm{UC}_2(t)}{1-t} b = \alpha k(k-1)(1-t)t^{k-2} b
\label{phi UC} \,.
\end{eqnarray}
An example of the potential $V^\mathrm{UC}(b,t,\alpha)$ for $k=3$ is plotted as a function of $b$ for different values of $t$ and for fixed $\alpha = 0.8$ in Figure \ref{Fig_potential}.

The times at which the trajectories cross $\Sigma_\mathrm c$ and $\Sigma_\mathrm s$ are obtained by solving (numerically) for $b$ and $t$ with fixed $\alpha$ the equations $\bigl\{ \bigl( {V^\mathrm{UC}}' = 0 \bigr) \wedge \bigl( {V^\mathrm{UC}}'' = 0 \bigr) \bigr\}$ and $\bigl\{ \bigl( {V^\mathrm{UC}}' = 0 \bigr) \wedge \bigl( V^\mathrm{UC} = 0 \bigr) \bigr\}$ (respectively).
The bounds (\ref{bound on phi(b,t,alpha)}) and (\ref{bound on phi'(b,t,alpha)}) obviously hold, since $\phi^\mathrm{UC}$ is simply $\rho_1 b$ and $\rho_1 \leq 1$, with the equal sign on the contradiction surface $\Sigma_\mathrm q$. Moreover, the denominator in (\ref{d alpha / dt clustering finale}) is $\de_\alpha \gamma' = k \{[t+b(1-t)]^{k-1}-t^{k-1}\} > 0$, and the denominator in (\ref{d alpha / dt sat/unsat final}) is $\de_\alpha \gamma$ which is also strictly positive. This ensures that $t_\mathrm c(\alpha)$ and $t_\mathrm s(\alpha)$ are strictly decreasing functions of $\alpha$.
The time at which contradictions are generated with probability 1 is obtained by solving $2 c_2^\mathrm{UC}(t) / (1-t) = 1$ for $t$ at fixed $\alpha$.
The plots of $t_\mathrm c(\alpha)$, $t_\mathrm s(\alpha)$ and $t_\mathrm q(\alpha)$ are shown in Figure~\ref{Fig t_c t_s and t_q for UC}.

\begin{figure}
 \centering
 \includegraphics[width=0.6\textwidth]{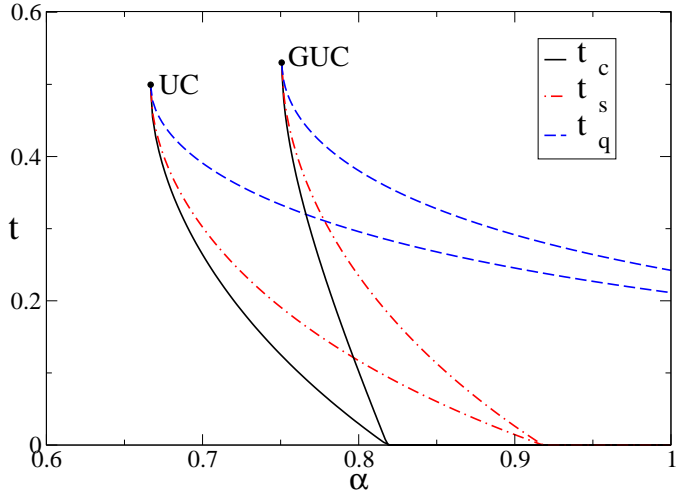}
 \caption{
  Times of crossing of $\Sigma_\mathrm c$, $\Sigma_\mathrm s$ and $\Sigma_\mathrm q$ for $k=3$ for UC and GUC. For $\alpha = \alpha_\mathrm c \simeq 0.818$ the initial formula is at the clustering transition and $t_\mathrm c = 0$ for both heuristics. The same happens with the \sat/\unsat\ transition at $\alpha = \alpha_\mathrm s \simeq 0.918$. As expected, $t_\mathrm c(\alpha)$ and $t_\mathrm s(\alpha)$ are single-valued. The fact that they are strictly decreasing means that for UC and GUC the denominators of (\ref{d alpha / dt clustering finale}) and (\ref{d alpha / dt sat/unsat final}) never change sign.
 }
 \label{Fig t_c t_s and t_q for UC}
\end{figure}

The largest value of $\alpha$ for which the algorithm finds a solution with finite probability (which I shall denote $\alpha_\mathrm h^\mathrm{UC}(k)$, the `h' standing for `heuristic') is the smallest value of $\alpha$ for which the trajectory crosses the \sat/\unsat\ transition surface $\Sigma_\mathrm s$. Alternatively, it can be computed as the smallest value of $\alpha$ for which the equation $2 c_2^\mathrm{UC}(t) / (1-t) = 1$ has a solution, which was done in Section \ref{Sec Algorithms}:
\begin{equation}
 \alpha_\mathrm h^\mathrm{UC}(k) = \frac 1 k \left( \frac{k-1}{k-2} \right)^{k-2} \,.
\end{equation}
For $k=3$ this is equal to $2/3$ and for large $k$ it goes as $e/k + O(k^{-2})$.

\subsubsection*{Analysis of GUC}

The analysis of GUC is slightly more complicated. The analysis
of Section \ref{Sec Algorithms} shows that the equations of motion are
\begin{equation}
 \frac{d c_j}{dt} = \frac{(j+1)c_{j+1} - j c_j}{1-t} - \delta_{j,j^*(t)} \left[ \frac 1 j - \frac{(j-1)c_{j}}{1-t} \right] \hspace{0.5cm} \bigl( j \geq j^*(t) \bigr)
\label{equations GUC}
\end{equation}
where $j^*(t)$ is the smallest value of $j$ such that $c_j(t) > 0$, assuming the initial condition $c_j(0) = \alpha \delta_{j,k}$. The interpretation of these equations is that GUC always assigns a variable appearing in the shortest clause (or possibly clauses) in the formula. As long as $j^* c_{j^*} / (1-t) \leq 1/(j^*-1)$ the rate at which clauses of length $j^*-1$ are generated is small enough that they can be removed, and the density of clauses of length $j^*-1$ remains 0; when this bound is violated, an extensive number of clauses of length $j^*-1$ accumulates, and $c_{j^*-1}$ becomes positive. I shall call $t^*(j)$ the time at which $c_j(t)$ becomes positive. When this happens, the value of $j^*$ is decreased by 1.
The equations (\ref{equations GUC}) therefore hold for $j \geq j^*(t)$, while $c_j(t) \equiv 0$ for all $j < j^*(t)$.

Even though it is in principle possible to solve (\ref{equations GUC}) exactly for any finite $k$, the solution becomes more and more complicated as $k$ increases, since it involves matching the solutions of different differential equations at $k-2$ points (at least for $\alpha$ large enough that $j^*$ reaches 2). I shall only give the example of $k=3$, for which one obtains:
\begin{eqnarray}
 c_3^\mathrm{GUC}(t) &=& \alpha (1-t)^3 \,, \\
 c_2^\mathrm{GUC}(t) &=& \frac 1 2 (1-t) \left\{ 3 \alpha \left[ 1 - (1-t)^2 \right] + \log(1-t) \right\} \,.
\end{eqnarray}

Notice that from (\ref{equations GUC}) it is clear that the $\rho_j(t)$ are all 0 except for two of them:
\begin{eqnarray}
 \rho_{j^*} &=& \frac 1 {j^*} - \frac{(j^*-1)c_{j^*}}{1-t} \,, \\
 \rho_{j^*-1} &=& \frac{j^* c_{j^*}}{1-t} \,.
\end{eqnarray}
For a fixed value of $j^* = \bar j$, $t$ varies between $t^*(\bar j)$ and $t^*(\bar j -1)$, and during this interval of time, $\bar j c_{\bar j}(t) / (1-t)$ varies between 0 and $1/(\bar j - 1)$, so that we have:
\begin{equation}
 \frac 1 {j^*(t)} \leq \rho_{j^*(t)} + \rho_{j^*(t)-1} \leq \frac 1 {j^*(t)-1} \,.
\label{bound rho_j* + rho_j*+1}
\end{equation}
It is easy to see that the bound expressed in (\ref{bound on phi(b,t,alpha)}) is respected (actually the previous inequality is even more stringent) and that the bound in (\ref{sum j rho_j}) is respected but saturated: $\sum_j j \rho_j(t) = 1$ at all times.

The crossing times of $\Sigma_\mathrm c$, $\Sigma_\mathrm s$ and $\Sigma_\mathrm q$ are computed solving numerically the equations obtained from the conditions (\ref{cond b*}), (\ref{cond clustering}) and (\ref{cond sat/unsat}), as for UC. The results are shown for $k=3$ in Figure~\ref{Fig t_c t_s and t_q for UC}. 

The largest value of $\alpha$ for which GUC succeeds with positive probability in finding a solution, $\alpha_\mathrm h^\mathrm{GUC}(k)$, can be found by looking for the value of $\alpha$ for which $\max_{t \in [0,1]} 2 c_2^\mathrm{GUC} / (1-t) = 1$. For $k=3$ this gives the equation $6 \alpha - \log(6 \alpha) = 3$, so that $\alpha_\mathrm h^\mathrm{GUC}(3) \simeq 0.750874$. Notice that this is larger than $\alpha_h^\mathrm{UC}(3)$, as could be expected.

\section{Bounds on the values of $\alpha$ for which poissonian heuristics can succeed}
\label{Sec_bounds}

I shall now discuss how the results of the previous Section on the general properties of poissonian heuristics are related to the phase diagram of \kxorsat, and in particular what consequences this relation has on the performance of poissonian heuristics in the various phases.

At the end of Section~\ref{Sec_potential} I have shown that the surfaces $\Sigma_\mathrm c$ and $\Sigma_\mathrm s$ intersect each other (I called the intersection \emph{critical surface} $\Sigma_\mathrm k$) and that $\Sigma_\mathrm c$, $\Sigma_\mathrm s$ and $\Sigma_\mathrm k$ are tangent to each other and to the contradiction surface $\Sigma_\mathrm q$. This a property of the phase diagram of \kxorsat\ which has nothing to do with specific DPLL heuristics. However, a continuity argument based on the fact that the trajectories generated by poissonian heuristics can cross the surfaces $\Sigma_\mathrm c$ and $\Sigma_\mathrm s$ at most once confirms it. The argument goes as follow.

For any heuristic of the poissonian class, there is a threshold $\alpha_\mathrm h(k)$ below which the heuristic finds a solution with positive probability and above which this probability vanishes. The heuristic fails with probability 1 if the (average) trajectory intersects the contradiction surface $\Sigma_\mathrm q$. Since for $\alpha < \alpha_\mathrm h$ the trajectory must not intersect $\Sigma_\mathrm q$ while for $\alpha > \alpha_\mathrm h$ it must, by continuity (of the trajectory and its derivatives, and of $\Sigma_\mathrm q$ and its derivatives) this implies that the trajectory corresponding to $\alpha_\mathrm h$ must be \emph{tangent} to $\Sigma_\mathrm q$.

In the same manner, since the trajectories can cross $\Sigma_\mathrm s$ at most once, if a trajectory enters the \unsat phase, it cannot escape from it, and the algorithm must fail. This means that for $\alpha < \alpha_\mathrm h$ the trajectories must not cross $\Sigma_\mathrm s$, while for $\alpha > \alpha_\mathrm h$ they must. As before, by continuity this implies that the trajectory corresponding to $\alpha_\mathrm h$ must be tangent to $\Sigma_\mathrm s$. The same argument can be made to show that it is also tangent to $\Sigma_\mathrm c$. 

Finally, since $\Sigma_\mathrm c$, $\Sigma_\mathrm s$ and $\Sigma_\mathrm q$ intersect on the critical surface $\Sigma_\mathrm k$ and the trajectory corresponding to $\alpha_\mathrm h$ must be tangent to all of them, without crossing any of them, this means that the trajectory must be tangent to each of them \emph{on} the critical surface $\Sigma_\mathrm k$. Therefore, $\Sigma_\mathrm c$, $\Sigma_\mathrm s$ and $\Sigma_\mathrm q$ are tangent to each other on $\Sigma_\mathrm k$.

Indeed, it is very simple to see that this argument is correct. The point of a trajectory generated by a poissonian heuristic which is closer to the contradiction surface $\Sigma_\mathrm q$ will verify the stationarity condition
\begin{equation}
 \frac d {dt} \frac {2 c_2(t)}{1-t} = \frac{2 \dot c_2(t)}{1-t} + \frac{2 c_2(t)}{(1-t)^2} = 0
\end{equation}
which, together with the equations of motion (\ref{differential equation for heuristic}) gives
\begin{equation}
 \frac{d c_2(t)}{dt} = \frac{3 c_3(t) - 2 c_2(t)}{1-t} - \rho_2(t) = - \frac{c_2(t)}{1-t} \,.
\end{equation}
The critical trajectory (i.e. the trajectory corresponding to $\alpha_\mathrm h$) will be such that the value of $2 c_2(t) / (1-t)$ at the maximum is 1. When this happens, $\rho_1(t) = 1$ so we must have $\rho_2(t) = 0$ (the heuristic only performs Unit Propagations), and we obtain
\begin{equation}
 \frac{3 c_3(t)}{1-t} = \frac 1 2
\end{equation}
which, together with $2 c_2(t) / (1-t) = 1$ is the equation of the critical surface $\Sigma_\mathrm k$ given in (\ref{Sigma_k}). As $2 c_2(t) / (1-t)$ is maximum in the point of intersection, the trajectory must be tangent to it.

This has a direct implication for the shape of the curves representing $t_\mathrm c(\alpha)$, $t_\mathrm s(\alpha)$ and $t_\mathrm q(\alpha)$: since each of these curves ends for the value of $(\alpha,t)$ that corresponds to the point where the trajectory is tangent to $\Sigma_\mathrm k$, the three curves must end in the same point (which I shall call \emph{critical point}) in the $(\alpha, t)$ plane, and they must be tangent to each other in the critical point. Since at the critical point the trajectory is on the contradiction surface, so that $\rho_1 = 2 c_2 / (1-t) = 1$, from (\ref{d alpha / dt sat/unsat final}) it is clear that $d t_\mathrm s / d \alpha$ diverges at the critical point, and since the three curves are tangent, they all have infinite derivative. This is clearly seen in Figure~\ref{Fig t_c t_s and t_q for UC} for UC and GUC with $k=3$. The value of $\alpha$ of the critical point is the largest value for which the heuristic succeeds with positive probability, i.e. $\alpha_\mathrm h(k)$.

We can now derive the main result of this Chapter, which follows in a straightforward manner from the previous discussion. The curve representing $t_\mathrm c (\alpha)$ starts at the point $(\alpha_\mathrm c(k), 0)$ and ends at the point $(\alpha_\mathrm h(k), t_\mathrm k)$. Moreover, $t_\mathrm c (\alpha)$ is a single valued function of $\alpha$, and its derivative is negative at $\alpha = \alpha_\mathrm c(k)$. This implies that 
\begin{equation}
 \alpha_\mathrm h(k) < \alpha_\mathrm c(k)
 \label{alpha h < alpha c}
\end{equation}
i.e. that poissonian heuristics fail with probability 1 in the clustered phase.

This result is, as far as I know, the first that relates the performance of a class of heuristics for DPLL with the properties of the phase diagram of the optimization problem.

\section{Optimality of GUC for large $k$}
\label{Sec_optimality GUC}

The result of the previous Section states that no poissonian heuristic for DPLL can succeed with positive probability in the clustered phase, i.e. for $\alpha > \alpha_\mathrm c(k)$. It is then natural to ask what is the maximum value of $\alpha$ which can actually be attained, and which heuristic reaches it, that is to say, what is the \emph{optimal} heuristic.

It is clear that the optimal heuristic will be the one which minimizes
\begin{equation}
 \Delta \alpha_\mathrm h \equiv \alpha_\mathrm c - \alpha_\mathrm h 
= \int_0^{t_\mathrm k} dt' \, \frac{d \alpha_\mathrm c(t')}{dt'} 
= - \int_0^{t_\mathrm k} dt' \, \left. \frac {1 - \phi'(b,t',\alpha)} {\de_\alpha \gamma'(b,t,\alpha)} \right|_{b=b^*_\mathrm c(t'), \alpha = \alpha_\mathrm c(t')}
\label{Delta alpha}
\end{equation}
where I used (\ref{d alpha / dt sat/unsat final}) and where $t_\mathrm k$ is the time coordinate of the critical point in the $(\alpha,t)$ plane, which will depend on the heuristic. Finding the optimal heuristic is a very difficult task: on one hand, the functions $\phi'(b,t,\alpha)$, $\gamma'(b,t,\alpha)$, $b^*_\mathrm c(t)$ and $\alpha_\mathrm c(t)$ have a highly non-trivial dependence on the parameters which characterize the heuristic, i.e. the probability functions $\{p_j(C_1,\dots,C_k)\}$; on the other hand, the quantity which must be minimized is an integral, which requires a \emph{functional} optimization.

I shall therefore discuss two more accessible results: first, that for finite $k$ GUC \emph{locally} minimizes the \emph{numerator} of (\ref{Delta alpha}); and second, that in the limit $k \to \infty$ GUC indeed is optimal, i.e. $\alpha_\mathrm h(k) \to \alpha_\mathrm c(k)$.

The first statement needs clarification: by locally optimize, I mean that on each point of the trajectory described by GUC, it minimizes the numerator in (\ref{Delta alpha}). This is a much weaker requirement than optimality, because a different trajectory, which is sub-optimal in some points, might turn out to be much better in some other points, and overall be better than GUC. And of course also because the denominator should be considered as well. However, I think this result is interesting because it sheds some light on why it is impossible for poissonian heuristics to penetrate the clustered phase.

Indeed, from the definition of $\phi$, which gives
\begin{equation}
 \phi'(b,t,\alpha) = \sum_{j=1}^k j \rho_j(t) b^{j-1} \,,
\end{equation}
and from the bound
\begin{equation}
 \sum_{j=1}^k j \rho_j(t) = \phi'(1,t,\alpha) \leq 1
 \label{bound j rho_j}
\end{equation}
it is clear that $\phi'$ will be maximized (and hence $1-\phi'$ will be minimized) by taking ``the largest possible $\rho_j$ for the smallest possible $j$''. This means that a heuristic which tries to minimize the numerator in the integrand that gives $\Delta \alpha$ should always select the variables to assign in the shortest available clauses, and this is exactly what GUC does. 

Moreover, I already noted at the end of Section~\ref{Sec_trajectories} that GUC saturates the bound (\ref{bound j rho_j}). This implies that GUC achieves the largest possible value of $\sum_j \rho_j$, which is the rate at which clauses are eliminated from the formula. Since this only happens through Unit Propagations, it also means that GUC achieves the highest possible rate of Unit Propagations per variables assigned, and therefore minimizes the fraction of variables that are assigned random values. I think this argument makes it at least plausible that GUC is actually the best poissonian heuristic.

A much stronger argument can be made to support the claim that GUC indeed is optimal in the limit $k \to \infty$. From (\ref{minus gamma dot}) and (\ref{bound rho_j* + rho_j*+1}) we have, integrating $dt$:
\begin{equation}
 \alpha - \int_0^t \frac {dt'} {j^*(t')-1} \leq -\gamma(1,t,\alpha) \leq \alpha - \int_0^t \frac {dt'} {j^*(t')} \,.
\end{equation}
This integral over $dt$ is equal to a sum over the values of $j$ between $k$ and $j^*(t)$,
\begin{equation}
 \alpha - \sum_{j=j^*(t)}^k \frac{t^*(j)-t^*(j+1)}{j-1} \leq -\gamma(1,t,\alpha) \leq \alpha - \sum_{j=j^*(t)}^k \frac{t^*(j)-t^*(j+1)}{j}
\end{equation}
since $j^*(t)$ is a step-like function, which is a constant $j^*(t) = \bar j$ for $t^*(\bar j) \leq t < t^*(\bar j-1)$.

It is reasonable to assume that, in the large $k$ limit,
\begin{equation}
 t^*(j)-t^*(j+1) = \frac 1 k + o(k^{-1})
\label{delta t*}
\end{equation}
for most values of $j$, i.e. for $j$ such that $0 < j/k < 1$. This assumption is well supported by numerical data for $k$ in the range $2^8$ to $2^{16}$, as we shall see later. 

Under this assumption, we obtain
\begin{equation}
 -\gamma(1,t,\alpha) = \alpha - \frac 1 k \sum_{j=j^*(t)}^k \frac 1 j \,.
\end{equation}
In order for the algorithm to generate a contradiction with probability 1, we must have $2c_2/(1-t) \geq 1$, and to have $c_2 > 0$, $j^*(t)$ must reach 2. So if $j^*$ always remains larger than 2, the algorithm must have a finite probability to succeed. If it does indeed succeed, it stops when $\gamma(1,t,\alpha) = 0$, since $\gamma(1,t,\alpha)$ is the number of clauses in the formula at time $t$. The smallest value of $\alpha$ for which the algorithm fails with probability 1 is therefore such that
\begin{equation}
 0 = \alpha_\mathrm h^\mathrm{GUC}(k) - \frac 1 k \sum_{j=2}^k \frac 1 j = \alpha_\mathrm h^\mathrm{GUC}(k) - \frac{\log k + O(1)}{k} \hspace{0.5cm} (\text{for } k \to \infty)
\end{equation}
where the term $O(1)$ in the numerator comes from the fact that it is possible that for a number of terms of order $o(k)$ the asymptotic expansion (\ref{delta t*}) doesn't hold. We obtain:
\begin{equation}
 \alpha_\mathrm h^\mathrm{GUC}(k) = \frac {\log k} k + O(k^{-1}) \hspace{0.5cm} (\text{for } k \to \infty) \,.
\end{equation}
This is the same scaling that is found for $\alpha_c(k)$ (see Table~\ref{Tab_threshold}), so that to the leading order in $k$
\begin{equation}
 \alpha_\mathrm h^\mathrm{GUC}(k) \sim \alpha_\mathrm c(k) \hspace{0.5cm} (\text{for } k \to \infty) \,.
\end{equation}

Let us now turn to the assumption (\ref{delta t*}). In order to verify it, we have performed a series of numerical simulations, in which the equations of motion of GUC are integrated by finite differences, for values of $k$ equal to the powers of two between $2^8$ and $2^{16}$. A finite-size scaling (with respect to $k$) of the results, shown in Figure~\ref{Fig finite size scaling}, 
is consistent with the scaling
\begin{equation}
 k \left[ t^*(j) - t^*(j+1) \right] = 1 + k^\nu \times f(j/k)
\label{scaling}
\end{equation}
where $f(x)$ is a function independent on $k$ and which goes as $x^{-\mu}$ for $x \to 0$. The values of $\mu$ and $\nu$ are found to be both equal to $1/2$. Integrating the scaling form (\ref{scaling}) with $\mu = \nu = 1/2$ one obtains that the first correction to the leading term $\log k / k$ in $\alpha_\mathrm h ^\mathrm{GUC}(k)$ is of order $1/k$, in agreement with the numerical estimates of $\alpha_\mathrm h ^\mathrm{GUC}(k)$ which give $\alpha_\mathrm h ^\mathrm{GUC}(k) \simeq \log k / k + 2.15/k$.

\begin{figure}
 \includegraphics[width=0.5\textwidth]{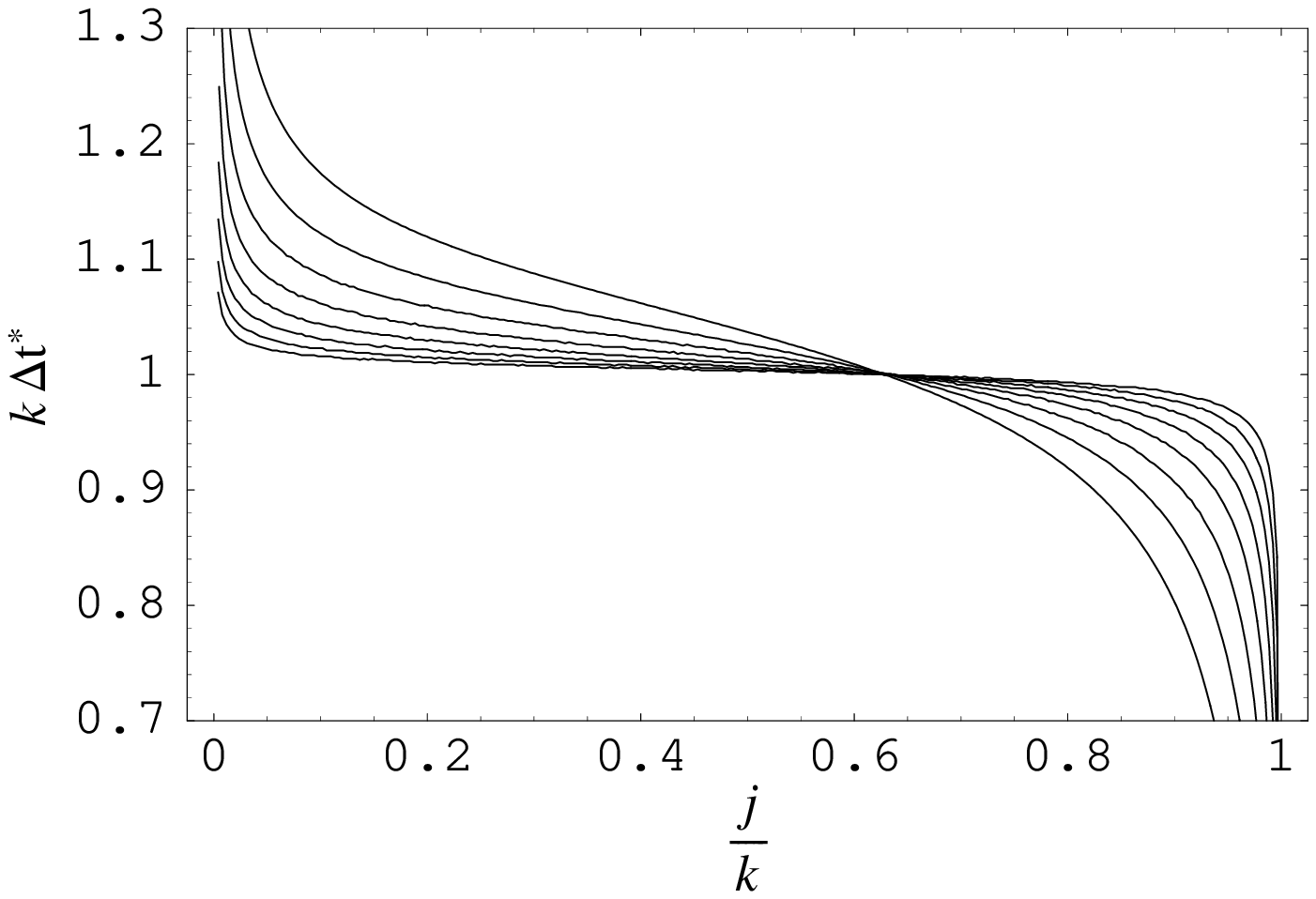}
 \includegraphics[width=0.5\textwidth]{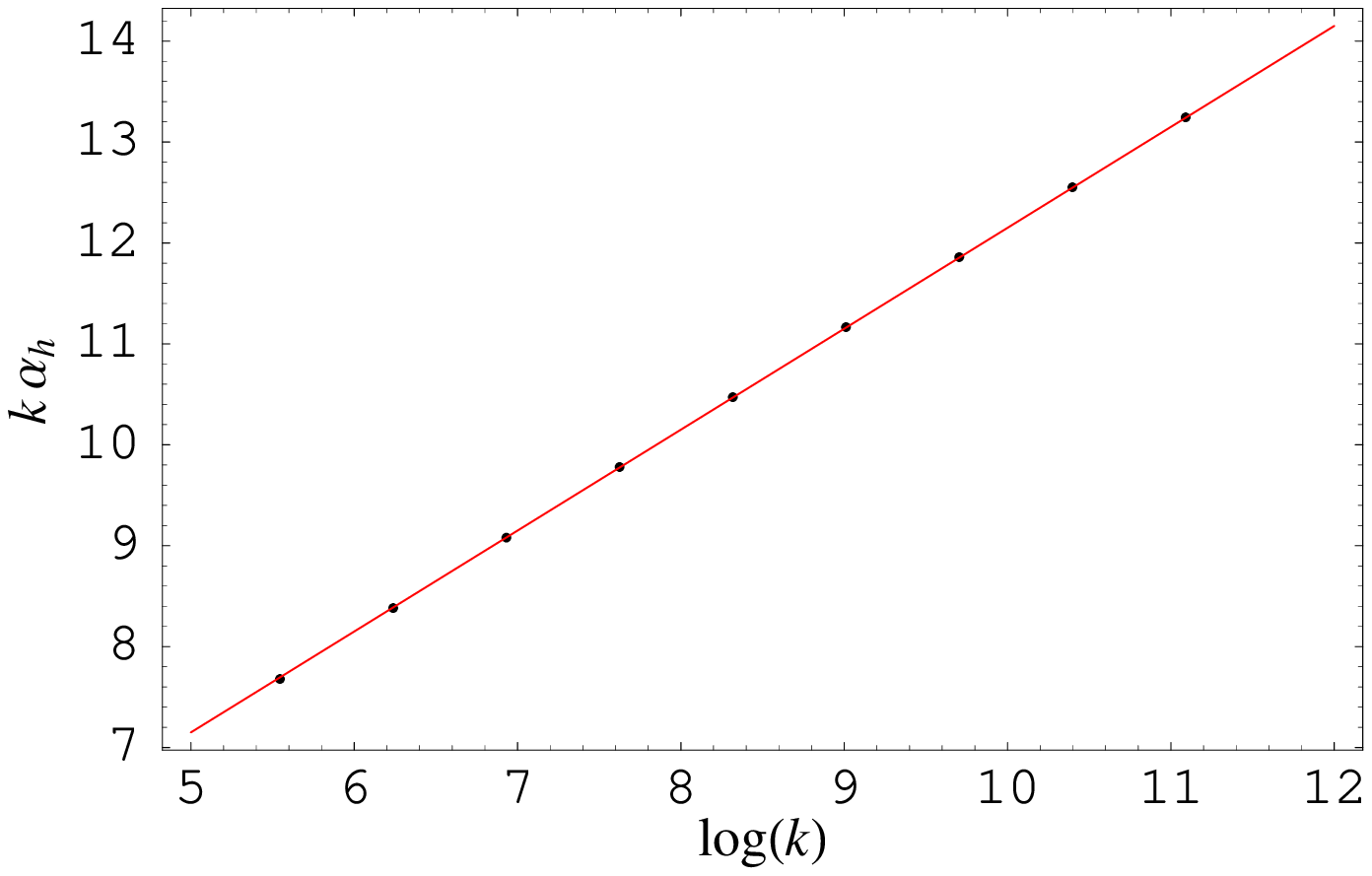} \\
 \includegraphics[width=0.5\textwidth]{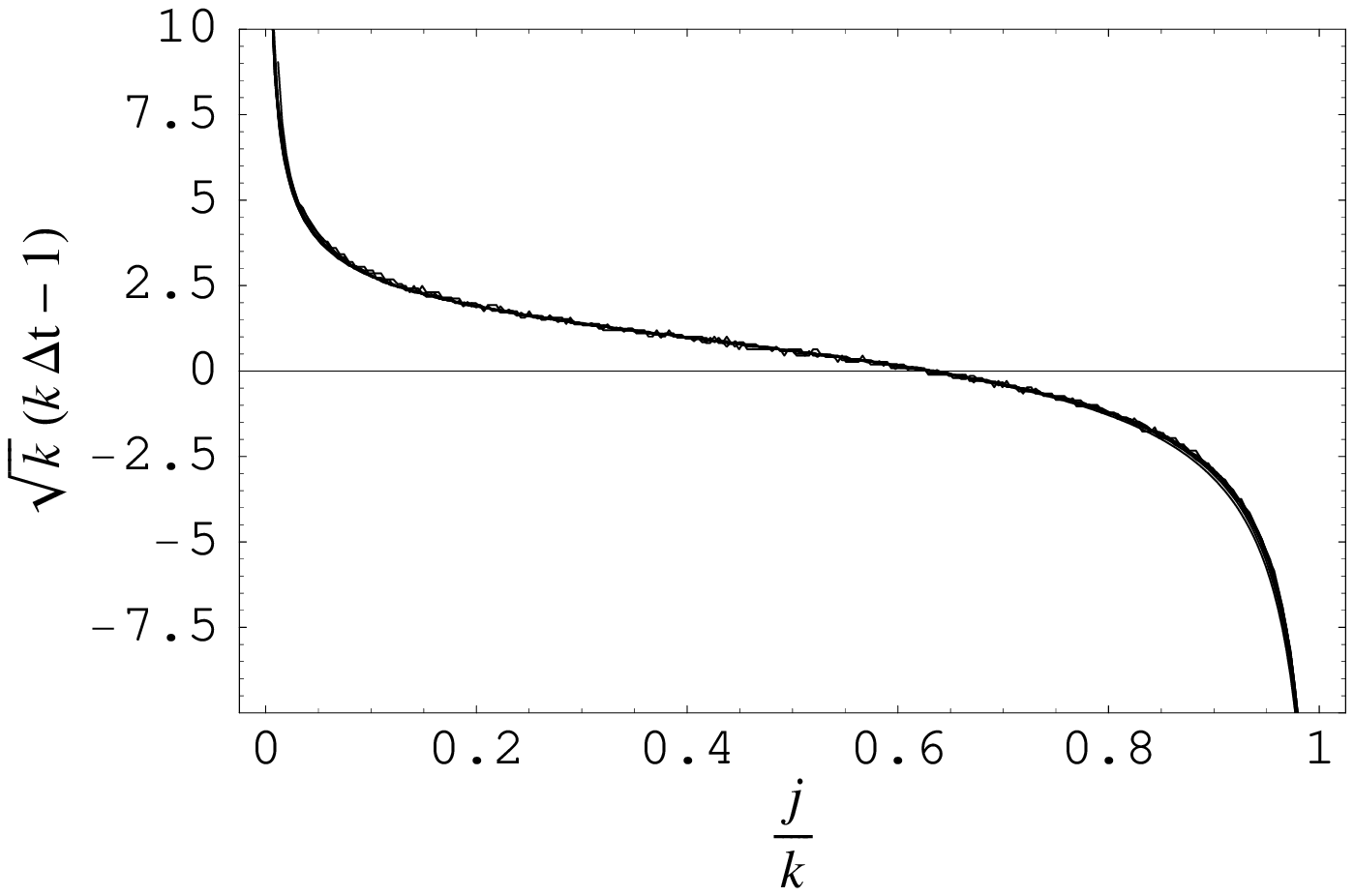}
 \includegraphics[width=0.5\textwidth]{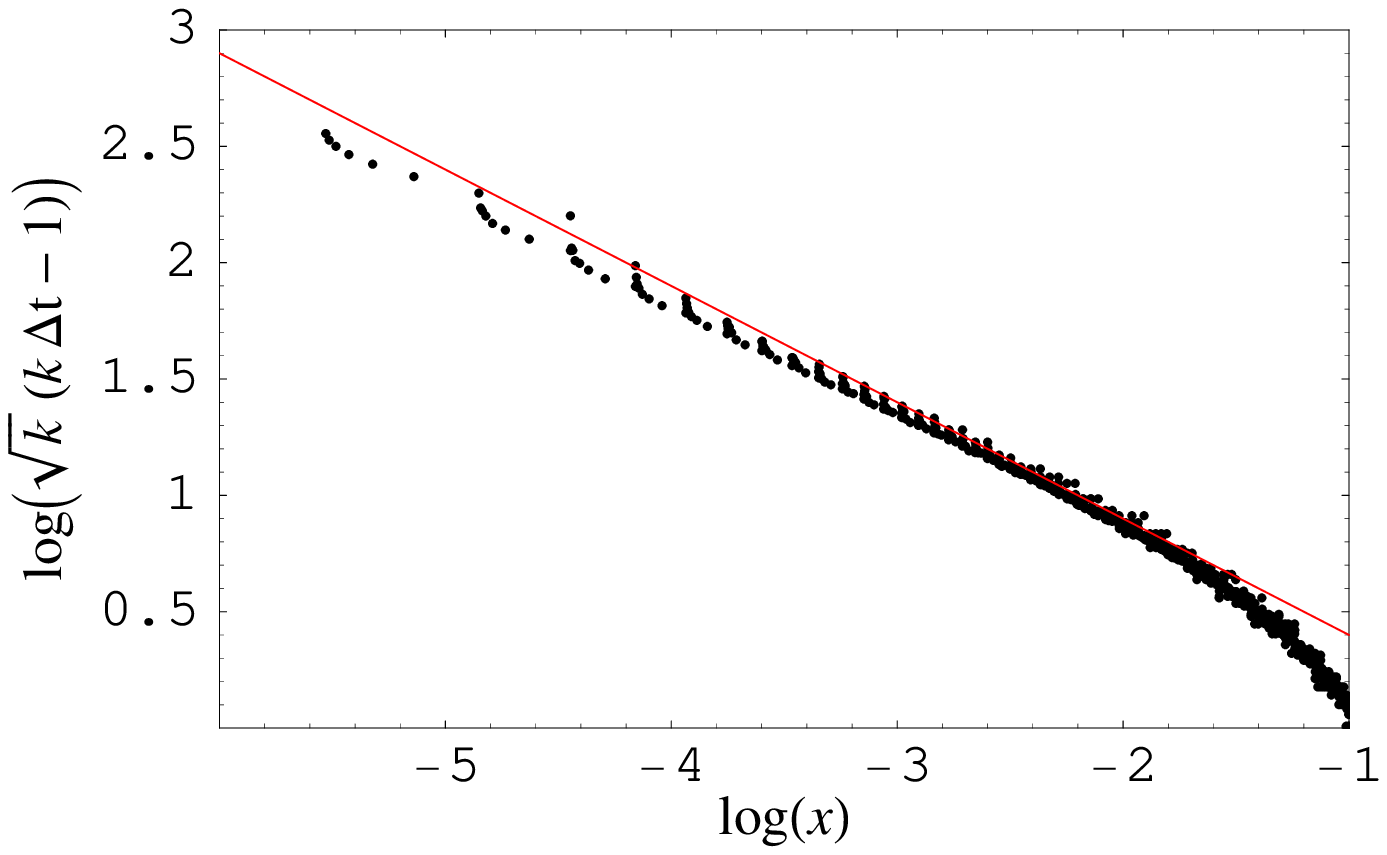}
 \caption{
  Finite size scaling results for GUC at large $k$. 
  \emph{Top~Left}~Each curve shows the values of $k[t^*(j)-t^*(j+1)]$ as a function of $j/k$ for $k=2^8,2^9,\dots,2^{16}$ (from the farthest to the closest curve to 1), and was obtained by integrating the equations of motion (\ref{equations GUC}) by finite differences. For each $k$, the value of $\alpha$ used is $\alpha_\mathrm h ^\mathrm{GUC}(k)$, determined as the value of $\alpha$ for which the maximum reached by $2 c_2(t) / (1-t)$ is 1. 
  \emph{Top~Right}~Data points of $\alpha_\mathrm h ^\mathrm{GUC}(k)$ versus $\log k / k + 2.15 / k$ (red line).
  \emph{Bottom~left}~The same data as above, plotted as $\{k\times[t^*(j)-t^*(j+1)]\}\times k^{1/2}$. The curves ``collapse'', showing $f(x)$ and confirming the value of $\nu = 1/2$. 
  \emph{Bottom~right}~By plotting the same curves on logarithmic scale it is easily seen that for $x$ close to 0 $f(x) \simeq x^{-\mu}$ with $\mu = 1/2$, corresponding to the slope of the red line.
 }
 \label{Fig finite size scaling}
\end{figure}

I believe that the above numerical results make a strong case supporting the assumption (\ref{delta t*}), and therefore the optimality of GUC.

\section{Conclusions and perspectives}
\label{Sec Conclusions 4}

In this Chapter, I have discussed some very general bounds on the performance of poissonian heuristics for DPLL for the solution of \kxorsat\ formul\ae\ and for that of its NP-complete extensions, called \uecsp. In particular, I have proved that such heuristics generate contradictions (i.e. fail) with probability 1 in the clustered phase of the problem.

A point of caution should be placed in the interpretation of this result: it is a very peculiar feature of \kxorsat\ that the clustering and freezing transitions coincide. What is found in general in other problems is that the clustering transition, where solutions form an exponential number of connected clusters that are well separated, and the freezing transition, where some variables take a constant value in all the solutions of a given cluster, are distinct. It is well known that in problems where these thresholds are distinct, it is the freezing transition that corresponds to the onset of hardness for known local algorithms. It can be argued that in \kxorsat\ too, what causes DPLL poissonian heuristics to fail, is the strong correlations between variables that are present in the frozen phase, rather than the separation of the clusters.
In view of this, it would be very interesting to understand what similar bounds could be obtained in problems where the two thresholds are distinct, and notably in \ksat.

Another interesting question concerns the extention to more general, \emph{non}-poissonian heuristics. In this regard, I have obtained some partial results that seem promising, even though a general theory is still far. More specifically, I have been able to solve the leaf removal equations for the case in which the mixed system to which it is applied is not poissonian, but instead is characterized by some arbitrary distribution of the number of occurrences. However, due to the complicated structure of the solution, it has resulted impossible so far to characterize the phase transitions in terms of a potential, which then would allow to derive some general properties of the trajectories, and possibly some bounds on the values of $\alpha$ for which solutions can be found. Some further work in this direction seems worth undertaking.
\chapter{Characterization of the solutions of $k$-SAT at large $\alpha$}
\label{Chap_ksat}

In this Chapter I shall discuss the properties of the solutions of random \ksat\ at large $\alpha$. This might seem oxymoronic, since at large $\alpha$ random \ksat\ formul\ae\ are \unsat\ with probability 1. The idea is precisely to restrict the formul\ae\ that are considered to those that, for a given large $\alpha$, are \sat, then to form an ensemble of these formul\ae\ with uniform weight, and study the properties of their solutions.

Apart from the intrinsic interest of the question, i.e. studying the properties of this particular ensemble of \ksat\ formul\ae, this problem is relevant because of some recent results by Feige and collaborators \cite{Feige02,Feige06}: for the first time (as far as I know), they have been able to relate the average case complexity of a satisfiability problem with the worst case complexity of another class of problems, thus bridging the gap between complexity theory and results derived from statistical mechanics methods.

Feige's result can be summarized as follows: under the assumption that there is no polynomial-time algorithm capable of recognizing \emph{every} \sat\ instance (and \emph{most} \unsat\ instances) of \ksat[3] for arbitrarily large (but bounded in $N$) values of $\alpha$, the approximation problem to several optimization problems (including min bisection, dense $k$-subgraph and max bipartite clique) is hard, i.e. non-polynomial in time in the worst case. The complexity class of the approximation problems considered by Feige was previously not known.

With this motivation, R\'emi Monasson, Francesco Zamponi and I have studied in \cite{Altarelli07} the problem of characterizing the solutions of \ksat[3] at large $\alpha$, with the objective of showing that a simple message-passing procedure is able to contradict a probabilistic version of Feige's assumption, in which ``every'' is substituted with ``with probability $p$'', for any (finite) value of $p$.

In the following Sections, I shall therefore present more in detail Feige's result and define the problem (Section~\ref{Sec Problem definition}); then I shall present the computation of the free energy of the uniform distribution of satisfiable \ksat[3]\ formul\ae, in Section~\ref{Sec Free energy}; in Section~\ref{Sec Cavity formalism} a similar result is derived from the cavity formalism; then, in Section~\ref{Sec Comparison of P_Sat and P_Plant} I shall compare the results obtained with those that are valid for a different ensemble of formul\ae, which was studied by Feige, and draw their algorithmic implications; I shall then comment, in Section~\ref{Sec Stability of RS free energy} on the stability of the RS solution of Sections~\ref{Sec Free energy} and \ref{Sec Cavity formalism}; finally, in Section~\ref{Sec Conclusion} I shall present and discuss the conclusions of this work.

\section{Problem definition and previously established results}
\label{Sec Problem definition}

I shall now define the problem I want to study, and give a brief overview of Feige's results, concerning on one hand the relation between the average-case complexity of \ksat[3] and the worst-case complexity of a class of approximation problems, and on the other hand the properties of a very simple message-passing algorithm, which on a particular ensemble of satisfiable \ksat[3] formul\ae\ has interesting properties (in view of the previous complexity result).

\subsection{Definition of the random ensembles}
\label{Par Definition of ensembles}

Let us consider random \ksat[3] formul\ae\ $\mathcal F$ involving $N$ boolean variables $\{x_1,\dots,x_N\}$ and $M = \alpha N$ clauses, with finite $\alpha$ (as $N \to \infty$). I shall denote assignments of the $N$ variables as $X \equiv \{x_i|i=1,\dots,N\} \in \{\true,\false\}^N$. Alternatively, I shall represent them as configurations of $N$ Ising spins $\sigma_i \in \{-1,1\}$, collectively denoted by $\sigma \equiv \{\sigma_i|i=1,\dots,N\}$, with $\sigma_i = 1$ corresponding to $x_i = \true$ and $-1$ to \false.

The Uniform Ensemble $\mathcal P_\mathrm{Unif}[\mathcal F]$ is obtained by giving the same weight to each possible formula $\mathcal F$. When $\alpha > \alpha_\mathrm s(3) \simeq 4.267$, the probability over $\mathcal P_\mathrm{Unif}[\mathcal F]$ that a formula $\mathcal F$ is \sat\ is 0: the overwhelming majority of formul\ae\ are \unsat.
It is therefore interesting to introduce two particular ensembles that include only those formul\ae\ that are \sat:

\begin{description}
 \item[Satisfiable Ensemble] $\mathcal P_\mathrm{Sat}$ is the ensemble of satisfiable formul\ae, with uniform weight. This is simply the restriction of $\mathcal P _\mathrm{Unif}$ to satisfiable formul\ae.
 \item[Planted Ensemble] Given an assignment $X$, the ensemble $\mathcal P_\mathrm{Plant}^X[\mathcal F]$ of \sat\ formul\ae\ ``planted on $X$'' is defined as the uniform ensemble of formul\ae\ that admit $X$ as a solution. The Planted Ensemble $\mathcal P_\mathrm{Plant}[\mathcal F]$ is obtained by averaging over $X$ with uniform weight for all possible configurations.
\end{description}

Notice that any satisfiable formula is present in both ensembles, but with different weights, as is easily seen from a simple computation:
for each clause involving $k$ literals, there is only one assignment of the corresponding $k$ variables that is not \sat. The number of formul\ae\ $\mathcal N_\mathrm f[X]$ that admit $X$ as a solution is therefore
\begin{equation}
 \mathcal N_\mathrm f [X] = \left[ \binom N k \left( 2^k - 1 \right) \right] ^ M \equiv \mathcal N_\mathrm f
\end{equation}
which is independent on $X$. 
The Planted Ensemble is then by definition
\begin{equation}
 \mathcal P_\mathrm{Plant}[\mathcal F] = \frac 1 {2^N} \sum_X \frac{\mathbb I[\mathcal F \text{ is satisfied by } X]}{\mathcal N_\mathrm f [X]} = \frac {\mathcal N_\mathrm s [\mathcal F]} {2^N \mathcal N_\mathrm f}
\end{equation}
where $\mathcal N_\mathrm s[\mathcal F]$ is the number of solutions admitted by $\mathcal F$. It is then clear that $\mathcal P_\mathrm{Plant}[\mathcal F]$ is not uniform, but proportional to the number of solutions of $\mathcal F$.

As we shall see in the following paragraphs, the two ensembles $\mathcal P_\mathrm{Sat}$ and $\mathcal P_\mathrm{Plant}$ appear in Feige's results.

\subsection{Hardness of approximation results}
\label{Par Hardness}

In this paragraph I shall give a very brief (and non-rigorous) overview of a theorem proved by Feige in \cite{Feige02}.

Feige considers a class of algorithms that take a \ksat[3] formula as an input and have two possible outputs: either \sat\ or \unsat. The algorithms in question need not be deterministic: for a given formula, it is admissible that the output be a random variable, whose distribution will then depend on the formula. Notice that, since there are two incompatible outputs, algorithms of this kind can give a wrong answer. However, we shall consider only \emph{asymmetric} algorithms, i.e. such that if the input formula is \sat\ then the output is \emph{always} \sat; on the other hand, it is admissible that if the input formula is \unsat\ the output be \sat, and we shall only require that the probability of this error be smaller than 1/2 (or some other finite constant, the actual value of which is unimportant).

He then examines the following
\begin{description}
 \item[Hypothesis 1] Even when $\alpha$ is an arbitrarily large constant (independent on $N$), there is no polynomial time algorithm that refutes most Random-\ksat[3] formul\ae\ and \emph{never} wrongly refutes a satisfiable formula.
\end{description}
This hypothesis states that no algorithm of the class described above can work in polynomial time \emph{on average} for \ksat[3] formul\ae\ drawn from the Uniform Ensemble $\mathcal P_\mathrm{Unif}$. In the statement of this Hypothesis, the crucial word \emph{never} refers both to the choice of the formula and to the random moves of the algorithm. According to the author, no algorithms are known to contradict it. Notice that numerical experiments demonstrate that as $\alpha$ grows beyond the \sat/\unsat\ transition threshold, \ksat\ becomes more ``easy'' (i.e. the average running time for refutation decreases). However, all known algorithms remain exponential time, and it is only the prefactor of the exponent which decreases. Therefore this observation does not contradict \hyp{1}. Also, notice that the fact that $\alpha$ is a constant independent of $N$ is crucial: polynomial time algorithms \emph{are} known for $\alpha \gg N^{1/2}$.

 In his paper Feige also considers a weaker form of this hypothesis, which has several advantages. The motivation for it is the following. For large $\alpha$, not only typical random formul\ae\ are \unsat, but the number of violated constraints becomes concentrated (relative to the Uniform Ensemble of formul\ae) around $M/8$, for \emph{every} assignment. Therefore, the formul\ae\ that are \emph{not typical} include all satisfiable formul\ae, and also all the formul\ae\ that admit at least one assignment which violates a number of clauses $\epsilon M$ with $0 < \epsilon < 1/8$. 
\begin{description}
 \item[Hypothesis 2] For every fixed $\epsilon \geq 0$, even when $\alpha$ is an arbitrarily large constant (independent on $N$), there is no polynomial time algorithm that on most Random-\ksat[3] formul\ae\ outputs \typ\ and \emph{never} outputs \typ\ on formul\ae\ with $(1-\epsilon)M$ satisfiable clauses.
\end{description}
In this case the algorithm considered has two possible outputs, \typ\ and \notyp, and again the admissible error is asymmetric. For $\epsilon = 0$ \hyp{2} reduces to \hyp{1}.

Notice that, despite the appearence, \hyp{1} implies \hyp{2} and therefore \hyp{2} is \emph{weaker} than \hyp{1}. In order to realize it, let me show that if \hyp{2} is violated, then \hyp{1} is also violated. Indeed, if \hyp{2} is violated, an algorithm exists which is able to identify formul\ae\ that have a fraction of satisfiable clauses larger than $1-1/8$. In most cases the output of this algorithm will be \typ, meaning that the fraction of satisfiable clauses is $1-1/8$; however, if the formula has a fraction of satisfiable clauses larger than $1-1/8$, it will be identified as such. Therefore, such an algorithm will output \typ\ most of the time, but it will never output \typ\ if the formula is \sat\ (and therefore has a fraction of satisfiable clauses larger than $1-1/8$), thus contradicting \hyp{1}.

The main result from \cite{Feige02} is the following
\begin{description}
 \item[Theorem 1] The existence of an algorithm able to approximate in polynomial time the solution to any of the following problems would contradict \hyp{2}: min bisection, dense $k$-subgraph, max bipartite clique (all within a constant approximation factor) and 2-catalog (within a factor $N^\delta$ where $N$ is the number of edges and $0 < \delta < 1$ some constant).
\end{description}
I shall not define these problems, which are well known in theoretical computer science and of little interest for the following\footnote{A definition is given in \cite{Feige02}}. It suffices me to say that their complexity class is not known. If \hyp{2} were proved to be true, as a consequence all these problems would be NP-hard, and this would be an interesting new result.

As I already mentioned, this theorem establishes a relation between the \emph{average-case complexity} of \ksat[3] at large $\alpha$ and the \emph{worst-case complexity} of some other problems. In this regard, it is a very striking result, and it opens the possibility of applying statistical mechanics methods to complexity theory.

Without any ambition to rigor, let me just sketch the proof of the theorem,
which is rather interesting. Let us define a problem $P$ as R-\ksat[3]-hard if
the existence of a polynomial time algorithm to solve $P$ would contradict
\hyp{2}. In particular, a problem is R-\ksat[3]-hard if it is possible to
reduce any instance of \ksat[3] to an instance of $P$ to which $A$ can be
applied, in such a way as to contradict \hyp{2}. Then, Feige proves that
several other boolean constraint satisfaction problems, and their optimization
versions, are R-\ksat[3]-hard.

More specifically, let us consider a boolean function over three variables, $f:
\, \{\true,\false\}^3 \rightarrow \{\true,\false\}$. The number of such
functions is $2^{2^3}$, most of which coincide up to renaming or negation of
the variables. For each of them, let us define as $t$ the number of possible
inputs, out of $2^3$, for which $f$ evaluates to \true, and $b$ (for
\emph{bias}) the number of possible inputs with an \emph{odd} number of \true\
values and for which $f$ evaluates to \true\ (or, if it is larger, the same
quantity with \emph{even} instead of \emph{odd}). Then, there are 13
\emph{distinct} such functions for which $2b > t$, including \textsc{and},
\textsc{or} and \textsc{xor}.

Consider a ``$3f$-clause'' involving 3 literals over $N$ variables and based on
any of these 13 functions $f$, and a random ``3$f$-formula'' made of $M=\alpha
N$ such clauses. Feige proves the following
\begin{description}
  \item[Theorem 2] It is R-\ksat[3]-hard to distinguish between those random
$3f$-formul\ae\ in which a fraction just over $t/8$ of the clauses are
satisfied, and those in which this fraction is just below $b/4$ (assuming
$\alpha$ is sufficiently large). In particular, this implies that it is
R-\ksat[3]-hard to approximate \textsc{max}-$3f$ within a constant factor
better than $t/2b$.
\end{description}
This theorem is very interesting in itself: it is here that the link between
the complexity of a decision problem (namely R-\ksat[3]) and that of an
approximation problem is established (even though, only for the average case).
The proof of Theorem 2 is straightforward but complicated, and I shall
omit it.

Feige then proves the following
\begin{description}
  \item[Proposition] For every $\epsilon > 0$, there is an $\alpha_\epsilon$ such that for any $\alpha > \alpha_\epsilon$ and $N$ large enough, whith probability 1 the following holds: \emph{every} set of $(1/8 + \epsilon)M$ clauses in a R-\ksat[3] formula with $M = \alpha N$ clauses contains at least $N+1$ distinct literals.
\end{description}
The crucial point, which will allow to establish a link between average-case and worst-case complexity, is that the proposition holds, with probability 1 over the choice of the \ksat[3] formula, for \emph{every} set of $(1/8 + \epsilon)M$ clauses of a given formula. The proof of this proposition is rather simple: given $N$ variables, corresponding to $2N$ literals, let us select a set $S$ containing $N$ literals. The probability that a random clause contains no literal from $S$ is $(1/2)^3$, and the probability that $m$ clauses out of $M$ contain no literals from $S$ is
\begin{equation}
  P_S(m) = \binom M m \left( \frac 1 2 \right)^{3m} \left[ 1 - \left( \frac 1 2 \right)^3 \right]^{M-m}
\end{equation}
which, for large $M$ and $m = \mu M$, is asymptotically
\begin{equation}
  P_S(m) \sim \exp \left\{ M \log 2 \left[ - \mu \log_2 \mu - (1-\mu) \log_2 (1-\mu) - 3\mu + (1-\mu) \log_2 (7/8)  \right] \right\} \equiv e ^ {M \phi(\mu)} \,.
\end{equation}
This probability is maximum for $\mu = 1/8$, and verifies the large deviations relation
\begin{equation}
  \mathbb P \left[ \mu = \frac 1 8 + \epsilon \right] \sim \exp \left[ \alpha N \phi''(1/8) \frac {\epsilon^2} 2 \right]
\end{equation}
with $\phi''(1/8) = - 64/7$.

Therefore, for any given $\epsilon > 0$, provided $\alpha > -3 \times 2 / \left[ \phi''(1/8) \epsilon^2 \right]$, we shall have
\begin{equation}
  \mathbb P \left[ \mu = \frac 1 8 + \epsilon \right] < 2^{-3N} \Rightarrow \mathbb P \left[ \mu < \frac 1 8 + \epsilon \right] > 1 - 2^{-3N} \,.
\end{equation}

More explicitly:
\begin{equation}
  \mathbb P \left[ \text{at least } (1/8 + \epsilon) M \text{ clauses out of } M \text{ contain no literal from } S \right] < 2^{-3N} \,.
\end{equation}
We can now use Boole's inequality,
\begin{equation}
  \mathbb P \left[ \bigcup_i A_i \right] \leq \sum_i \mathbb P[A_i]
\end{equation}
and write, for all the possible subsets $S$ of $N$ literals out of $2N$,
\begin{eqnarray}
  && \mathbb P \left[ \text{at least } (1/8 + \epsilon) M \text{ clauses out of } M \text{ contain no literal from } \bigcup S \right] < \sum_S 2^{-3N} \\
  &\Leftrightarrow& \mathbb P \left[ \text{at least } (1/8 + \epsilon) M \text{ clauses out of } M \text{ contain no literal from \emph{any} set of } N \text{ literals} \right] < 2^{-N} \nonumber \\
\end{eqnarray}
since the number of possible sets $S$ is less than $2^{2N}$. This statement is equivalent to the one in the Proposition: every set of $(1/8 + \epsilon)M$ clauses contains at least $N+1$ literals, with probability 1 over the choice of the formula from which the clauses are taken.

The proof of Theorem 1 then proceeds as follows, for each of the graph-based problems $P$ listed in the enunciate. Some 3\textsc{and}-formula $\mathcal F$ with $M = \alpha N$ clauses in $N$ variables is mapped to a graph $\mathcal G$ by a suitable construction. The actual constructions vary with the specific problem $P$ and I shall omit them. Let us make the case of min bisection for concreteness. The Proposition is used to prove that if $\mathcal F$ has at most $(1/8+\epsilon)M$ satisfiable clauses, then the corresponding $\mathcal G$ has a cut of width at least $(1-\epsilon)M$; while if $\mathcal F$ has at least $(1/4-\epsilon)M$ satisfiable clauses, then the corresponding $\mathcal G$ has a cut of width $3(1/4+\epsilon)M$. This means that if it is possible to approximate min bisection on \emph{every} instance within a factor $3/4$, then it is possible to compute the approximate bisection, and from the approximate value it will be possible to distinguish the two cases (i.e. of \emph{typical} 3\textsc{and}-formul\ae\ with $(1/8+\epsilon)M$ satisfiable clauses vs. \emph{non-typical} 3\textsc{and}-formul\ae\ with $(1/4-\epsilon)M$ satisfiable clauses). Because of Theorem 2, this contradicts \hyp{2}, and thus proves Theorem 1.

\subsection{Performance of \wprop\ on the planted distribution}
\label{Par WP on planted}

The problem of establishing or refuting \hyp{1} and/or \hyp{2} was tackled by Feige and collaborators in \cite{Feige06}. That paper makes a step forwards in the direction of refutation, but does not achieve to prove it in general.

The authors consider a simple message passing procedure, called Warning Propagation (WP). Given a \ksat[3] formula $\mathcal F$ and the factor graph $\mathcal G$ representing it, two kinds of messages are defined for each edge in $\mathcal G$, i.e. for each pair $(\mathcal C_a, x_i)$ where $\mathcal C_a$ is a clause and $x_i$ a variable appearing in it: \emph{clause-to-variable messages} $u_{a \to i}$ are binary variables equal to 0 or 1; \emph{variable-to-clause messages} $h_{i \to a}$ are integer variables (positive, negative or null). The following update rule is defined:
\begin{equation}
 \left\{ 
  \begin{aligned}
     h_{i \to a} &= \sum_{b \in \de_+ i \backslash a} u_{b \to i} \ - \!\!\! \sum_{b \in \de_- i \backslash a} u_{b \to i} \,, \\
     u_{a \to i} &= \prod_{j \in \de a \backslash i} \mathbb I [h_{j \to a} < 0]
  \end{aligned}
 \right.
\label{WP rule}
\end{equation}
where $\de a$ is the set of variables appearing in clause $\mathcal C_a$, the back-slash denotes privation, $\de_+ i$ is the set of clauses in which variable $x_i$ appears non-negated, and $\de_- i$ is the set of clauses in which it appears negated.

WP is defined as the following algorithm, taking a \ksat[3] formula $\mathcal F$ as input and returning a \emph{partial} assignment $X$ as output:
\begin{algorithmic}
 \Procedure{Warning Propagation}{$\mathcal F$}
 \State{Construct the factor graph $\mathcal G$ representing $\mathcal F$}
 \State{Randomly initialize the clause-to-variable messages $\{u_{a \to i}\}$ to 0 or 1}
 \Repeat
  \State{Randomly order the edges of $\mathcal G$}
  \State{Update the messages $h_{i \to a}$ and $u_{a \to i}$ in the selected order according to the rule (\ref{WP rule})}
 \Until{No message changes in the update}
 \State{Compute a partial assignment $X$ based on $\{h_{i \to a}\}$:}
 \If{$\sum_{a \in \de_+ i} h_{i \to a} - \sum_{a \in \de_- i} h_{i \to a} > 0$}
  \State{$x_i = \true$}
 \ElsIf{$\sum_{a \in \de_+ i} h_{i \to a} - \sum_{a \in \de_- i} h_{i \to a} < 0$}
  \State{$x_i = \false$}
 \Else
  \State{$x_i$ is unassigned}
 \EndIf
 \State{Return $X$}
 \EndProcedure
\end{algorithmic} 
Notice that some variables in $X$ will be unassigned at the end of WP. 

The main result proved in \cite{Feige06} is the following
\begin{description}
 \item[Theorem 2] For any assignment $Y$ and any formula $\mathcal F$ from the ensemble $\mathcal P_\mathrm{Plant}^{Y}[\mathcal F]$ planted on $Y$ with large enough $\alpha$ (but constant in $N$), the following is true with  probability $1 - e^{-O(\alpha)}$ over the choice of the formula and the random moves of WP:
  \begin{enumerate}
    \item WP$(\mathcal F)$ converges after at most $O(\log N)$ iterations
    \item The fraction of variables assigned in $X$ is $1 - e^{-O(\alpha)}$, and for each of them $x_i = y_i$ (the value it takes in the planted assignment $Y$)
    \item The formula obtained by simplifying $\mathcal F$ with the values assigned in $X$ can be satisfied in time $O(N)$
  \end{enumerate}
\end{description}

\subsection{Discussion of the known results and problem definition}
\label{Par Problem definition}

Theorem 2 establishes that WP has some properties of the algorithm described in \hyp{1}, but with some important differences, as I shall discuss in this paragraph.

First, WP is a constructive algorithm, but it is not complete: it is possible that it never converges (i.e. that the loop goes on for ever); however, if it does converge, it provides an assignment which can be easily checked. One can set a fixed maximum number of iterations $\mathcal N_\mathrm i$ and stop the execution if it is reached; the output will then be \unsat, and this will possibly be wrong. If, on the contrary, an assignment is returned (and it is checked to be satisfying), the output will be \sat, and this will surely be true.

Therefore WP is an asymmetric algorithm, which never outputs \sat\ to an \unsat\ formula, but which sometimes outputs \unsat\ to a \sat\ formula. 
The algorithm described in \hyp{1} is different in this regard, as it must \emph{never} return \unsat\ to a \sat\ formula. 

Second, the statements in Theorem 2 hold in probability for formul\ae\ drawn from the Planted Ensemble, while in \hyp{1} the Uniform Ensemble is considered.

The conclusion which can be drawn is that Theorem 2 refutes the following modified
\begin{description}
 \item[Hypothesis $\boldsymbol{1_p}$ Planted] Even when $\alpha$ is an arbitrarily large constant (independent on $N$), there is no polynomial time algorithm that refutes most Random-\ksat[3] formul\ae\ \emph{from the Planted Ensemble} $\mathcal P_\mathrm{Plant}$, and outputs \sat\ \emph{with probability} $p$ on a \ksat[3] formula which is satisfiable.
\end{description}
The differences relative to \hyp{1} are written in italics in \hyp{$1_p$~Planted}: the distribution of formul\ae\ is the Planted Ensemble instead of the Uniform one, and satisfiable formul\ae\ are recognized with probability $p$ instead of always.

The question I shall try to answer in the rest of this Chapter is if it is possible to make further progress towards the refutation of \hyp{1}, and in particular if the convergence of WP can be established for formul\ae\ drawn from he Satisfiable Ensemble $\mathcal P_\mathrm{Sat}$. This is equivalent to proving it for the Uniform Ensemble, since $\mathcal P_\mathrm{Sat}$ is the restriction of $\mathcal P_\mathrm{Unif}$ to satisfiable formul\ae, and for formul\ae\ that are not \sat\ it is admissible for the algorithm to give wrong answers (i.e. not to converge).

The main conclusion that we shall reach is that this is indeed true, and that the following
\begin{description}
 \item[Hypothesis $\boldsymbol{1_p}$] Even when $\alpha$ is an arbitrarily large constant (independent on $N$), there is no polynomial time algorithm that refutes most Random-\ksat[3] formul\ae\ $\mathcal P_\mathrm{Plant}$, and outputs \sat\ \emph{with probability} $p$ on a \ksat[3] formula which is satisfiable.
\end{description}
is wrong for any $p < 1$. A similarly probabilistic version of \hyp{2} will also be refuted.

\section{Free energy of the uniform distribution of satisfiable formul\ae}
\label{Sec Free energy}

In this Section, I shall apply the replica formalism for diluted systems described in Paragraph \ref{Par Replica method for diluted} to a spin glass problem which is equivalent to Random-\ksat[3], in order to derive the properties of the formul\ae\ in $\mathcal P_\mathrm{Sat}$ and of their solutions.

The following computation follows the one presented in \cite{Monasson97}, the main difference being the introduction (in Paragraph \ref{Par Chemical potential}) of a ``chemical potential'' that will permit to select only the formul\ae\ that are satisfiable.

\subsection{Replicated partition function of \ksat}
\label{Par Replicated Z}

In this Section, we shall use the representation of an assignment $X$ as a collection of $N$ Ising spins, $\sigma \equiv \{\sigma_1,\dots,\sigma_N\}$. For a given \ksat\ formula $\mathcal F$ and a given configuration $\sigma$ we define the energy function
\begin{equation}
 E_\mathcal F(\sigma) = \sum_{i=1}^M \mathbb I[\sigma \text{ verifies } \mathcal C_i]
\label{Energy}
\end{equation}
where $\mathcal C_i$ is the $i^\text{th}$ clause in $\mathcal F$. This energy is simply the number clauses in $\mathcal F$ that are violated by $\sigma$.

The partition function is defined as
\begin{equation}
 Z_\mathcal F(\beta) = \sum_\sigma e^{-\beta E_\mathcal F(\sigma)} 
 = \sum_\sigma \prod_{i=1}^M z_i(\sigma) 
\end{equation}
where $z_i(\sigma) \equiv \exp \{-\beta \ \mathbb I [\sigma \text{ verifies } \mathcal C_i]\}$.
The average of the replicated partition function over the choice of the formula from the Uniform Ensemble $\mathcal P_\mathrm{unif}$, which I shall denote by an overline, is
\begin{equation}
 \overline{ Z(\beta)^n } \equiv \sum_{\mathcal F} \mathcal P_\mathrm{Unif}[\mathcal F] Z_\mathcal F(\beta)^n 
 = \overline{ \sum_{\sigma^1,\dots,\sigma^n} \prod_{i=1}^M z_i(\sigma^1) \cdots z_i(\sigma^n) }
\end{equation}
where $\sigma^a$ is the $N$-component configuration of replica $a$.

Since the literals appearing in each clause are extracted independently on the other clauses, the average over the choice of the formula reduces to the average over the literals appearing in a clause, raised to the power $M$:
\begin{equation}
 \overline{ Z(\beta)^n }  
 = \sum_{\sigma^1,\dots,\sigma^n} \left[ \overline{ z(\sigma^1) \cdots z(\sigma^n) } \right] ^ M \,.
\label{Replicated Z average}
\end{equation}
Let us consider a term in the sum, corresponding to a given $\boldsymbol \sigma \equiv (\sigma^1,\dots,\sigma^n)$. It is the average over the choice of the literals appearing in the clause of a product over the replica index $a$ of a quantity which is 1 if the clause considered is satisfied by replica $a$ and $e^{-\beta}$ otherwise. Let us denote by $i_j$ the index of the $j^\text{th}$ literal in the clause, and by $q_j$ a variable which is $-1$ if it is negated and 1 otherwise. We have:
\begin{equation}
 \overline{ z(\sigma^1) \cdots z(\sigma^n) }
 = \binom N k ^ {-1} \sum_{i_1 < \cdots < i_k} ^ {1,N} \frac 1 {2^k} \sum_{ q_1, \dots, q_k} ^ {\{-1,1\}} \prod_{a=1}^n \left\{ 1 + \left( e^{-\beta} -1 \right) \prod_{j=1}^k \delta(\sigma_{i_j}^a, q_j) \right\} 
\label{z^1 z^n average}
\end{equation}
where the $\delta$ is a Kronecker function. In the following we shall consider the limit $N \to \infty$, and in view of this we can neglect the constraint of the $k$ indices being different and approximate the binomial with $N^{k}$. 

The product over $a$ appearing in (\ref{z^1 z^n average}) is a function of the replicated configurations $\vec \sigma_{i_j}$ at the sites $\{i_1,\dots,i_k\}$. Since we are averaging over the choice of the sites, it is convenient to introduce
\begin{equation}
 \rho(\vec \tau | \boldsymbol \sigma) \equiv \frac 1 N \sum_{i=1}^N \prod_{a=1}^n \delta(\tau^a, \sigma_i^a)
\label{Def rho tau sigma}
\end{equation}
which is the fraction of sites that, for a given $\boldsymbol \sigma$, are equal to the $n$-component configuration $\vec \tau$. We then have
\begin{equation}
 \overline{ z(\sigma^1) \cdots z(\sigma^n) }
 = \sum_{\vec \tau_1, \dots, \vec \tau_k} \rho(\vec \tau_1 | \boldsymbol \sigma) \cdots \rho(\vec \tau_k | \boldsymbol \sigma) \, \mathcal E (\vec \tau_1, \dots, \vec \tau_k)
\end{equation}
where 
\begin{equation}
 \mathcal E (\vec \tau_1, \dots, \vec \tau_k) \equiv \frac 1 {2^k} \sum_{q_1,\dots,q_k}^{\{-1,1\}} \prod_{a=1}^n \left\{ 1 + \left( e^{-\beta} -1 \right) \prod_{j=1}^k \delta(\tau_j^a, q_j) \right\} \,.
\label{Def effective coupling}
\end{equation}

The replicated partition function (\ref{Replicated Z average}) is then
\begin{equation}
 \overline{ Z(\beta) ^ n} = \sum_{\boldsymbol \sigma} \exp \left\{ M \log \sum_{\vec \tau_1, \dots, \vec \tau_k} \rho(\vec \tau_1 | \boldsymbol \sigma) \cdots \rho(\vec \tau_k | \boldsymbol \sigma) \, \mathcal E (\vec \tau_1, \dots, \vec \tau_k) \right\} \,.
\end{equation}
We can introduce the function $c(\vec \tau)$ and multiply the previous expression by the functional integral
\begin{equation}
 \int_0^1 \delta c(\cdot) \, \delta \bigl[ c(\vec \tau) - \rho(\vec \tau | \boldsymbol \sigma) \bigr] = 1 
\end{equation}
(where the integrand is a functional Dirac distribution), to obtain
\begin{eqnarray}
 \overline{ Z(\beta) ^ n} &=& \int_0^1 \delta c(\cdot) \, \exp \left\{ \alpha N \log \sum_{\vec \tau_1, \dots, \vec \tau_k} c(\vec \tau_1) \cdots c(\vec \tau_k) \, \mathcal E (\vec \tau_1, \dots, \vec \tau_k) \right\} 
 \sum_{\boldsymbol \sigma} \delta \bigl[ c(\cdot) - \rho(\cdot | \boldsymbol \sigma) \bigr] \,.
 \nonumber \\ 
\label{Replicated Z average 2}
\end{eqnarray}
The sum over $\boldsymbol \sigma$ is the number of $n$-replicated $N$-sites configurations such that for any $\vec \tau$, the fraction of sites that have a replicated configuration $\vec \tau$ is equal to $c(\vec \tau)$. For each of the possible values of $\vec \tau$ one has to choose the $N c(\vec \tau)$ sites that will have $\vec \tau$ as their replicated configuration, and the number of ways to do it is the multinomial coefficient:
\begin{equation}
 \sum_{\boldsymbol \sigma} \delta \bigl[ c(\cdot) - \rho(\cdot | \boldsymbol \sigma) \bigr] = \frac{N!}{\prod_{\vec \tau} \left[N c(\vec \tau)\right]!}
 \sim \exp \left[ -N \sum_{\vec \tau} c(\vec \tau) \log c(\vec \tau) \right] 
\label{Entropy of c}
\end{equation}
to the leading order as $N \to \infty$.

The ``physical'' interpretation of the previous results is the following: the function $c(\cdot)$ is the order parameter of our theory, the term which multiplies $N$ in the exponent of (\ref{Replicated Z average 2}) is the effective energy expressed in terms of $c(\cdot)$, and (\ref{Entropy of c}) is the (microcanonical) entropy of $c(\cdot)$. This follows exactly the scheme traced in Paragraph \ref{Par Replica method for diluted}. Notice, moreover, that the ``physical'' inverse temperature $\beta$ only appears in the definition of $\mathcal E$, which is the effective interaction strength, and that the parameter which plays the role of the inverse temperature in the effective theory is $\alpha$. 

\subsection{Free energy and replica symmetric ansatz}
\label{Par Replica symmetric free energy}

We can write (\ref{Replicated Z average 2}) in terms of an effective free energy density
\begin{equation}
\mathscr F[c(\cdot), n, \beta, \alpha] \equiv - \sum_{\vec \tau} c(\vec \tau) \, \log c(\vec \tau) + \alpha \log \sum_{\vec \tau_1, \dots, \vec \tau_k} c(\vec \tau_1) \cdots c(\vec \tau_k) \, \mathcal E (\vec \tau_1, \dots, \vec \tau_k)
\label{Free energy}
\end{equation}
as the functional integral
\begin{equation}
 \overline{ Z(\beta) ^ n} = \int_0^1 \delta c(\cdot) \, e^{N \mathscr F[c(\cdot), n, \beta, \alpha]} \,,
\end{equation}
which in the thermodynamic limit $N \to \infty$ can be evaluated by saddle point. The free energy density, defined as
\begin{equation}
 f(\beta, \alpha) \equiv \frac 1 \beta \lim_{N \to \infty} \frac 1 N \lim_{n \to 0} \frac 1 n \log \overline{ Z(\beta) ^ n}
\end{equation}
is then equal to the limit for $n \to 0$ of the extremum value of (\ref{Free energy}) over $c(\cdot)$,
\begin{equation}
 f(\beta, \alpha) = \lim_{n \to 0} \frac 1 {\beta n} \extremum_{c(\cdot)} \mathscr F[c(\cdot), n, \beta, \alpha] \,.
\label{Saddle point equation 1}
\end{equation}
Notice however that, as usual with replica calculations, the order of the two limits $N \to \infty$ and $n \to 0$ has been reversed, which has no \emph{a priory} justification.

In order to compute the extremum of the effective free energy, some assumption must be made on the form of the function $c(\cdot)$. The replica symmetric \emph{ansatz} considers only functions that are symmetric in the replica index, of the form:
\begin{equation}
 c(\vec \tau) = \gamma \left( \sum_{a=1}^n \tau^a \right) \,.
\end{equation}
Under this assumption, a convenient parameterization of the function $\gamma(\cdot)$ is in terms of the auxiliary function $R(h)$
\begin{equation}
 c(\vec \tau) = \int_{-\infty}^\infty dh \, R(h) \frac{\exp \left\{ \frac {\beta h} 2 \sum_{a=1}^n \tau^a \right\}} {\left[ 2 \cosh (\beta h / 2) \right] ^ n} \,.
\label{Def R(h)}
\end{equation}
A few remarks are in order. First, we expect the function $\gamma(\cdot)$ to be even, because in (\ref{Def effective coupling}) we are summing over the values of $q_j$, and $\overline{ Z(\beta)^n}$ is therefore invariant under $\vec \tau \to - \vec \tau$. This implies that $R(h)$ must also be even. Second, $c(\vec \tau)$ is normalized to 1 (it is the fraction of sites that have replicated configuration $\vec \tau$), so also $R(h)$ must be normalized,
\begin{equation}
 \int_{-\infty}^\infty R(h) \, dh = 1 \,.
\label{Normalization R(h)}
\end{equation}
Third, the equation (\ref{Def R(h)}) defining $R(h)$ is, apart from the factor in the denominator, a Laplace transform, so that $R(h)$ is indeed well defined.
Finally, notice that the expression multiplying $R(h)$ in (\ref{Def R(h)}) is the Gibbs weight of a system of $n$ Ising spins $\tau^a$ in a uniform magnetic field $h$ at temperature $\beta$. Since the physical interpretation\footnote{This is true for the function $c(\cdot)$ which extremizes the free energy density. The physical interpretation of $h$ and $R(h)$ therefore holds only for the $R(h)$ corresponding to the extremum.} of $c(\vec \tau)$ is the fraction of sites having the replicated configuration $\vec \tau$, that is to say, the probability that the configuration $\vec \tau$ is observed, the interpretation of $h$ is indeed that of a magnetic field acting on the spins, and the interpretation of $R(h)$ is that of the probability distribution of the values of the field $h$. This observation motivates the introduction of $R(h)$.

\subsection{Selection of satisfiable formul\ae\ by means of a ``chemical potential''}
\label{Par Chemical potential}

So far, the computation has been performed for any $\beta$, and nothing in it ensures that only satisfiable formul\ae\ are considered in the average: the ensemble we are considering is $\mathcal P_\mathrm{Unif}[\mathcal F]$ instead of $\mathcal P_\mathrm{Sat}[\mathcal F]$. I shall now introduce a method which allows to restrict the ensemble to $\mathcal P_\mathrm{Sat}[\mathcal F]$.

\subsubsection*{General strategy}

In all generality, for systems with discrete configurations $\mathscr C$ and discrete non-negative energy $E(\mathscr C) \in \{E_0, E_1, \dots\}$, the partition function $Z(\beta)$ can be written as
\begin{equation}
 Z(\beta) = \sum_{\mathscr C} e^{-\beta E(\mathscr C)} = \sum_i g_i e^{-\beta E_i}
\end{equation}
where $g_i$ is the number of configurations with energy $E_i$. In the presence of disorder, the values of $\{E_i\}$ depend on the sample. The average over disorder of the replicated partition function is then
\begin{equation}
 \overline{ Z(\beta)^n } = \overline{ \left[g_0 e^{-\beta E_0} + g_1 e^{-\beta E_1} + \cdots \right] ^ n } \,.
\label{Zn per nu}
\end{equation}

In order to compute the free energy of the model, the limit $n \to 0$ must be taken, and if one is interested in the low temperature behavior of the model, the question of the value of the product $\beta n \equiv \nu$ rises.

Normally, what is needed is the low temperature behavior of the average of the free energy over all values of the disorder, and $n$ must go to 0 \emph{before} sending $\beta \to \infty$, which corresponds to $\nu = 0$. In our case, however, we would like to select the values of the disorder parameters that minimize the energy of the system, and to restrict the average to these values. Let us see what happens when taking $\beta \to \infty$ and $n \to 0$ with finite $\nu$. We can formally develop the multinomial in  (\ref{Zn per nu}) to obtain
\begin{eqnarray}
 \overline{ Z(\beta)^n } 
 &=&  \overline{ g_0^n e^{-\nu E_0} } + n \ \overline{ g_0^{n-1} e^{-\beta (n-1) E_0} \left[ g_1 e^{-\beta E_1} + \cdots \right] } \, + \cdots \nonumber \\
 &=& \overline{ g_0^n e^{-\nu E_0} } + n \ \overline{ g_0^{n-1} g_1 e^{-\nu E_0} \, e^{\beta (E_0 - E_1)} } \, + \cdots 
\end{eqnarray}
where each term after the first has a factor $e^{\beta(E_0 - E_i)}$ which makes it vanish, so that only the first term contributes. Since $g_0$ is independent on $n$, and $n \to 0$ we remain with
\begin{equation}
  \overline{ Z(\beta)^n } \simeq \overline{ e^{-\nu E_0} } \,.
\end{equation}
We see that the consequence of taking $\nu > 0$ is to include in the computation of the replicated partition function, for a given realization of the disorder, only the lowest energy configurations. 

The energy $E_0$ is the \emph{extensive} energy of the ground state of the system for a given realization of the disorder, which can be regarded as a random variable over the distribution of disorder. Let us denote by $\omega(\epsilon)$ the large deviations function of the distribution of the energy density $\epsilon = E_0 / N$ of the ground state, i.e.
\begin{equation}
 \mathbb P[E_0 = N \epsilon] = e^{N \omega(\epsilon) + o(N)} \,.
\end{equation}
It is reasonable to expect that $\omega(\epsilon)$ will be a negative convex function (i.e. $\omega''(\epsilon) < 0$), vanishing in its maximum. Let's assume that this is the case. For large $N$ we shall have
\begin{equation}
   \overline{ Z(\beta)^n } 
 \simeq \int d\epsilon \, e^{N \left[ \omega(\epsilon) - \nu \epsilon \right] } \simeq e^{N \varphi(\nu)}
\label{Zn as function of nu}
\end{equation}
where
\begin{equation}
 \varphi(\nu) \equiv \max_{\epsilon} \left[ \omega(\epsilon) - \nu \epsilon \right]
\end{equation}
is the Legendre transform of $\omega(\epsilon)$, provided the convexity assumption on $\omega$ holds (which can be verified \emph{a posteriori}).

The integral in (\ref{Zn as function of nu}) will be dominated by the contribution from the value of $\epsilon$ which maximizes the exponent, which is given by 
\begin{equation}
 \epsilon_0(\nu) = -\de_\nu \varphi(\nu) \,.
\label{Average energy}
\end{equation}
The partition function computed in (\ref{Zn as function of nu}) is therefore averaged only on those values of disorder that give a ground state energy equal to $N\epsilon_0(\nu)$:  the parameter $\nu$ allows to restrict the distribution of the disorder to some subset with a well defined ground state energy. In this regard, it plays a role similar to a chemical potential in thermodynamics.

Let us now turn to the application of this program to compute the replica symmetric free energy of \ksat\ over the Satisfiable ensemble $\mathcal P_\mathrm{Sat}[\mathcal F]$. The strategy will be to substitute the replica symmetric \emph{ansatz} (\ref{Def R(h)}) for $c(\cdot)$ in the free energy (\ref{Free energy}), and take the limits $\beta \to \infty$ and $n \to 0$ with finite $\nu = \beta n$, to obtain a free energy functional depending on $\nu$, analogous to $\varphi(\nu)$ in (\ref{Zn as function of nu}), and which will have a functional dependence on $R(h)$. Then to derive the saddle point equation corresponding to (\ref{Saddle point equation 1}) and which will determine $R(h)$, and solve them for generic $\nu$. We shall compute the average ground state energy as a function of $\nu$, as in (\ref{Average energy}), and find the value of $\nu$ corresponding to zero energy, which will select satisfiable formul\ae. The equilibrium distribution of fields $R(h)$ over the Satisfiable Ensemble will finally allow us to characterize the solutions.

\subsubsection*{Entropic term}

The entropic term of (\ref{Free energy}),
\begin{equation}
 \mathscr S[c(\cdot)] \equiv - \sum_{\vec \tau} c(\vec \tau) \log c(\vec \tau)
\end{equation}
can be computed by means of the following identity:
\begin{equation}
 x \log x = \left. \frac{d x^{p+1} }{d p} \right|_{p=0} \,.
\end{equation}
We obtain:
\begin{equation}
 \mathscr S[c(\cdot)] = - \left. \sum_{\vec \tau} \frac {d} {dp} c(\vec \tau)^{p+1} \right|_{p=0} = - \left. \frac {d} {dp} \sum_{\vec \tau} c(\vec \tau)^{p+1} \right|_{p=0}
\label{Entropic term 1}
\end{equation}
with
\begin{eqnarray}
 \sum_{\vec \tau} c(\vec \tau)^{p+1} &=& \sum_{\vec \tau} \left\{ \int_{-\infty}^\infty dh \, R(h) \frac {\exp \left[\frac {\beta h} 2 \sum_{a=1}^n \tau^a \right]}{[2 \cosh \frac {\beta h} 2 ]^n} \right\}^{p+1}  \\
 &=& \int_{-\infty}^\infty dh_1 \cdots dh_{p+1} \, R(h_1) \cdots R(h_{p+1}) \left[ \frac {2 \cosh \left[ \frac \beta 2 \sum_{j=1}^{p+1} h_j \right]} {2 \cosh \frac{\beta h_1} 2 \cdots \cosh \frac{\beta h_{p+1}} 2 } \right] ^n \,.
\end{eqnarray}
We can now multiply by
\begin{equation}
 1 = \int_{-\infty}^\infty d \hat x \, \delta \left( \hat x - \sum_{j=1}^{p+1} h_j \right) = \int_{-\infty}^\infty \frac{dx \, d \hat x}{2 \pi} e^{i x \left( \hat x - \sum_{j=1}^{p+1} h_j \right)} 
\end{equation}
to obtain
\begin{equation}
 \sum_{\vec \tau} c(\vec \tau)^{p+1} = \int_{-\infty}^\infty \frac{dx \, d \hat x}{2 \pi} e^{i x \hat x} \left( 2 \cosh \frac {\beta \hat x} 2 \right)^n \left[ \int_{-\infty}^\infty dh \, \frac {R(h) e^{-ixh}}{\left( 2 \cosh \frac {\beta h} 2 \right)^n} \right] ^{p+1} \,.
\end{equation}
By taking the derivative as in (\ref{Entropic term 1}) we find
\begin{equation}
 \mathscr S[c(\cdot)] = - \int_{-\infty}^\infty \frac{dx \, d \hat x}{2 \pi} e^{i x \hat x} \left( 2 \cosh \frac {\beta \hat x} 2 \right)^n \phi(x) \log \phi(x)
\end{equation}
where 
\begin{equation}
 \phi(x) \equiv \int_{-\infty}^\infty dh \, \frac {R(h) e^{-ixh}}{\left( 2 \cosh \frac {\beta h} 2 \right)^n} \,.
\end{equation}

In the limit $\beta \to \infty$, $n \to 0$ with finite $\nu = \beta n $ we have
\begin{equation}
 \lim_{n \to 0} \left[ 2 \cosh \frac {\nu h} {2 n} \right]^n = e^{\nu \frac {|h|} 2}
\label{Limit of cosh beta h}
\end{equation}
and
\begin{equation}
 \mathscr S[c(\cdot)] = - \int_{-\infty}^\infty \frac{dx \, d \hat x}{2 \pi} e^{i x \hat x + \nu \frac {\left| \hat x \right|} 2} \phi(x) \log \phi(x)
\end{equation}
with
\begin{equation}
 \phi(x) = \int_{-\infty}^\infty dh \, e^{-i x h - \nu \frac {|h|} 2} R(h) \,.
\label{Def phi(x)}
\end{equation}

\subsubsection*{Energetic term}

For the second term in (\ref{Free energy}), we have 
\begin{eqnarray}
 \mathscr E[c(\cdot)] &\equiv& \alpha \log \sum_{\vec \tau_1, \dots, \vec \tau_k} c(\vec \tau_1) \cdots c(\vec \tau_k) \, \mathcal E (\vec \tau_1, \dots, \vec \tau_k) \\
 &=& \alpha \log \sum_{\vec \tau_1, \dots, \vec \tau_k} c(\vec \tau_1) \cdots c(\vec \tau_k) \, \exp \left\{- \beta \sum_{a=1}^n \prod_{j=1}^k \delta (\tau_j^a , 1) \right\}
\label{Energetic term 1}
\end{eqnarray}
where I have simplified the expression (\ref{Def effective coupling}) of the effective coupling $\mathcal E$ taking profit from the sum over $\vec \tau_i$.
Substitution of replica symmetric \emph{ansatz} (\ref{Def R(h)}) gives
\begin{align}
 \mathscr E[c(\cdot)]  & = \alpha \log \sum_{\vec \tau_1, \dots, \vec \tau_k}  \int_{-\infty}^\infty dh_1 \cdots dh_k \, R(h_1) \cdots R(h_k) \times \nonumber \\
 & \hspace{0.5cm} \times \frac{\exp \left[\beta \frac {h_1} 2 \sum_{a=1}^n \tau_1^a \right]}{\left( 2 \cosh \frac {\beta h_1} 2 \right)^n} \cdots \frac{\exp \left[\beta \frac {h_k} 2 \sum_{a=1}^n \tau_1^a \right]}{\left( 2 \cosh \frac {\beta h_k} 2 \right)^n} \, \exp \left\{- \beta \sum_{a=1}^n \prod_{j=1}^k \delta (\tau_j^a , 1) \right\} \\
 &= \alpha \log \int_{-\infty}^\infty dh_1 \cdots dh_k \, \frac {R(h_1)} {\left( 2 \cosh \frac {\beta h_1} 2 \right)^n} \cdots \frac {R(h_k)} {\left( 2 \cosh \frac {\beta h_k} 2 \right)^n} \, \times \nonumber \\
 & \hspace{0.5cm} \times \left\{ \sum_{\vec \tau} \exp \ \beta \left[ \sum_{j=1}^k \frac {h_j} 2 \tau_j - \mathbb I [\tau_1 = \cdots = \tau_k = 1 ] \right] \right\}^n \,.
\label{Energetic term 2}
\end{align}

As $\beta \to \infty$ the sum over $\tau$ is dominated by the term which maximizes the square parenthesis in (\ref{Energetic term 2}), while the hyperbolic cosines are given by (\ref{Limit of cosh beta h}), so that:
\begin{equation}
 \mathscr E[c(\cdot)] = \alpha \log \int_{-\infty}^\infty dh_1 \cdots dh_k \, R(h_1) \cdots R(h_k) \, e^{\nu \Phi(\mathbf h)}
\end{equation}
with $\mathbf h \equiv (h_1,\dots,h_k)$ and 
\begin{eqnarray}
 \Phi(\mathbf h) &=& \max_{\tau \in \{-1,1\}^k} \frac 1 2 \sum_{j=1}^k \left( \tau_j h_j - |h_j| \right) - \mathbb I[\tau, \mathbf 1]  \\
 &=& \left\{
 \begin{aligned}
  & \! - \min \{1, h_1, \dots, h_k \} && \text{ if } h_j > 0 \ \forall j \\
  & 0 && \text{ otherwise}
 \end{aligned}
 \right. \,. \label{Def Phi(x)}
\end{eqnarray}

The free energy functional we obtain, putting $\mathscr E$ and $\mathscr S$ together, is:
\begin{equation}
 \mathscr F[R(\cdot),\nu,\alpha] = - \int_{-\infty}^\infty \frac{dx \, d \hat x}{2 \pi} e^{i x \hat x + \nu \frac {\left| \hat x \right|} 2} \phi(x) \log \phi(x) + \alpha \log \int_{-\infty}^\infty dh_1 \cdots dh_k \, R(h_1) \cdots R(h_k) \, e^{\nu \Phi(\mathbf h)}
\label{Free energy nu}
\end{equation}
with $\phi(x)$ and $\Phi(\mathbf h)$ defined in (\ref{Def phi(x)}) and (\ref{Def Phi(x)}).

\subsection{Saddle point equations}
\label{Par Saddle point equations}

We are now in position to derive the saddle point equations that will determine $R(h)$ from the extremality condition (\ref{Saddle point equation 1}) for $\mathscr F[R(\cdot),\nu,\alpha]$, subject to the normalization condition (\ref{Normalization R(h)}), which we write as
\begin{align}
 & \frac \delta {\delta R(\cdot)} \left\{ \mathscr F[R(\cdot),\nu,\alpha] + \lambda \left[ \int R(h) dh - 1 \right] \right\} = 0  \\
 \Leftrightarrow \ & 
 - \int_{-\infty}^\infty \frac{dx \, d \hat x}{2 \pi} e^{i x \hat x + \frac 1 2 \nu \left| \hat x \right| - i x h -  \frac 1 2 \nu \left|  h \right|} \left[ 1 + \log \phi(x) \right] + \nonumber \\ 
 & + \frac {\alpha k} {\mathscr D[R(\cdot)]} \int_{-\infty}^\infty dh_2 \cdots dh_k \, R(h_2) \cdots R(h_k) \, e^{\nu \Phi(h,h_2,\dots,h_k)} + \lambda = 0
\label{Saddle point 2}
\end{align}
where 
\begin{equation}
 \mathscr D[R(\cdot)] \equiv \int_{-\infty}^\infty dh_1 \cdots dh_k \, R(h_1) \cdots R(h_k) \, e^{\nu \Phi(\mathbf h)}
\label{Def D[R]}
\end{equation}
and where it should be noted that the integral over the fields $h_j$ in (\ref{Saddle point 2}) starts with $h_2$: both terms are functions of $h$, and they must be identically null.

In principle, the symmetry condition $R(h) = R(-h)$ should also be imposed, by means of a second Lagrange multiplier. However, it suffices to restrict the range over which (\ref{Saddle point 2}) defines $R(h)$ to positive values of $h$ and define $R(-h) \equiv R(h)$ with $h > 0$. 

Because of the definition (\ref{Def Phi(x)}) of $\Phi(\mathbf h)$, it is convenient to write the integral over the fields in (\ref{Saddle point 2}) over $\mathbb R^+$ only. This can be done by noticing that if one of the $h_i$ is negative then $\Phi(\mathbf h) = 0$, so that
\begin{align}
 & \int_{-\infty}^\infty dh_2 \cdots dh_k \, R(h_2) \cdots R(h_k) \, e^{\nu \Phi(h,h_2,\dots,h_k)} \nonumber \\
 & = \int_{-\infty}^\infty dh_2 \cdots dh_k \, R(h_2) \cdots R(h_k) \, + \nonumber \\
& \hspace{0.5cm} + \int_0^\infty dh_2 \cdots dh_k \, R(h_2) \cdots R(h_k) \, \left( e^{\nu \Phi(h,h_2,\dots,h_k)} - 1 \right) \\
 & = 1 - \frac 1 {2^{k-1}} + \int_0^\infty dh_2 \cdots dh_k \, R(h_2) \cdots R(h_k) \, e^{- \nu \min\{1,h,h_2,\dots,h_k\} } 
\end{align}
because of the normalization and the symmetry of $R(h)$.
We now multiply by the identity
\begin{equation}
 \int_{-\infty}^\infty \frac {dy \, d \hat y} {2 \pi} e^{i y \left[ \hat y - \min\{1,h_2,\dots,h_k\} \right] } = 1
\end{equation}
to obtain
\begin{align}
 & \int_{-\infty}^\infty dh_2 \cdots dh_k \, R(h_2) \cdots R(h_k) \, e^{\nu \Phi(h,h_2,\dots,h_k)} \nonumber \\
 & = 1 - \frac 1 {2^{k-1}} + \int_{-\infty}^\infty \frac {dy \, d \hat y} {2 \pi} e^{- \nu \min \{h, \hat y\} - i y \hat y} \times \nonumber \\
 & \hspace{0.5cm} \times \int_0^\infty dh_2 \cdots dh_k \, R(h_2) \cdots R(h_k) \, e^{ i y \min\{1,h_2,\dots,h_k\} } \,.
\end{align}
Notice that
\begin{equation}
 \min \{h, \hat y\} = \frac 1 2 \Bigl[ h + \hat y - | h - \hat y |  \Bigr]
\end{equation}
so the exponent in the first integral of the previous equation can be written as
\begin{equation}
 - i y \hat y - \frac 1 2 \nu (h + \hat y) + \frac 1 2 \nu | h - \hat y | 
\end{equation}
and changing the integration variables to $x = y - \frac i 2 \nu $ and $\hat x = h - \hat y$ we obtain
\begin{align}
 & \int_{-\infty}^\infty dh_2 \cdots dh_k \, R(h_2) \cdots R(h_k) \, e^{\nu \Phi(h,h_2,\dots,h_k)} \nonumber \\
 & = 1 - \frac 1 {2^{k-1}} + \int_{-\infty}^\infty \frac {dx \, d \hat x} {2 \pi} e^{i x \hat x + \frac 1 2 \nu |\hat x| - i x h - \frac 1 2 \nu h } \times \nonumber \\
 & \hspace{0.5cm} \times \int_0^\infty dh_2 \cdots dh_k \, R(h_2) \cdots R(h_k) \, e^{  i (x + \frac i 2 \nu) \min\{1,h_2,\dots,h_k\} } \,.
\end{align}
The exponent in the integral over $dx \, d \hat x$ is the same as in the first term of (\ref{Saddle point 2}), and 
\begin{equation}
 \int_{-\infty}^\infty \frac {dx \, d \hat x} {2 \pi} e^{i x \hat x + \frac 1 2 \nu |\hat x| - i x h - \frac 1 2 \nu h } = \int_{-\infty}^\infty d \hat x \, \delta ( \hat x - h ) e^{ \frac 1 2 \nu | \hat x | - \frac 1 2 \nu h }= 1
\end{equation}
since $h > 0$, so we can collect all the terms in (\ref{Saddle point 2}) under the same integral. Let us define the following functions
\begin{eqnarray}
 K(h,x) &=& \int_{-\infty}^\infty \frac {d \hat x}{2 \pi} e^{i x \hat x + \frac 1 2 \nu \left| \hat x \right| - i x h -  \frac 1 2 \nu \left|  h \right|} \, , \\
 Q(x) &=& \int_0^\infty dh_2 \cdots dh_k \, R(h_2) \cdots R(h_k) \, e^{ i x \min\{1,h_2,\dots,h_k\} }
\end{eqnarray}
in terms of which the saddle point equation (\ref{Saddle point 2}) becomes
\begin{equation}
 \int_{-\infty}^\infty dx \, K(h,x) \left\{ - [ 1 + \log \phi(x) ] + \frac {\alpha k} {\mathscr D[R(\cdot)]} \left[ 1 - \frac 1 {2^{k-1}} + Q \left( x + \frac i 2 \nu \right) \right] + \lambda \right\} = 0 \,.
\end{equation}

A solution to this equation is obtained if the curly bracket vanishes identically. In that case, inverting (\ref{Def phi(x)}) we obtain
\begin{eqnarray}
 R(h) &=& \int_{-\infty}^\infty \frac {dx}{2 \pi} \, e^{ixh + \frac 1 2 \nu h} \phi(x) \\
 &=& \int_{-\infty}^\infty \frac {dx}{2 \pi} \exp \left\{ ixh + \frac 1 2 \nu h - 1 +  \frac {\alpha k} {\mathscr D[R(\cdot)]} \left[ 1 - \frac 1 {2^{k-1}} + Q \left( x + \frac i 2 \nu \right) \right] + \lambda \right\} \,.
\label{Saddle point 3}
\end{eqnarray}

\subsection{Distribution of fields}
\label{Par Distribution of fields}

We are now in position to determine the distribution of fields $R(h)$ that satisfies the saddle point equation (\ref{Saddle point 3}). Since this is a functional equation, it's resolution is greatly simplified by making some assumption on the form of the function. I shall consider the following \emph{ansatz} for $R(h)$, 
\begin{equation}
 R(h) = \sum_{p = -\infty}^\infty r_p \, \delta( h - p )
\label{Integer R(h)}
\end{equation}
where only integer values of $h$ are considered. I shall later prove that a more general form in which fractional values are considered reduces to this, suggesting that this is the only solution. 

With this assumption (\ref{Saddle point 3}) becomes an equation for the coefficients $\{r_p\}$. Let us begin from (\ref{Def D[R]}) by computing 
\begin{eqnarray}
 \mathscr D[R(\cdot)] &=& \int_{-\infty}^\infty dh_1 \cdots dh_k \, R(h_1) \cdots R(h_k) \, e^{\nu \Phi(\mathbf h)} \\
 &=& \sum_{p_1,\dots,p_k}^{-\infty, \infty} r_{p_1} \cdots r_{p_k} \times 1 + \sum_{p_1,\dots,p_k}^{1, \infty} r_{p_1} \cdots r_{p_k} \left( e^{-\nu} - 1 \right) \\
 &=& 1 + \left( \frac {1-r_0} 2 \right)^k \left(e^{-\nu}-1\right) \ \ \equiv \ \ D
\label{Integer D}
\end{eqnarray}
where I used the fact that, for integer $\{h_j\}$, if some $h_j$ is negative or null then $\Phi(\mathbf h)$ is 0 and otherwise it is 1, and the symmetry and normalization of $R(h)$ which imply that $r_p = r_{-p}$ and $\sum_{p=-\infty}^\infty r_p = 1$.

Similarly for the term in $Q$ in the exponent of (\ref{Saddle point 3}), which can be written
\begin{eqnarray}
 Q \left( x + \frac i 2 \nu \right)
 &=& \int_0^\infty dh_2 \cdots dh_k \, R(h_2) \cdots R(h_k) \cos \left[x \min\{1,h_2,\dots,h_k\} \right] \, e^{ - \frac 1 2 \nu \min\{1,h_2,\dots,h_k\} } \nonumber \\ \\
 &=& \sum_{p_2,\dots,p_k}^{0,\infty} r_{p_2} \cdots r_{p_k} \times 1 + \sum_{p_2,\dots,p_k}^{1,\infty} r_{p_2} \cdots r_{p_k} \left[ \cos(x) \, e^{-\frac 1 2 \nu} - 1 \right] \\
 &=& \left( \frac {1+r_0} 2 \right) ^{k-1} + \left( \frac {1-r_0} 2 \right)^{k-1} \left[\cos(x) \, e^{-\frac 1 2 \nu} - 1 \right] \\
 &=& A + B \cos(x)
\label{Integer Q}
\end{eqnarray}
with
\begin{eqnarray}
 A &\equiv& \left( \frac {1+r_0} 2 \right)^{k-1} - \left( \frac {1-r_0} 2 \right)^{k-1} , 
\label{Def A} \\
 B &\equiv& \left( \frac {1-r_0} 2 \right)^{k-1} e^{ - \frac 1 2 \nu }  \,.
\label{Def B}
\end{eqnarray}

Substituting (\ref{Integer D}) and (\ref{Integer Q}) into the saddle point equation (\ref{Saddle point 3}) gives
\begin{equation}
 R(h) = e^{\lambda' + \frac 1 2 \nu |h|} \int_{-\infty}^\infty \frac {dx}{2 \pi} \cos(xh) \exp \left\{ \alpha k \left[ \frac A D + \frac B D \cos(x) \right] \right\}
\end{equation}
where $h$ can be positive or negative, and 
\begin{equation}
 \lambda' = \lambda - 1 + \frac {\alpha k} D \left[ 1 - \frac 1 {2^{k-1}} \right] \,.
\end{equation}
This form is compatible with the \emph{ansatz} (\ref{Integer R(h)}), since it vanishes unless $h$ is an integer, and we can invert it to obtain
\begin{eqnarray}
 r_p &=& e^{\lambda' + \frac 1 2 \nu |h| + \alpha k \frac A D} \int_{-\pi}^\pi \frac {dx}{2 \pi} e^{ixp} \exp \left[ \alpha k \frac B D \cos(x) \right] \\
 &=&  e^{\lambda' + \frac 1 2 \nu |h| + \alpha k \frac A D} I_p \left( \alpha k \frac B D \right)
\end{eqnarray}
where $I_p(x)$ is the modified Bessel function of integer order $p$.
The value of $\lambda'$ is determined by the normalization of $R(h)$, and we obtain
\begin{equation}
 r_p = \frac {  e^{\frac 1 2 \nu |p|} I_p \left(\alpha k \frac B D \right) } { \sum_{q=-\infty}^\infty e^{\frac 1 2 \nu |q| } I_q \left(\alpha k \frac B D \right) } \,.
\label{Solution for r_p}
\end{equation}
In this formula, $B$ depends on $r_0$. It is therefore an equation for $r_0$ and, once solved for $r_0$, and identity for all other values of $p$.

\subsection{Ground state energy}
\label{Par Ground state energy}

Having obtained the explicit expression of the equilibrium distribution $R(h)$, we can compute the average value of the ground state energy density $\epsilon_0 (\nu)$ for general $\nu$. 

Following (\ref{Average energy}), we write from the form of the free energy density functional (\ref{Free energy nu})
\begin{eqnarray}
 \epsilon_0(\nu) &=& -\frac \de {\de \nu} \mathscr F[R(\cdot),\nu,\alpha] \\
 &=& \frac 1 2 \int_{-\infty}^\infty \frac {dx \, d \hat x} {2 \pi} e^{ ix \hat x + \frac 1 2 \nu | \hat x |} \left\{ | \hat x | \phi(x) \log \phi(x) - \left[ 1 + \log \phi(x) \right] \int_{-\infty}^\infty dh \, e^{-ixh - \frac 1 2 \nu |h| } \, |h| \, R(h) \right\} + \nonumber \\
 && - \alpha \int_{-\infty}^\infty dh_1 \cdots dh_k \frac {R(h_1) \cdots R(h_k)} {\mathscr D[R(\cdot)]} \Phi(\mathbf h) e^{\nu \Phi(\mathbf h)} \,.
\end{eqnarray}

The integrals $dx \, d \hat x$ can be eliminated by means of the saddle point conditions (\ref{Saddle point 3}) and (\ref{Saddle point 2}), which give
\begin{gather}
 \log \phi(x) = \lambda' + \alpha k \int_0^\infty dh_2 \cdots dh_k \frac {R(h_2) \cdots R(h_k)} {\mathscr D[R(\cdot)]} e^{\left( ix - \frac 1 2 \nu \right) \min\{1,h_2,\dots,h_k\} } \\
 \int_{-\infty}^\infty \frac {dx \, d \hat x} {2 \pi} e^{i x \hat x + \frac 1 2 \nu | \hat x |} \left[ 1 + \log \phi(x) \right] e^{-ixh - \frac 1 2 |h|}
 = \lambda \alpha k \int_0^\infty dh_2 \cdots dh_k \frac {R(h_2) \cdots R(h_k)} {\mathscr D[R(\cdot)]} e^{\nu \Phi(h,h_2,\dots,h_k)}
\end{gather}
from which we obtain
\begin{eqnarray}
 \epsilon_0(\nu) &=& -\int_0^\infty h R(h) \, dh + \frac {\alpha k} 4 \int_0^\infty dh_2 \cdots dh_k \frac {R(h_2) \cdots R(h_k)} {\mathscr D[R(\cdot)]} \min \{1,h_2,\dots,h_k\} + \nonumber \\
 && + \alpha \int_0^\infty dh_1 \cdots dh_k \frac {R(h_1) \cdots R(h_k)} {\mathscr D[R(\cdot)]} \left[ \frac k 2 \min \{1,h_2,\dots,h_k\} + (1-k) \min \{1,h_1,\dots,h_k\} \right] \times \nonumber \\
 && \times e^{-\nu \min \{1,h_1,\dots,h_k\} } \,.
\label{Epsilon_0(nu) 1}
\end{eqnarray}

This expression is valid independently on the form of $R(h)$. For the \emph{ansatz} (\ref{Integer R(h)}) we have
\begin{eqnarray}
 \epsilon_0(\nu) &=& - \sum_{p=1}^\infty p \, r_p + \frac {\alpha k} 4 \sum_{p_2,\dots,p_k}^{1,\infty} \frac {r_{p_2} \cdots r_{p_k}} D + \frac {\alpha k} 4 r_0 \sum_{p_2,\dots,p_k}^{1,\infty} \frac {r_{p_2} \cdots r_{p_k}} D + \nonumber \\
 && + \frac {\alpha k} 2 \sum_{p_1,\dots,p_k}^{1,\infty} \frac {r_{p_1} \cdots r_{p_k}} D e^{-\nu \min \{1,p_1\}} + \alpha(1-k) \sum_{p_1,\dots,p_k}^{1,\infty} \frac {r_{p_1}\cdots r_{p_k}} D e^{-\nu} \\
 &=& - \sum_{p=1}^\infty p \, r_p + \frac {\alpha k} {2 D} \left( \frac {1-r_0} 2 \right)^{k-1} \left( \frac {1+r_0} 2 \right) + \frac \alpha D \left[ \frac k 2 +(1-k) \right] \left( \frac {1-r_0} 2 \right)^k e^{-\nu} \\
 &=& -\sum_{p=1}^\infty p \, r_p + \frac {\alpha k} 2 \frac B D \left( \frac {1+r_0} 2 \right) e^{\nu/2} + \alpha \left( 1 - \frac k 2 \right) \frac B D \left( \frac {1-r_0} 2 \right) e^{-\nu/2}
 \label{Epsilon_0(nu) 2}
\end{eqnarray}
where the term corresponding to $p_1 = 0$ in the first term of the second line of (\ref{Epsilon_0(nu) 1}) has an extra factor $1/2$ coming from the integral $\int_0^\infty \delta(x) dx$. The sum in the last expression can be computed as
\begin{equation}
 \sum_{p=1}^\infty p \, r_p = - \frac \de {\de \nu} \log \mathscr I \left(\alpha k \frac B D, \nu \right)
\end{equation}
where 
\begin{equation}
 \mathscr I (x, \nu) \equiv \sum_{p=-\infty}^\infty e^{-\frac 1 2 \nu |p|} I_p(x) = 2 e^{x \cosh (\nu/2)} - I_0(x) - 2 \sum_{p=1}^\infty e^{-\nu/2} I_p(x)
\end{equation}
converges very fast for $\nu > 0$.

\subsubsection*{Large $\nu$ expansion}

I am going to show that the condition $\epsilon_0(\nu) = 0$, which corresponds to the selection of satisfiable formul\ae\ from the ensemble $\mathcal P _\mathrm{Sat}[\mathcal F]$, is obtained for $\nu \to \infty$.

Let me denote $\varepsilon = e^{-\nu}$ and, to first order in $\varepsilon$
\begin{eqnarray}
 G &\equiv& \frac {\alpha k} 2 \frac B D e^{\nu / 2} = \frac {\alpha k} 2 \frac { \left( \frac {1-r_0} 2 \right) ^{k-1} } {1 - \left( \frac {1-r_0} 2 \right) ^ k (1 -\varepsilon) } = \mathscr G \left[ 1 - \varepsilon \frac {2 \mathscr G }{\alpha k} \frac {1-r_0} 2 \right] \,, \\
 \mathscr G &\equiv& \mathscr G(r_0) \equiv \frac {\alpha k} 2 \frac { \left( \frac {1-r_0} 2 \right) ^{k-1} } {1 - \left( \frac {1-r_0} 2 \right) ^ k } \,.
\label{Def G}
\end{eqnarray}
The Bessel functions $I_p(x)$ can be expanded for small $x$ and $p \geq 0$ as
\begin{equation}
 I_p(x) = \frac {x^p} {2^p p!} \left[ 1 + \frac {x^2} {4 (p+1)} + O(x^4) \right]
\end{equation}
and $I_{-p}(x) = I_p(x)$. Since from the definition (\ref{Def B}) of $B$ we have that it is $O(\varepsilon^{1/2})$ while from (\ref{Integer D}) we have $D = O(1)$, we can expand
\begin{eqnarray}
 \mathscr I \left(\alpha k \frac B D, \nu \right) &=& \sum_{p=-\infty}^\infty e^{\frac 1 2 \nu |p|} I_p \left(\alpha k \frac B D \right) \\
 &=& \sum_{p=-\infty}^\infty e^{\frac 1 2 \nu |p|} \frac { \left( \frac {\alpha k} 2 \frac B D \right)^{|p|} } { 2^{|p|} |p|!} \left[ 1 + \frac {\left( \frac {\alpha k} 2 \frac B D \right)^2} { 4(|p|+1) } + O(\varepsilon^2) \right] \\
 &=& \sum_{p=-\infty}^\infty \frac {G^{|p|}} {|p|!} \left[ 1 + \varepsilon \frac{G^2}{|p|+1} + O(\varepsilon^2) \right] \\
 &=& 2 e^G -1 + \varepsilon \left( 2 G e^G - G^2 - 2G \right) + O(\varepsilon^2) \,.
\end{eqnarray}

We can then write, in the equation (\ref{Solution for r_p}) for $r_0$, to the leading order in $\varepsilon$:
\begin{eqnarray}
 r_0 &=& \frac { I_0 \left( 2 G e^{\nu / 2} \right) } {2 e^G -1 + \varepsilon \left( 2 G e^G - G^2 - 2G \right) + O(\varepsilon^2) } \\
 &=& \frac { 1 + \varepsilon G^2 + O(\varepsilon^2) } {2 e^G -1 + \varepsilon \left( 2 G e^G - G^2 - 2G \right) + O(\varepsilon^2) } \\
 &=& \frac 1 {2 e^{\mathscr G} - 1 } \left\{ 1 + \varepsilon \mathscr G^2 + \frac \varepsilon {2 e^{\mathscr G} - 1} \left[ 2 e^\mathscr G \frac {2 \mathscr G}{\alpha k} \frac {1-r_0} 2 - 2 \mathscr G e^ \mathscr G + \mathscr G^2 + 2 \mathscr G \right] + O(\varepsilon^2) \right\} \\
 &\equiv& F_0(r_0) + \varepsilon F_1(r_0) + O(\varepsilon^2) \,.
\label{r_0 at large nu}
\end{eqnarray}
Let me define 
\begin{eqnarray}
 \rho_0 &=& \lim_{\nu \to \infty} r_0 \,, \\
 \rho_1 &=& \lim_{\nu \to \infty} \frac {r_0 - \rho_0} \varepsilon
\end{eqnarray}
so that $r_0 = \rho_0 + \varepsilon \rho_1 + o(\varepsilon)$. The value of $\rho_0$ is determined by the equation
\begin{equation}
 \rho_0 = \frac 1 {2 e^{\mathscr G(\rho_0)} -1} \,.
\label{rho_0 replica}
\end{equation}
The value of $\rho_1$ is obtained by developing (\ref{r_0 at large nu}) around $\rho_0$:
\begin{equation}
 \rho_0 + \varepsilon \rho_1 = \rho_0 + F'_0(\rho_0) \varepsilon \rho_0 + \varepsilon F_1(\rho_0) 
\end{equation}
which gives
\begin{equation}
 \rho_1 = \frac {F_1(\rho_0)} {1-F'_0(\rho_0)}
\end{equation}

In order to write the average ground state energy for large $\nu$ we also need to compute
\begin{eqnarray}
 \sum_{p=1}^\infty p \, e^{\frac 1 2 \nu p} I_p \left( \frac {\alpha k} 2 \frac B D \right) &=& \sum_{p=1}^\infty \frac {G^p}{(p-1)!} \left[ 1 + \frac {\varepsilon G^2} {p+1} + O(\varepsilon^2) \right] \\
 &=& G e^G + \varepsilon G \left( 1 - e^G + G e^G \right) + O(\varepsilon^2) \,.
\end{eqnarray}

Using these expansions in the expression for the average ground state energy (\ref{Epsilon_0(nu) 2}) we obtain, after some algebra,
\begin{eqnarray}
 \epsilon_0(\nu) &=& - \frac { G e^G + \varepsilon G \left( 1 - e^G + G e^G \right) } {2 e^G -1 + \varepsilon \left( 2 G e^G - G^2 - 2G \right) } + G \frac {1 + r_0} 2 + \varepsilon \left( \frac 2 k -1 \right) G \frac {1-r_0} 2 \\
 &=& - \mathscr G e^ \mathscr G \rho_0 \left\{ 1 - \varepsilon \frac {2 \mathscr G^2} {\alpha k} \frac{1-\rho_0}2 \left[ \frac 1 {\mathscr G} - \rho_0 \right] + \varepsilon \left[ e^{-\mathscr G} - 1 + \mathscr G \right] - \varepsilon \rho_0 \left[ 2 \mathscr G e^{\mathscr G} - \mathscr G^2 - 2 \mathscr G \right] \right\} + \nonumber \\
 && + \mathscr G \rho_0 e^{\mathscr G} \left[ 1 - \varepsilon \frac {2 \mathscr G} {\alpha k} \frac {1-\rho_0} 2 \right] + \varepsilon \left( \frac 2 k - 1 \right) \mathscr G \frac {1-\rho_0} 2 \\
 &=& \varepsilon \mathscr G e^{\mathscr G} \rho_0 \left\{ - \frac { \mathscr G^2 \rho_0 (1-\rho_0)}{\alpha k} - \left( e^{-\mathscr G} -1 + \mathscr G \right) + \rho_0 \left( 2 \mathscr G e^{\mathscr G} - \mathscr G^2 - 2 \mathscr G \right) + \frac 1 {\rho_0 e^{\mathscr G} } \left( \frac 2 k -1 \right) \frac {1-\rho_0} 2 \right\} \nonumber \\
\label{GS energy}
\end{eqnarray}
where everything except $\varepsilon$ is $O(1)$ as $\nu \to \infty$. Notice that the term in $\rho_1$ does not contribute to the first order result in the end. 

The conclusion of this somewhat tedious calculation is that 
\begin{equation}
 \epsilon_0(\nu) \mathop{\sim}_{\nu \to \infty} e^{-\nu} \,.
\end{equation}
Therefore, in order to obtain the equilibrium distribution of fields for formul\ae\ extracted from the Satisfiable Ensemble $\mathcal P_\mathrm{Sat}[\mathcal F]$, it is sufficient to take the limit $\nu \to \infty$ in (\ref{Solution for r_p}), giving
\begin{eqnarray}
 \rho_0 &=& \frac 1 {2 e^{\mathscr G(\rho_0)}-1} \,, 
 \label{Sol rho_0} \\
 \rho_p &\equiv& \lim_{\nu \to \infty} r_p = \frac {\mathscr G(\rho_0)^{|p|}} {|p|!} \frac 1 {2 e^{\mathscr G(\rho_0)}-1} \hspace{1cm} (p \neq 0)
 \label{Sol rho_p}
\end{eqnarray}
where $\mathscr G(\rho_0)$ is defined by (\ref{Def G}) as
\begin{equation}
 \mathscr G(\rho_0) = \frac {\alpha k} 2 \frac {\left( \frac {1-\rho_0} 2 \right) ^{k-1} } {1 - \left( \frac {1-\rho_0} 2 \right)^k } \,.
\label{G(rho_0)}
\end{equation}

For any $k$ and $\alpha$ it is easy to solve (\ref{Sol rho_0}) to find $\rho_0$, and then use it to compute all other $\rho_p$, thus completely defining the distribution of fields $R(h)$. In the following we shall see that this is sufficient to characterize some very interesting properties of the solutions. We shall also return on the two \emph{ansatz} we made to derive these results: the replica symmetric form (\ref{Def R(h)}) of $c(\cdot)$, and the integer-only form of $R(h)$ in (\ref{Integer R(h)}), in the Section \ref{Sec Stability of RS free energy} about the stability of the solution.

\section{Cavity formalism for the fields distribution}
\label{Sec Cavity formalism}

The results of the previous Section can be obtained in a rather more straightforward way, at the price of some more assumptions.

Let us consider a formula over $N-1$ variables, and let us add a new variable, which will appear in $\ell_+$ new clauses as a non-negated literal, and in $\ell_-$ as a negated one. For random formul\ae\ from the Uniform Ensemble, $\ell_+$ and $\ell_-$ will be random variables with independent poissonian distribution
\begin{equation}
 p_{L}(\ell) = \frac {(\alpha' k / 2)^\ell}{\ell!} e^{-\alpha' k / 2}
\end{equation}
where $\alpha'$ is some constant that we shall determine later.

Let us denote by $1 - \rho_0$ the probability that an ``old'' variable is constrained, i.e. if it changes value some existing clause will be violated. Then, the new variable will be constrained if and only if all the $k-1$ other variables in the clause are constrained, and if they appear with the ``wrong'' sign in the new clause. The probability for this to happen is
\begin{equation}
 q = \left( \frac {1-\rho_0} 2 \right) ^ {k-1} \,.
\label{Def q}
\end{equation}

The number of clauses that contain the new variable $x$ or its negation $\bar x$ and which constrain them, which I shall denote $m_+$ and $m_-$ respectively, will be independent random variables with distribution
\begin{eqnarray}
 p_M(m) &=& \sum_{\ell = m} ^ \infty p_L(\ell) \binom \ell m q^m (1-q)^{\ell-m} \\
 &=& \frac{(\alpha' k q / 2)^m}{m!} e^{-\alpha' k q /2} \,.
\label{Distribution p_M}
\end{eqnarray}
I shall also introduce a weighted distribution, in which $\ell$ is the weight, for later use:
\begin{eqnarray}
 p^\mathrm w_M(m) &=& \sum_{\ell = m} ^ \infty \ell p_L(\ell) \binom \ell m q^m (1-q)^{\ell-m} \\
 &=& p_M(m) \left[ m + \frac {\alpha' k} 2 (1-q) \right] 
\label{Distribution p_M^w}
\end{eqnarray}
(notice that this is \emph{not} normalized, since $\sum_{m=0}^\infty p^\mathrm w_M(m) = \alpha' k / 2$).

The $m_+$ clauses that constrain $x$ will be satisfied if $x = \true$, while the $m_-$ clauses that constrain $\bar x$ will be satisfied if $x = \false$. The minimal increase in energy after the addition of $x$ to the formula is therefore
\begin{equation}
 \Delta E = \min\{m_+,m_-\} \,.
\end{equation}
Let me define the ``magnetic field'' $h$ as the difference $m_+ - m_-$. Both $\Delta E$ and $h$ will be random variables, with joint distribution
\begin{equation}
 P(\Delta E,h) = \sum_{m_+ = 0}^\infty p_M(m_+) \sum_{m_- = 0}^\infty p_M(m_-) \delta_{\Delta E,\min \{m_+,m_-\}} \delta_{h,m_+-m_-} \,.
\end{equation}

In the spirit of Paragraph \ref{Par Chemical potential}, I am going to weight each possible new formula with a factor $e^{- \nu \Delta E}$. The probability that the new variable is subject to a field $h = p \in \mathbb Z$ is then
\begin{equation}
 r_p (\nu) = \frac {\sum_{\Delta E \geq 0} P(\Delta E,p) e^{-\nu \Delta E} } { \sum_{\Delta E \geq 0} \sum_{m=-\infty}^\infty P(\Delta E, m) e^{-\nu \Delta E} } \,.
\end{equation}

In order to restrict the computation to satisfiable formul\ae, let us take the limit $\nu \to \infty$, so that only formul\ae\ with $\Delta E = 0$ contribute. The probability that the new variable has zero field (i.e. that it is not constrained) is then
\begin{equation}
 \rho_0 = \lim_{\nu \to \infty} r_0(\nu) = \frac { P(0,0) } {\sum_{m=-\infty}^{\infty} P(0,m) } = \frac 1 {2 e^{\alpha' k q / 2} - 1}
\label{rho_0 cavity}
\end{equation}
and since $q$ is a function of $\rho_0$ defined in (\ref{Def q}), this expression is an equation which determines $\rho_0$.

If we had no restrictions on the clauses added to the formula, their average number would be $\alpha k$. However, we are restricting the ensemble to satisfiable formul\ae\ only: some of the potential new clauses will have to be rejected, because they would make the formula \unsat, and the average number of clauses effectively added will be
\begin{eqnarray}
 \left< l_+ + l_- \right> &=& \frac { \sum_{m_+,m_-}^{0,\infty} \left[ p^\mathrm w_M(m_+) p_M(m_-) + p_M(m_+) p^\mathrm w_M(m_-) \right] \delta_{0,\min\{m_+,m_-\}} } { \sum_{m=0}^\infty P(0,m) } \\
 &=& \alpha' k \left[ 1 - \left( \frac {1-\rho_0} 2 \right) ^ k \right] \,.
\end{eqnarray}

In order for $\alpha$ to be the clause to variable ratio of the formula, we must impose
\begin{equation}
 \alpha = \alpha ' \left[ 1 - \left( \frac {1-\rho_0} 2 \right) ^ k \right]
\label{alpha' a}
\end{equation}
which determines $\alpha'$.

Multiplying on both sides by $k q / 2$ and recalling the definition of $q$ we obtain
\begin{equation}
 \frac{\alpha' k q}{2} = \frac {\alpha k} 2 \frac { \left( \frac {1 - \rho_0} 2 \right) ^{k-1} } { 1 - \left( \frac {1 - \rho_0 } 2 \right) ^ k }
\end{equation}
which, compared with (\ref{G(rho_0)}), gives
\begin{equation}
 \mathscr G(\rho_0 ) = \frac {\alpha' k q} 2 \,.
\end{equation}

The equation (\ref{rho_0 cavity}) for $\rho_0$ is then 
\begin{equation}
 \rho_0 = \frac 1 { 2 e^{\mathscr G (\rho_0)} - 1}
\end{equation}
which is exactly the same as (\ref{rho_0 replica}).

Notice that the distribution that we have computed is the distribution of the \emph{cavity} fields, i.e. the fields acting on the new variable and generated by the old ones. \emph{A priori} this distribution is different from that of the real fields, which include the effect of the new clauses on the values of the old variables (and therefore of the fields they induce). The distribution we are interested in is the one of the real fields, which is what we have computed by means of the replica calculation, not the distribution of cavity fields. However, it can be shown that these two distributions coincide in the case when they are poissonian. I shall now prove that this is indeed so, at least in the limit of large $\alpha$.

The generating function $\mathfrak g(x)$ of the distribution of variable occurrences $\ell_+ + \ell_-$ over satisfiable formul\ae, i.e. such that $\min \{m_+,m_-\} = 0$, can be computed as
\begin{eqnarray}
 \mathfrak g(x) &=& \sum_{m_+ = 0}^\infty \sum_{\ell_+ = m_+}^\infty p_L(\ell_+) \binom {\ell_+} {m_+} q^{m_+} (1-q)^{\ell_+-m_+} \times \nonumber \\
 && \times \sum_{m_- = 0}^\infty \sum_{\ell_- = m_-}^\infty p_L(\ell_-) \binom {\ell_-} {m_-} q^{m_-} (1-q)^{\ell_--m_-} \times x^{\ell_+ + \ell_-} \delta_{0, \min\{m_+,m_-\}} \\
 &=& e^{ \alpha' k (x-1) (1-q) } \frac {2 e^{\alpha' k x q / 2} - 1} {2 e^{\alpha' k q / 2} - 1 } \,.
\end{eqnarray}

For $\alpha \to \infty$ we see from (\ref{rho_0 cavity}) that $\rho_0 \to 0$ and from (\ref{alpha' a}) that $\alpha' = O(\alpha)$, so that
\begin{equation}
 \mathfrak g(x) = e^{\alpha' k (x-1) (1- q/2)} + e^{-O(\alpha)} = e^{\alpha k (x-1)}
\end{equation}
which is the generating function of a poissonian distribution of parameter $\alpha k$.

The conclusion of this Section is that the interpretation of the field $h$ as the number of clauses that are violated if a variable is flipped is correct, and the distribution of fields $R(h)$ is the distribution over the variables and the formul\ae\ from $\mathcal P_\mathrm{Sat}$ of their values.

\section{Comparison of  $\mathcal P_\mathrm{Sat}$ and $\mathcal P_\mathrm{Plant}$ at large $\alpha$}
\label{Sec Comparison of P_Sat and P_Plant}

I am now going to use the distribution of fields computed in Section (\ref{Sec Free energy}) to show that, for $\alpha \to \infty$, the statistical properties of formul\ae\ extracted from $\mathcal P_\mathrm {Sat}$ coincide with those of formul\ae\ from $\mathcal P_\mathrm{Plant}$.

For $\alpha \to \infty$ the solution to (\ref{Sol rho_0}), (\ref{Sol rho_p}) and (\ref{G(rho_0)}) is
\begin{eqnarray}
 \rho_0 &=& \frac 1 {2 e^{\gamma} - 1} \,, 
\label{rho_0 large alpha}\\
 \rho_p &=& \frac { \gamma ^ {|p|} } { |p|! } \frac 1 {2 e^\gamma - 1} \,,
\label{rho_p large alpha} \\
 \gamma &\equiv& \mathscr G(0) = \frac {\alpha k}{2^k - 1} \,.
\label{Gamma large alpha}
\end{eqnarray}

Since $\gamma = O(\alpha)$, this means that the fraction of variables that are not constrained is $\rho_0 = e^{-O(\alpha)}$: the solutions to a satisfiable formula at large $\alpha$ are all very similar to each other. Moreover, the average value of the fields is $O(\gamma) = O(\alpha)$, so the constrained variables have strong fields that force them to the correct assignment. 

\subsection{Distribution of fields}
\label{Par Comparison fields}

I shall now compute the distribution of fields for formul\ae\ extracted from the Planted Ensemble $\mathcal P_\mathrm{Plant}$.

Let us consider a configuration $X$, and a random clause $\mathcal C$ satisfied by $X$. If one variable $x_i$ is flipped, what is the probability $q$ that $\mathcal C$ is no longer satisfied? It is the product of the probability that $\mathcal C$ contains $x_i$, which is $k/N$, times the probability that all the other literals in the clause have been chosen with the wrong sign, which is $1/(2^k-1)$
\begin{equation}
 q = \frac k N \frac 1 {2^k-1} \,.
\end{equation}
The number $p$ of such clauses will be a random variable, with a binomial distribution $P(p)$ of parameter $q$
\begin{equation}
 P(p) = \binom M p q^p (1-q)^{M-p} \,.
\end{equation}
For $N \to \infty$ this reduces to a poissonian of parameter $\alpha k / (2^k-1)$, which is $\gamma$ defined in (\ref{Gamma large alpha}),
\begin{equation}
 P(p) \mathop{\sim}_{N \to \infty} e^{-\gamma} \frac {\gamma ^ p}{p!} \,.
\end{equation}

In a random configuration $X$, half the variables will be \true, giving rise to positive fields, and half will be \false, giving negative fields. The distribution of fields, i.e. of the number of satisfied clauses that are violated if a variable is flipped, with the plus sign if that variable is \true\ and minus otherwise, is
\begin{equation}
 \rho_p^\mathrm{Plant} = \delta_{p,0} \, e^{-\gamma} + (1-\delta_{p,0})  \frac 1 2 e^{-\gamma}\frac {\gamma^{|p|}}{|p|!} \,.
\end{equation}

Comparing with (\ref{rho_p large alpha}) we see that the two distributions of fields corresponding to the Satisfiable Ensemble at large $\alpha$ and to the Planted Ensemble differ by terms $e^{-O(\alpha)}$.

\subsection{Correlation between field and number of occurrences}
\label{Par Comparison correlations}

Not only the typical magnitude of the fields in formul\ae\ from $\mathcal P_\mathrm{Sat}$ is of order $\alpha$ at large $\alpha$, but it is correlated to a bias in the distribution of the relative number of occurrences of variables and their negations, as I shall prove with the following computation.

In order for a formula to be satisfiable, there must be no variable that receive contradictory messages, i.e. which is constrained by some clauses to be \true\ and by some other to be \false. If we assume that the field on the variable is $h>0$, this means that the number $m_-$ of clauses that constrain it to be \false\ must be 0, while the number $m_+$ of clauses that constrain it to be \true\ will be positive or null.

Let us denote by $\left< \ell_+ \right>_{h>0}$ the average number of occurrences of such a variable in clauses where it is not negated, and by $\left< \ell_- \right>_{h>0}$ the corresponding number for its negation. These will be random variables whose distribution can be expressed in terms of (\ref{Distribution p_M}) and (\ref{Distribution p_M^w}) as
\begin{equation}
 \left< \ell \, \right>_{h>0} = \frac { \sum_{m_+ \geq 1} p_M^\mathrm w(m_+) \, p_M(0) } { \sum_{m_+ \geq 1} p_M(m_+) \, p_M(0) }
\label{Average ell for E=0}
\end{equation}
where in the numerator $p_M(0)$ is the probability that the number of clauses sending a negative message to the variable is 0, $p_M^\mathrm w(m_+)$ is proportional to the average number of occurrences of the variable conditioned on the message it receives being positive, and the denominator is a normalization.

Using the explicit distributions (\ref{Distribution p_M}) and (\ref{Distribution p_M^w}) we have
\begin{eqnarray}
 \left< \ell_+ \right>_{h>0} &=& \frac { \sum_{m_+ \geq 1} \frac{(\alpha' k q / 2)^{m_+}}{m_+!} e^{-\alpha' k q /2} \left[ m_+ + \frac {\alpha' k} 2 (1-q) \right] \times e^{-\alpha' k q /2} } { \sum_{m_+ \geq 1} \frac{(\alpha' k q / 2)^{m_+}}{m_+!} e^{-\alpha' k q /2} \times e^{-\alpha' k q /2} } \\
 &=& \frac {\alpha' k} 2 \left[ \frac {1-(1-q)e^{-\mathscr G}} {1-e^{-\mathscr G}} \right] \\
 &=& \frac {\alpha k} 2 \frac 1 {1-2^{-k}} + e^{-O(\alpha)} \,, \\
 \left< \ell_- \right>_{h>0} &=& \frac { \sum_{m_+ \geq 1} \frac{(\alpha' k q / 2)^{m_+}}{m_+!} e^{-\alpha' k q /2} \times e^{-\alpha' k q /2} \left[ \frac {\alpha' k} 2 (1-q) \right] } { \sum_{m_+ \geq 1} \frac{(\alpha' k q / 2)^{m_+}}{m_+!} e^{-\alpha' k q /2} \times e^{-\alpha' k q /2} } \\
 &=& \frac {\alpha' k} 2 (1-q) \\
 &=& \frac {\alpha k} 2 \frac {1 - 2^{-(k-1)}} {1-2^{-k}} + e^{-O(\alpha)}
\end{eqnarray}
from which we obtain the average value of the \emph{bias}
\begin{equation}
 \frac { \left< \ell_+ \right>_{h>0} - \left< \ell_- \right>_{h>0} } { \left< \ell_+ \right>_{h>0} + \left< \ell_- \right>_{h>0} } = \frac 1 {2^k-1} + e^{-O(\alpha)} \,.
\label{bias}
\end{equation}

Therefore variables with positive field appear more frequently non-negated than negated. Of course, the opposite is true for variables with negative field.

The same computation can be easily performed for formul\ae\ from the Planted Ensemble. Given a configuration $X$ and $k$ indices of variables composing a clause, out of the $2^k$ possible choices of the negations of the corresponding literals only $2^k-1$ will give satisfied clauses. If a variable $x$ is \true\ in $X$, then the number of satisfied clauses in which it appears non-negated is $2^{k-1}$, corresponding to the random choices of the signs of the other literals; the number of clauses in which it appears negated, however, will be smaller, as at least one of the other literals must have the proper sign to satisfy the clause, giving $2^{k-1}-1$ possible choices.

Since the average numbers of occurrences of $x$ and $\bar x$ are proportional to these probabilities, we shall have
\begin{eqnarray}
 \frac { \left< \ell_+ \right>_\mathrm{Plant} - \left< \ell_- \right>_\mathrm{Plant} } { \left< \ell_+ \right>_\mathrm{Plant} + \left< \ell_- \right>_\mathrm{Plant} } &=& \frac { 1 / {2^{k-1}} - 1 / \left(2^{k-1}-1\right) } { 1 / {2^{k-1}} + 1 / \left(2^{k-1}-1\right) } \\
 &=& \frac 1 {2^k -1} \,.
\end{eqnarray}

Comparing with (\ref{bias}), we see that the distribution of the bias in the Planted Ensemble is the same as in the Satisfiable Ensemble at large $\alpha$, up to terms $e^{-O(\alpha)}$.


\subsection{Finite energy results}
\label{Par Finite energy results}

The results of the two previous paragraphs extend to formul\ae\ with small positive energy, i.e. which are \emph{not} satisfiable.

The average value of the ground state energy, given by (\ref{GS energy}), greatly simplifies for large $\alpha$, giving
\begin{equation}
 \epsilon_0(\nu) = \frac \gamma k \left[ 1 + O(\gamma^2 e^{-\gamma}) \right] \, e^{-\nu}
\label{Finite energy}
\end{equation}
with $\gamma = O(\alpha)$ defined in (\ref{Gamma large alpha}).

The computation of the bias (\ref{Average ell for E=0}) can be generalized to finite large values of $\nu$ by including positive values of $m_-$, weighted with a factor $e^{-\nu m_-}$. To first order in $e^{-\nu}$ only $m_- = 1$ contributes and we have
\begin{eqnarray}
  \left< \ell \, \right>_{h>0} &=& \frac { \sum_{m_+ \geq 1} p_M^\mathrm w(m_+) \, \sum_{0 \leq m_- < m_+} p_M(m_-) \, e^{-\nu m_-} } { \sum_{m_+ \geq 1} p_M(m_+) \, \sum_{0 \leq m_- < m_+} p_M(m_-) \, e^{-\nu m_-} } \\
 &=&  \frac { \sum_{m_+ \geq 1} p_M^\mathrm w(m_+) \, p_M(0) + \sum_{m_+ \geq 2} p_M^\mathrm w(m_+) \, p_M(1) \, e^{-\nu} } { \sum_{m_+ \geq 1} p_M(m_+) \, p_M(0) + \sum_{m_+ \geq 2} p_M^\mathrm w(m_+) \, p_M(1) \, e^{-\nu} } + O(e^{-2\nu}) \,.
\end{eqnarray}
Computing the sums as for (\ref{Average ell for E=0}), we obtain
\begin{equation}
  \frac { \left< \ell_+ \right>_{h>0} - \left< \ell_- \right>_{h>0} } { \left< \ell_+ \right>_{h>0} + \left< \ell_- \right>_{h>0} } = \frac 1 {2^k-1} - \frac {\alpha k} {2 (2^k-1)^2} e^{-\nu} + O(\alpha^{-1}) + O(e^{-2\nu}) \,,
\end{equation}
where we can use (\ref{Finite energy}) to eliminate $e^{-\nu}$ and obtain
\begin{equation}
  \frac { \left< \ell_+ \right>_{h>0} - \left< \ell_- \right>_{h>0} } { \left< \ell_+ \right>_{h>0} + \left< \ell_- \right>_{h>0} } = \frac 1 {2^k-1} \left[ 1 - \epsilon_0 k2^k \left( \frac 1 2 - \frac 1 {2^{k+1}-2} - \frac{2^k} {\alpha k} \right) \right] + o(\epsilon_0)\,.
\end{equation}

We see that as long as $\epsilon_0 \ll 2^{-k} / k$ the bias remains of the same order as for satisfiable formul\ae.

\subsection{Algorithmic implications}
\label{Par Algorithmic implications}

In this section, I have shown that the distribution of fields $\rho_p$ and the average bias obtained for formul\ae\ extracted from the Planted Ensemble coincides with those for formul\ae\ extracted from the Satisfiable Ensemble for large enough $\alpha$, and that this extends to finite energy formul\ae\ from the Uniform Ensemble, provided the energy is $\epsilon_0 \ll 2^{-k}/k$.

The demonstration of \cite{Feige06} of the convergence of WP is based on the following facts, which are proved for the Planted Ensemble:
\begin{itemize}
 \item At large $\alpha$, typical formul\ae\ have a large \emph{core}, i.e. a set of variables that take the same value in all the solutions to the formula. The fraction of variables that are not in the core is $e^{-O(\alpha)}$. 
 \item The cavity fields corresponding to core variables and computed for satisfying assignments are of order $O(\alpha)$.
 \item Even for random assignments, the cavity fields are of order $O(\alpha)$. This is due to the fact that the value of core variables in satisfying assignments is correlated to a bias in the relative number of occurrences of the variable and its negation in the formula.
\end{itemize}

As we have seen in this Section, each of these properties holds as well for formul\ae\ drawn from the Satisfiable Ensemble $\mathcal P_\mathrm{Sat}$, provided $\alpha$ is large enough. This supports the conclusion that the convergence of WP should extend to $\mathcal P_\mathrm{Sat}$. I therefore claim that \hyp{$1_p$}, formulated at the end of Paragraph \ref{Par Problem definition}, is refuted by WP for any $p > 0$.

Moreover, a probabilistic version of \hyp{2} states that
\begin{description}
 \item[Hypothesis $\boldsymbol{2_p}$] For every fixed $\epsilon \geq 0$, even when $\alpha$ is an arbitrarily large constant (independent on $N$), there is no polynomial time algorithm that on most Random-\ksat[3] formul\ae\ outputs \typ\ and  outputs \notyp\  with probability $p$ on formul\ae\ with $(1-\epsilon)M$ satisfiable clauses.
\end{description}
The finite energy results of Paragraph \ref{Par Finite energy results} support the claim that \hyp{$2_p$} is refuted by WP for any $p > 0$ provided $\epsilon \ll 2^{-k}/k$.

\section{Stability of the RS free energy}
\label{Sec Stability of RS free energy}

The conclusions of the previous sections are based on two \emph{ansatz}: that the order parameter $c(\cdot)$ has the replica symmetric form (\ref{Def R(h)}), and that the distribution of fields $R(h)$ is non zero only for integer values of the fields, in (\ref{Integer R(h)}).

In this Section, I shall support the claim that these two \emph{ansatz} are correct. In order to do this, I shall prove that more general solutions for the saddle point equations that determine $R(h)$, which are non zero for fractional values of $h$, reduce to the \emph{ansatz}, i.e. that the non-integer contributions vanish. Then I shall prove that the eigenvalues of the stability matrix of the saddle point equations computed for the replica symmetric form of $c(\cdot)$ are all negative for large enough $\alpha$ and $\nu \to \infty$. This does not prove that the \emph{ansatz} corresponds to a global minimum, but only to a \emph{local} one. In order to rule out the existence of other solutions to the saddle point equations, I shall prove that two real replicas of the formula necessarily have the same distribution of fields, and therefore must be in the same thermodynamic state, which is therefore unique.

\subsection{Solutions with non-integer fields}
\label{Par Stability non-integer fields}

Instead of the integer valued \emph{ansatz} of (\ref{Integer R(h)}), let us assume that $R(h)$ takes the more general form
\begin{equation}
 R(h) = \sum_{p=-\infty}^\infty r_p \, \delta \left( h - \frac p q \right)
\end{equation}
where $q$ is an integer larger than 1. Substituting this assumption in the saddle point equations (\ref{Saddle point 3}) gives the following \emph{functional} equation
\begin{equation}
 \sum_{p=-\infty}^{\infty} r_p \cos \left( x \frac p q \right) e^{- \frac {\nu |p|} {2q} } = \exp \left[ \mu + \alpha k \sum_{j=1}^q A_j \cos \left( x \frac p q \right) e^{- \frac { \nu j } { 2 q} } \right]
\label{Fractional r_p}
\end{equation}
which must be true for any $x$, where $\mu$ is a constant, and where
\begin{eqnarray}
 A_1 &\equiv& \frac { w^{k-1} - (w - r_1)^{k-1} } { 1 - w^k } \,, \\
 A_j &\equiv& \frac { (w - r_{j-1})^{k-1} - (w - r_j)^{k-1} } { 1 - w^k } \hspace{0.5cm} (1 < j < q) \,, \\
 A_q &\equiv& \frac { (w - r_{p-1})^{k-1} } { 1 - w^k } \,, \\
 w &\equiv& \frac {1 - r_0} 2 \,.
\end{eqnarray}

The value of $\mu$ can be determined by taking $x = i \nu / 2$ and then sending $\nu \to \infty$, which gives
\begin{equation}
 \sum_{p = -\infty}^\infty r_p \, \frac { 1 + \delta_{p,0} } 2 = \exp \left( \mu + \frac {\alpha k} 2 \sum_{j=1}^q A_j \right) \,.
\end{equation}
By taking instead $x=0$ and sending $\nu \to \infty$ one also obtains that
\begin{equation}
 r_0 = e^\mu \,.
\end{equation}
Combining these two identities, we obtain an equation for $r_0$:
\begin{equation}
 r_0 = \frac 1 {2 \exp \left[ \frac {\alpha k} 2 \frac {w^{k-1}} {1-w^k} \right] - 1 } \,.
\end{equation}

Notice that this is exactly the same equation (\ref{Sol rho_0}) and (\ref{G(rho_0)}) that we have obtained with the \emph{ansatz} of integer fields (\ref{Integer R(h)}). 

For $j=1$ we have from (\ref{Fractional r_p}):
\begin{equation}
 r_1 = r_0 \frac {A_1} 2 
\end{equation}
which can be written as
\begin{equation}
 r_1 = \frac {r_0} 2 \frac { w^{k-1} - (w - r_1)^{k-1} } { 1 - w^k } \,.
\label{Fractional r_1}
\end{equation}
Notice that $r_1 = 0$ is a solution of this equation. The derivative with respect to $r_1$ of the right hand side is
\begin{equation}
 \frac{r_0} 2 \frac {(k-1) (w-r_1)^{k-2}} {1-w^k} \,.
\end{equation}
When $\alpha$ is large, $r_0 = e^{-O(\alpha)}$ and $w = 1/2 - e^{-O(\alpha)}$. The possible range of value of $r_1$ goes from 0 to $w$ (which is the probability of the field being positive, and therefore must be larger than or equal to $r_1$). For large enough $\alpha$ this derivative is much smaller than $1$ for any of the possible values of $r_1$, and therefore there cannot be another solution to (\ref{Fractional r_1}).

A similar argument can be constructed for any of the coefficients $r_p$ corresponding to fractional values of the field, showing that only integer values are admissible among rationals.
Of course, this doesn't prove that other distributions $R(h)$ satisfying the saddle point equations and involving irrational fields cannot exist, but it is a rather strong indication that the \emph{ansatz} (\ref{Integer R(h)}) is correct.

\subsection{Eigenvalues of the stability matrix}
\label{Par Stability eigenvalues}

The stability matrix of the free energy (\ref{Free energy}) is defined as its second derivative,
\begin{equation}
 \mathbf M _{\vec \sigma \vec \tau} = \frac { \de^2 \mathscr F } { \de c(\vec \sigma) \, \de c(\vec \tau) }
\end{equation}
which gives:
\begin{eqnarray}
 \mathbf M _{\vec \sigma \vec \tau} &=& - \frac 1 {c(\sigma)} \delta_{\vec \sigma , \vec \tau} + \frac { \alpha k (k-1) \sum_{\vec \sigma_3 \cdots \vec \sigma_k} c(\vec \sigma_3) \cdots c(\vec \sigma_k) \, \mathcal E (\vec \sigma, \vec \tau, \vec \sigma_3, \dots, \vec \sigma_k) } { \sum_{\vec \sigma_1 \cdots \vec \sigma_k} c(\vec \sigma_1) \cdots c(\vec \sigma_k) \, \mathcal E (\vec \sigma_1, \dots, \vec \sigma_k) } + \nonumber \\
 && - \frac { \alpha k^2 \sum_{\vec \sigma_2 \cdots \vec \sigma_k} c(\vec \sigma_2) \cdots c(\vec \sigma_k) \, \mathcal E (\vec \sigma, \vec \sigma_2, \dots, \vec \sigma_k) \sum_{\vec \sigma'_2 \cdots \vec \sigma'_k} c(\vec \sigma'_2) \cdots c(\vec \sigma'_k) \, \mathcal E (\vec \tau, \vec \sigma'_2, \dots, \vec \sigma'_k) } { \left[ { \sum_{\vec \sigma_1 \cdots \vec \sigma_k} c(\vec \sigma_1) \cdots c(\vec \sigma_k) \, \mathcal E (\vec \sigma_1, \dots, \vec \sigma_k) } \right]^2 } \,. \nonumber \\
\label{Stability M 1}
\end{eqnarray}

The solution  $c(\cdot)$ of the saddle point equations, given by the equations (\ref{Def R(h)}), (\ref{Integer R(h)}) and (\ref{Sol rho_p}), can be written as
\begin{equation}
 c(\vec \sigma) = \frac 1 {2 e^{\mathscr G}-1} \left\{ \exp \left[ \mathscr G e^{-\frac {\nu (1-s)} 2} \right] + \exp \left[ \mathscr G e^{-\frac {\nu(1+s)} 2} \right] - 1 \right\}
\end{equation}
where 
\begin{equation}
 s \equiv \frac 1 n \sum_{a=1}^n \sigma^a
\end{equation}
and $\mathscr G$ is defined in (\ref{G(rho_0)}). In the limit $\nu \to \infty$ this reduces to 
\begin{equation}
 c(\vec \sigma) = \frac 1 { 2 e^{\mathscr G} - 1 } \left[ e^\mathscr G \delta_{|s|,1} + \left(1 - \delta_{|s|,1} \right) \right] \,.
\end{equation}
For large $\alpha$ this further simplifies, as $\mathscr G = O(\alpha)$ so that
\begin{equation}
 c(\vec \sigma) = \frac 1 2 \, \delta_{|s|,1} + e^{-O(\alpha)} \,.
\end{equation}

We can now compute the sums that appear in the expression of $\mathbf M$. In order to do this, let me recall the definition of the effective interaction $\mathcal E$ from (\ref{Def effective coupling}):
\begin{eqnarray}
 \mathscr A_k &\equiv& \sum_{\vec \sigma_1 \cdots \vec \sigma_k} c(\vec \sigma_1) \cdots c(\vec \sigma_k) \, \mathcal E (\vec \sigma_1, \dots, \vec \sigma_k) \\
 &=& \sum_{\vec \sigma_1, \dots, \vec \sigma_k} c(\vec \sigma_1) \cdots c(\vec \sigma_k) \frac 1 {2^k} \sum_{q_1, \dots, q_k}^{\{-1,1\}} \exp \left\{- \beta \sum_{a=1}^n \prod_{j=1}^k \delta (\sigma_j^a , q_j) \right\} \,.
\end{eqnarray}
In the limit $\beta \to \infty$, only the terms where the exponent vanish contribute. The value of $\mathcal E$ is then $2^{-k}$ times the number of $k$-component vectors $\underline v$ such that for any $j=1,\dots,k$ and any $a = 1,\dots,n$ we have $v_j \neq \sigma_j^a$. Since the only $\vec \sigma$ that have a non-vanishing $c(\vec \sigma)$ are $\{\sigma_a = 1 \ \ (\forall a = 1,\dots,n)\}$ and $\{\sigma_a = -1 \ \ (\forall a = 1,\dots,n)\}$, these $n$ conditions are actually identical, and only one (out of the possible $2^k$) vector $\underline v$ is excluded.

The sum over the $k$ vectors $\vec \sigma_j$ therefore has $2^k$ terms (corresponding to the 2 possible values of $\vec \sigma_j$), each of which has a factor $2^{-k}$ from the product of the $c(\cdot)$'s, and a factor $2^{-k} \times (2^k -1)$ from the $\mathcal E$, so that
\begin{equation}
 \mathscr A_k = 2^k \times \frac 1 {2^k} \times \frac 1 {2^k} (2^k - 1) 
 = 1 - \frac 1 {2^k} \,.
\end{equation}

In a very similar way,
\begin{eqnarray}
 \mathscr A_{k-1}(\vec \sigma) &\equiv& \sum_{\vec \sigma_2 \cdots \vec \sigma_k} c(\vec \sigma_2) \cdots c(\vec \sigma_k) \, \mathcal E (\vec \sigma, \vec \sigma_2, \dots, \vec \sigma_k) \\
 &=& 2^{k-1} \times \frac 1 {2^{k-1}} \times \frac 1 {2^k} \left[ 2^k - \left(2-\delta_{|s|,1}\right) \right] \\
 &=& 1 - \frac 1 {2^{k-1}} + \frac {\delta_{|s|,1}}{2^k} \,,
\end{eqnarray}
since if $|s|=1$ all the columns in the matrix $\boldsymbol \sigma$ will be equal (and only one vector $\underline v$ will be excluded), while if $|s| < 1$ there will be two column values (and correspondingly 2 vectors $\underline v$ excluded).

Finally,
\begin{eqnarray}
 \mathscr A_{k-2}(\vec \sigma, \vec \tau) &\equiv& \sum_{\vec \sigma_3 \cdots \vec \sigma_k} c(\vec \sigma_3) \cdots c(\vec \sigma_k) \, \mathcal E (\vec \sigma, \vec \tau, \vec \sigma_3, \dots, \vec \sigma_k) \\
 &=& 2^{k-2} \times \frac 1 {2^{k-2}} \times \frac 1 {2^k} \left[ 2^k - A(\vec \sigma, \vec \tau) \right] \\
 &=& 1 - \frac {A(\vec \sigma, \vec \tau)} {2^k}
\end{eqnarray}
where $A(\vec \sigma, \vec \tau)$ counts the number of different pairs, among the possible four which are $(1,1)$, $(1,-1)$, $(-1,1)$, $(-1,-1)$, that actually occur in the set $\{(\sigma^a,\tau^a) | a = 1, \dots, n\}$.

We can then substitute these expression in (\ref{Stability M 1}) to obtain, up to terms of order $e^{-O(\alpha)}$,
\begin{eqnarray}
 \mathbf M _{\vec \sigma \vec \tau} &=& - \frac { 2 \, e^\mathscr G \, \delta_{\vec \sigma, \vec \tau} } { e^\mathscr G \, \delta_{|s|,1} + \left(1-\delta_{|s|,1} \right) } + \frac {\alpha k (k-1) \left[ 2^k - A(\vec \sigma, \vec \tau) \right] } {2^k-1} + \\
 && - \frac{ \alpha k^2 \left[ 2^k - 2 + \delta_{|s|,1} \right] \left[ 2^k - 2 + \delta_{|t|,1} \right] } { (2^k-1)^2 }
\end{eqnarray}
where $t$ is defined for $\vec \tau$ as $s$ for $\vec \sigma$.
This matrix is invariant under the exchange of replica indices, and therefore it can be block-diagonalized in subspaces of well-defined replica symmetry. 

In order to take into account the normalization constraint
\begin{equation}
 \sum_{\vec \sigma} c(\vec \sigma) = 1 \,,
\end{equation}
it is convenient to decompose the dependency of $\mathscr F[c(\cdot)]$ in two, writing
\begin{equation}
 \mathscr F[c(\vec \sigma)] \equiv \mathscr F' \left[1-\sum_{\vec \sigma} c'(\vec \sigma), c'(\vec \sigma) \right]
\end{equation}
with $c'(\vec \sigma) = c(\vec \sigma)$ for every $\vec \sigma$ except $\vec \sigma = \vec 1 \equiv (1,\dots,1)$, and 0 otherwise, and where $\mathscr F'$ is the functional defined by the previous identity.
The stability matrix of $\mathscr F'$ is then
\begin{eqnarray}
 \mathbf M'_{\vec \sigma \vec \tau} &=& \mathbf M_{\vec \sigma \vec \tau} - \mathbf M'_{\vec \sigma \vec 1} - \mathbf M'_{\vec 1 \vec \tau} + \mathbf M'_{\vec 1 \vec 1} \\
 &=& - \frac{2 \, e^{\mathscr G} \, \delta_{\vec \sigma \vec \tau} }{ e^\mathscr G \, \delta_{|s|,1} + \left(1-\delta_{|s|,1} \right) } - 2 - \frac { \alpha k (k-1) } { 2^k - 1 } \left[ 1 + A(\vec \sigma, \vec \tau) - A(\vec 1, \vec \tau) - A(\vec \sigma, \vec 1) \right] + \nonumber \\
 && - \frac { \alpha k^2 \left[\delta_{|s|,1} - 2 \right] \left[ \delta_{|t|,1} - 2 \right] } { (2^k-1)^2 } \,.
\end{eqnarray}

In non-symmetric subspaces, $|s| \neq 1 \neq |t|$, and the previous equation becomes
\begin{equation}
 \mathbf M'_{\vec \sigma \vec \tau} = - 2 \, e^{\mathscr G} \, \delta_{\vec \sigma \vec \tau} - 2 - \frac { \alpha k (k-1) } { 2^k - 1 } \left[ 1 + A(\vec \sigma, \vec \tau) - A(\vec 1, \vec \tau) - A(\vec \sigma, \vec 1) \right] -
\frac {4 \alpha k^2} {(2^k-1)^2} \,.
\end{equation}
The diagonal terms of this matrix are of order $O(e^\alpha)$, while the off-diagonal terms are of order $O(\alpha)$. The contribution of the off-diagonal terms to the eigenvalues will be given by $2^n$ terms, each of $O(\alpha)$. Since the contribution of the diagonal terms to the eigenvalues is of order $O(e^\alpha)$ and it is negative, this ensures that for large enough $\alpha$ all the eigenvalues will be negative.
In this subspaces, the replica symmetric solution is therefore a local maximum.

In the symmetric subspace, all the diagonal elements of $ \mathbf M'_{\vec \sigma \vec \tau}$ are of order $O(e^\alpha)$, \emph{except} the term corresponding to $\vec \sigma = \vec \tau = - \vec 1$: for this term, the exponential contributions vanish. However, we can then write
\begin{equation}
 \mathbf M'_{\vec \sigma \vec \tau} = - 2 \, e^{\mathscr G} \, \delta_{\vec \sigma \vec \tau} + \mathbf V_{\vec \sigma \vec \tau}
\end{equation}
and treat $\mathbf V$ as a perturbation. For  $\vec \sigma = \vec \tau = - \vec 1$ the matrix element of $\mathbf V$ is
\begin{equation}
 \mathbf V_{- \vec 1 , \, - \vec 1} =  \mathbf M'_{- \vec 1 , \, - \vec 1} = \mathbf M_{- \vec 1 , \, - \vec 1} - \mathbf M_{- \vec 1 , \, \vec 1} - \mathbf M_{\vec 1 , \, - \vec 1} +\mathbf M_{\vec 1 , \, \vec 1} 
\end{equation}
and from(\ref{Stability M 1}) this is equal to $-4$, so it is negative.

The conclusion of this analysis is that, for $\alpha$ large enough, all the eigenvalues of the stability matrix of $\mathscr F$, computed for the $c(\cdot)$ which satisfies the replica symmetric saddle point equations, are negative, and therefore that this solution is \emph{locally} stable.

\subsection{Uniqueness of the solution}
\label{Par Stability from cavity}

The conclusion from the previous Paragraph cannot rule out the existence of other solutions to the saddle point equations, which could possibly be the true \emph{global} maximum of $\mathscr F$. I shall now provide an argument supporting that the saddle point equations have a unique solution, which is therefore the one found in Section \ref{Sec Free energy}.

Let us consider two real replicas of the system, i.e. two identical satisfiable formul\ae. I shall indicate by $\alpha$ the thermodynamic state of the first replica, and by $\beta$ that of the second one (the context will make it obvious when $\alpha$ refers to the clause to variable ratio of the formula). I want to study, with the cavity method, the joint probability for a variable of having a positive, negative or null field in the two states $\alpha$ and $\beta$, which I shall denote by the following quantities:
\begin{equation}
\begin{array}{ccc}
 p_{++}^{\alpha \beta} & p_{+0}^{\alpha \beta} & p_{+-}^{\alpha \beta} \\
 p_{0+}^{\alpha \beta} & p_{00}^{\alpha \beta} & p_{0-}^{\alpha \beta} \\
 p_{-+}^{\alpha \beta} & p_{-0}^{\alpha \beta} & p_{--}^{\alpha \beta}
\end{array}
\end{equation}
What I want to prove is that for large $\alpha$:
\begin{itemize}
 \item The off-diagonal terms become negligible, so that the fields are equal in the two states for most variables;
 \item The term $p_{00}^{\alpha \beta}$ is much smaller than $p_{++}^{\alpha \beta}$ and $p_{--}^{\alpha \beta}$.
\end{itemize}
The consequence of these two properties will be that most variables will be constrained to take the same values in the two states $\alpha$ and $\beta$, which is therefore a single, unique, thermodynamic state.

\subsubsection*{Distribution of the number of messages}

Let us assume that a new variable is added to the formula, appearing non-negated in $l_+$ clauses and negated in $l_-$ clauses. These will be two independent random variables with identical poissonian distribution of parameter $\alpha' k / 2$, where $\alpha'$ is some constant which will be determined later.

A clause will send a message to a variable (that is, it will constrain it) if all the other variables in the clause are constrained (that is, have a non-zero field) and appear with the ``wrong'' sign in the clause, which happens with probabilities
\begin{eqnarray}
 q^\alpha &=& \left[ \frac{ 1 - (p_{0+}^{\alpha \beta} + p_{00}^{\alpha \beta} + p_{0-}^{\alpha \beta})}{2} \right] ^{k-1} \,, 
\label{Self alpha} \\
 q^\beta &=& \left[ \frac{ 1 - (p_{+0}^{\alpha \beta} + p_{00}^{\alpha \beta} + p_{-0}^{\alpha \beta})}{2} \right] ^{k-1}
\label{Self beta}
\end{eqnarray}
respectively in the states $\alpha$ and $\beta$.

For a given $l_+$, the probability that in the state $\alpha$ the number clauses sending a message to the new variable is $m_+^\alpha$ is equal to 
\begin{eqnarray}
 p_M^\alpha \left( m_+^\alpha | l_+ \right) = \binom {l_+}{m_+^\alpha} \left( q^\alpha \right)^{m_+^\alpha} \left( 1 - q^\alpha \right)^{l-m_+^\alpha}
\end{eqnarray}
and identical distributions are valid for $m_-^\alpha$ for fixed $l_-$, and for the corresponding quantities in the state $\beta$.

The number of occurrences $l_+$ must be the same in the two states (the replicas are identical), and must be larger than $m_+^\alpha$ and $m_+^\beta$. The joint distribution of $m_+^\alpha$ and $m_+^\beta$ is obtained by summing over the allowed values of $l_+$:
\begin{eqnarray}
 p_M^{\alpha \beta}(m_+^\alpha, m_+^\beta) &=& \sum_{l_+ = \max(m_+^\alpha, m_+^\beta)}^\infty \frac{1}{(l_+)!} \left( \frac{\alpha' k}{2} \right)^{l_+} e^{-\alpha' k /2} \times \binom {l_+} {m_+^\alpha} \left( q^\alpha	\right)^{m_+^\alpha} \left( 1 - q^\alpha \right)^{l_+-m_+^\alpha} \nonumber \\
 && \times \binom {l_+} {m_+^\beta} \left( q^\beta	\right)^{m_+^\beta} \left( 1 - q^\beta \right)^{l_+-m_+^\beta}
\end{eqnarray}
and similarly for the negative messages.

The joint probability of all messages is given by the product of the distributions of positive and negative messages, since they are independent:
\begin{equation}
 \mathbb P[m_+^\alpha, m_+^\beta, m_-^\alpha, m_-^\beta] = p_M^{\alpha \beta}(m_+^\alpha, m_+^\beta) \times p_M^{\alpha \beta}(m_-^\alpha, m_-^\beta) \,.
\end{equation}
The values of $\{p_{++}^{\alpha \beta}, \cdots, p_{--}^{\alpha \beta}\}$ are obtained from this distribution by summing over the appropriate ranges the values of $m_\pm$.

\subsubsection*{Selection of satisfiable formul\ae}

In order to have a satisfiable formula, no variable must receive contradictory messages. This means that the ranges to be considered in the sums to compute  $\{p_{++}^{\alpha \beta}, \cdots, p_{--}^{\alpha \beta}\}$ must be the following:
\begin{eqnarray}
 p_{00}^{\alpha \beta} &:& p_M^{\alpha \beta} (0,0) \times p_M^{\alpha \beta} (0,0) \\
 p_{++}^{\alpha \beta} &:& p_M^{\alpha \beta} (m_+^\alpha,m_+^\beta) \times p_M^{\alpha \beta} (0,0) \\
 p_{--}^{\alpha \beta} &:& p_M^{\alpha \beta} (0,0) \times p_M^{\alpha \beta} (m_-^\alpha,m_-^\beta) \\
 p_{+-}^{\alpha \beta} &:& p_M^{\alpha \beta} (m_+^\alpha,0) \times p_M^{\alpha \beta} (0,m_-^\beta) \\
 p_{-+}^{\alpha \beta} &:& p_M^{\alpha \beta} (0,m_+^\beta) \times p_M^{\alpha \beta} (m_-^\alpha,0) \\
 p_{0+}^{\alpha \beta} &:& p_M^{\alpha \beta} (0,m_+^\beta) \times p_M^{\alpha \beta} (0,0) \\
 p_{0-}^{\alpha \beta} &:& p_M^{\alpha \beta} (0,0) \times p_M^{\alpha \beta} (0,m_-^\beta) \\
 p_{+0}^{\alpha \beta} &:& p_M^{\alpha \beta} (m_+^\alpha,0) \times p_M^{\alpha \beta} (0,0) \\
 p_{-0}^{\alpha \beta} &:& p_M^{\alpha \beta} (0,0) \times p_M^{\alpha \beta} (m_-^\alpha,0)
\end{eqnarray}
where all the $m_\pm$ are positive, and must be summed between 1 and infinity.

I therefore define:
\begin{eqnarray}
 S_0 &\equiv& p_M^{\alpha \beta}(0,0)^2 \,, \\
 S_1 &\equiv& \sum_{m_+^\alpha, m_+^\beta = 1}^\infty p_M^{\alpha \beta} (m_+^\alpha, m_+^\beta) \ p_M^{\alpha \beta} (0,0)
 \ \ = \sum_{m_-^\alpha, m_-^\beta = 1}^\infty p_M^{\alpha \beta} (0,0) \ p_M^{\alpha \beta} (m_-^\alpha, m_-^\beta) \,, \\
 S_2 &\equiv& \sum_{m_+^\alpha=1}^\infty \sum_{m_-^\beta=1}^\infty p_M^{\alpha \beta} (m_+^\alpha,0) \ p_M^{\alpha \beta} (0,m_-^\beta)
 \ \ = \sum_{m_-^\alpha=1}^\infty \sum_{m_+^\beta=1}^\infty p_M^{\alpha \beta} (0,m_+^\beta) \ p_M^{\alpha \beta} (m_-^\alpha,0) \,, \\
 S_3 &\equiv& \sum_{m_+^\beta = 1}^\infty p_M^{\alpha \beta}(0,m_+^\beta) p_M^{\alpha \beta} (0,0)
 \ \ = \sum_{m_-^\beta = 1}^\infty p_M^{\alpha \beta}(0,0) p_M^{\alpha \beta} (0,m_-^\beta) \,, \\
 S_3' &\equiv& \sum_{m_+^\alpha = 1}^\infty p_M^{\alpha \beta}(m_+^\alpha,0) 
 p_M^{\alpha \beta} (0,0)
 \ \ = \sum_{m_-^\alpha = 1}^\infty p_M^{\alpha \beta}(0,0) p_M^{\alpha \beta} (m_-^\alpha,0) \,, \\
\mathcal{N} &\equiv& S_0 + 2 S_1 + 2 S_3 + 2 S_3' \,,
\end{eqnarray}
so that
\begin{eqnarray}
 p_{00}^{\alpha \beta} &=& \frac{S_0}{\mathcal N} \,, \\
 p_{++}^{\alpha \beta} &=& \frac{S_1}{\mathcal N} \ = p_{--}^{\alpha \beta} \,,\\
 p_{+-}^{\alpha \beta} &=& \frac{S_2}{\mathcal N} \ = p_{-+}^{\alpha \beta} \,, \\
 p_{0+}^{\alpha \beta} &=& \frac{S_3}{\mathcal N} \ = p_{0-}^{\alpha \beta} \,, \\
 p_{+0}^{\alpha \beta} &=& \frac{S_3'}{\mathcal N} \ = p_{-0}^{\alpha \beta} \,.
\end{eqnarray}

All these sums are computed by inverting the order of the sums over $l_\pm$ and $m_\pm$ and adding the term corresponding to $m_\pm = 0$, for example
\begin{eqnarray}
 \sum_{m_+^\alpha,m_+^\beta = 1}^\infty \ \sum_{l_+ = \max(m_+^\alpha , m_+^\beta)}^\infty \longrightarrow \ \ \sum_{l_+ = 0}^{\infty} \ \sum_{m_+^\alpha, m_+^\beta = 0}^{l_+} - \text{ terms with }  m_+=0 \,.
\end{eqnarray}
This gives:
\begin{eqnarray}
 S_0 &=& \exp \left\{ - \alpha' k \left[ 1 - (1-q^\alpha) (1-q^\beta) \right] \right\} \\
 S_1 &=& \exp \left\{ - \frac{\alpha' k}{2} \left[ 1 - (1-q^\alpha)(1-q^\beta) \right] \right\} \nonumber \times \left\{ 1 - \exp \left[ - \frac{\alpha' k}{2} q^\alpha \right]- \exp \left[ - \frac{\alpha' k}{2} q^\beta \right] \right\} + \nonumber \\
 && + \exp \left\{ - \alpha' k \left[ 1 - (1-q^\alpha)(1-q^\beta) \right] \right\} \\
 S_2 &=& \left\{ \exp \left[ - \frac{\alpha' k}{2} q^\alpha \right] - \exp \left[ - \frac{\alpha' k}{2} \left(1 - (1-q^\alpha)(1-q^\beta) \right) \right] \right\} \nonumber \\
 && \times \left\{ \exp \left[ - \frac{\alpha' k}{2} q^\beta \right] - \exp \left[ - \frac{\alpha' k}{2} \left(1 - (1-q^\alpha)(1-q^\beta) \right) \right] \right\} \\
 S_3 &=& \exp \left\{ - \frac{\alpha' k}{2} \left[ 1 - (1-q^\alpha)(1-q^\beta) + q^\alpha \right] \right\} - \exp \left\{ - \alpha' k \left[ 1 - (1-q^\alpha)(1-q^\beta) \right] \right\} \\
 S_3' &=& \exp \left\{ - \frac{\alpha' k}{2} \left[ 1 - (1-q^\alpha)(1-q^\beta) + q^\beta \right] \right\} - \exp \left\{ - \alpha' k \left[ 1 - (1-q^\alpha)(1-q^\beta) \right] \right\}
\end{eqnarray}
\begin{eqnarray}
 \mathcal N &=& 2 \exp \left\{-\frac{\alpha' k} 2 \left[1-(1-q^\alpha)(1-q^\beta)\right]\right\} \times \left\{1-\exp\left[-\frac{\alpha' k} 2 q^\alpha \right]		-\exp\left[-\frac{\alpha' k} 2 q^\beta \right] \right\} + \nonumber \\
 && + 2 \exp \left\{-\frac{\alpha' k} 2 (q^\alpha + q^\beta)\right\}  +\exp\left\{-\frac{\alpha' k} 2 \left[ 1 - (1-q^\alpha)(1-q^\beta) \right] \right\}
\label{Normalization N}
\end{eqnarray}

The self-consistency equations (\ref{Self alpha}) and (\ref{Self beta}) are then
\begin{eqnarray}
 q^\alpha &=& \left\{ \frac{1}{2} \left[ 1 - \frac{S_0 + 2 S_3}{\mathcal N} \right] \right\}^{k-1} \,, \\
 q^\beta &=& \left\{ \frac{1}{2} \left[ 1 - \frac{S_0 + 2 S_3'}{\mathcal N} \right] \right\}^{k-1} \,.
\end{eqnarray}
Notice that these equations are coupled, as $S_0$, $S_3$ and $S_3'$ contain both $q^\alpha$ and $q^\beta$.

\subsubsection*{Solution of the self-consistency equations}

These equations have four fixed points, of which for $\alpha \to \infty$ only one is stable. To see it, I consider that as $\alpha \to \infty$, also $\alpha' \to \infty$ (I shall verify this later). Then, keeping only the leading exponential term in $\alpha'$,
\begin{eqnarray}
 S_0 & \ll & S_3, \ S'_3 \,, \\
 S_3 & \sim & \exp \left\{ - \frac{\alpha' k}{2} \left[ 1 - (1-q^\alpha)(1-q^\beta) + q^\alpha \right] \right\} \,, \\
 S'_3 & \sim & \exp \left\{ - \frac{\alpha' k}{2} \left[ 1 - (1-q^\alpha)(1-q^\beta) + q^\beta \right] \right\} \,, \\
 \mathcal N & \sim & 2 \exp\left\{-\frac{\alpha' k} 2 \left[ 1 - (1-q^\alpha)(1-q^\beta) \right] \right\} \,.
\end{eqnarray}
The self consistency equations then decouple:
\begin{eqnarray}
 q^\alpha & = & \left\{ \frac 1 2 \left[ 1 - \exp \left(-\frac{\alpha' k} 2 q^\alpha \right) \right] \right\} ^{k-1} \,, \\
 q^\beta & = & \left\{ \frac 1 2 \left[ 1 - \exp \left(-\frac{\alpha' k} 2 q^\beta \right) \right] \right\} ^{k-1} \,.	
\end{eqnarray}
These equations are identical. Each admits two solutions: one for $q \simeq 0$, and one for $q \simeq 1/2^{k-1}$ (of course, $q=0$ is also a solution, but a trivial one). The solution close to 0 is
\begin{equation}
 q_0 = \left( \frac{\alpha' k} 4 \right) ^ {-\frac{k-1}{k-2}} + \dots
\end{equation}
and it is unstable, since the derivative of the right hand side is larger than 1. The other solution is
\begin{eqnarray}
 q^* = \frac 1 {2^{k-1}} \left\{ 1 - (k-1) \exp \left[-\frac{\alpha' k} {2^k} \right] \right\} + \dots
\end{eqnarray}
and this solution is stable. Therefore, for $\alpha \to \infty$ we shall have $q^\alpha = q^\beta = q^*$.

The computation of $\alpha'$ as a function of $\alpha$ is similar as the one I've shown in Section \ref{Sec Cavity formalism}. We must impose that the average total number of occurrences of the new variable be
\begin{equation}
 \left< l_+ + l_- \right> _\mathrm{Sat} = \alpha k \,.
\end{equation}
The distribution of $(l_+,l_-)$ conditioned on the formula being satisfiable is obtained by summing over the values of $m_\pm$ that give no contradictions, i.e.
\begin{eqnarray}
 P_\mathrm{Sat}(l_+,l_-) &=& \frac 1 {\mathcal N} \Biggl\{ P_M^{\alpha \beta}(0,0|l_+) P_M^{\alpha \beta}(0,0|l_-) + \sum_{m_+^\alpha, m_+^\beta = 1}^{l_+} P_M^{\alpha \beta}(m_+^\alpha,m_+^\beta|l_+) P_M^{\alpha \beta}(0,0|l_-) + \nonumber \\
 &&  + \dots + \sum_{m_-^\alpha=1}^{l_-} P_M^{\alpha \beta}(0,0|l_+)   P_M^{\alpha \beta}(m_-^\alpha,0|l_-) \Biggr\} \\
 &=&  \frac {1}{\mathcal N} \frac {1}{(l_+)!(l_-)!} \left( \frac {\alpha' k}{2} \right) ^ {l_++l_-} e^{-\alpha'k} \times \left[(1-q^\alpha)^{l_++l_-}- (1-q^\alpha)^{l_+}-1-q^\alpha)^{l_-}\right] \times \nonumber \\
 && \times\left[(1-q^\beta)^{l_++l_-}-(1-q^\beta)^{l_+}- (1-q^\beta)^{l_-}\right] 
\end{eqnarray}
where the normalization factor $\mathcal N$ is the one from (\ref{Normalization N}). We obtain:
\begin{eqnarray}
\left< l_+ + l_- \right> _\mathrm{Sat} &=& \alpha' k \times \frac 1 {\mathcal N} \times e^{-\alpha' k} \times \nonumber \\
 && \times \Biggl\{ (1-q^\alpha)(1-q^\beta) \exp \left[ \alpha' k (1-q^\alpha)(1-q^\beta) \right] + \nonumber \\
 && +(2-q^\alpha-q^\beta) \exp \left[ \frac{\alpha' k}2 (2-q^\alpha-q^\beta) \right] + \nonumber \\
 && + \left[1+(1-q^\alpha)(1-q^\beta)\right] \exp\left[ \frac{\alpha' k}2 [1+(1-q^\alpha)(1-q^\beta)] \right] + \nonumber \\
 && - (1-q^\alpha)(2-q^\beta) \exp \left[ \frac{\alpha' k} 2 (1-q^\alpha)(2-q^\beta) \right] + \nonumber \\
 && - (2-q^\alpha)(1-q^\beta) \exp \left[ \frac{\alpha' k} 2 (2-q^\alpha)(1-q^\beta) \right] \Biggr\} \,.
\end{eqnarray}
For $\alpha \to \infty$ we shall have $q^\alpha = q^\beta = q^*$, and the leading order term in the numerator is the one containing $ 1 + (1-q^*)^2$:
\begin{equation}
 \left< l_+ + l_- \right> _\mathrm{Sat} \sim \alpha' k \times \frac 1 {\mathcal N} \times \left[1+(1-q^*)^2\right] \exp\left\{-\frac{\alpha' k} 2 \left[1-(1-q^*)^2 \right] \right\} \,,
\end{equation}
with
\begin{equation}
 \mathcal N \sim 2 \exp\left\{-\frac{\alpha' k} 2 \left[1-(1-q^*)^2 \right] \right\}
\end{equation}
so that
\begin{equation}
 \left< l_+ + l_- \right> _\mathrm{Sat} = \frac 1 2 \alpha' k \left[1+(1-q^*)^2\right] + e^{-O(\alpha')}
\end{equation}
and
\begin{equation}
 \alpha' = \frac {2 \alpha} {1+(1-q^*)^2} + e^{-O(\alpha)} \,.
\end{equation}

\subsubsection*{Uniqueness of the state}

The joint probabilities are given, for large $\alpha$, by
\begin{eqnarray}
 p_{00}^{\alpha \beta} &=& \frac {S_0}{\mathcal N} \sim \frac 1 2 \exp \left\{ - \frac{\alpha' k} 2 \left[ 1 - \left(1- \frac 1 {2^{k-1}}\right)^2 \right]\right\} \,, \\
 p_{++}^{\alpha \beta} \ \ = \ \ p_{--}^{\alpha \beta} &=& \frac {S_1} {\mathcal N}  \sim \frac 1 2 - e^{-O(\alpha)} \,, \\
 p_{0+}^{\alpha \beta} \ \ = \ \ p_{0-}^{\alpha \beta} \ \ = \ \ p_{+0}^{\alpha \beta} \ \ = \ \ p_{-0}^{\alpha \beta} &=& \frac {S_3} {\mathcal N} \sim \frac 1 2 \exp \left[ - \frac{\alpha' k} {2^k} \right] \,.
\end{eqnarray}

This confirms that the off-diagonal terms are exponentially suppressed, and that $p_{00}^{\alpha \beta} \ll p_{++}^{\alpha \beta}, \, p_{--}^{\alpha \beta}$. Apart from a fraction of variables of order $e^{-O(\alpha)}$ we see that the variables are constrained and must take the same value in the two states $\alpha$ and $\beta$, so that there is actually only one unique state.

The solution to the saddle point equations that we found in Section \ref{Sec Free energy} is therefore unique.

\section{Discussion of the results and conclusion}
\label{Sec Conclusion}

In Paragraph \ref{Par Algorithmic implications} I have drawn the conclusion of this work: that the proof of convergence of WP provided in \cite{Feige06} for formul\ae\ extracted from the Planted Ensemble can be extended to formul\ae\ extracted from the Satisfiable Distribution. As we have seen, this contradicts a probabilistic version of \hyp{2}. There are two questions that remain open and deserve attention.

The first regards Feige's complexity result. Theorem 1 was based on a deterministic form of \hyp{2}, which is weaker than the probabilistic version refuted by the previous results. It would be very interesting to understand whether the hypotheses of Theorem 1 can be relaxed, and some conclusion reached on the basis of the refutation of \hyp{$2_p$}.

Even more interesting, from the physicist's point of view, is the second question. The above discussion for \ksat\ can be easily extended to other models, such as \kxorsat. The characterization of the solutions to large $\alpha$ satisfiable formul\ae\ in terms of the distribution of fields can be repeated, with similar results: that a fraction $1-e^{-O(\alpha)}$ of the variables are constrained to take a unique value in all the solutions, and that the fields acting on the variables are of order $O(\alpha)$. However, there is a crucial distinction between \ksat\ and \kxorsat: the correlation between the sign of the field acting on a variable and a bias in the number of occurrences between it and its negation, which is present in \ksat, cannot be present in \kxorsat\ for obvious symmetry reasons. Since this is a crucial ingredient of the convergence of WP, it should not be expected to apply to \kxorsat. It would then be very interesting to find an algorithm which identifies satisfiable \kxorsat\ formul\ae\ at large $\alpha$, and to understand the implications this would have on Theorem 1.

\end{part}

\pagestyle{empty}
\chapter*{Acknowledgements}

I would have never been able to start this work \textemdash\ let alone complete it \textemdash\ without the support and help of many persons, to whom I am deeply indebted and grateful, and whom I wish to thank: Susanna Federici, to whom this work is dedicated; my family and friends, and especially Giulia, Luca and Valentina, for their love and support; Irene and Andrea, for their crucial initial encouragement; Giorgio, R\'emi and Francesco, who taught me all I know in this field, and whom I have now the privilege to consider friends; the Laboratoire de Physique Th\'eorique at the Ecole Normale Superieure in Paris, for its warm welcome, and especially Simona, Nicolas Sourlas who accepted to be my official cotutor, as well as Guilhem and Andrea; and finally, to Silvio Franz, who accepted to referee this thesis.
\addcontentsline{toc}{chapter}{Acknowledgements}
\chapter*{List of notations}

\begin{tabular}{ll}
\hspace{2cm}       & \\
$\equiv$           & Identical to \\
$\sim$             & Asymptotically equal to, leading order in asymptotic expansions\\
$\simeq$           & Approximately equal to \\
$n \div m$         & Integer division of $n$ by $m$ \\ 
$\mathbb P[\cdot]$ & Probability \\ 
$\mathbb E[\cdot]$ & Expected value \\ 
$\mathbb I[\mathsf{event}]$ & Indicator function of $\mathsf{event}$, equal to 1 if $\mathsf{event}$ is true and 0 otherwise \\
$\vee$             & Logical OR \\
$\wedge$           & Logical AND \\
$\oplus$           & Logical XOR \\
$\left| \mathscr S \right|$ & Cardinality of set $\mathscr S$ \\
$\left< \cdot \right>$ & Thermodynamic average \\ 
$\overline{O}$     & Average over disorder of $O$ \\
$i,j,k,\dots$      & Site indices from 1 to $N$  \\
$a,b,c,\dots$      & Replica indices from 1 to $n$ \\
$\sigma_i$         & Individual spin \\
$\sigma$           & $N$-component spin configuration \\
$\boldsymbol \sigma$ & Replicated $N \times n$ spin configuration \\
$\sigma^a$         & $N$-component spin configuration of replica $a$ \\
$\vec \sigma_i$    & $n$-component spin configuration on site $i$ \\
$\vec \sigma, \vec \tau$ & Generic $n$-component spin configurations \\
$\sigma_i^a$       & Value of spin on site $i$ for replica $a$ \\
$\alpha$           & Ratio between number of clauses $M$ and number of variables $N$ in a boolean \\
                   & constraint satisfaction problems \\
$\alpha_\mathrm s$ & Threshold value for \sat/\unsat\ transition \\
$\alpha_\mathrm c$ & Threshold value for clustering transition \\
$\alpha_0$         & Lower bound on $\alpha_s$ from the second moment  inequality \\
$\alpha_\mathrm h$ & Largest value of $\alpha$ for which a poissonian DPLL heuristic succeeds with positive\\
                   & probability \\
$\Sigma_\mathrm c$ & Clustering transition surface \\
$\Sigma_\mathrm s$ & \sat/\unsat\ transition surface \\
$\Sigma_\mathrm k$ & Critical surface (i.e. intersection of $\Sigma_\mathrm c$ and $\Sigma_\mathrm s$) \\
$\Sigma_\mathrm q$ & Contradiction surface \\
$\mathcal F$       & \ksat\ formula \\
$\mathcal P_\mathrm{Unif}[\mathcal F]$ & Uniform measure over random formul\ae \\
$\mathcal P_\mathrm{Sat}[\mathcal F]$ & Uniform measure over satisfiable formul\ae \\
$\mathcal P_\mathrm{Unif}[\mathcal F]$ & Planted measure over random formul\ae 
\end{tabular}

\begin{tabular}{ll}
\hspace{2cm}       & \\
$c(\vec \sigma)$   & Fraction of sites with replicated configuration $\vec \sigma$, functional order parameter \\
$R(h)$             & Distribution of fields, functional order parameter equivalent to $c(\vec \sigma)$ \\
$\mathscr F$       & Free energy density functional \\
$\nu$              & ``Thermodynamic potential'', $\nu \equiv \beta n$ as 	$\beta \to \infty$ and $n \to 0$ \\
$\epsilon_0(\nu)$  & Ground state energy density of formul\ae\ conditioned on $\nu$ \\
$r_p$              & Weight of $R(h)$ over $h=p \in \mathbb Z$ \\
$I_p(x)$           & Modified Bessel function of integer order \\
$\rho_p$           & Limit of $r_p$ for $\nu \to \infty$
\end{tabular}

\addcontentsline{toc}{chapter}{Bibliography}
\newpage

\end{document}